\documentclass[a4paper,twoside]{article}
\newcommand{\affil}[1]{$^{\rm #1}$}
\usepackage[authoryear]{natbib}

\bibpunct{(}{)}{;}{a}{}{,}
\usepackage{graphicx}
\usepackage{sectsty}
\sectionfont{\large}
\subsectionfont{\normalsize}
\date{} 
\title{\large\bf\flushleft Spatially Resolved Spectroscopy of Starburst and Post--Starburst Galaxies in The Rich z$\sim 0.55$ Cluster CL~0016+16$^{1}$}

\author{\parbox{\textwidth}{\flushleft
\vspace{-0.5cm}
{\it Michael B. Pracy\affil{A,C}, Warrick J. Couch\affil{A}, Harald Kuntschner\affil{B}}\\
\vspace{0.4cm}
{\small \affil{A}\,Centre for Astrophysics and Supercomputing, Swinburne University of Technology, P.O. Box 218, Hawthorn, VIC, 3122, Australia}\\
{\small \affil{B}\,Space Telescope European Coordinating Facility, European Southern Observatory, Karl-Schwarzschild Strasse 2, 85748, Garching, Germany}\\
{\small \affil{C}\,Email:mpracy@astro.swin.edu.au}}}
\begin{document}
\twocolumn[
\begin{changemargin}{.8cm}{.5cm}
\begin{minipage}{.9\textwidth}
\vspace{-1cm}
\maketitle
%
%
\small{\bf Abstract:}
We have used the Low Resolution Imaging Spectrograph (LRIS) on the W.M. Keck I telescope
to obtain spatially resolved spectroscopy of a small sample of six `post--starburst'  and 
three `dusty--starburst' galaxies in the rich cluster CL~0016+16 at $z=0.55$.  We use this
to measure radial profiles of the H$\delta$ and OII$\lambda 3727$ lines which are diagnostic
probes of the mechanisms that give rise to the abrupt changes in star--formation rates in these galaxies.
In the post--starburst sample we are unable to detect any radial gradients in the H$\delta$ line equivalent width
-- although one galaxy exhibits a gradient from one side of the galaxy to the other. The absence of H$\delta$
gradients in these galaxies is consistent with their production via interaction with the intra--cluster medium, 
however, our limited spatial sampling prevents  us from drawing robust conclusions. All members of the sample
have early type morphologies, typical of post--starburst galaxies in general, but lack the high incidence of tidal tails
and disturbances seen in local field samples. This argues against a merger origin and adds weight to a scenario
where truncation by the intra--cluster medium is at work. The post--starburst spectral signature is consistent over the 
radial extent probed with no evidence of OII$\lambda 3727$ emission
and strong H$\delta$ absorption at all radii i.e. the post-starburst classification is not an aperture effect. 
In contrast the 'dusty--starburst' sample shows a tendency for  a central concentration
of OII$\lambda 3727$ emission. This is most straightforwardly interpreted as the 
consequence of a central starburst. However, other possibilities exist such as a non-uniform dust 
distribution (which is expected in such galaxies) and/or  a non-uniform starburst age distribution. 
The sample exhibit late type and irregular morphologies.


\medskip
\medskip
\end{minipage}
\end{changemargin}
]
\small
\footnotetext[1]{The data presented herein were obtained at the W.M. Keck Observatory, which is operated as a scientific partnership 
among the California Institute of Technology, the University of California and the National Aeronautics and Space Administration. The 
Observatory was made possible by the generous financial support of the W.M. Keck Foundation.}

\section{Introduction}
Some of the first direct evidence for the redshift evolution
of galaxy populations came from observations of galaxies residing
in the cores of rich galaxy clusters. In the local universe, galaxies
in these environments show little sign of active, or recent, star--formation.
However the photometric observations of \citet{butcher78,butcher84} of cluster
galaxies at intermediate redshift revealed a large increase in the fraction
of blue galaxies present. Spectroscopic studies have established the increased
blue fraction is the result of an increase in the fraction of galaxies which are
actively forming stars in clusters at these redshifts 
\citep[e.g.][]{dressler82,dressler83,dressler92,couch87,couch94,poggianti99}. 
This implies that significant and, somewhat, rapid evolution has taken place in 
these environments over the last 3--5\,Gyrs.

In addition to the \citet{butcher78} effect, spectroscopic observations of intermediate
redshift clusters also resulted in the discovery of a new class of galaxy which lacked optical 
emission lines but had strong Balmer absorption lines \citep{dressler83}. The absence
of emission lines implies no ongoing star--formation whilst the strong Balmer absorption lines
requires a substantial contribution of light from A-stars. Such a spectroscopic signature requires
a strong episode of star--formation which has been rapidly truncated within the past $\sim 1$\, Gyr 
\citep{couch87,poggianti99}. These galaxies represent a population of `post--starburst galaxies' 
(also referred to as k+a, a+k or E+A galaxies) which are observed in the midst of rapid evolution 
in their star-formation properties and may mark the transition of a star-forming disk galaxy
into a quiescent spherical system \citep{caldwell96,zabludoff96}. 
Another class of galaxies with strong H$\delta$ absorption lines are the e(a) galaxies, which
possess strong H$\delta$ absorption in conjunction with mild [OII]$\lambda 3727$ emission 
\citep{dressler99}. 
Whilst it is possible to produce such a spectrum by having a small amount of residual star--formation 
in a post--starburst galaxy, the evidence points to these galaxies being dust enshrouded 
starbursts \citep{poggianti99,poggianti00}. These galaxies are commonly 
observed ($\sim$ 75\,per cent of cases) to be undergoing mergers or strong 
interactions \citep{poggianti00}. At intermediate redshift e(a) galaxies are 
common in both clusters and the field \citep{poggianti99,dressler99,dressler04,poggianti09}
suggesting that the presence of a dusty starburst in not caused by the cluster environment.
At these redshifts the frequency of E+A galaxies in the cluster environment is strongly
enhanced \citep{dressler99,tran04,poggianti09} and argues that a cluster-specific mechanism is
involved. This environmental dependence  (and the overall frequency) 
of e(a) galaxies precludes all of them being precursors to the E+A phase \citep{poggianti09}; 
however, they may represent the progenitors of some of the E+A population.

The physical mechanisms that drive the galaxy evolution occurring in rich clusters 
remains an unresolved issue. 
There are a plethora of possibilities including: major galaxy 
mergers \citep{mihos96,bekki05}, unequal mass mergers
\citep{bekki01}, galaxy interactions \citep{bekki05}, interactions with the cluster 
tidal field \citep{bekki99},
galaxy harassment \citep{moore96} and interaction with the hot intra--cluster 
gas \citep{gunn72,dressler83,bothun86}.
In the local field there is mounting evidence that the formation mechanism for 
post starburst galaxies involves
galaxy--galaxy mergers and interactions 
\citep{zabludoff96,norton01,blake04,yamauchi05,goto08,yang08,pracy09}. 
Such galaxies are rare in all environments at the present epoch and the cluster environment is not strongly 
preferred \citep{blake04}. However, at intermediate redshift the frequency
of these galaxies is strongly enhanced in the cluster environment 
\citep{dressler99,tran04,poggianti99} and this hints that
a cluster specific mechanism is at work, at least in part,  with ram pressure stripping being 
a favored candidate 
\citep{poggianti99,poggianti09}. 

Understanding the progenitor populations of post-starburst galaxies at 
low redshift has been progressed significantly via
spatially resolved optical spectroscopy \citep{norton01, pracy09}. 
The internal kinematics and distribution of the young and old
stellar populations convey important information about the processes that resulted 
in the starburst and its cessation. Tidal 
interactions and in particular major galaxy mergers are expected to result in a centrally-concentrated 
burst of star--formation and hence a strong enhancement in the Balmer line
absorption in the central regions of the galaxy during the subsequent post-starburst
phase \citep{noguchi88,barnes91,mihos92,mihos96,bekki05,bournaud08}. In contrast, uniform truncation of 
star--formation in a star-forming galaxy, as would be expected from ram-pressure stripping by 
the intra--cluster medium (ICM), should result in a more distributed young stellar 
population and hence Balmer absorption signature \citep{rose01,bekki05,pracy05}. In addition, 
these two different scenarios are expected to produce systems that are quite
different kinematically: If major galaxy mergers are the trigger, then they would 
(in general) destroy any disk rotation and the remnant 
should be pressure-supported. In contrast, other less violent mechanisms will allow rotation 
to persist in the remnant galaxy \citep{bekki05,pracy09}. In this context the
results from spatially resolved spectroscopic studies of E+A galaxies at
low redshift \citep{norton01, pracy09} have been divided as to the trigger mechanism. 
Norton et al's (2001) long-slit
spectroscopy of a significant number of the E+A's from the original \citet{zabludoff96}
sample, found in the majority of cases that the young post-starburst population was 
centrally-concentrated and its kinematics was consistent with it being pressure-supported, 
indicative of a major-merger origin. \citet{pracy09}'s two-dimensional integral field unit 
(IFU) spectroscopy of a similarly selected sample of nearby E+A galaxies, on the other hand,
found them all to have no detectable gradient in Balmer absorption line strength
across their central regions and exhibit significant rotation, pointing to either
a minor merger or some other less violent formation mechanism.

At higher redshifts ($z>0.1$), the situation is equally unclear, due largely to the
dearth of spatially resolved spectroscopic information on E+A galaxies, which at
these earlier epochs mostly reside in rich clusters. The only major such study to
be undertaken is that of \citet{pracy05}, who obtained IFU spectroscopy of 12 E+A galaxies
in the rich cluster, AC~114, at $z\sim 0.32$. They found quite diverse behaviour in the 
spatial distribution of the young stellar populations within their sample and interpreted 
this as the result of both major mergers and truncation by the intra--cluster 
medium as playing important roles in making up the cluster post-starburst population. 

In order to more thoroughly investigate and understand the mechanisms responsible
for E+A formation in intermediate redshift clusters, we have commenced an observational
program aimed at significantly increasing the amount of high quality spatially resolved
spectroscopy of E+A galaxies in these systems. In this paper we present the initial
pilot observations made for this program, where we exploited the excellent image
quality and light-gathering power of the 10m W. M. Keck I telescope to obtain
spatially resolved spectroscopy of E+A galaxies (as well as their
"e(a)" progenitors) in the rich cluster CL~0016+16 at $z=0.55$. Although at this
redshift such observations are very challenging in being able to sufficiently
resolve, spatially, the distribution and kinematics of the young stellar population 
that are key to discriminating between different formation mechanisms, we
demonstrate that in the very best (0.4\,arcsec) seeing conditions this is possible.

The plan of the paper is as follows: In the next section we describe the
selection of our E+A targets, their observation with the Low Resolution Imaging
Spectrograph (LRIS), and the reduction of the data. We present our results in 
Section 3, included in which is new information on the morphologies of our target
galaxies derived from deep Hubble Space Telescope images. In Section 4 we 
describe detailed modelling we have undertaken to understand the impact of
the smearing effects of seeing on our observations and our ability to
discern spatial structure. We discuss our finding and summarise our conclusions
in the final section. Throughout the paper we convert from observed to physical 
units assuming  a $\Omega_{M}=0.3$, $\Omega_{\Lambda}=0.7$ and $h=0.7$ cosmology, 
noting that at the redshift of CL~0016+16, 1\,arcsecond
\ corresponds to $\sim$6.4\,kpc.

\section{Observations and data reduction}

\subsection{Input catalogue}
CL~0016+16 is a cluster rich in post-starburst galaxies, with a total of 10
objects identified as "k+a", "a+k", "a+k/k+a" in the spectroscopic studies of
\citet{dressler92} and \citet{dressler99}. These studies also identified a
small number of (3) dusty starburst galaxies -- denoted "e(a)" by \citet{dressler99} -- 
which may well be the progenitors of the E+A population; these objects were
also included in our initial pool of targets. All these objects were brighter
than $r_{tot}$=22.25\,mag, this being the magnitude limit of the original
spectroscopy.

The final selection of targets for observation was made as part of the
slit mask design process for LRIS, which was carried out using the {\em AUTOSLIT3}
program provided by the Keck Observatory. This program was first used in its 
interactive mode to select targets and assign slits to them, and then used
to automatically grow the slits in an optimum way so as to use the full area
of the LRIS detectors while at the same time work within the geometrical
constraints imposed by the surface density of the targets, the minimum slit
length requirements (set to 5.0\,arcsec), and the length of the spectra on 
the detector. Of highest priority in this process was to select as many E+A 
types as possible, and then to assign any remaining slits to the e(a) galaxies.
Another important feature of the {\em AUTOSLIT3} program that was used was
the ability to tilt the slits with respect to the spatial axis. This
allowed us, in a few cases, to better align the slit with the major axis of
the galaxy, which is highly desirable if rotation is to be measured.

Despite this extra degree of freedom in being able to tilt the slits (up 
to $\pm$30\,degs relative to the spatial axis), the process of selecting targets 
and assigning slits to them was, 
in practice, highly constrained. Even with the use of two different masks, 
only 6 of the 10 E+A galaxies within the cluster could be observed, and 
the slit orientation with respect to the galaxies' major axis was often
far from optimum. The details of the final sample of objects that were observed
are listed in Table \ref{tab:targets}, where the first six columns contain the 
identification number, the Right Ascension, the Declination, the redshift, 
the apparent magnitude $r_{tot}$, and the $r-i$ colour, all taken from 
\citet{dressler92}. The three subsequent columns give the morphology
of each galaxy (taken from \citet{dressler99}), its approximate orientation
(face-on, edge-on, or ambiguous), and its spectral classification (also
from \citet{dressler99}). As can be seen from the table, all three of the
e(a) galaxies in CL~0016+16 were included in our final sample.

\begin{table*}
\begin{center}
\caption{\label{tab:targets}Target galaxies and their properties}
\begin{tabular}{llllllllll}\hline
Object id &   RA(J2000)   & Dec(J2000)  &   z    &  $r_{\rm tot}$  &  $r-i$      &  morph       & orientation &  spec type & spec type \\
         &               &             &           &               &             &             &  & (D99) & (this work) \\ \hline
DG\_106  &  00 18 36.19  & 16 25 7.0   & 0.5306    &  22.2         &  0.20       &    Sb      & face--on    &    a+k           &  a+k/k+a         \\
DG\_115  &  00 18 32.47  & 16 25 10.1  & 0.5485   &  21.5         &  0.51        &    S0       & face--on    &    a+k/k+a        &  k+a         \\
DG\_134  &  00 18 36.81  & 16 25 17.7  & 0.5525    &  22.2         &  0.05        &    Sa      & face--on    &    a+k          &  k+a/a+k         \\
DG\_181  &  00 18 30.94  & 16 25 41.7  & 0.5628    &  21.6         &  0.47        &    Sab     & face--on    &    k+a         &  k+a         \\
DG\_338  &  00 18 36.01  & 16 26 52.1  & 0.5570    &  22.2         &  0.42       &    Sd/irr  & ambiguos    &    e(a)          &  e(c)/e(a)        \\
DG\_352  &  00 18 31.58  & 16 26 56.6  & 0.5374    &  22.0         &  0.47        &    S0      & edge--on    &    k+a/a+k       &  k+a         \\
DG\_356  &  00 18 29.50  & 16 26 57.9  & 0.5379    &  22.2         &  0.33        &    Sd      & edge--on    &    e(a)          &  e(a)/e(c)        \\
DG\_371  &  00 18 38.95  & 16 27 03.7  & 0.5630    &  20.8         &  0.57        &    Sc      & face--on    &    k+a/e(a)      &  k+a/e(c)/e(a)         \\
DG\_411  &  00 18 32.97  & 16 27 22.3  & 0.5339    &  21.9        &  0.34        &    Irr     & ambiguous    &    k+a           &  k+a/a+k         \\ \hline
\end{tabular}
\end{center}
\end{table*}

\subsection{LRIS spectroscopy}
The observations were carried out on the W.M. Keck I telescope on the first 
half of the night of 2008 October 31, using the red camera on LRIS. The 600/7500
grism was used for our spectroscopy, which with the 0.75\,arcsec slit width used
for both masks, yielded a spectral resolution of 4.5\,\AA. This grism provides
a wavelength window of $\sim$2620\AA\, which was centred at $\lambda = 7010$\AA\, 
thereby covering the range $5700\leq\lambda\leq 8320$\AA\, which corresponds
to the interval 3680--5370\AA\ in the cluster rest frame. The spatial sampling
along the slit with this configuration was 0.21\,arcsec per pixel. 

Equal time was spent observing through each of 
the two masks designed for our observations (see above). Several objects 
(DG\_134, DG\_181, DG\_352, DG\_356, DG\_371) 
were observed through both masks. The total integration time for each mask was 2\,hours 
taken in $4\times 1800$\,s exposures and bounded by flat and arc-lamp observations 
on each side. The seeing was $\sim 0.4$\, arcseconds throughout the observations
taken with the first mask (hereafter Mask~1), and then it gradually increased to 
$\sim 0.8$\, arcseconds over the course of the observations taken with
the second mask (hereafter Mask~2).

\subsection{Data reduction}
The data were reduced using the automated LRISredux pipeline written
by Matt Auger (priv comm). The pipeline consists of a set of Python routines which
reduces the data to the point where each mask observation has been 
transformed to a set of 2-dimensional spectra (one for each slit)
and resampled onto a distortion free lambda,y plane. 
Spatially resolved spectra along the slit are then extracted along individual rows.
The spatial pixel scale is $\sim 0.21$\,arcsec/pixel, which at the cluster redshift, corresponds to $\sim 1.3$\,kpc/pixel.
We extract the central (spatial) spectrum individually and then extract and add two
adjacent spatial spectra moving outward from the centre until the signal-to-noise ratio becomes
insufficient. In the same process the individual exposures are combined by identifying the
same spatial bins in each exposure using the slit profiles. The result is a
set of 1 dimensional spectra for each target corresponding to different spatial
intervals along the slit. The final signal-to-noise ratio varies by object and
radial position. The minimum and maximum mean continuum signal-to-noise ratios
of the spatially resolved spectra used are 2.0\,\AA$^{-1}$ and 11.0\,\AA$^{-1}$, respectively.
Note: the minimum value corresponds to the outskirts of an emission line galaxy where we
get a much better signal-to-noise ratio in the emission lines. The minimum continuum signal-to-noise ratio 
in an absorption line system is 2.5\,\AA$^{-1}$. The median and standard 
deviation are 6.7\,\AA$^{-1}$ and  2.8\,\AA$^{-1}$. We also perform an extraction of the full spatial 
extent of each galaxy to obtain an integrated spectrum.

\subsection{Archival HST imaging}
The cluster CL~0016+16 has been extensively imaged with HST using its 
Advanced Camera for Surveys (ACS). A mosaic of images covering a region of 5\,arcmin by
5\,arcmin centred on the cluster and taken in both the F775W and  F606W filters
(hereafter referred to as R and V, respectively) was acquired as part of the GTO
program for this instrument (PI = H. Ford). The processed versions of these images 
were retrieved from the Hubble Legacy Archive to provide high spatial resolution photometry
and morphological information for the galaxies studied here. In order to provide a
visual indication of the morphology of each galaxy, postage-stamp images made from the
HST ACS F775W frames are shown in the first columns of Figures \ref{fig:psb}-\ref{fig:emmorph}. In Figures \ref{fig:psb} and \ref{fig:em}, 
these images are also used to show the orientation of each galaxy relative to the
LRIS slit, the latter being shown by the thick horizontal line. 

\subsection{Integrated spectra}
\begin{figure*}
  \begin{center}
    \begin{minipage}{0.95\textwidth}
      \includegraphics[width=4.6cm, angle=90, trim=0 0 0 0]{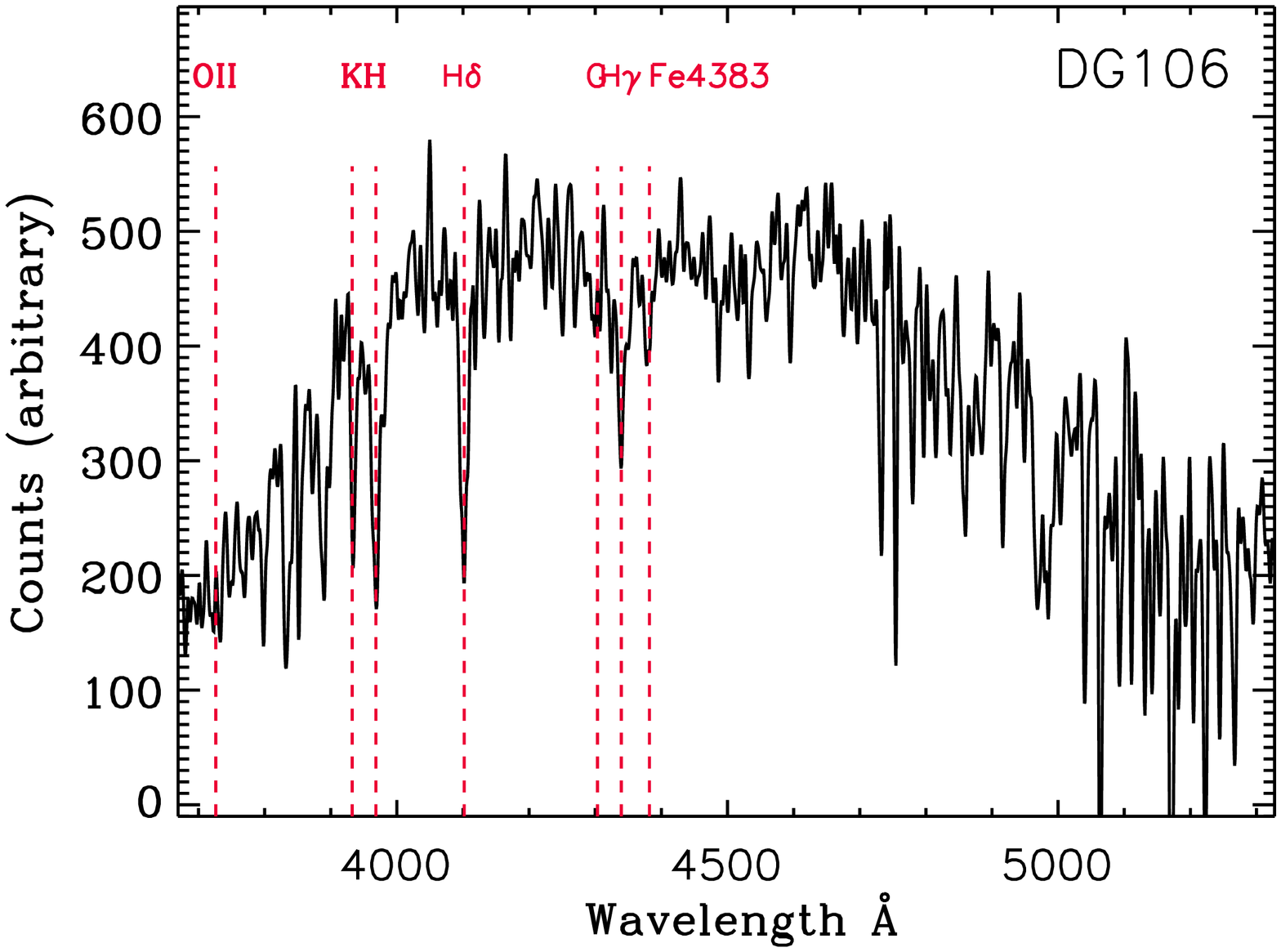}
      \includegraphics[width=4.6cm, angle=90, trim=0 0 0 0]{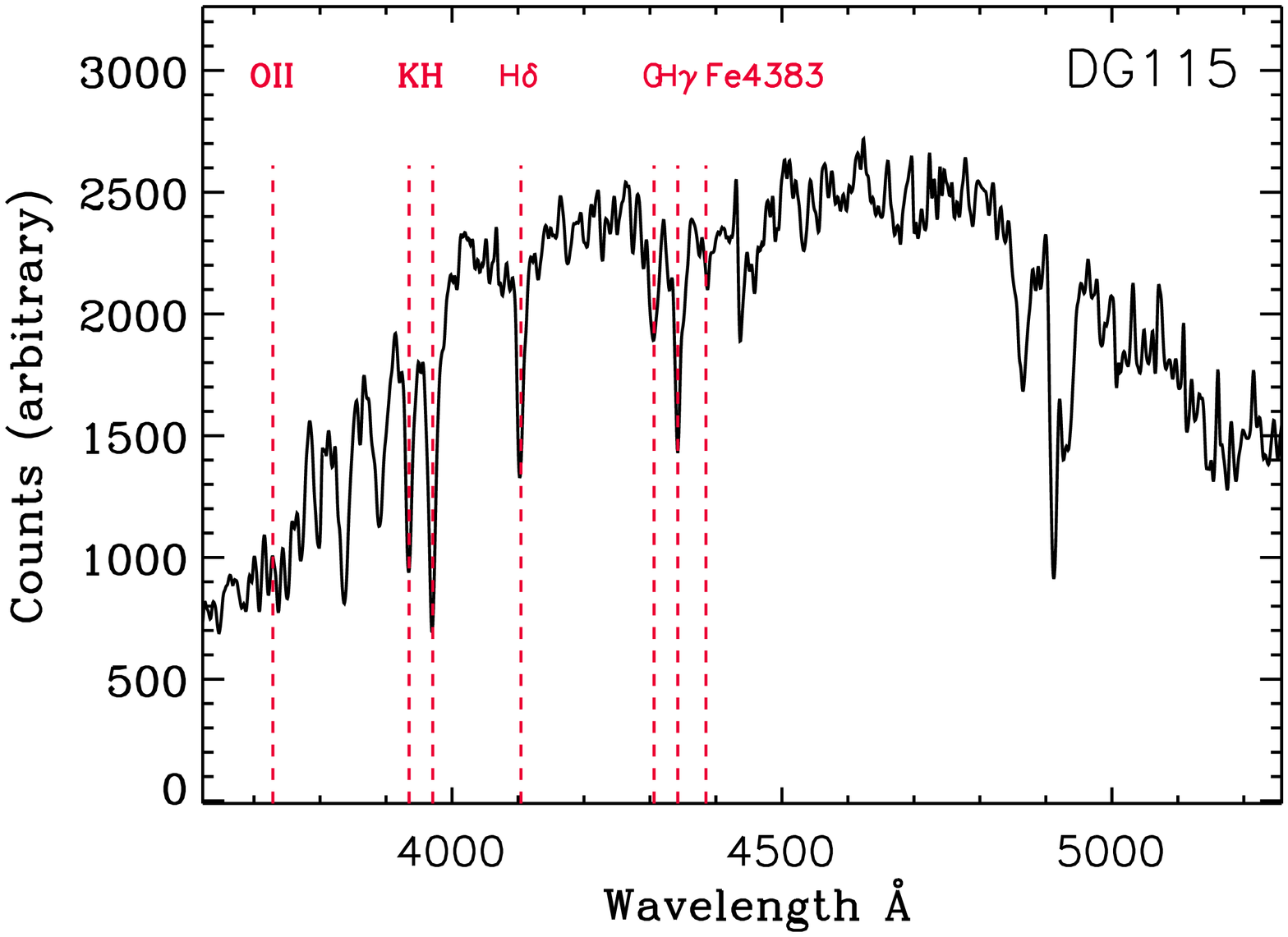}
    \end{minipage}
    \begin{minipage}{0.95\textwidth}
      \includegraphics[width=4.6cm, angle=90, trim=0 0 0 0]{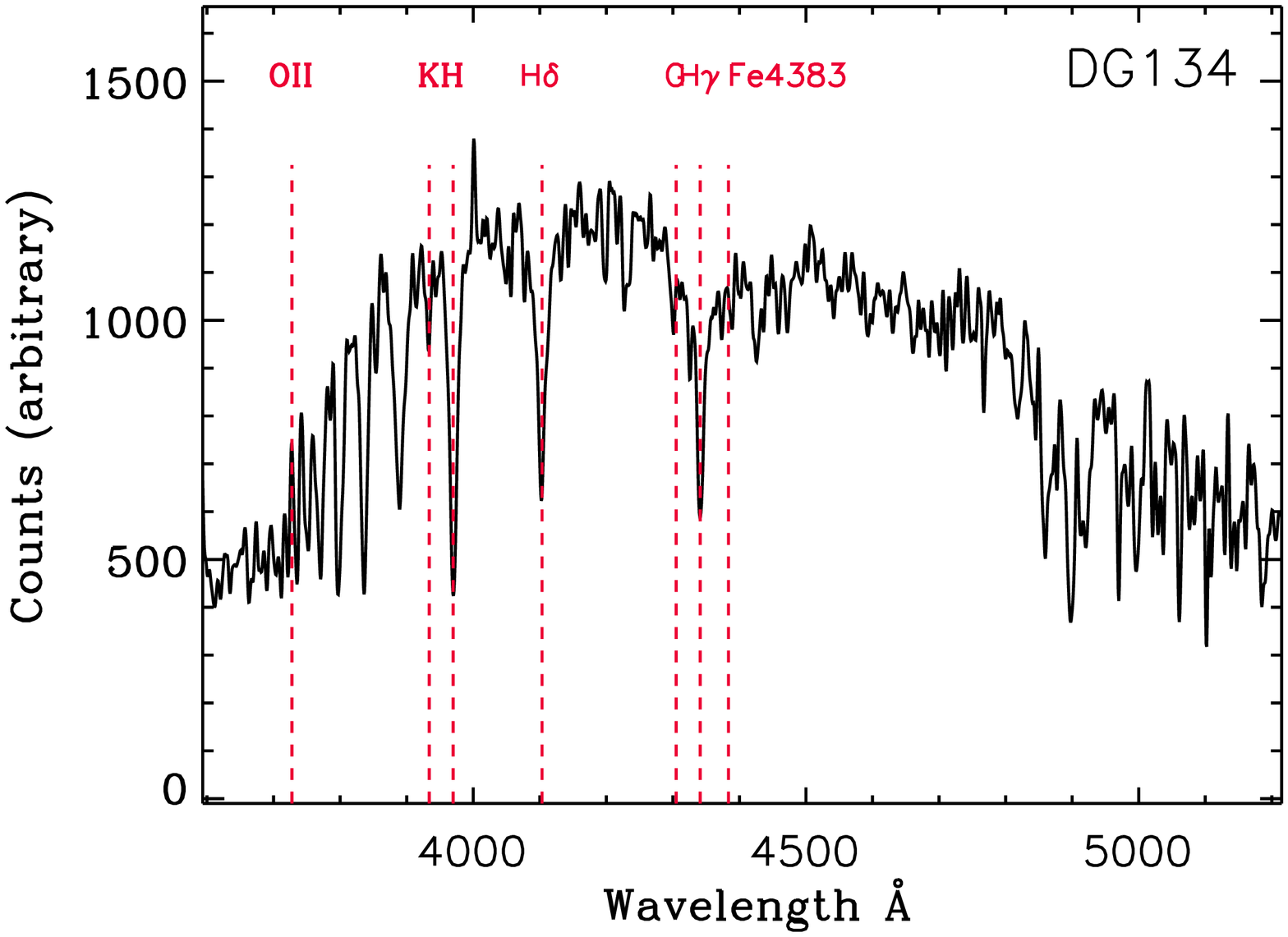}
      \includegraphics[width=4.6cm, angle=90, trim=0 0 0 0]{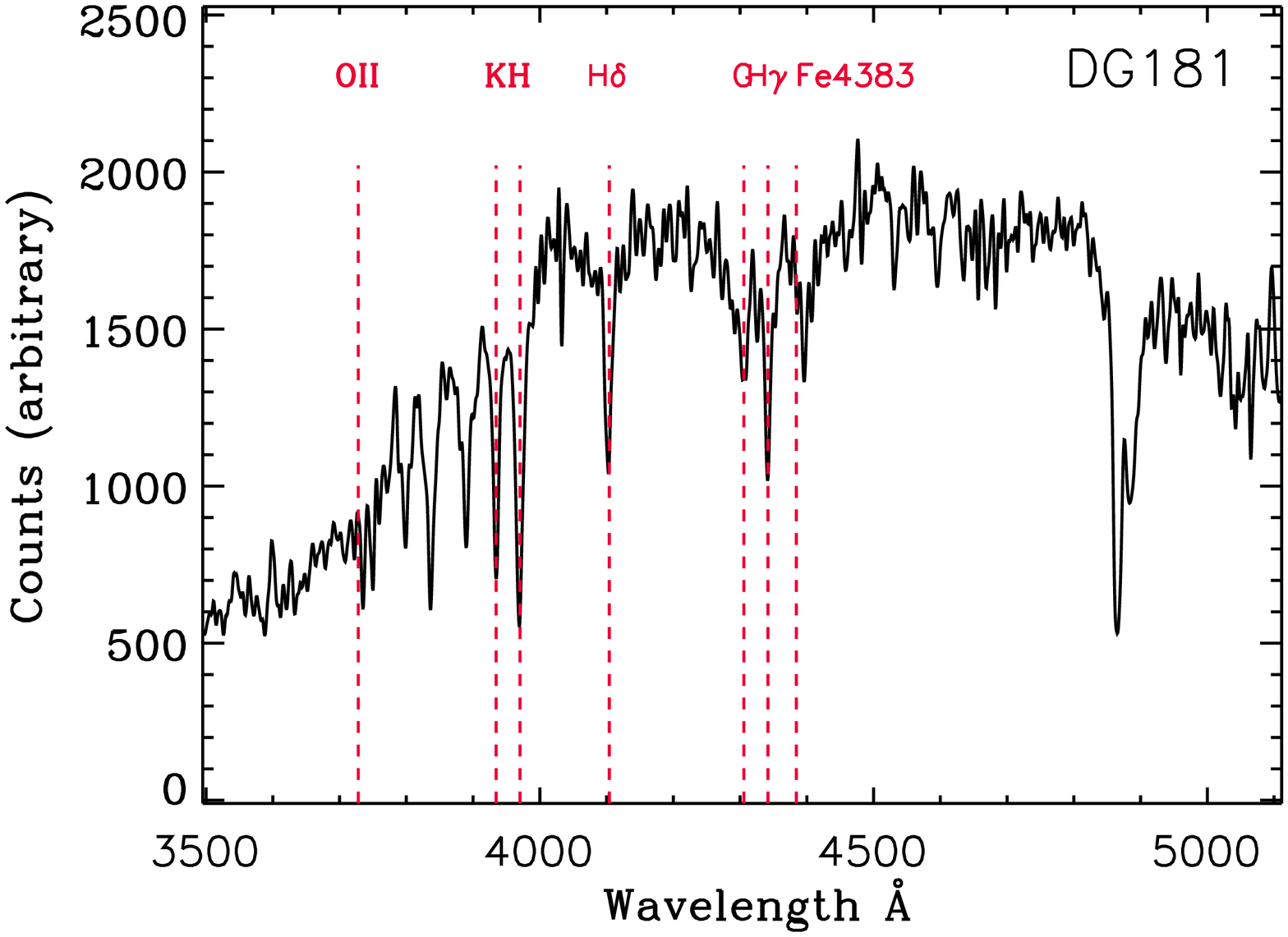}
    \end{minipage}
    \begin{minipage}{0.95\textwidth}
      \includegraphics[width=4.6cm, angle=90, trim=0 0 0 0]{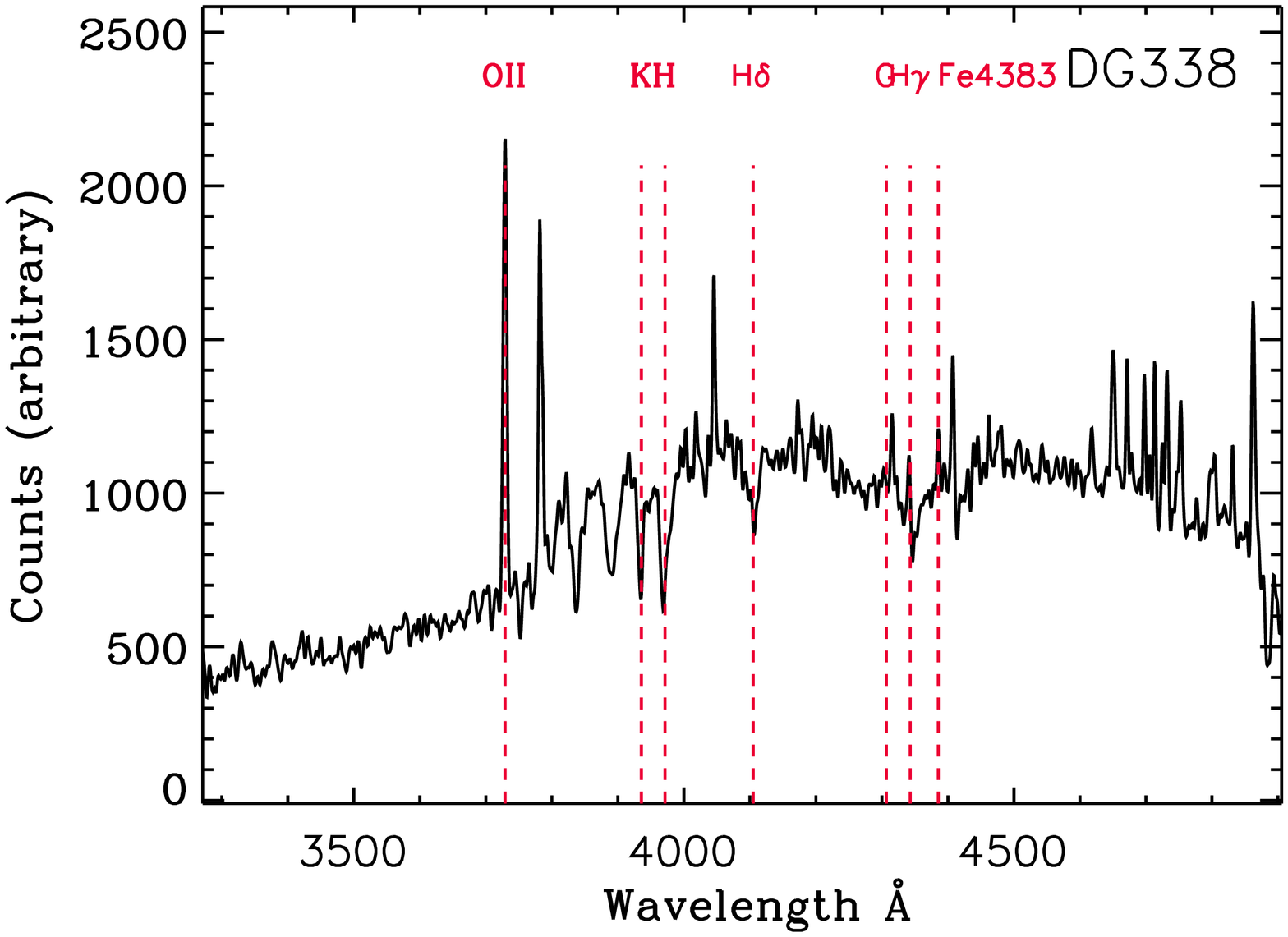}
      \includegraphics[width=4.6cm, angle=90, trim=0 0 0 0]{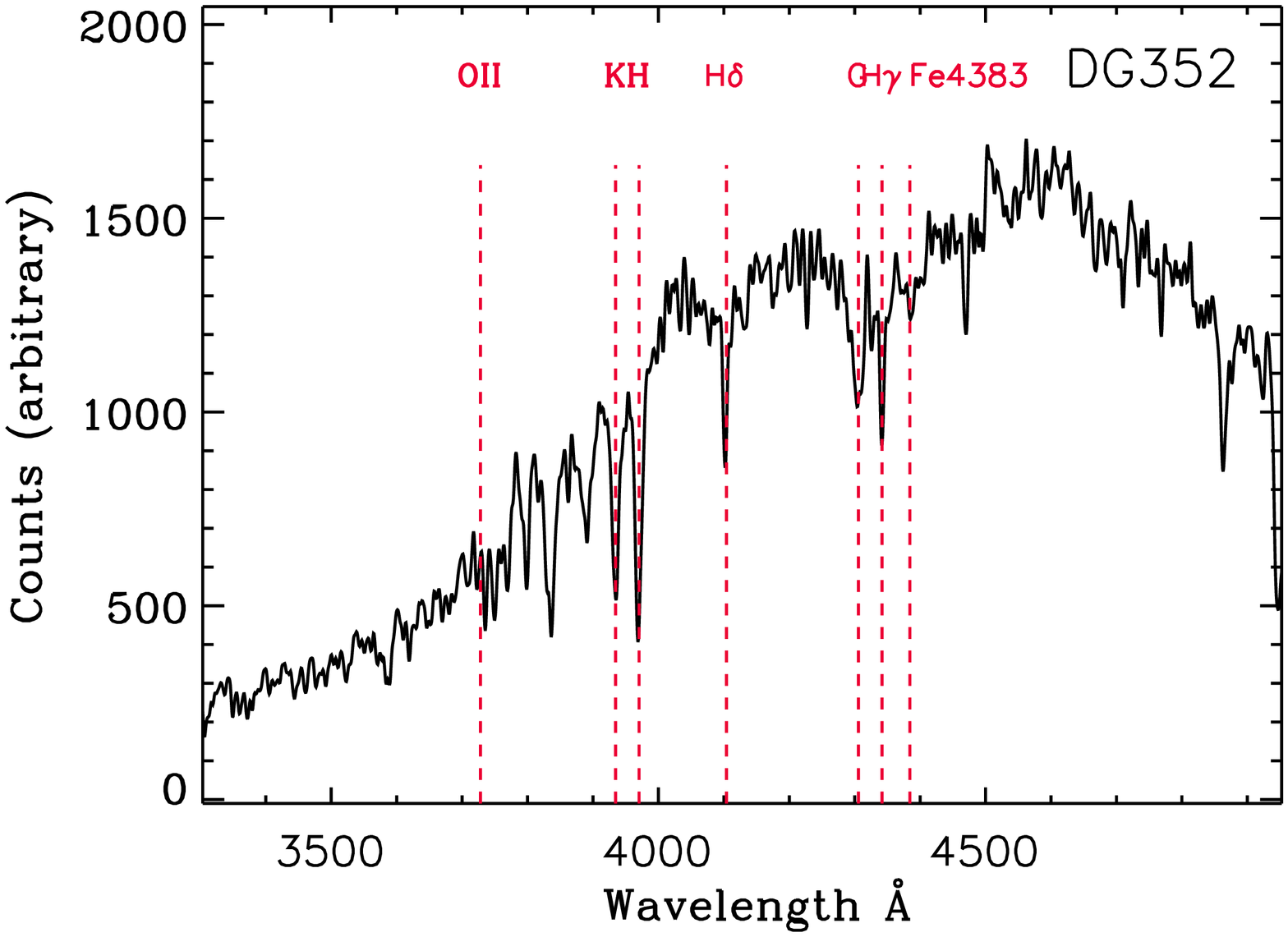}
    \end{minipage}
    \begin{minipage}{0.95\textwidth}
      \includegraphics[width=4.6cm, angle=90, trim=0 0 0 0]{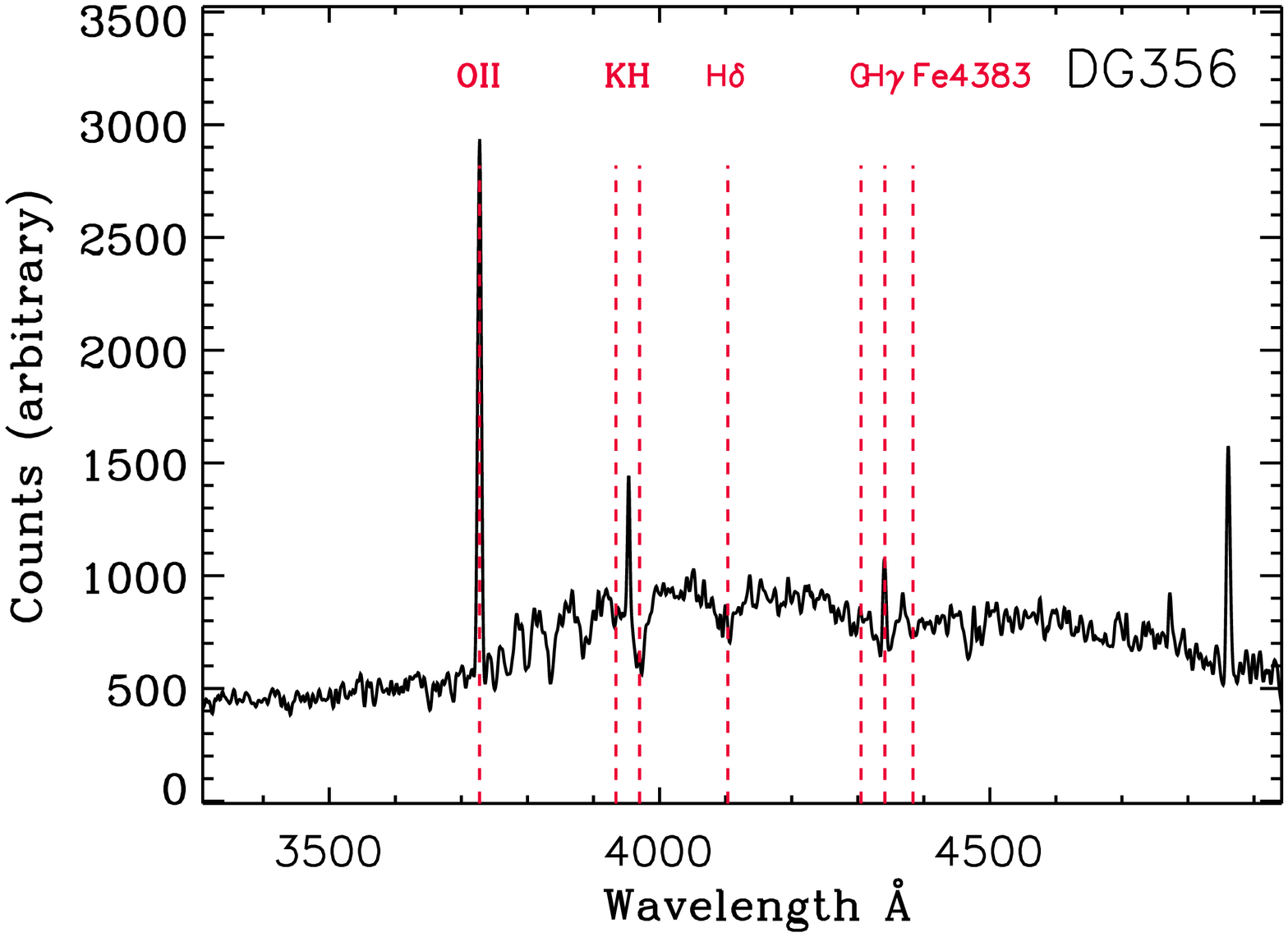}
      \includegraphics[width=4.6cm, angle=90, trim=0 0 0 0]{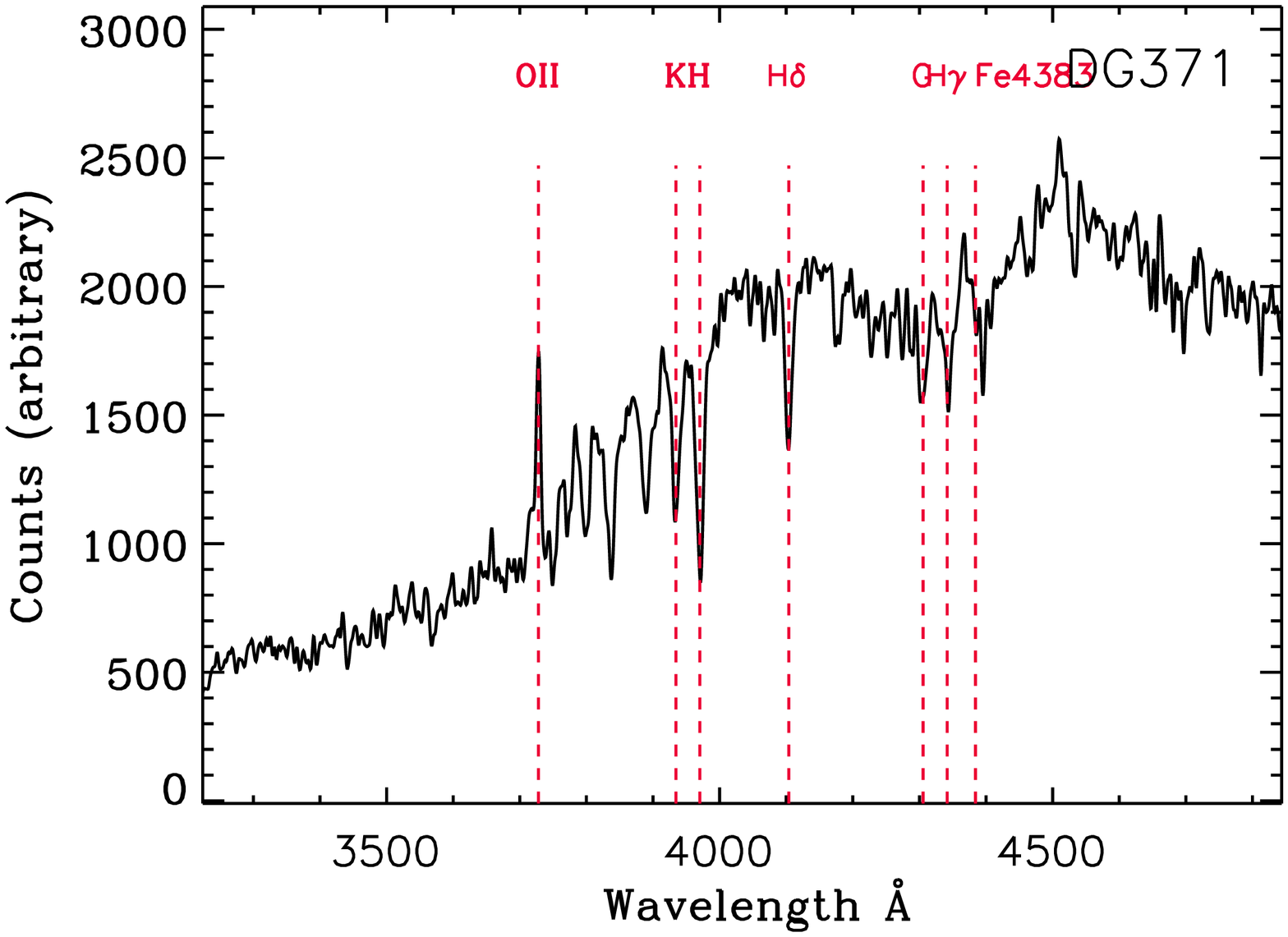}
    \end{minipage}
    \begin{minipage}{0.95\textwidth}
      \includegraphics[width=4.6cm, angle=90, trim=0 0 0 0]{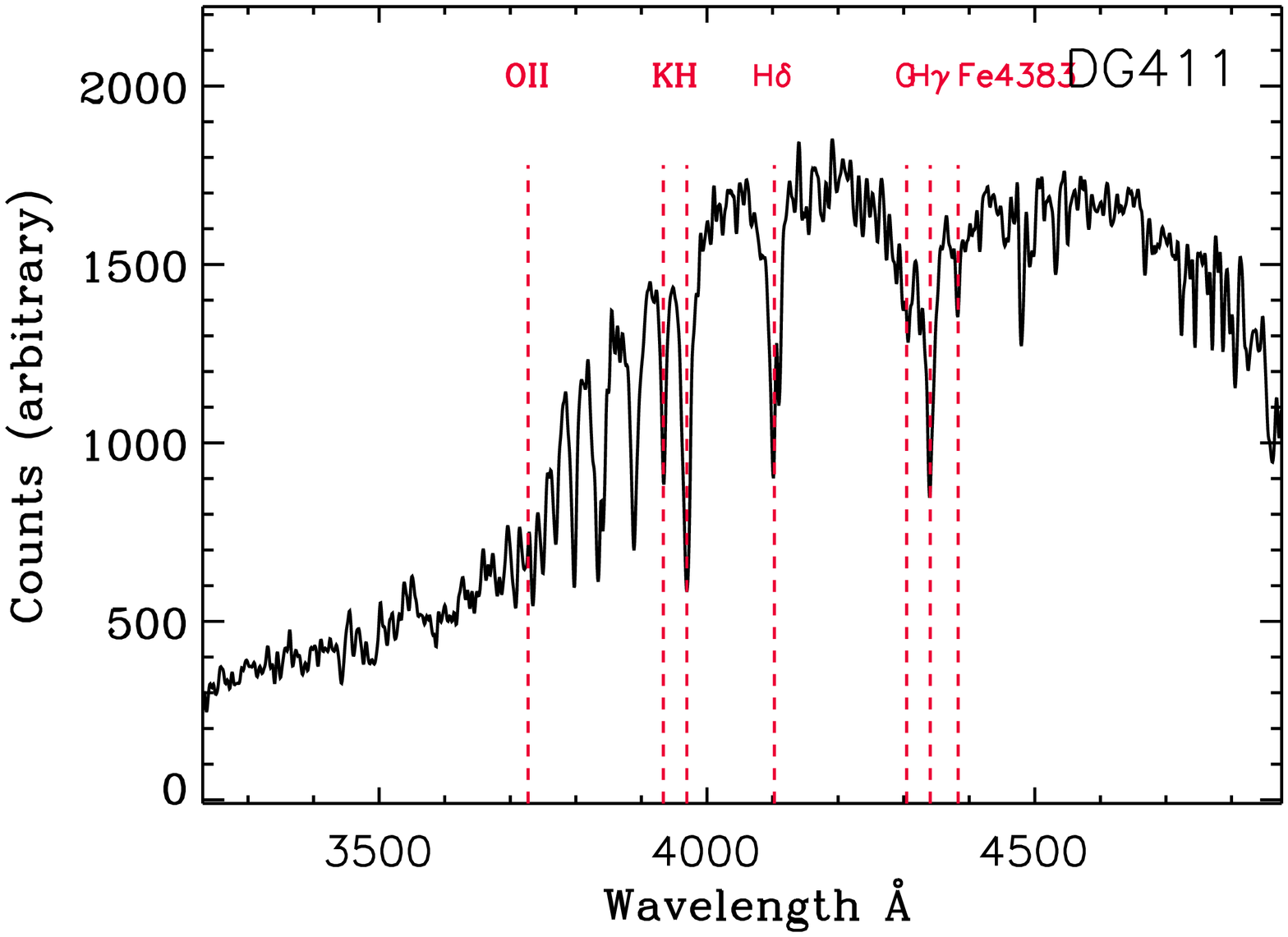}
    \end{minipage}
  \end{center}
  \caption{The integrated spectra of our sample in order of increasing \citet{dressler92} id number from top left
to bottom right: DG\_106; DG\_115; DG\_134; DG\_181; DG\_338; DG\_352; DG\_356; DG\_371; DG\_411}
\label{fig:intspec}
\end{figure*}

\begin{figure}
\begin{center}
\includegraphics[width=5.9cm, angle=90, trim=0 0 0 0]{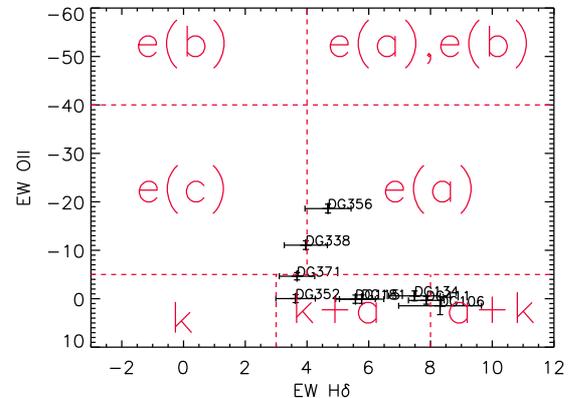}
\caption{\label{fig:spectype} The position of our sample galaxies in the equivalent width [OII]$\lambda 3727$ 
versus H$\delta$ plane which is used to define the spectral class.}
\end{center}
\end{figure}

The integrated spectrum for each galaxy in our sample is shown in Figure \ref{fig:intspec}. Of the sample
of nine galaxies three have clear [OII]$\lambda 3727$ emission lines whilst the remaining six have 
absorption line spectra with strong Balmer lines. We measure line indices from the spectra using the 
standard flux summing technique where the area in the line is measured by summing the counts in a bandpass 
centered on the line and the continuum is estimated in two flanking bandpasses either side of the line. 
For the H$\delta$ line we use the standard lick index definition for H$\delta_{\rm A}$ \citep{worthey97} 
and for the [OII]$\lambda 3727$ line we define the bands as: 3700--3720\AA\, (blue continuum),  3720--3734\AA\, 
(line) and 3734--3754\AA\, (red continuum). In Figure \ref{fig:spectype} we plot the positions of our 
sample in the equivalent width [OII]$\lambda 3727$ versus equivalent width H$\delta_{\rm A}$ plane.
The regions in this plane that define the different spectral classes of \cite{dressler99} are overlaid. 
Six of our galaxies sit neatly in the post--starburst (k+a and a+k) regions of 
this plane\footnote{we shall hereafter refer to the k+a and a+k types collectively as E+A galaxies}, and 
two lie in the star-forming regions [e(a) and e(c)]. One galaxy, DG\_371, occupies a position on the boundary 
of these two regions. The integrated spectrum 
of DG\_371 in Figure \ref{fig:intspec} reveals a highly significant detection of [OII]$\lambda 3727$ 
and the spectrum is very similar to that of the other two e(a) galaxies (DG\_356 and DG\_338). 
With high quality spectra like those shown in Figure \ref{fig:intspec}, it is reasonable to choose 
a tighter equivalent width constraint on the [OII]$\lambda 3727$ line in delimiting emission and absorption 
line systems; indeed an equivalent width of 2.5\AA\, is a more common 
definition \citep[e.g.][]{zabludoff96,blake04,goto07}. We therefore treat DG\_371 as an e(a)  system. 
In the last column of Table \ref{tab:targets}, we include our new spectral classifications, 
where we give multiple spectral classifications for a single object when it's 1$\sigma$ error bar overlaps
more than one region. While there are minor differences between our classifications and the 
original ones of \citet{dressler99} (listed in the previous column), there is generally good
agreement overall. 

\begin{figure}
\begin{center}
\includegraphics[width=5.9cm, angle=90, trim=0 0 0 0]{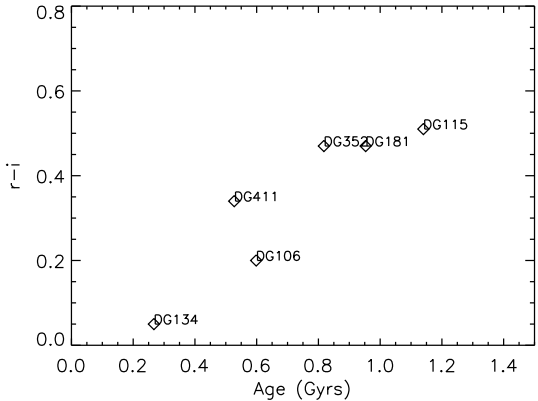}
\caption{\label{fig:agecol} $r-i$ colour of the E+A sample versus the time since
truncation for the post starburst galaxies.}
\end{center}
\end{figure}

The spatial distribution of the strength of absorption lines  such as H$\delta$ in E+A galaxies
is not expected to be constant but to depend on the time that has elapsed since the truncation of the 
starburst \citep[]{pracy05,bekki05}. It is, therefore, useful to have
an estimate for the age of the young stellar population in our E+A sample. In order to estimate the 
age of the young stellar population, we fit the E+A galaxy spectra with a set of single-age single-metallicity
stellar population synthesis models \citep{vazdekis07}. The fitting is performed using
a penalized pixel fitting algorithm \citep{cappellari04}.
The templates used range in age from 0.1\,Gyr to
2\,Gyr with age spacing of $\sim 0.1$\,Gyr. We also include two older templates of ages $\sim 5$\,Gyr
and $\sim 8$\,Gyrs to account for the underlying old population -- although we note that E+A light at
these wavelengths is very much dominated by the young stellar population \citep{pracy09}. We then
calculate a pseudo age of the young population by taking the weighted age of the best fitting 
young single age stellar population model templates. This method of estimating the age of the
young stellar population has been shown to be consistent with other age diagnostics \citep{pracy09}.
In Figure \ref{fig:agecol} we show our age estimates plotted against $r-i$ colour for our E+A sample.
The ages range from $\sim 0.2$ to $\sim 1.2$\,Gyrs -- consistent with their E+A status. As expected
there is a strong correlation of age with galaxy colour \citep[e.g.][]{couch87}.

\section{Results}

\subsection{Post-starburst galaxies}
Six of the nine galaxies in our sample have integrated spectra indicative of post--starburst
galaxies with strong Balmer absorption lines but lacking optical emission lines: DG\_106, 
DG\_115, DG\_134, DG\_181, DG\_352, and DG\_411. We now present the results of our
spatially resolved spectroscopy of these objects, as well as our reanalysis of their
morphologies based on HST imaging. 

\subsubsection{Spatially resolved spectroscopy}
\begin{figure*}
   \begin{center}
     \begin{minipage}{0.95\textwidth}
\hspace{-1.2cm}
         \includegraphics[width=3.6cm, angle=0]{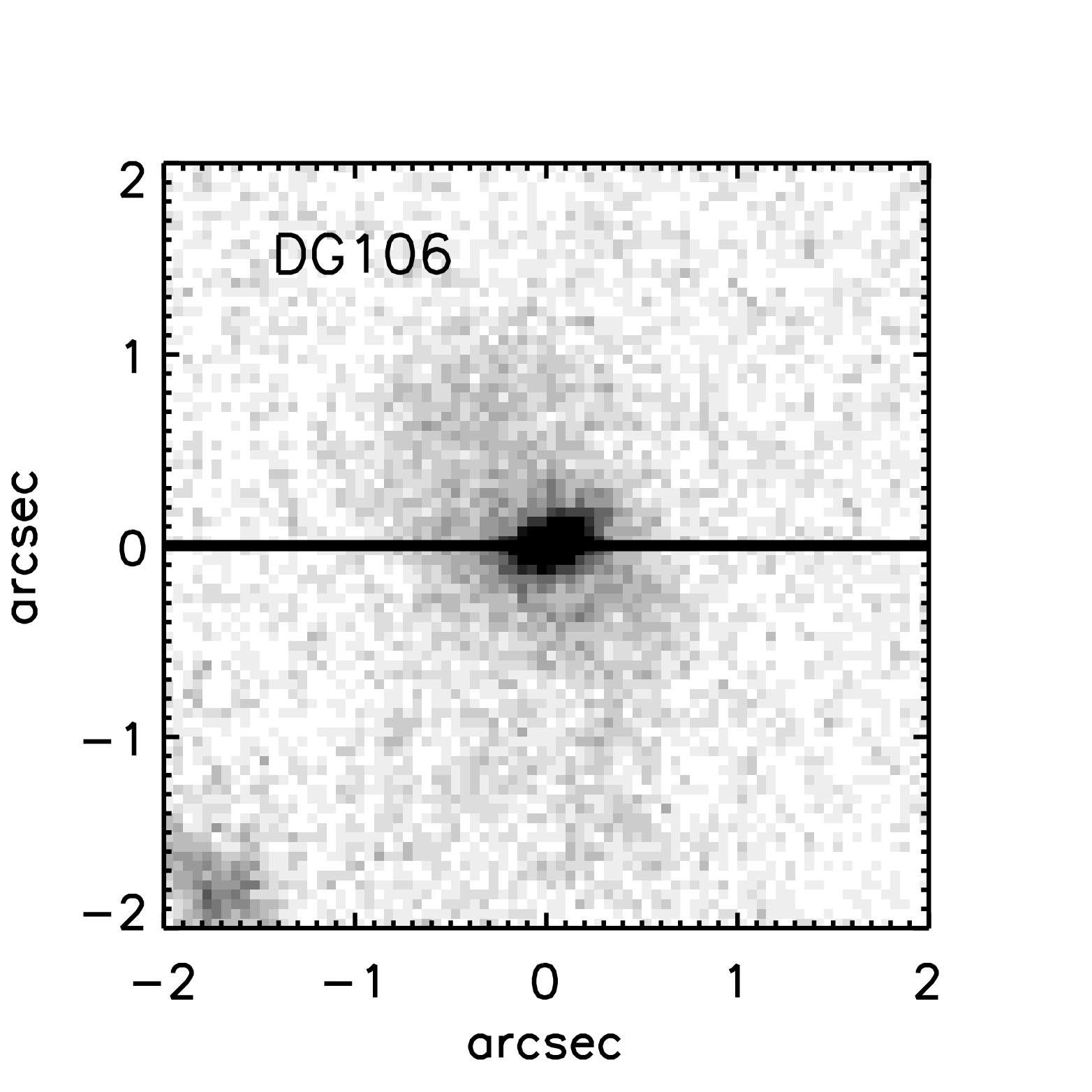}
\hspace{-0.8cm}
        \includegraphics[width=3.4cm, angle=90]{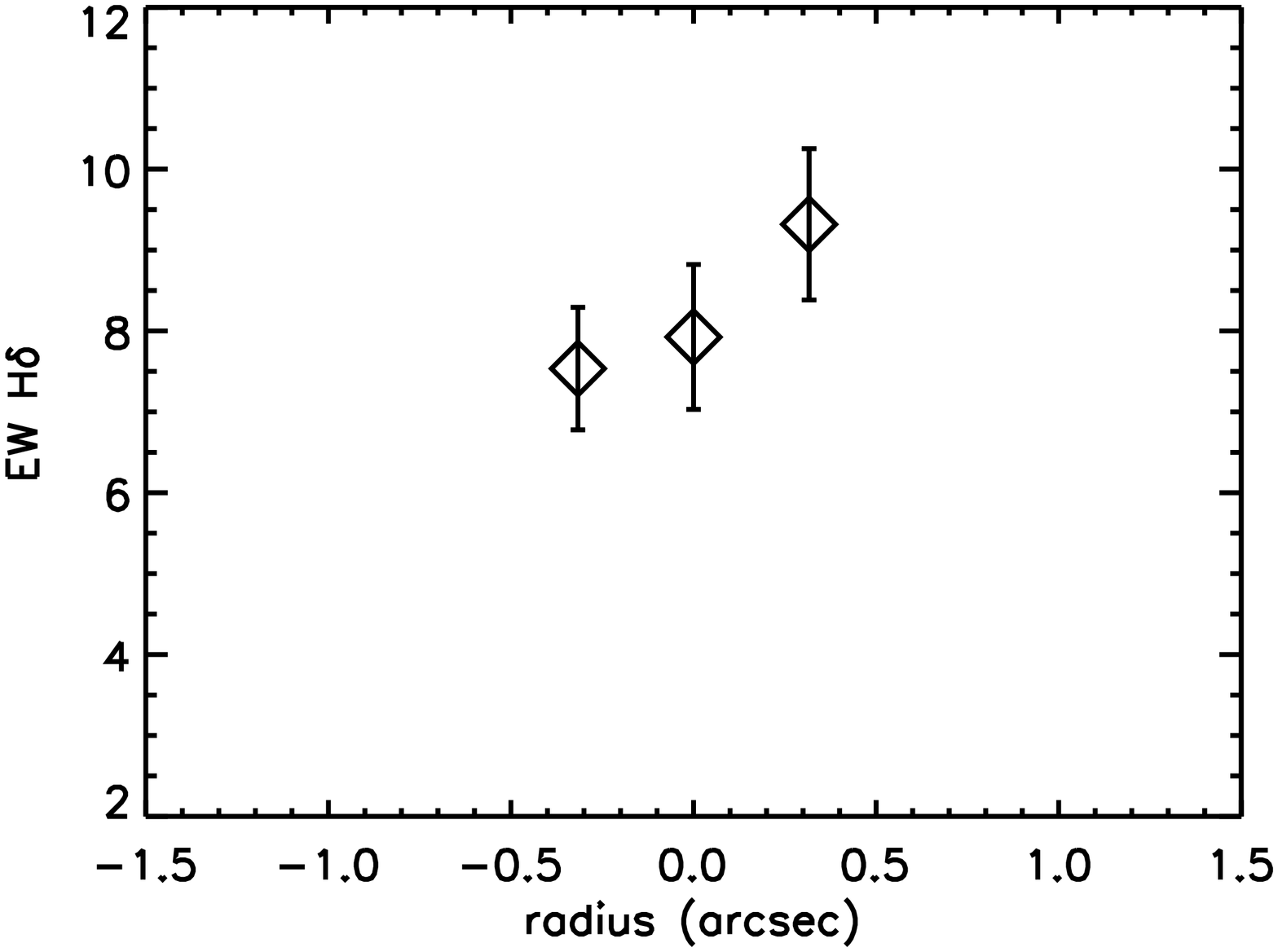}
\hspace{-0.6cm}
         \includegraphics[width=3.4cm, angle=90]{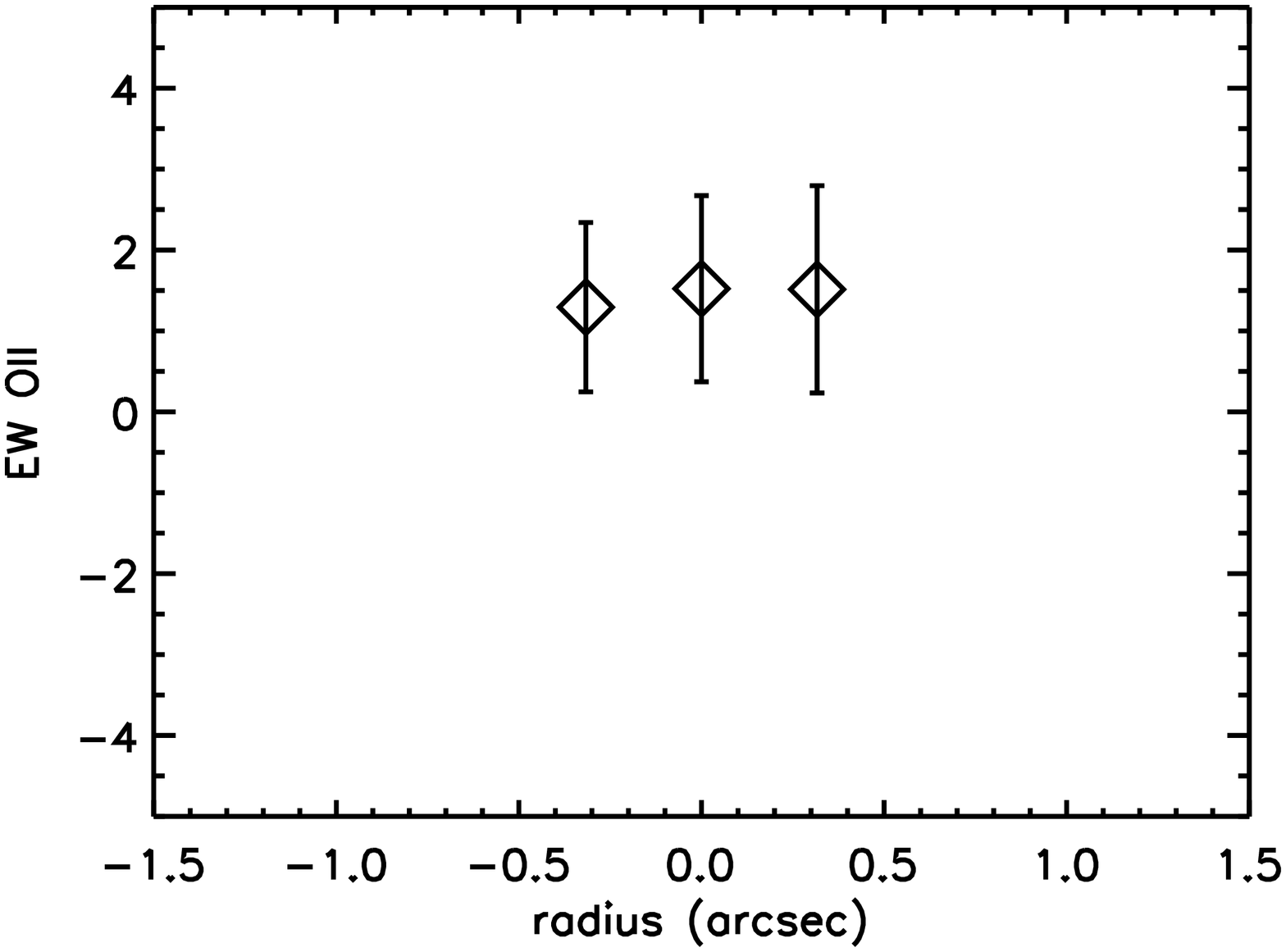}
\hspace{-0.6cm}
         \includegraphics[width=3.4cm, angle=90]{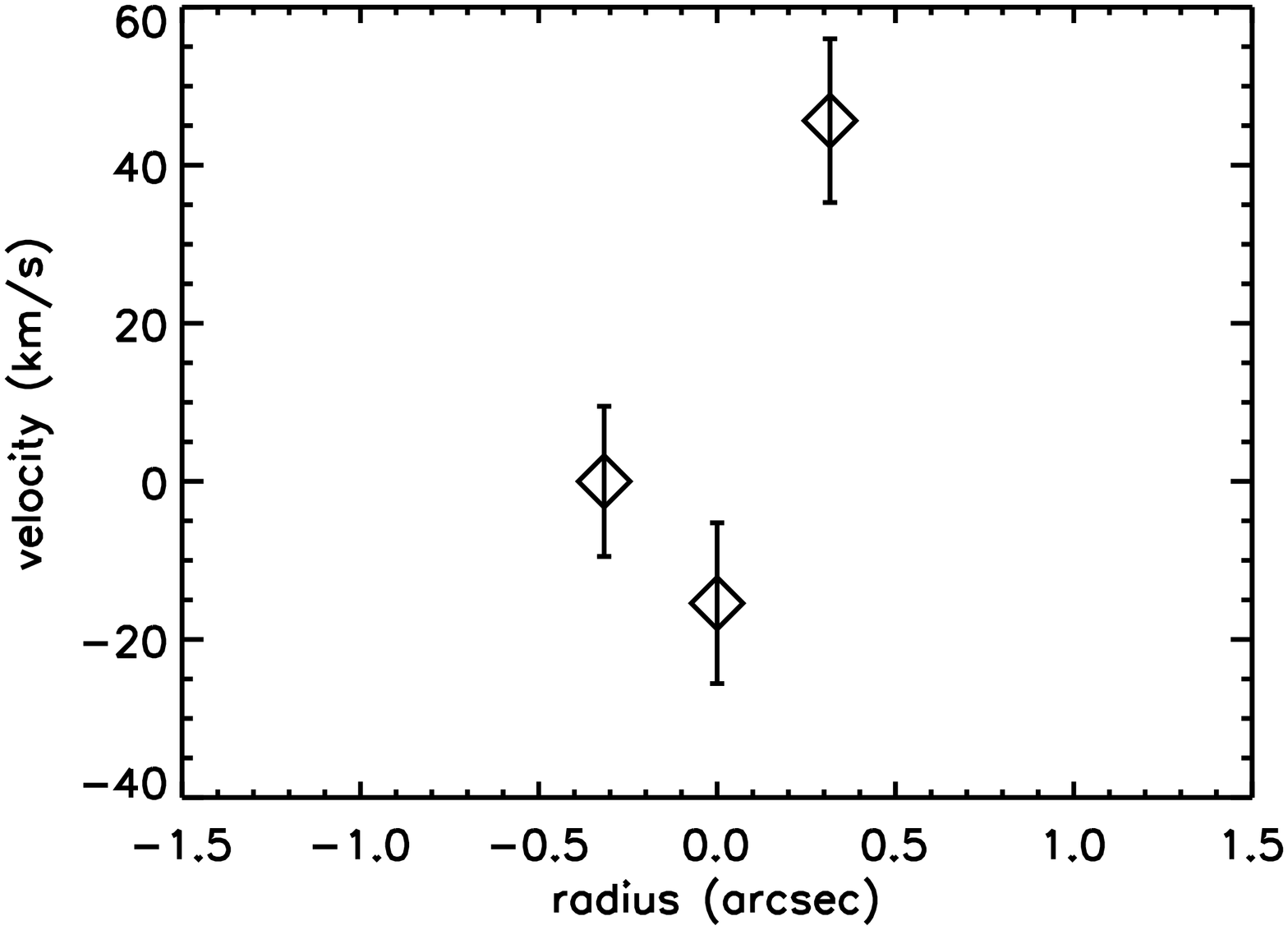}
      \end{minipage}
       \begin{minipage}{0.95\textwidth}
\hspace{-1.2cm}
         \includegraphics[width=3.6cm, angle=0, trim=0 0 0 0]{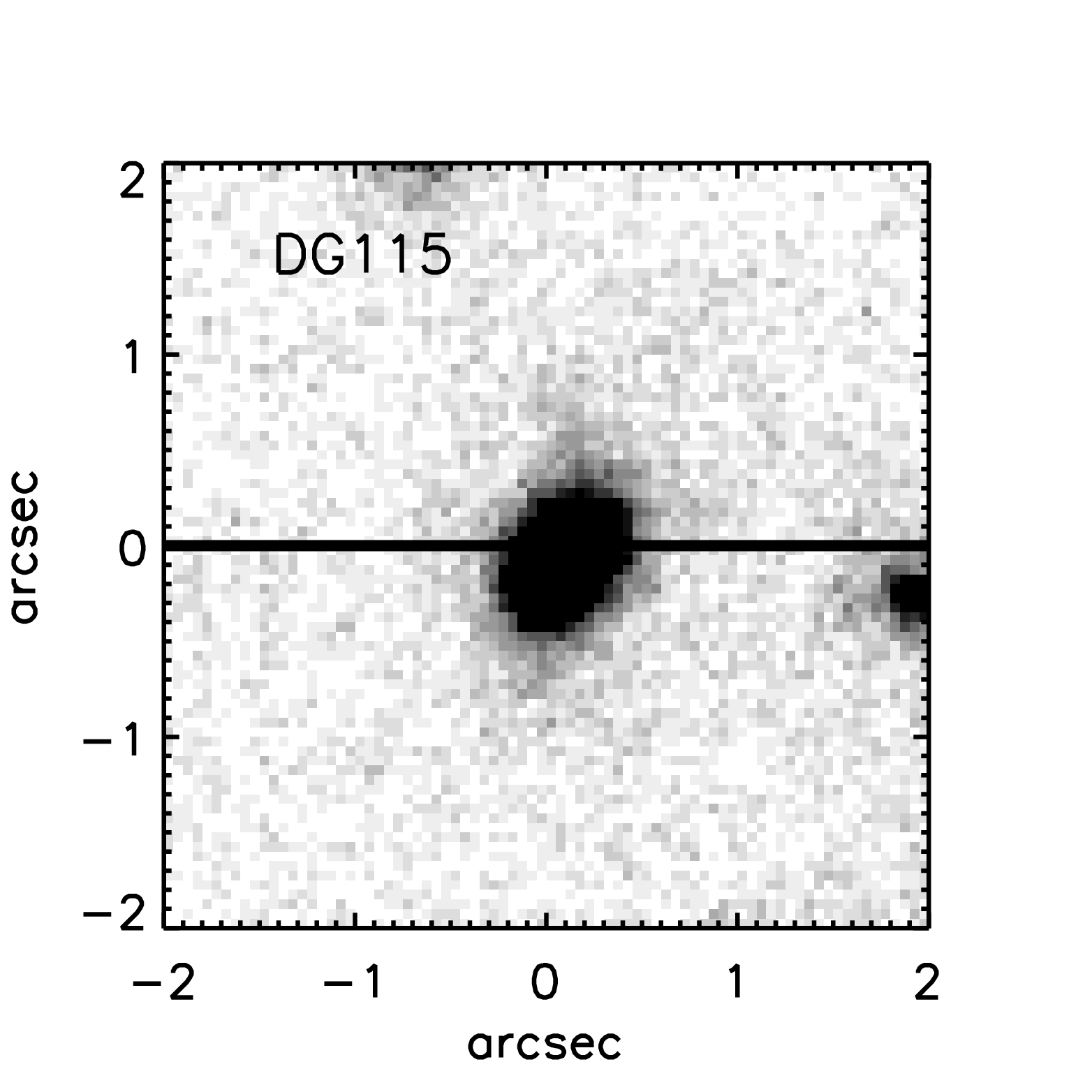}
\hspace{-0.8cm}
        \includegraphics[width=3.4cm, angle=90, trim=0 0 0 0]{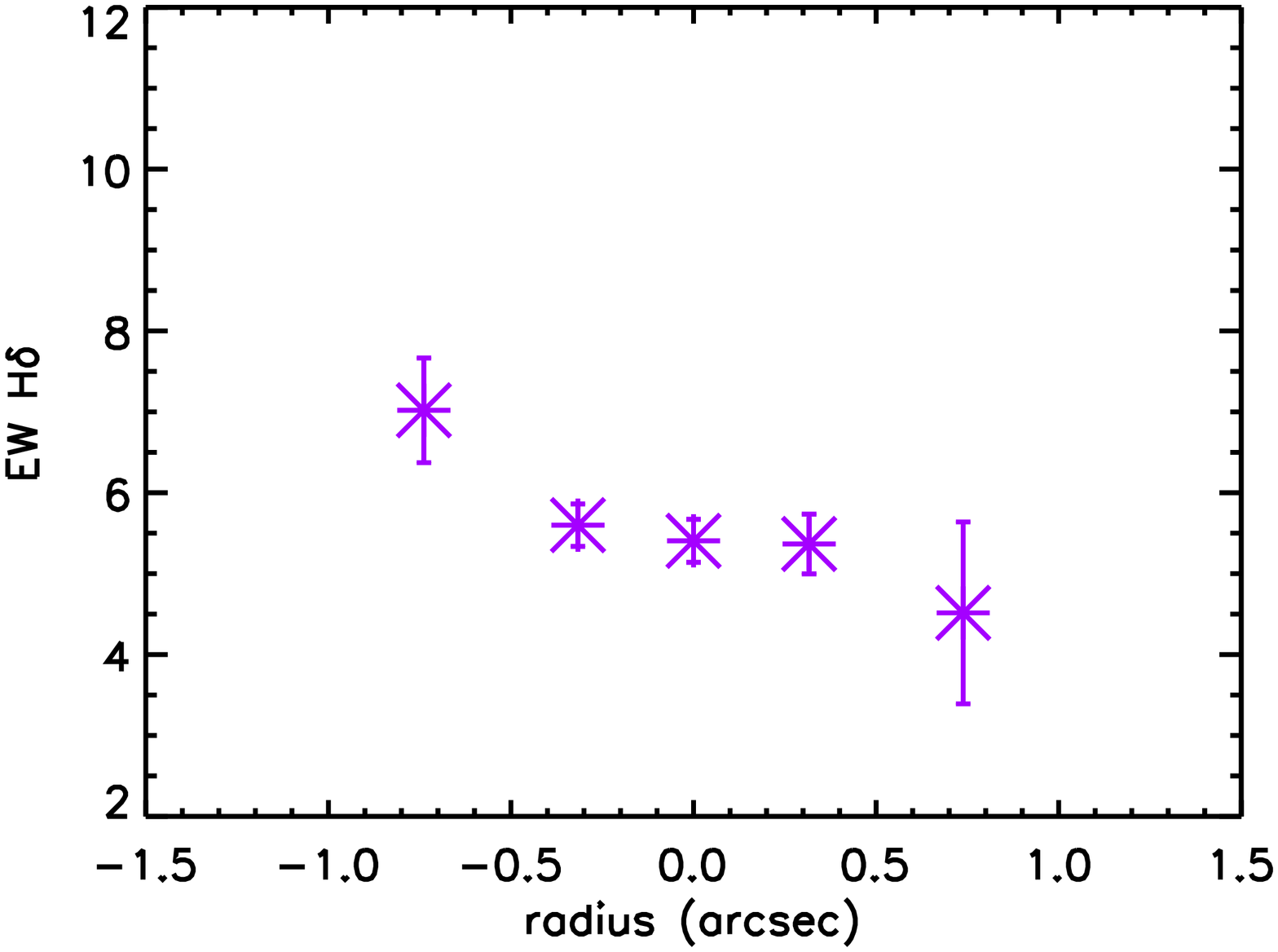}
\hspace{-0.6cm}
         \includegraphics[width=3.4cm, angle=90, trim=0 0 0 0]{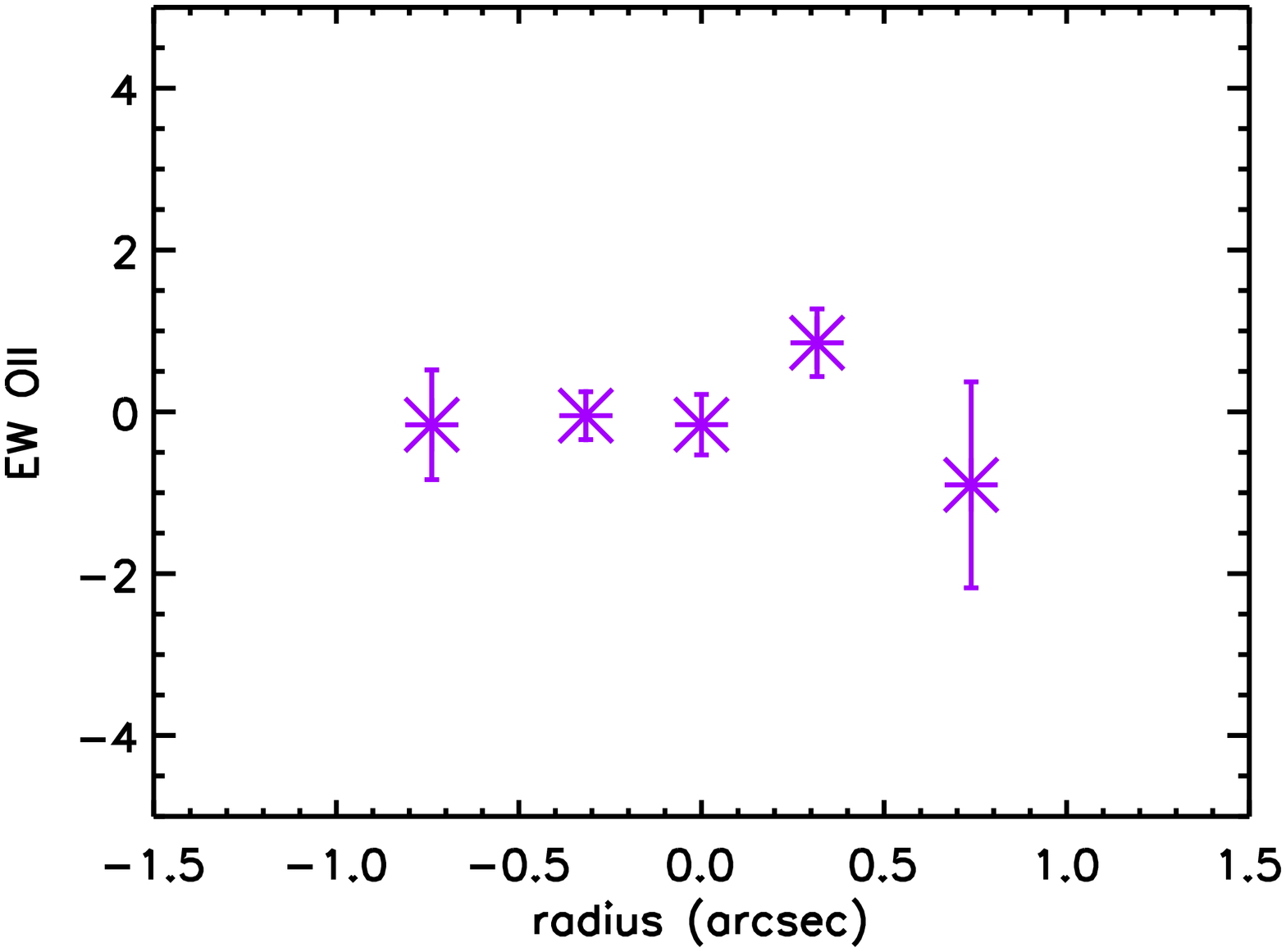}
\hspace{-0.6cm}
         \includegraphics[width=3.4cm, angle=90, trim=0 0 0 0]{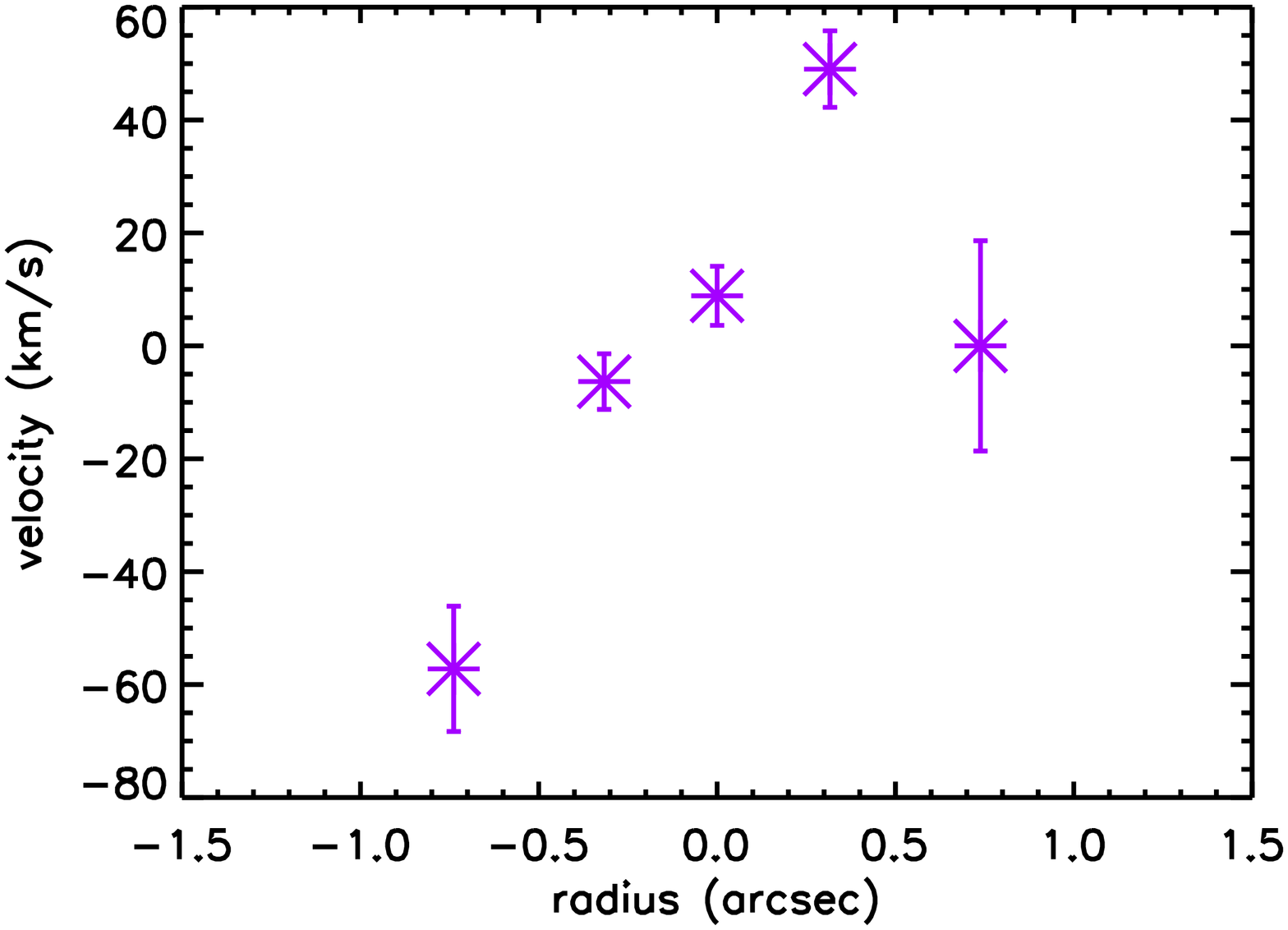}
      \end{minipage}
       \begin{minipage}{0.95\textwidth}
\hspace{-1.2cm}
         \includegraphics[width=3.6cm, angle=0, trim=0 0 0 0]{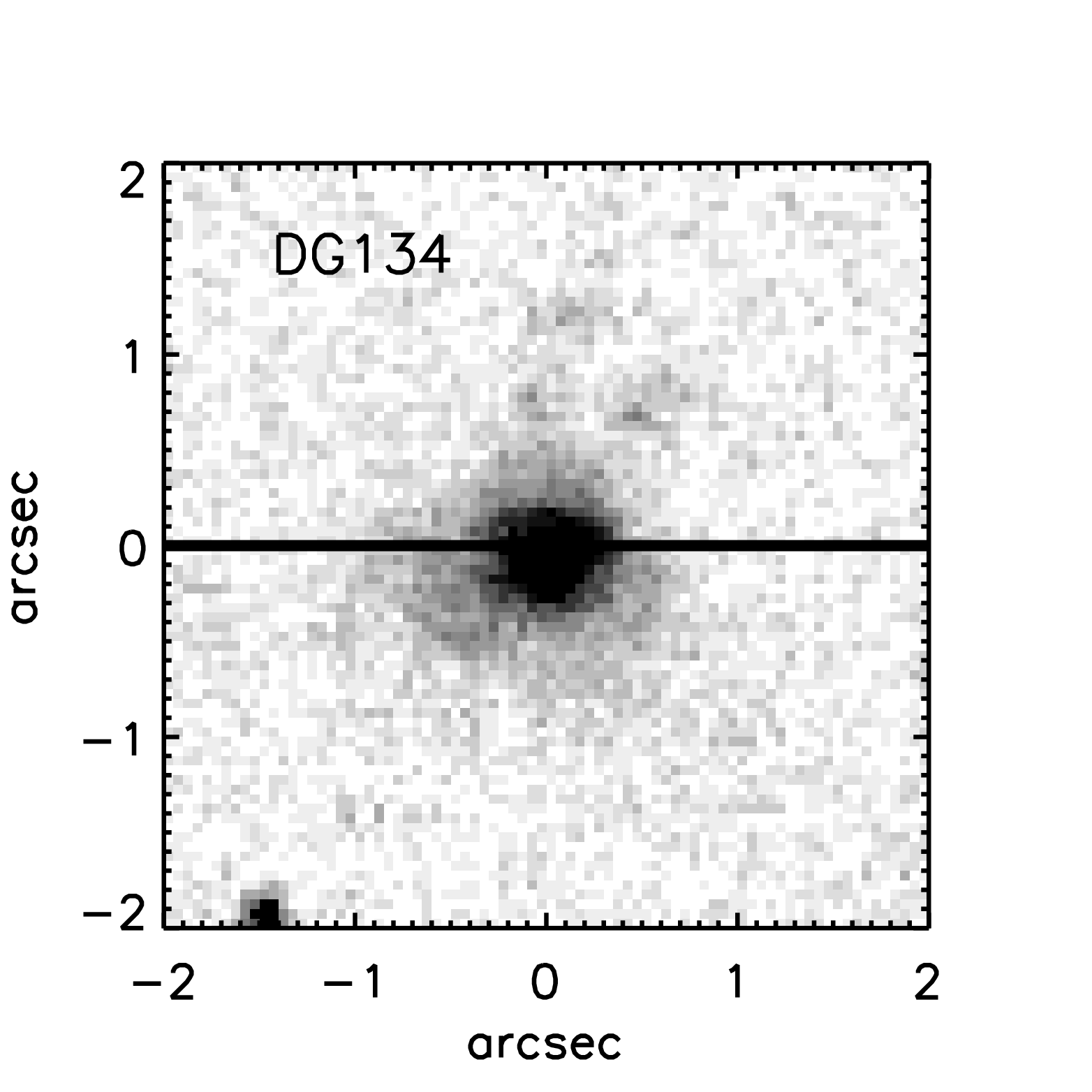}
\hspace{-0.8cm}
        \includegraphics[width=3.4cm, angle=90, trim=0 0 0 0]{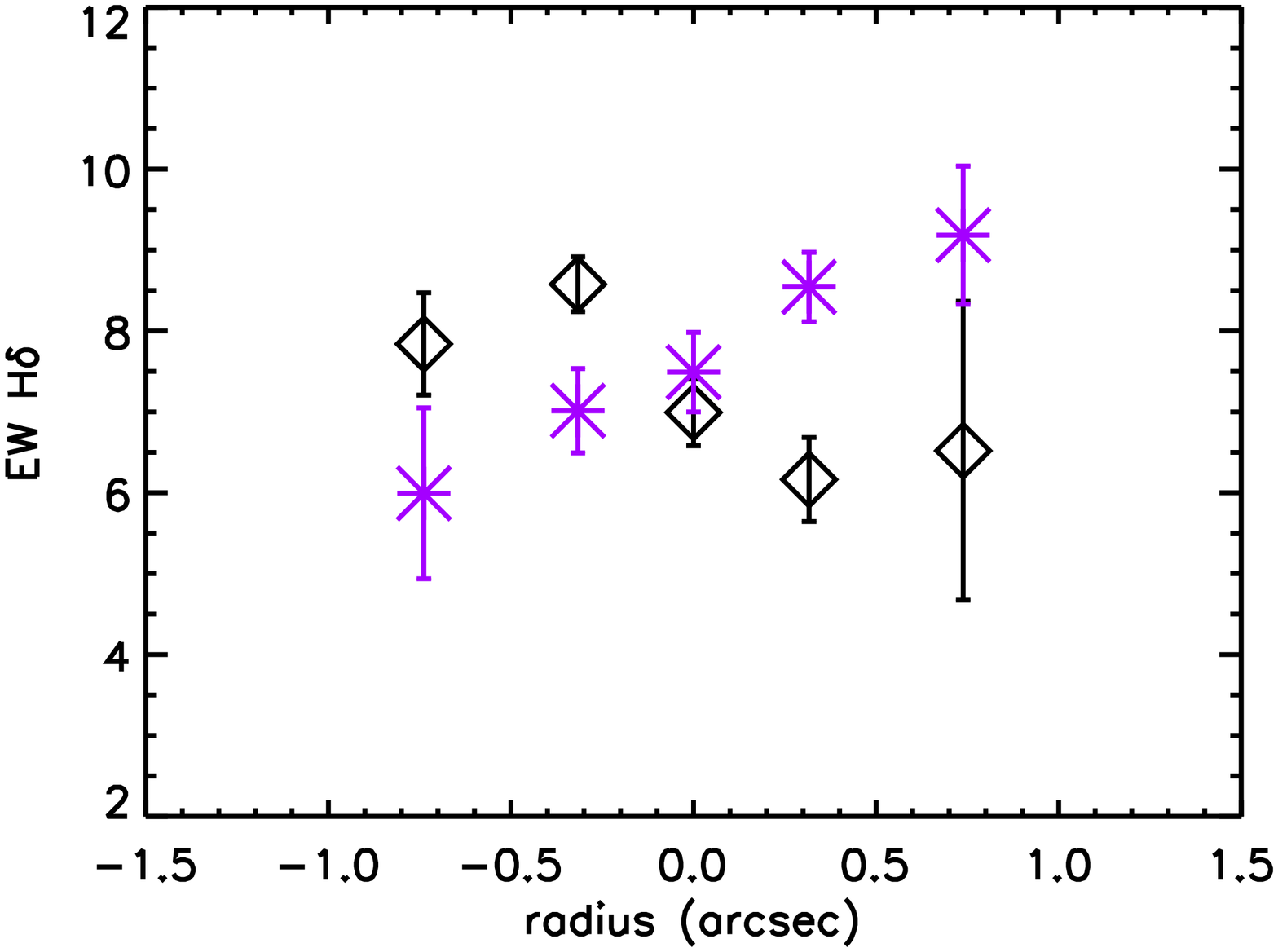}
\hspace{-0.6cm}
         \includegraphics[width=3.4cm, angle=90, trim=0 0 0 0]{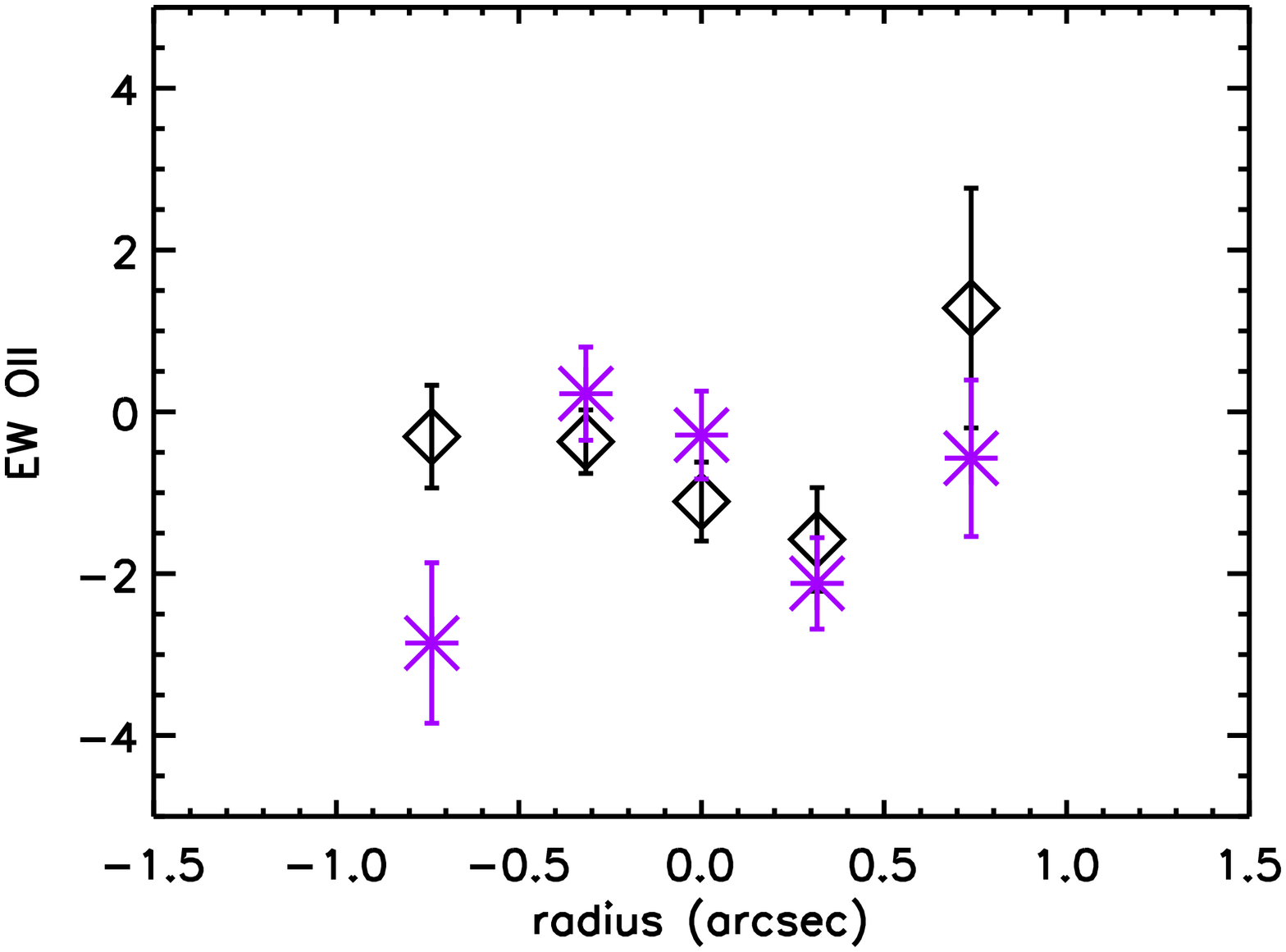}
\hspace{-0.6cm}
         \includegraphics[width=3.4cm, angle=90, trim=0 0 0 0]{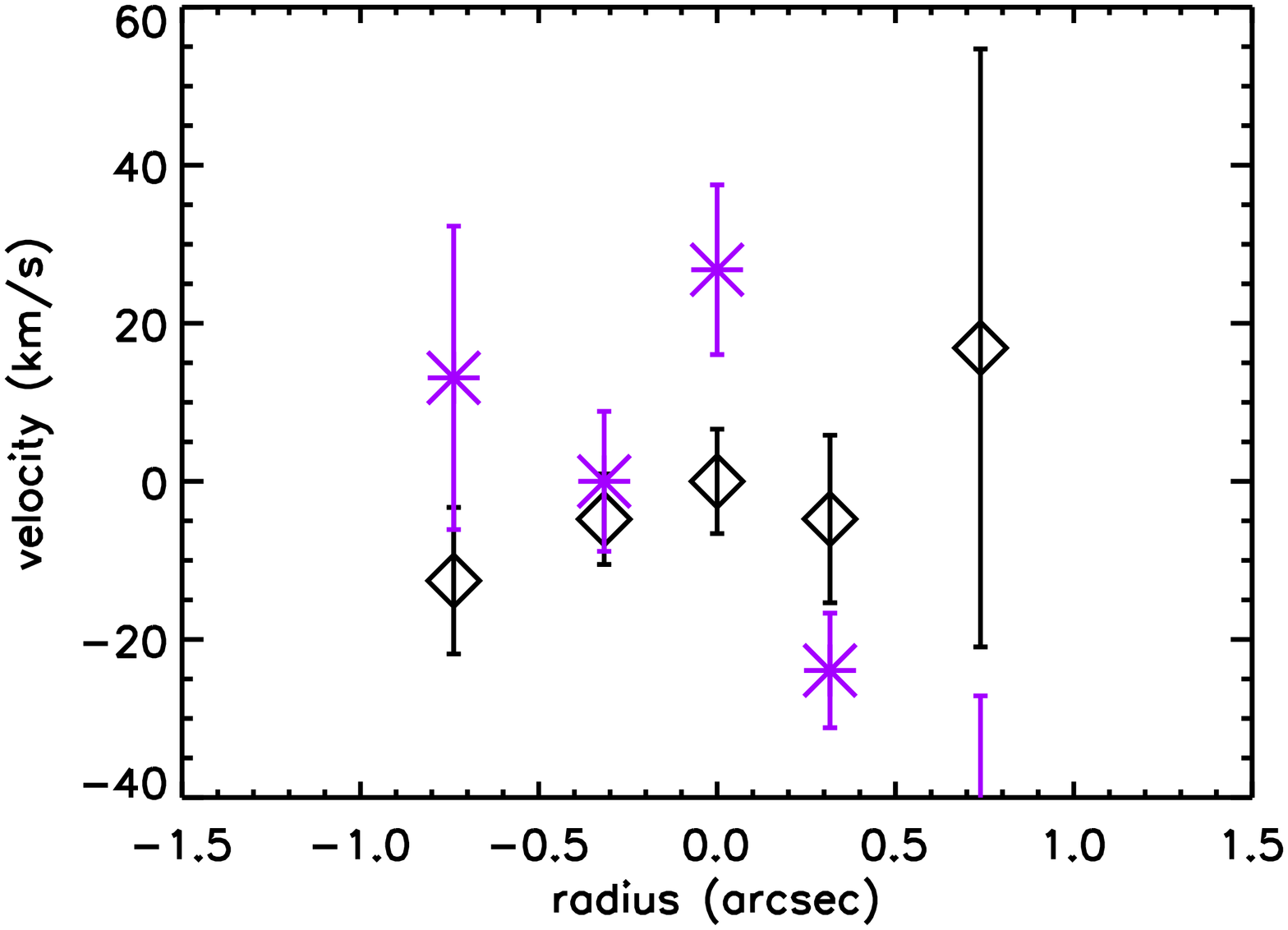}
      \end{minipage}
       \begin{minipage}{0.95\textwidth}
\hspace{-1.2cm}
         \includegraphics[width=3.6cm, angle=0, trim=0 0 0 0]{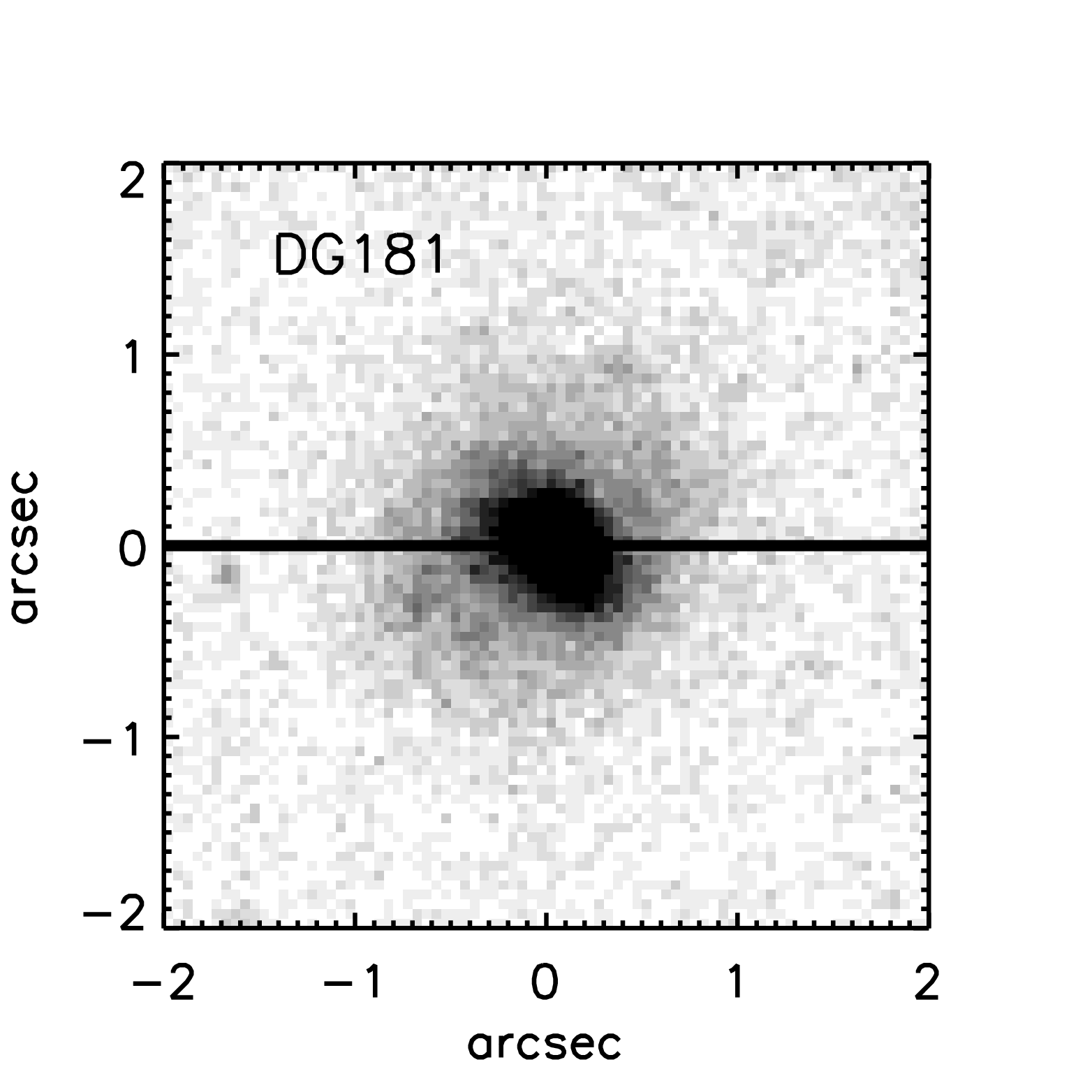}
\hspace{-0.8cm}
        \includegraphics[width=3.4cm, angle=90, trim=0 0 0 0]{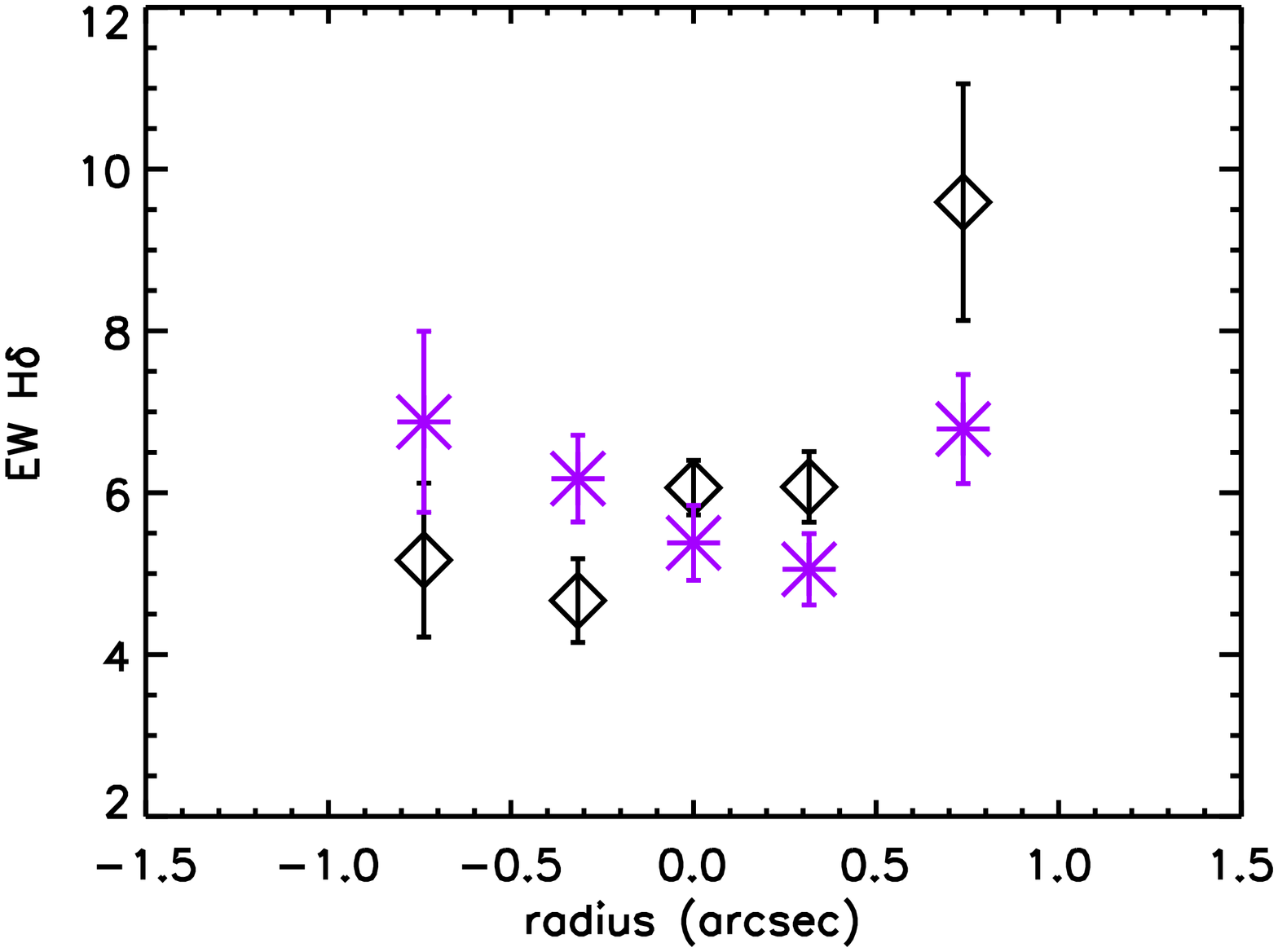}
\hspace{-0.6cm}
         \includegraphics[width=3.4cm, angle=90, trim=0 0 0 0]{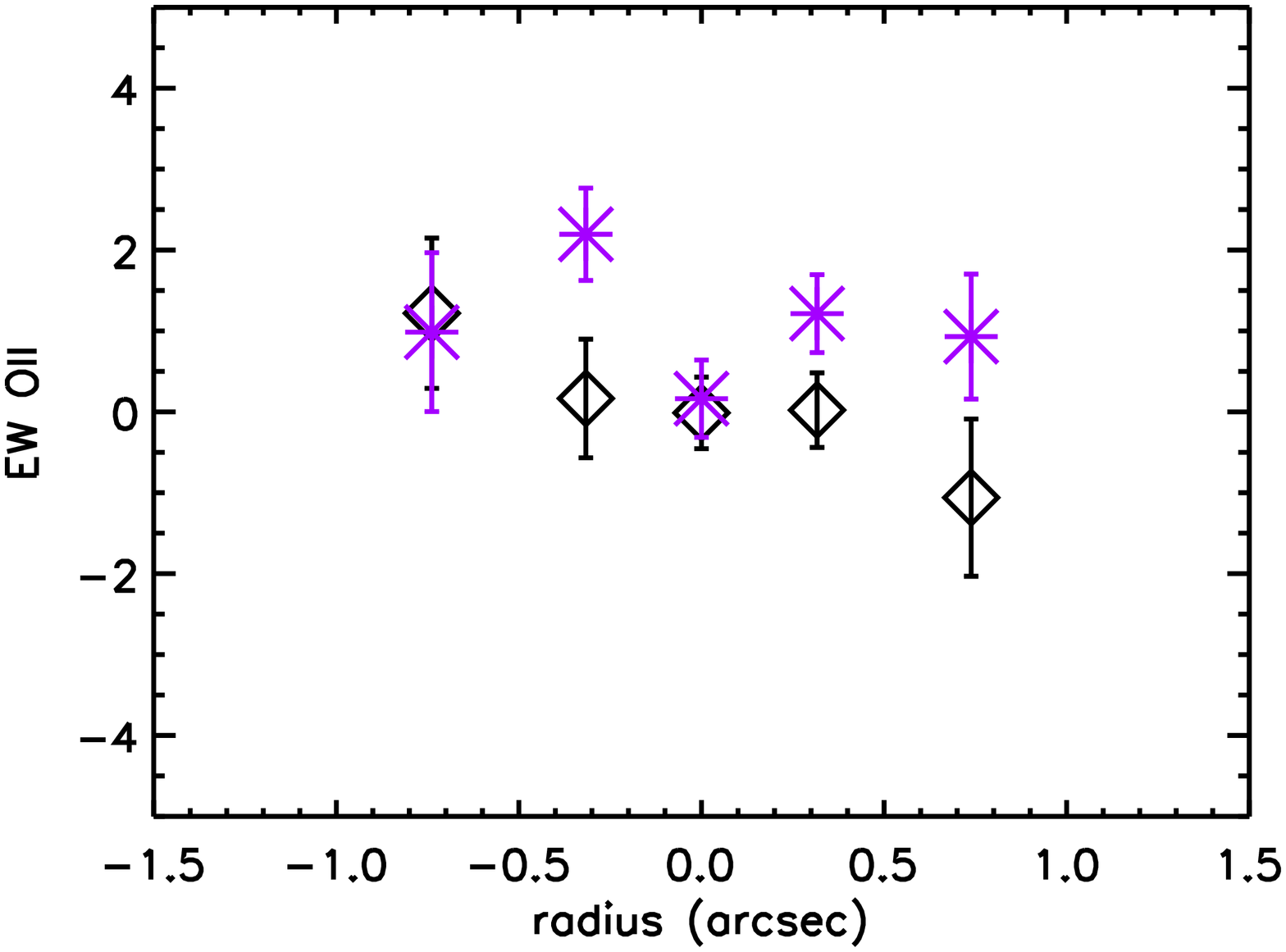}
\hspace{-0.6cm}
         \includegraphics[width=3.4cm, angle=90, trim=0 0 0 0]{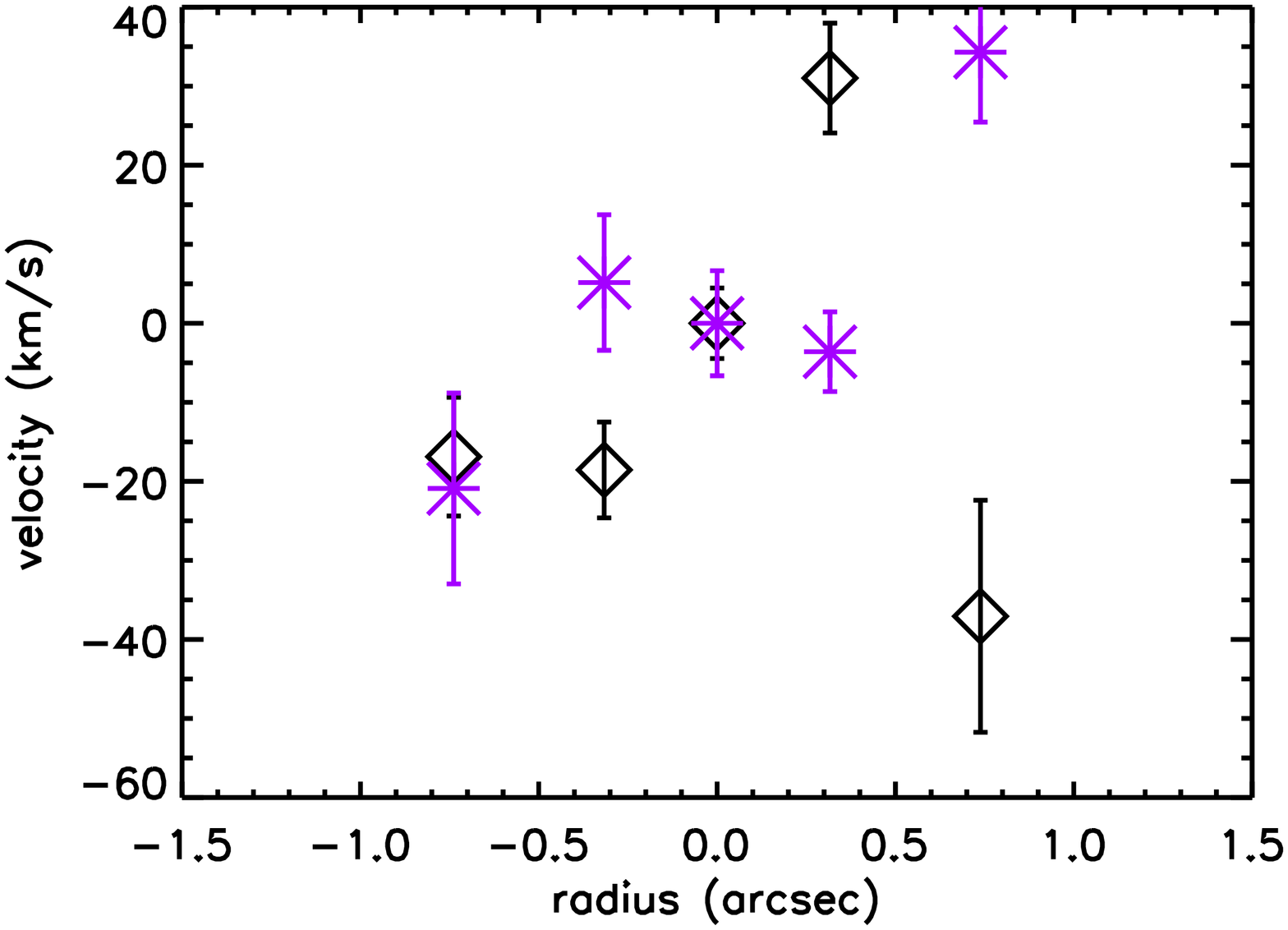}
      \end{minipage}
       \begin{minipage}{0.95\textwidth}
\hspace{-1.2cm}
         \includegraphics[width=3.6cm, angle=0, trim=0 0 0 0]{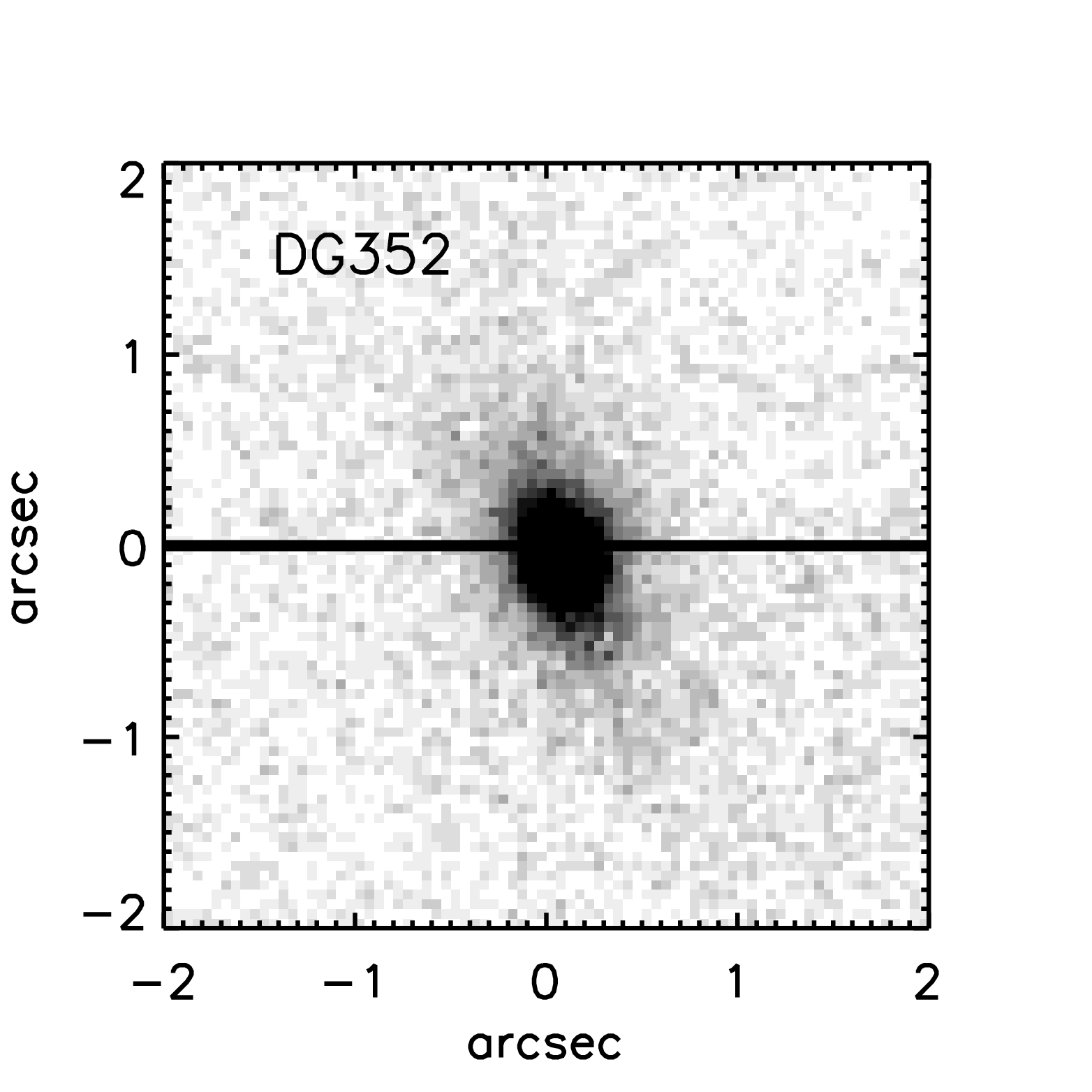}
\hspace{-0.8cm}
        \includegraphics[width=3.4cm, angle=90, trim=0 0 0 0]{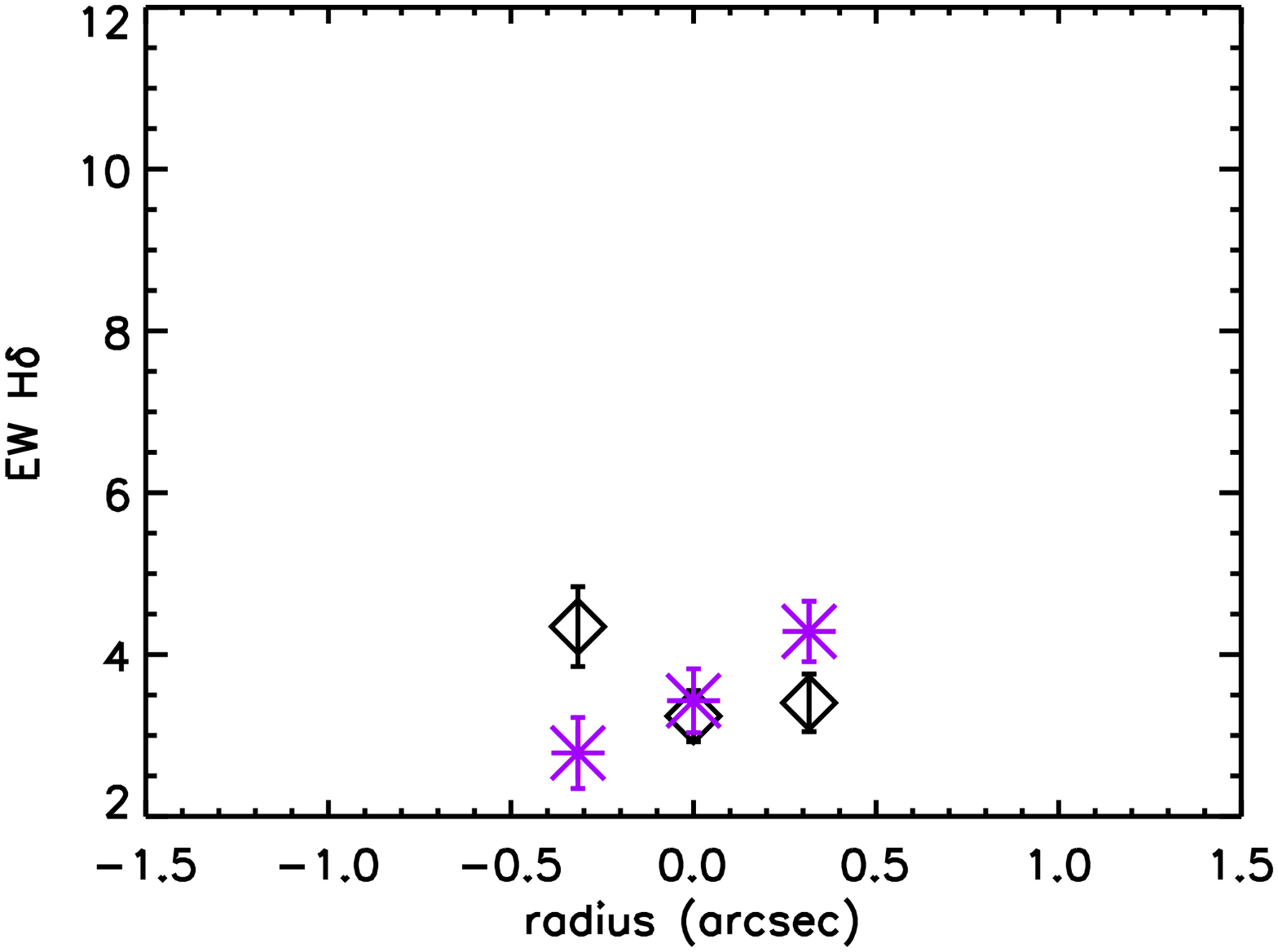}
\hspace{-0.6cm}
         \includegraphics[width=3.4cm, angle=90, trim=0 0 0 0]{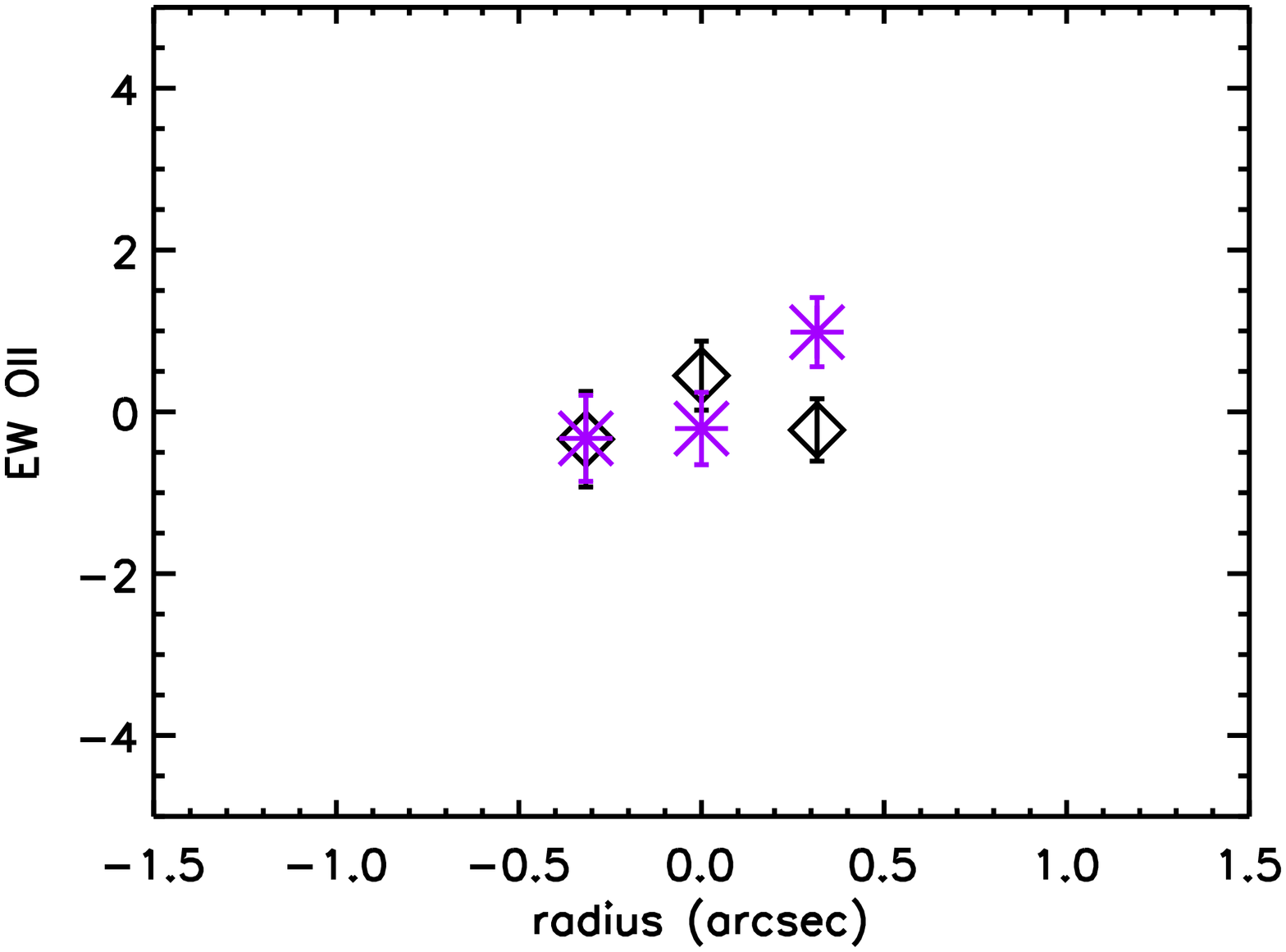}
\hspace{-0.6cm}
         \includegraphics[width=3.4cm, angle=90, trim=0 0 0 0]{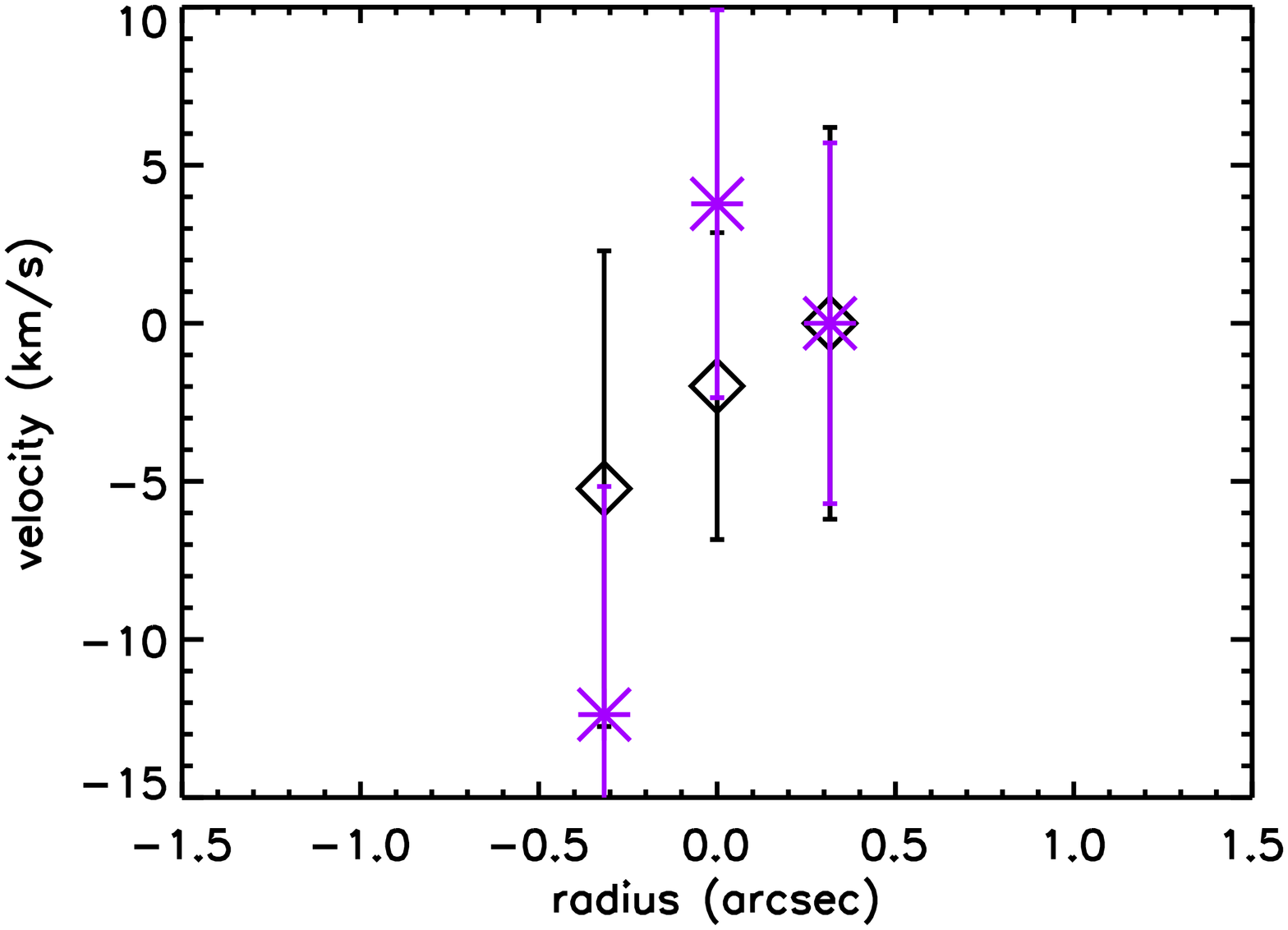}
      \end{minipage}
       \begin{minipage}{0.95\textwidth}
\hspace{-1.2cm}
         \includegraphics[width=3.6cm, angle=0, trim=0 0 0 0]{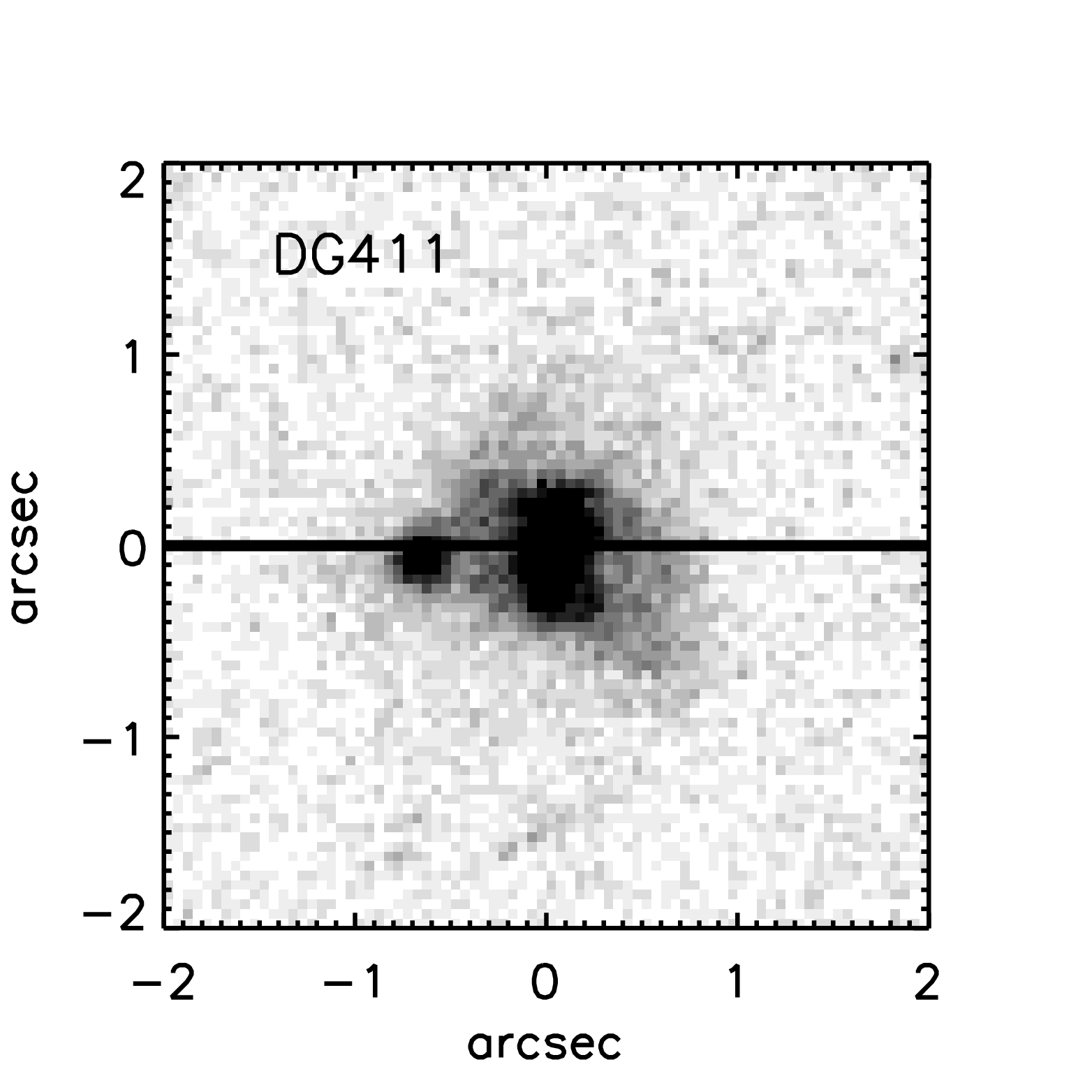}
\hspace{-0.8cm}
        \includegraphics[width=3.4cm, angle=90, trim=0 0 0 0]{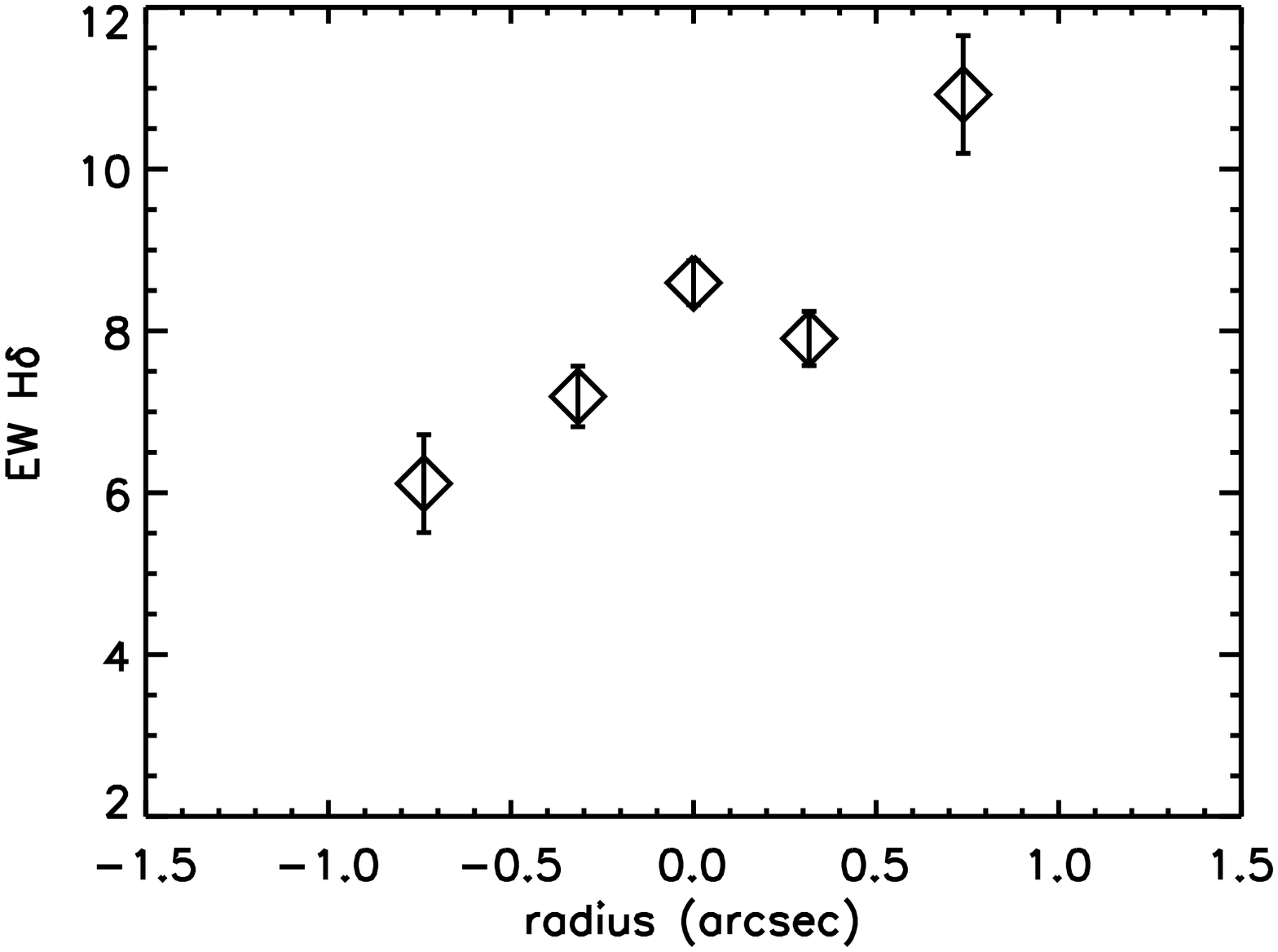}
\hspace{-0.6cm}
         \includegraphics[width=3.4cm, angle=90, trim=0 0 0 0]{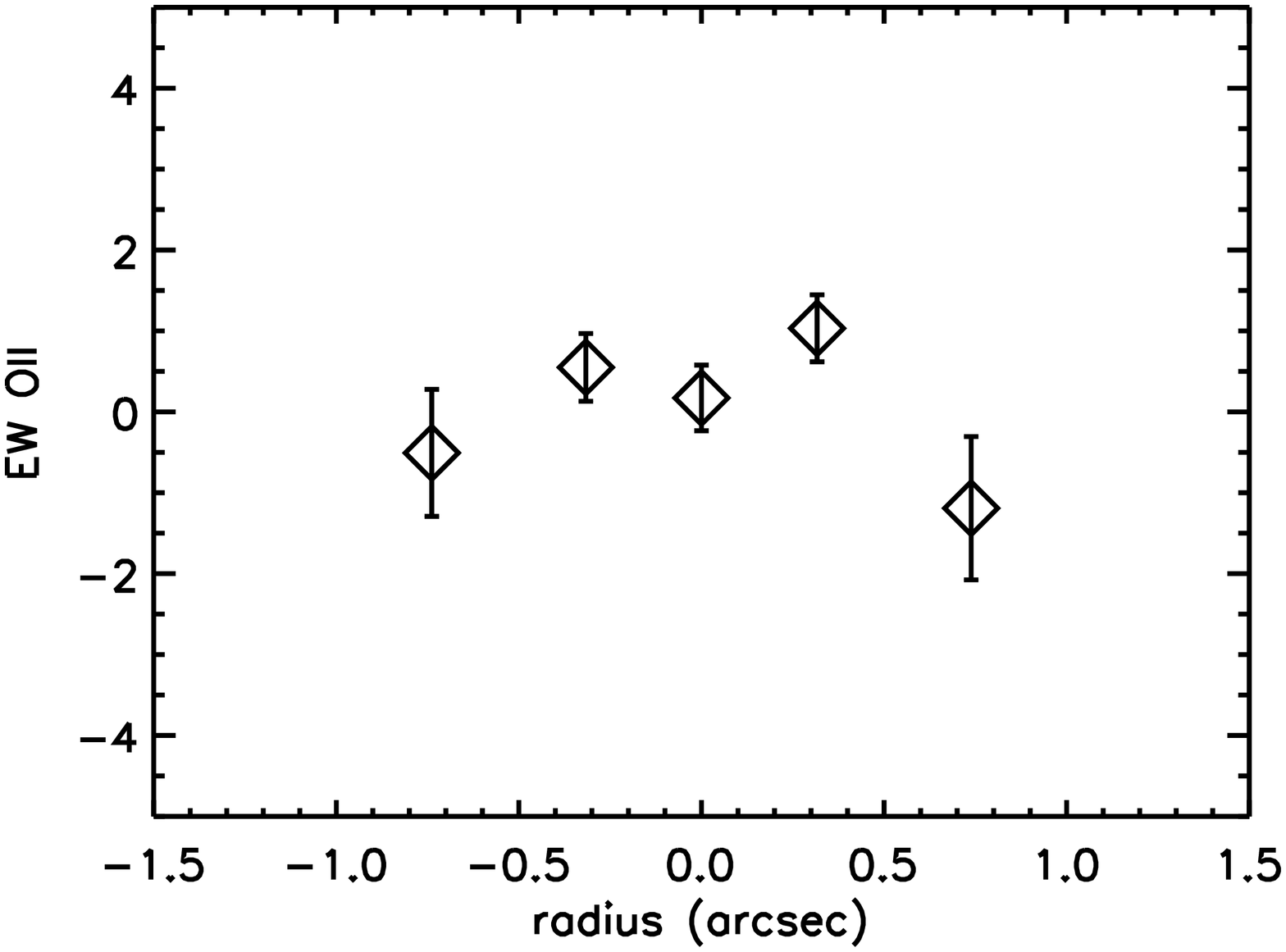}
\hspace{-0.6cm}
         \includegraphics[width=3.4cm, angle=90, trim=0 0 0 0]{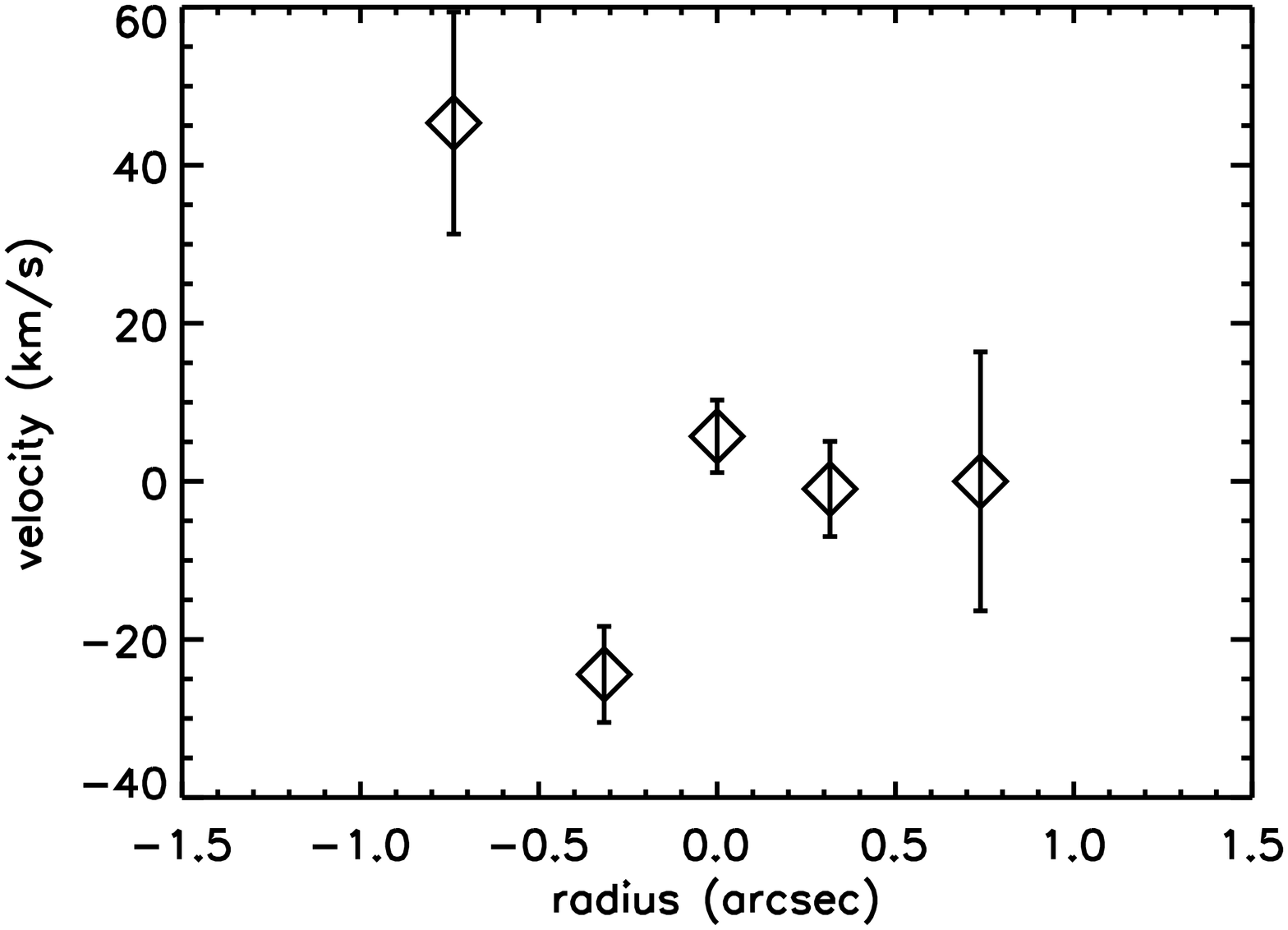}
      \end{minipage}
\end{center}
\caption{The post--starburst galaxies. From {\it left to right}: HST postage stamp image taken 
with the F775W filter and the angle of the slit superimposed; radial EW[H$\delta_{\rm A}$] profile; 
radial equivalent width [OII]$\lambda 3727$ profile; and projected streaming velocity profile along the slit. Mask 1 observations
are plotted as {\it black} symbols and Mask 2 observations are  plotted as {\it blue} symbols. At the redshift of the cluster
1\,arcsecond corresponds to $\sim 6.4$\,kpc.}
\label{fig:psb}
\end{figure*}

In the columns 2--4 of Figure \ref{fig:psb} we show the 
radial profiles along the slit of the H$\delta$ equivalent width, the [OII]$\lambda 3727$ 
equivalent width, and the streaming velocity for our sample of post-starburst
galaxies, respectively. Measurements made from the Mask~1 
observations are plotted as {\it black symbols} and measurements made from the Mask~2 observations 
are plotted as {\it blue symbols}. Several targets were observed in both masks, which allows a 
check on the repeatability of our measurements. The Mask~1 observations were taken in better 
(0.4\,arcsec) seeing, resulting in better delivered spatial resolution. Seeing has the effect of smearing out any 
radial gradients as the data points become more correlated as the image quality worsens. 
In principle this means it may be possible to detect radial trends in Mask~1 data that 
cannot be seen in Mask~2.  

The extracted radial data only extend over a few spatial resolution elements. This combined
with the moderate signal--to--noise of the line index measurements makes it 
difficult to detect weak radial variations (see Section 4 for a more detailed discussion). 
The equivalent width profiles show no evidence of any central enhancement (or deficit) or 
any radial gradients (either positive or negative), with the H$\delta$ equivalent width 
being uniformly strong across all the galaxies and there being no sign of [OII]$\lambda 3727$ emission 
at any radii. The exception is DG\_411 which shows evidence for a gradient in H$\delta$ 
equivalent width from one side of the galaxy to the other. This galaxy also has a nearby 
companion evident in the HST image. There is some evidence for rotation in two of the 
sample galaxies (DG\_115 and DG\_181); however, given the sometimes non-optimal 
orientations of the slit and mostly face--on nature of the sample, we would not expect 
to see significant rotation in all cases. 

\subsubsection{Morphological properties}
\begin{figure*}
   \begin{center}
\vspace{-1cm}
     \begin{minipage}{0.95\textwidth}
\hspace{-1.3cm}
         \includegraphics[width=4.0cm, angle=0]{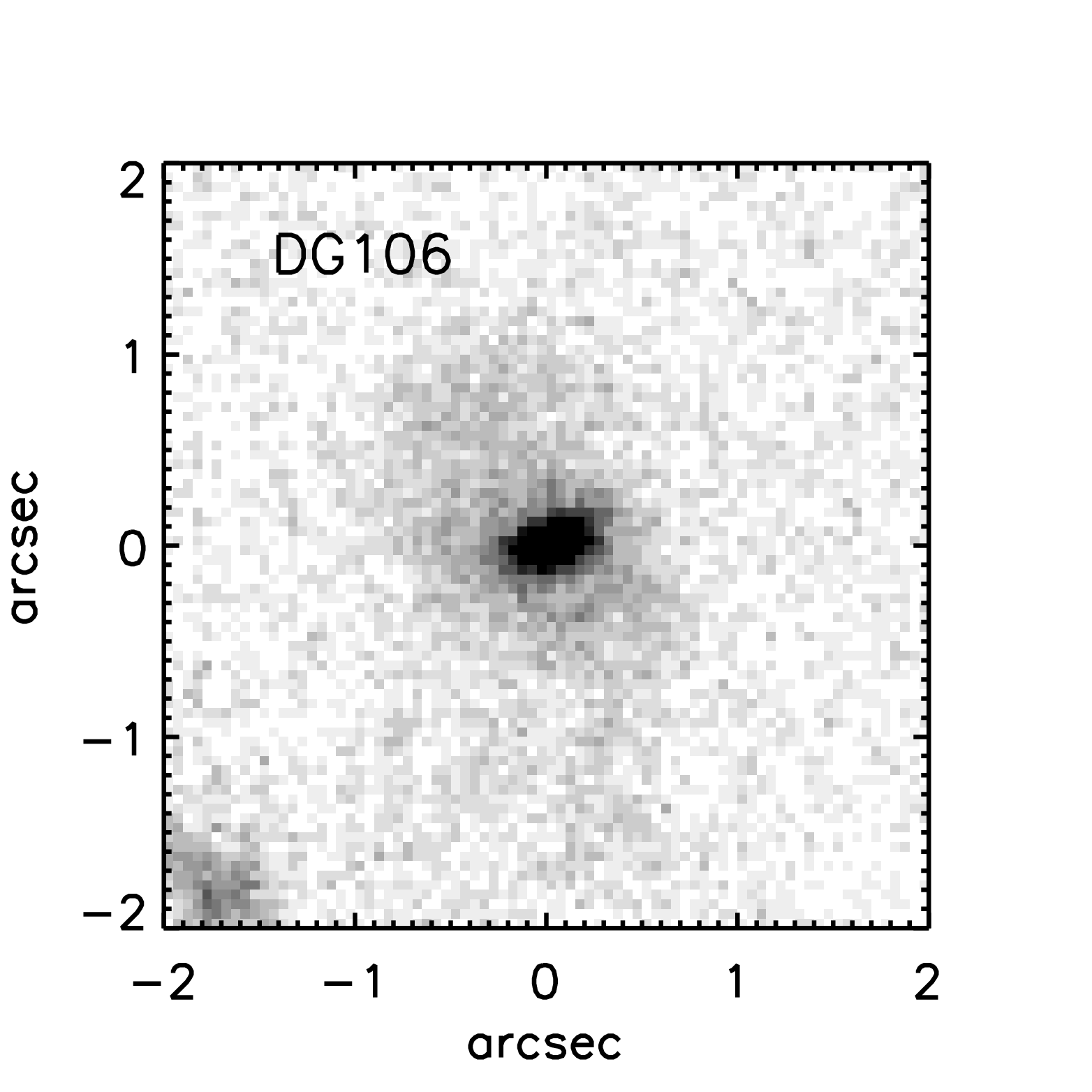}\hspace{-0.7cm}
         \includegraphics[width=4.0cm, angle=0]{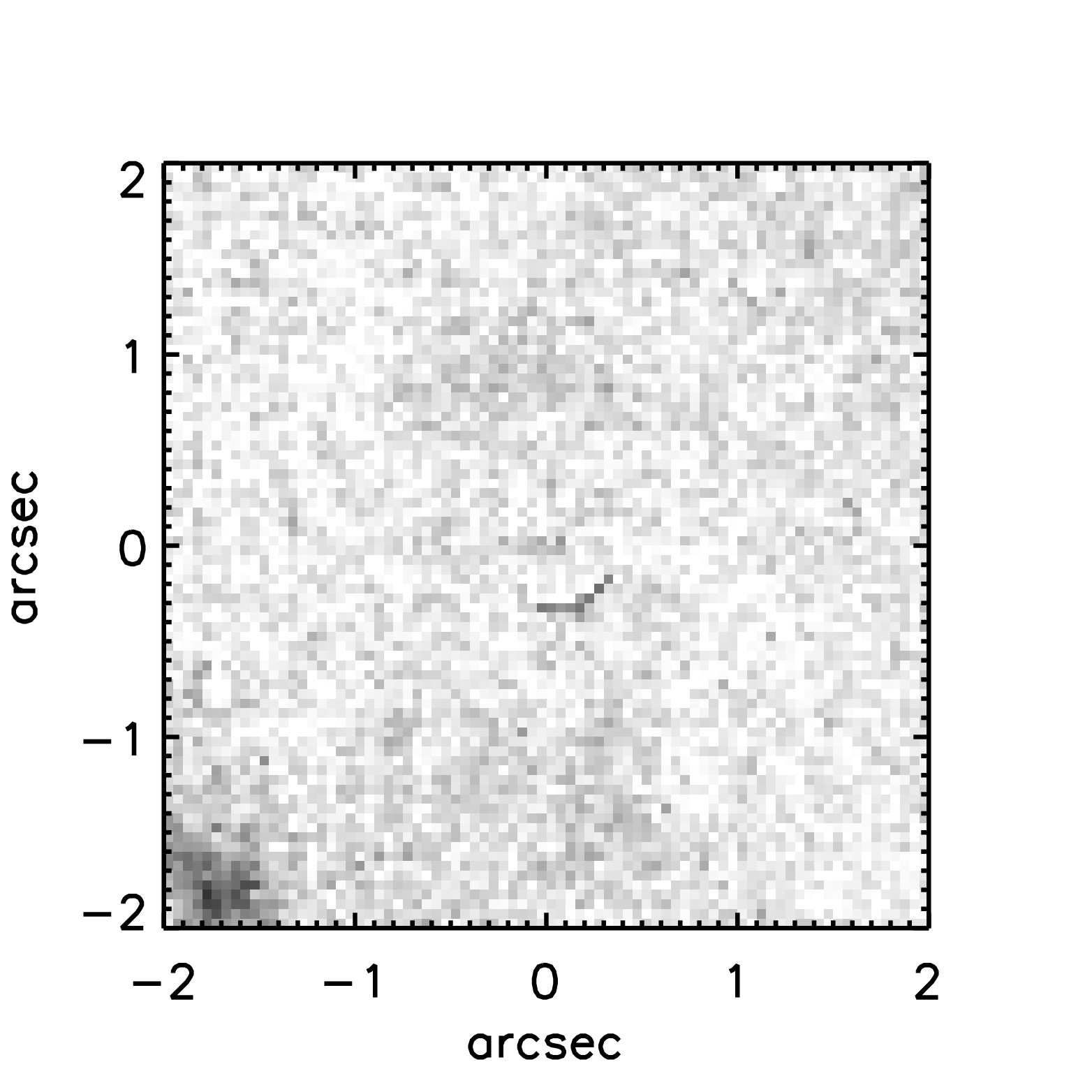}\hspace{-0.7cm}
         \includegraphics[width=3.6cm, angle=90]{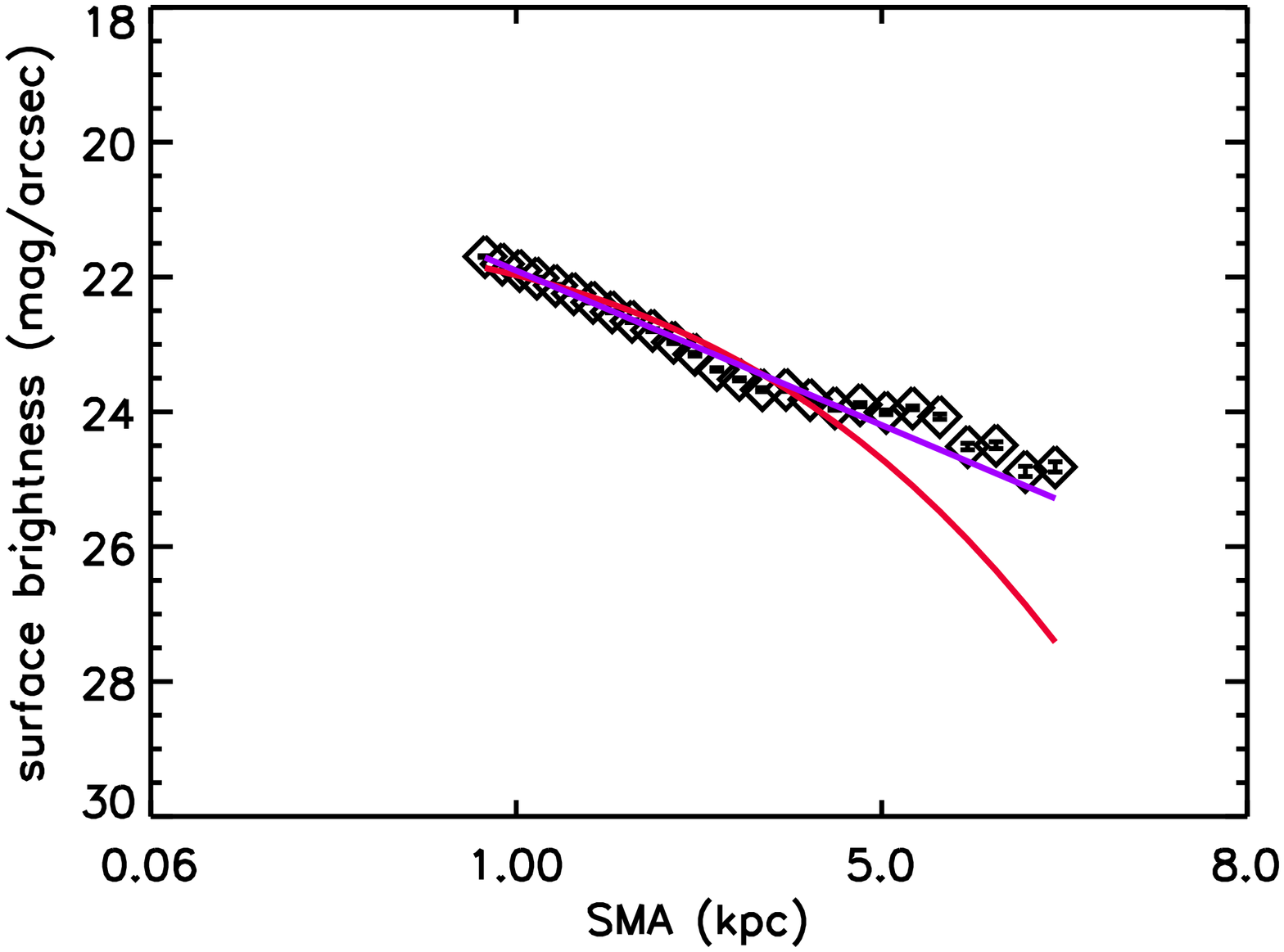}\hspace{-0.7cm}
         \includegraphics[width=5.4cm, angle=0]{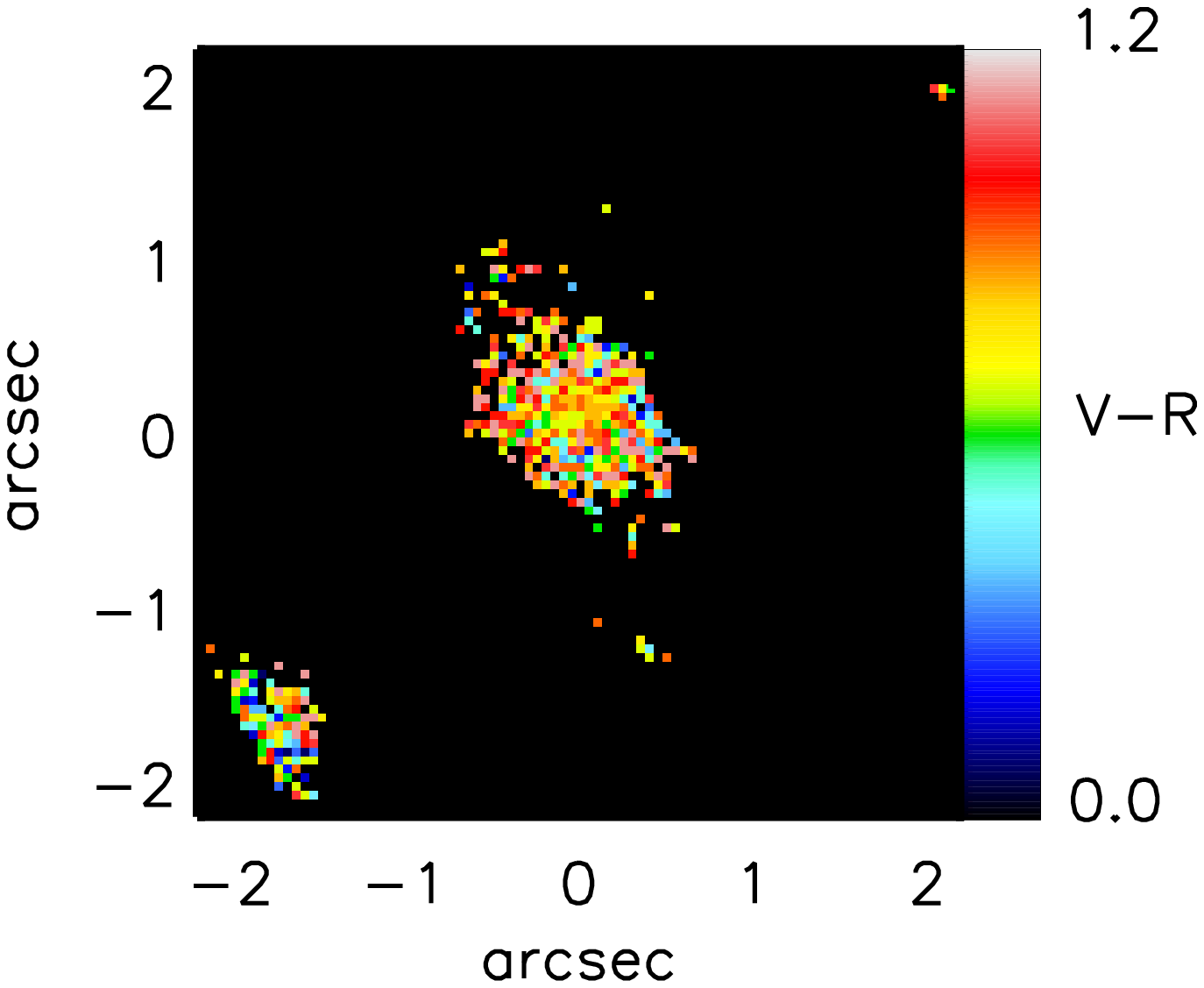}
      \end{minipage}
       \begin{minipage}{0.95\textwidth}
\hspace{-1.3cm}
         \includegraphics[width=4.0cm, angle=0, trim=0 0 0 0]{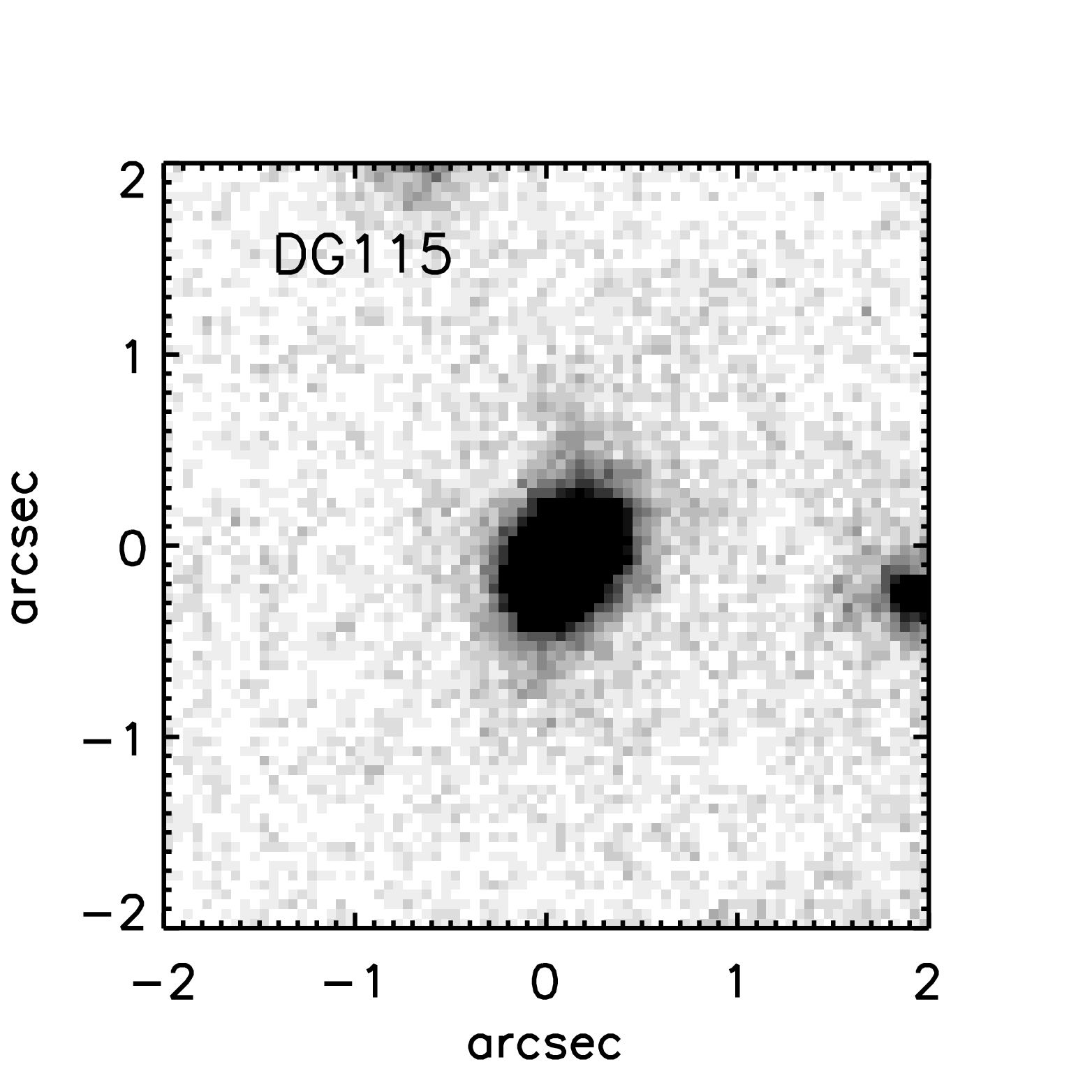}\hspace{-0.7cm}
         \includegraphics[width=4.0cm, angle=0, trim=0 0 0 0]{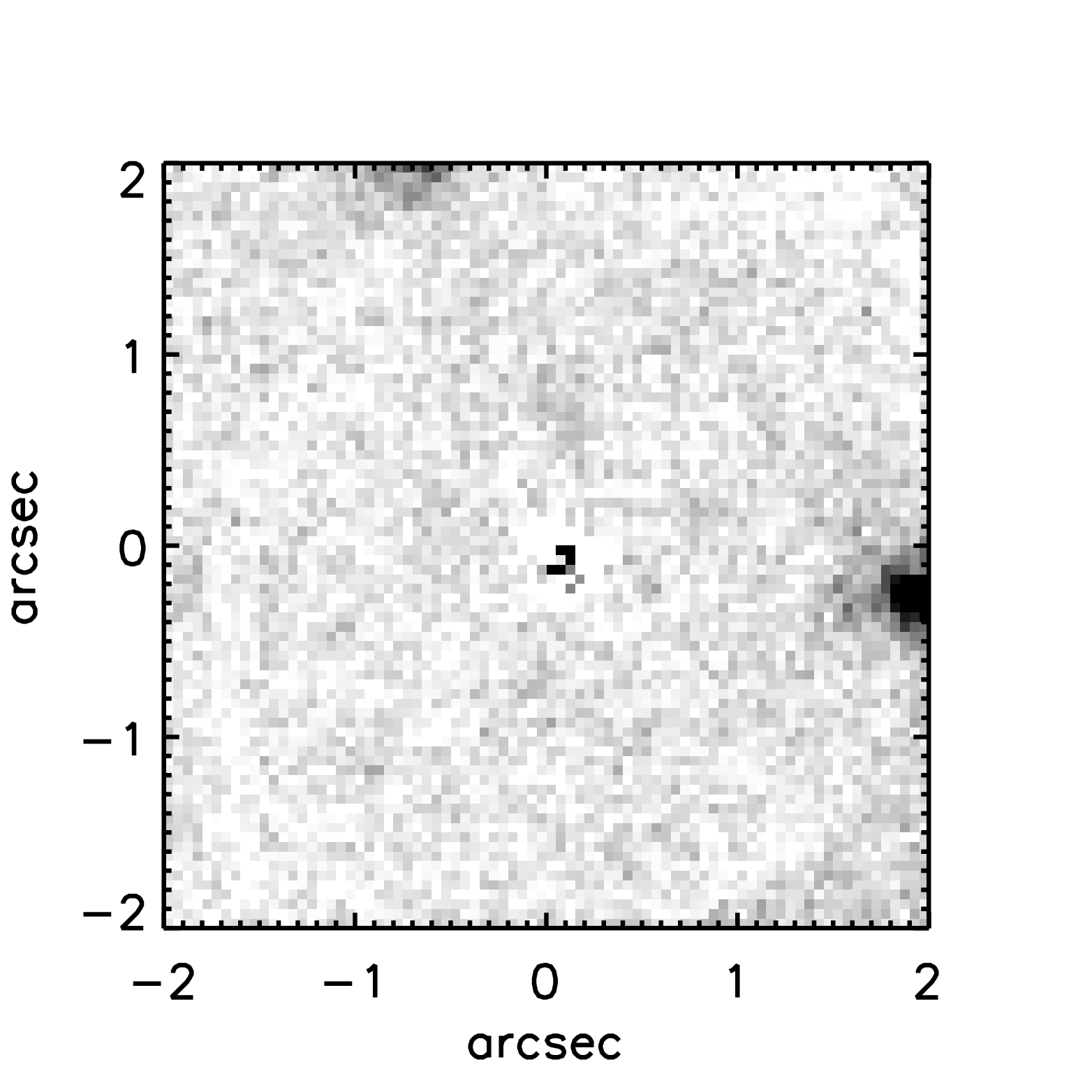}\hspace{-0.7cm}
         \includegraphics[width=3.6cm, angle=90, trim=0 0 0 0]{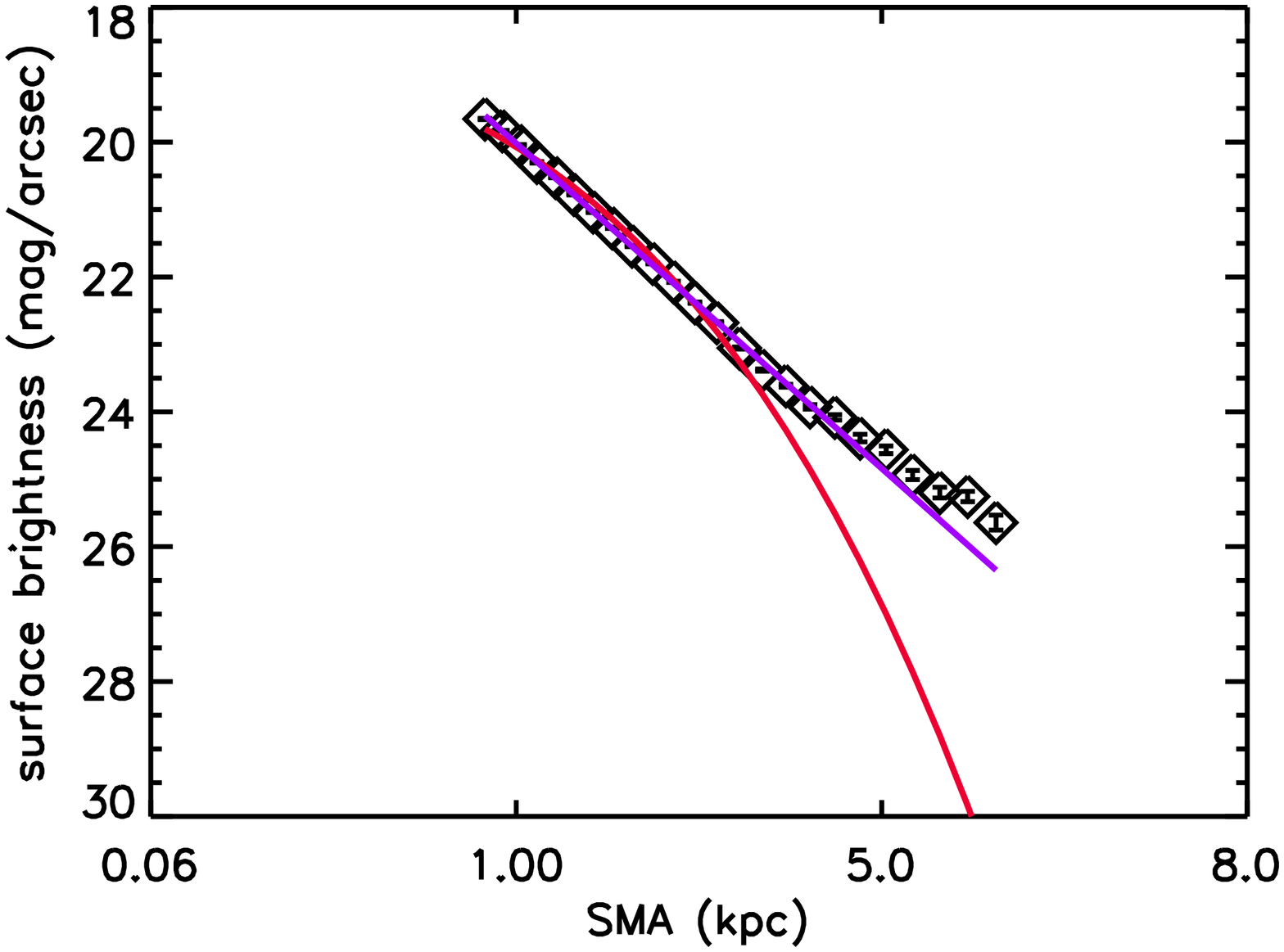}\hspace{-0.7cm}
         \includegraphics[width=5.4cm, angle=0, trim=0 0 0 0]{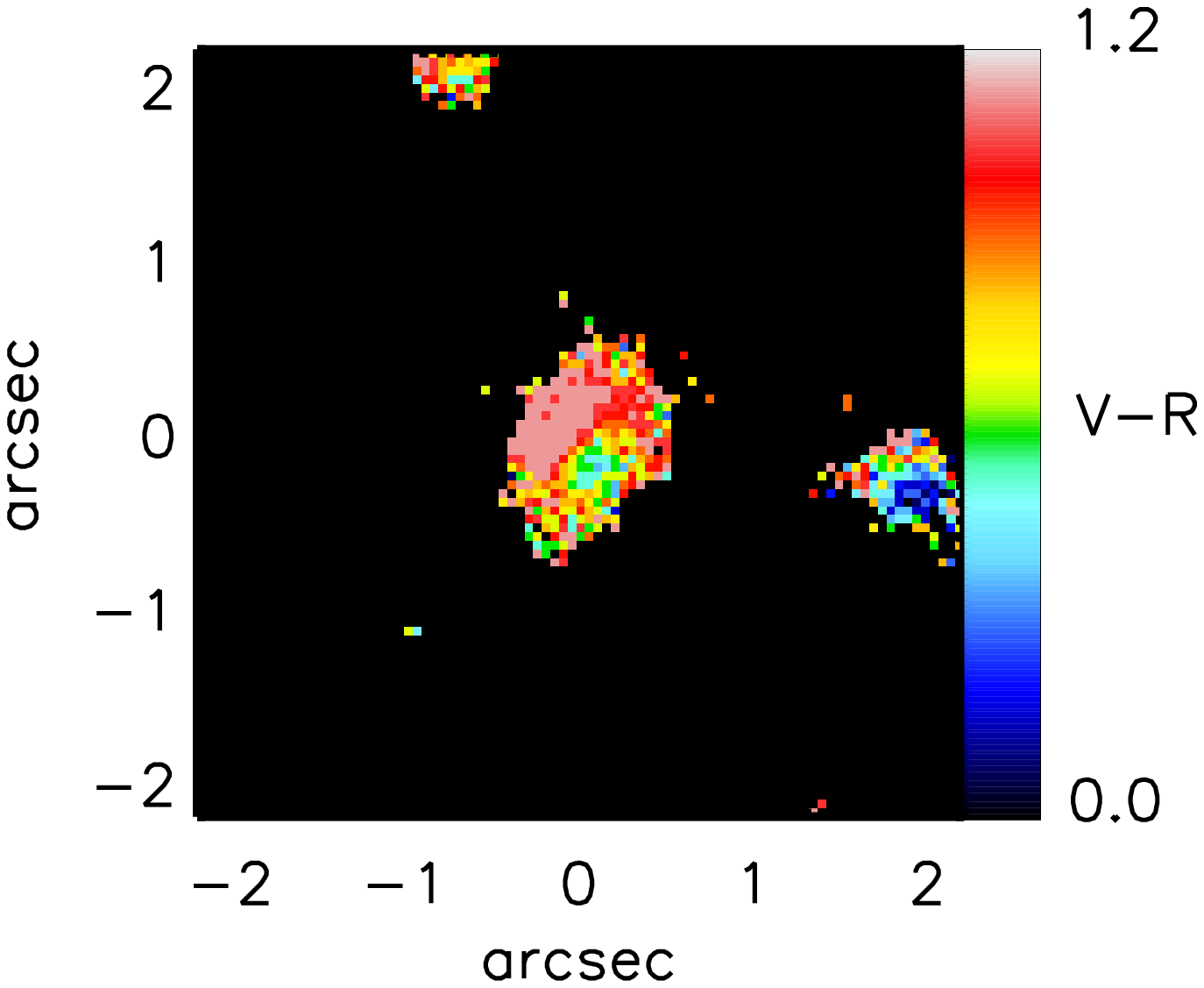}\hspace{-0.7cm}
       \end{minipage}
       \begin{minipage}{0.95\textwidth}
\hspace{-1.3cm}
        \includegraphics[width=4.0cm, angle=0, trim=0 0 0 0]{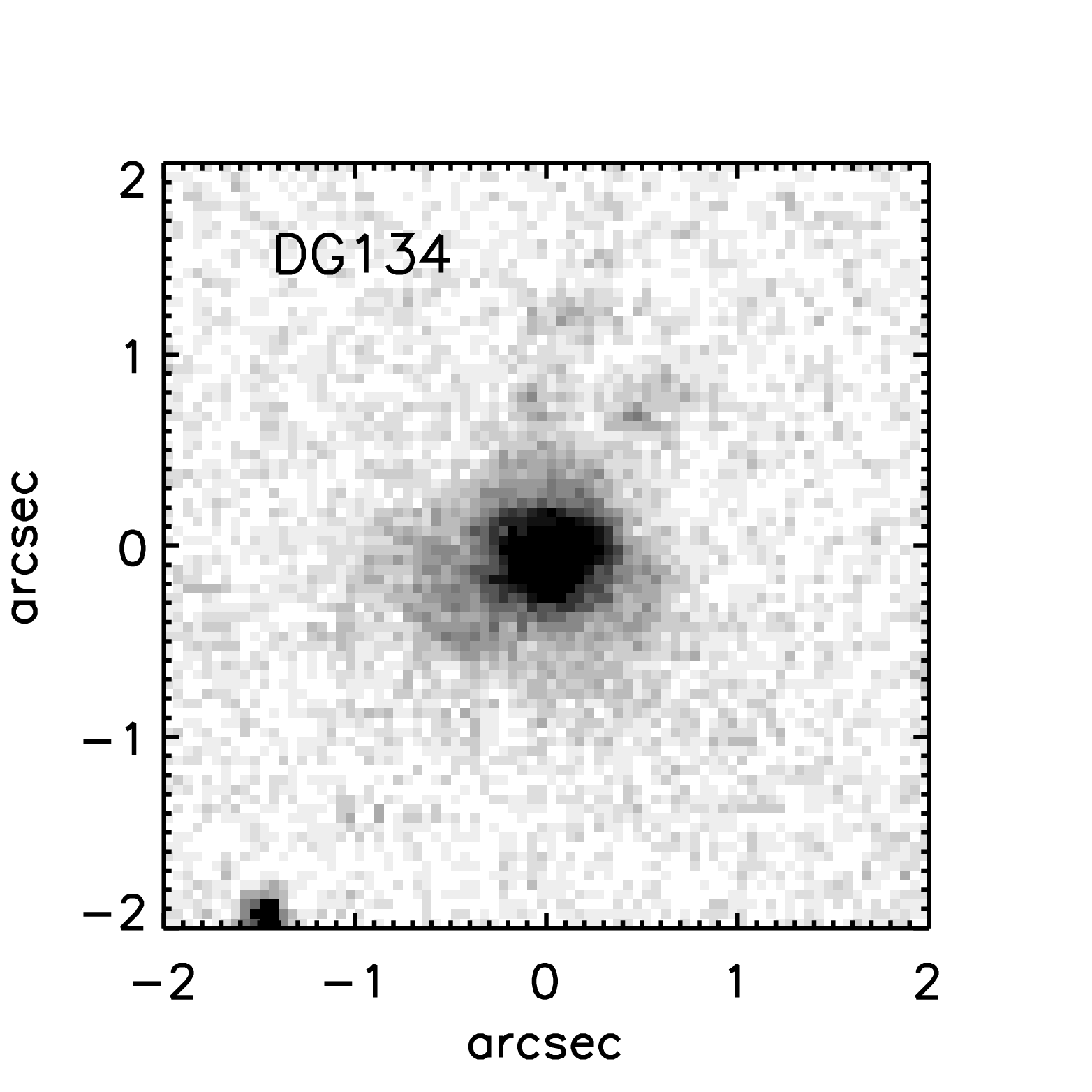}\hspace{-0.7cm}
        \includegraphics[width=4.0cm, angle=0, trim=0 0 0 0]{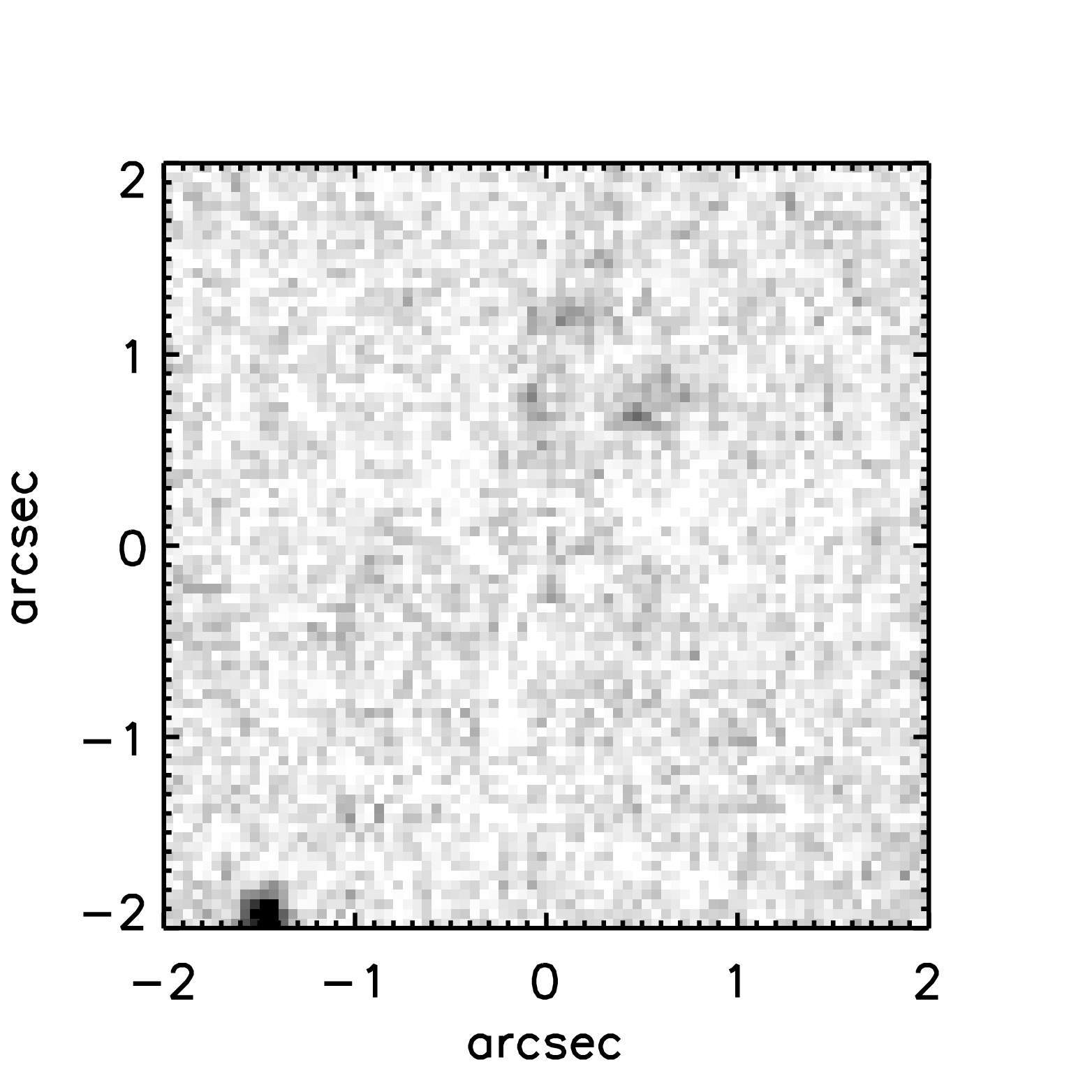}\hspace{-0.7cm}
         \includegraphics[width=3.6cm, angle=90, trim=0 0 0 0]{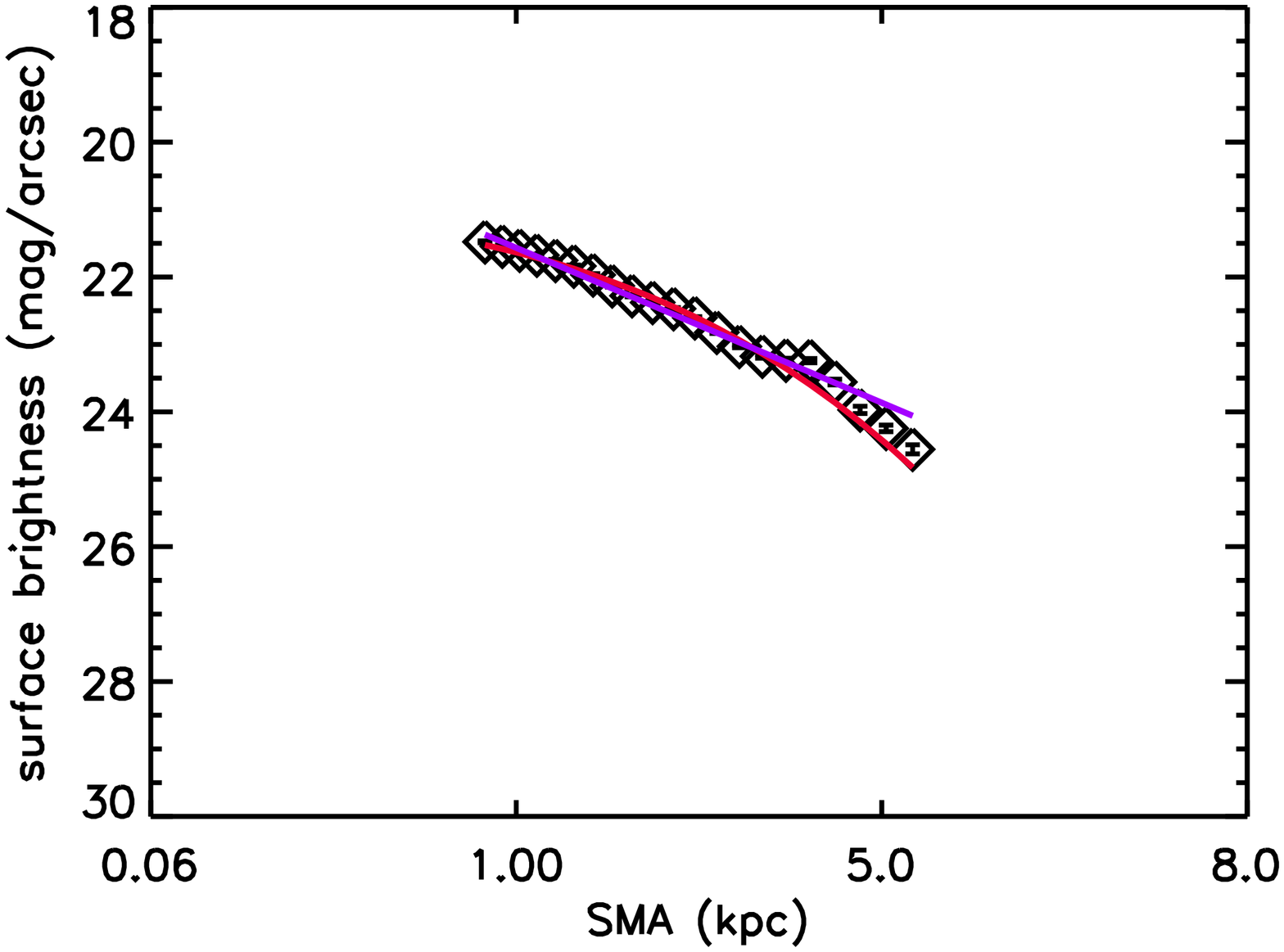}\hspace{-0.7cm}
        \includegraphics[width=5.4cm, angle=0, trim=0 0 0 0]{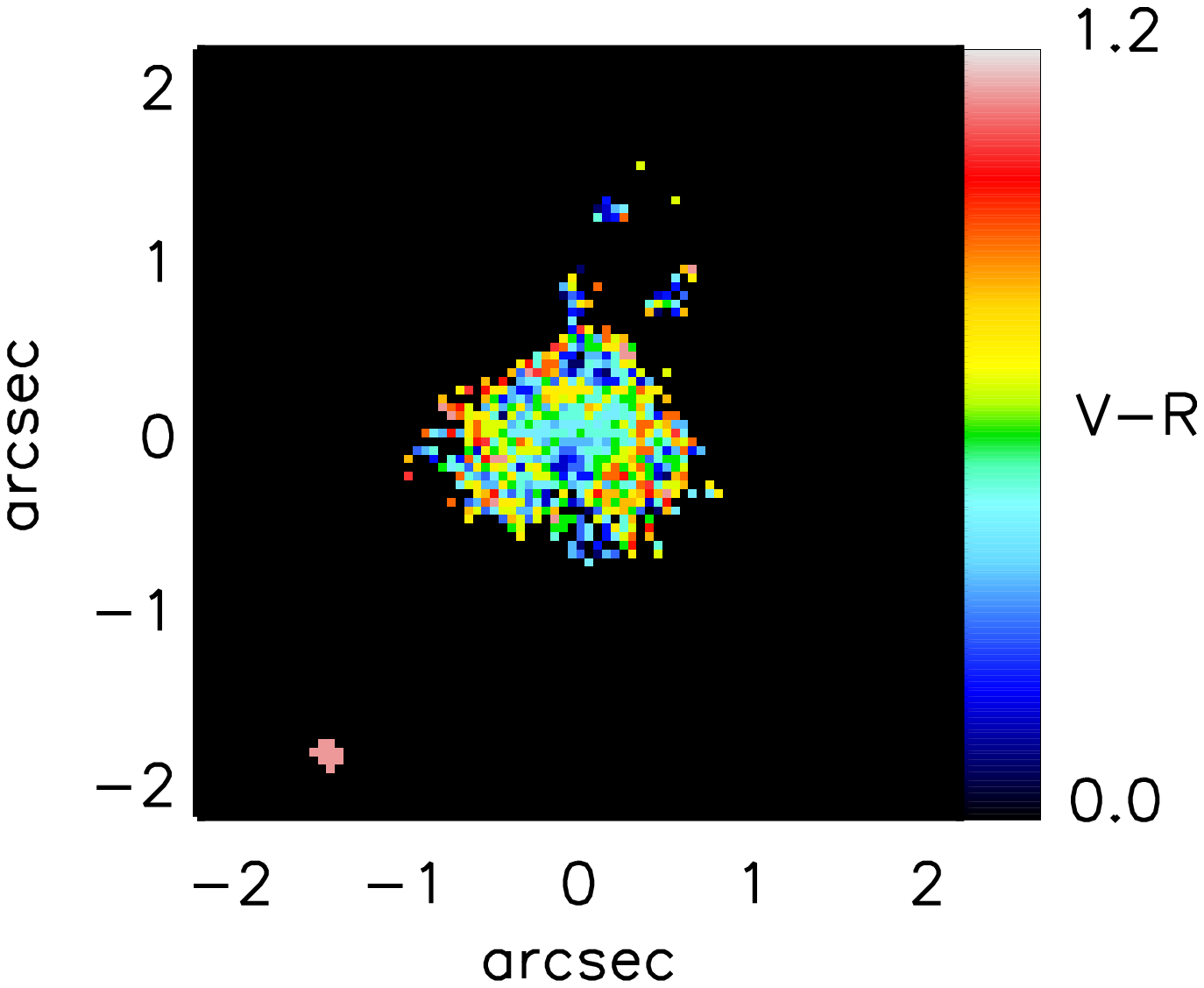}\hspace{-0.7cm}
      \end{minipage}
       \begin{minipage}{0.95\textwidth}
\hspace{-1.3cm}
         \includegraphics[width=4.0cm, angle=0, trim=0 0 0 0]{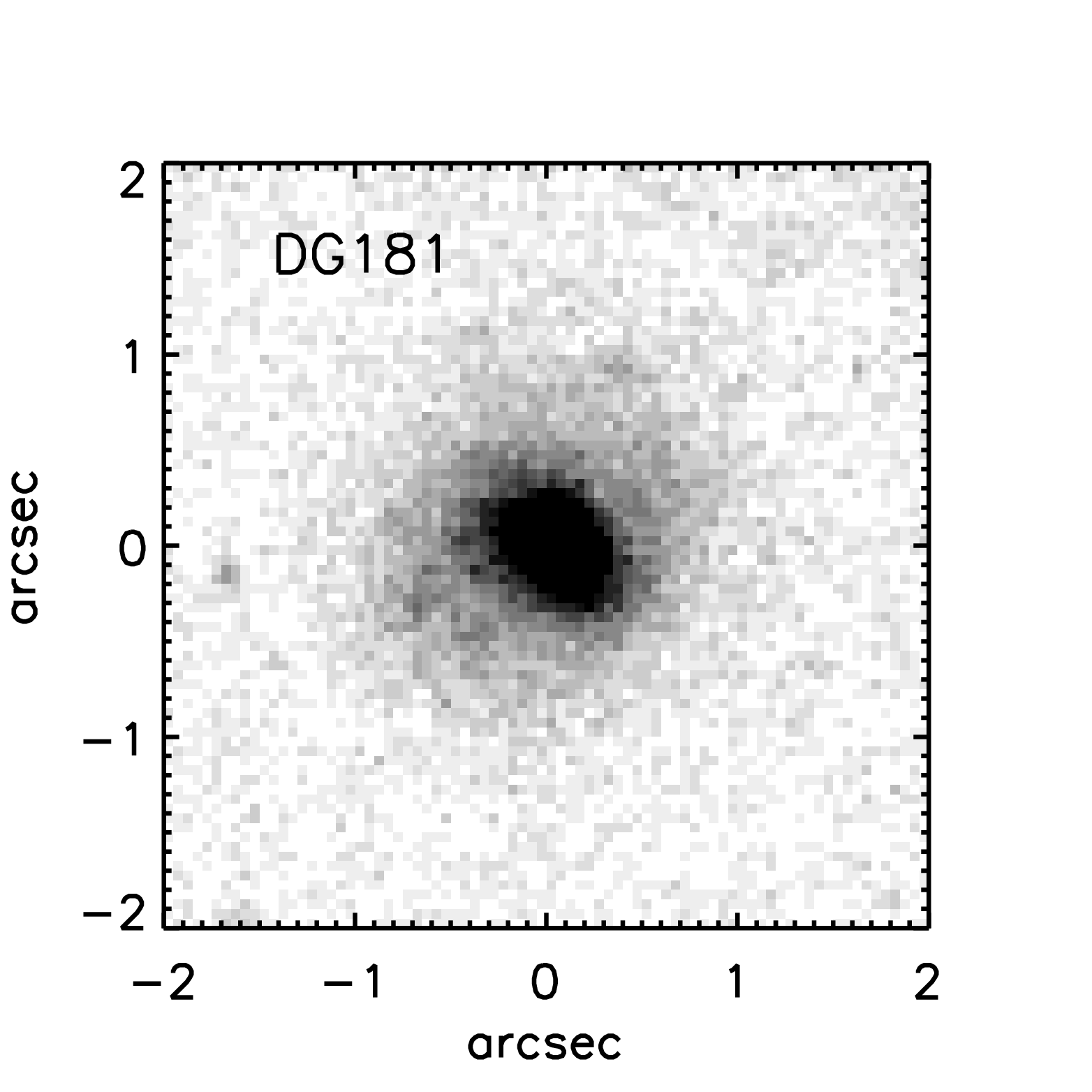}\hspace{-0.7cm}
         \includegraphics[width=4.0cm, angle=0, trim=0 0 0 0]{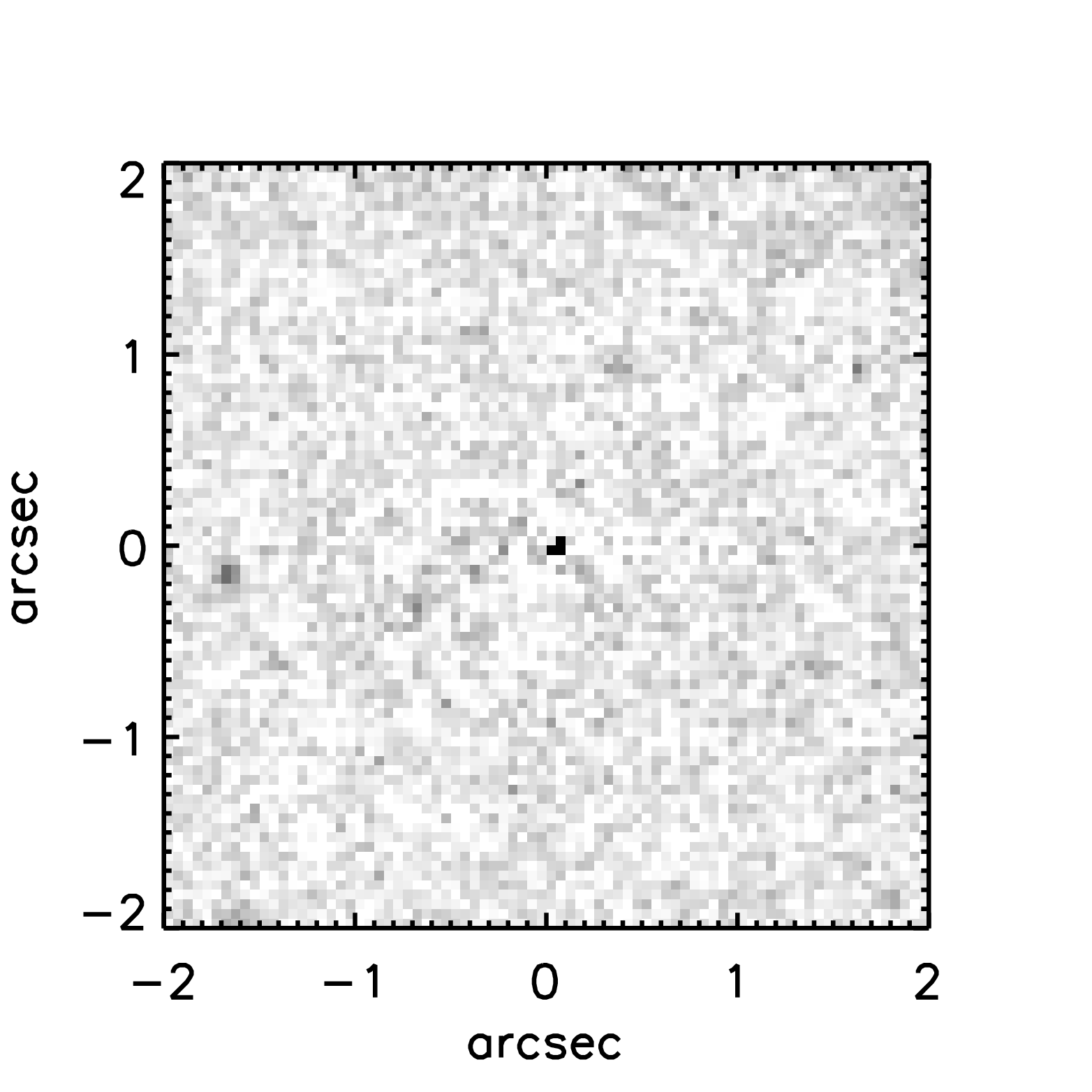}\hspace{-0.7cm}
         \includegraphics[width=3.6cm, angle=90, trim=0 0 0 0]{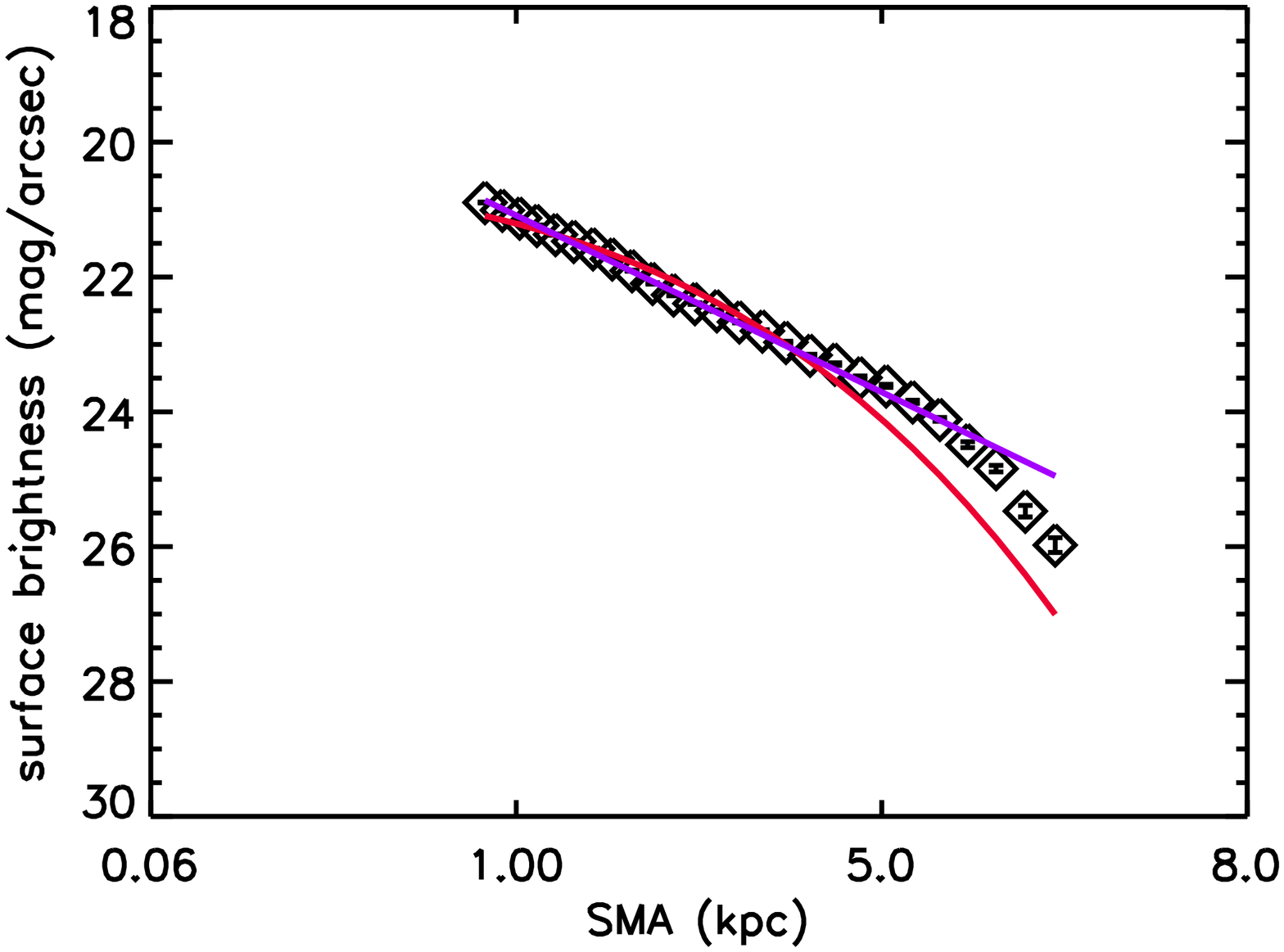}\hspace{-0.7cm}
         \includegraphics[width=5.4cm, angle=0, trim=0 0 0 0]{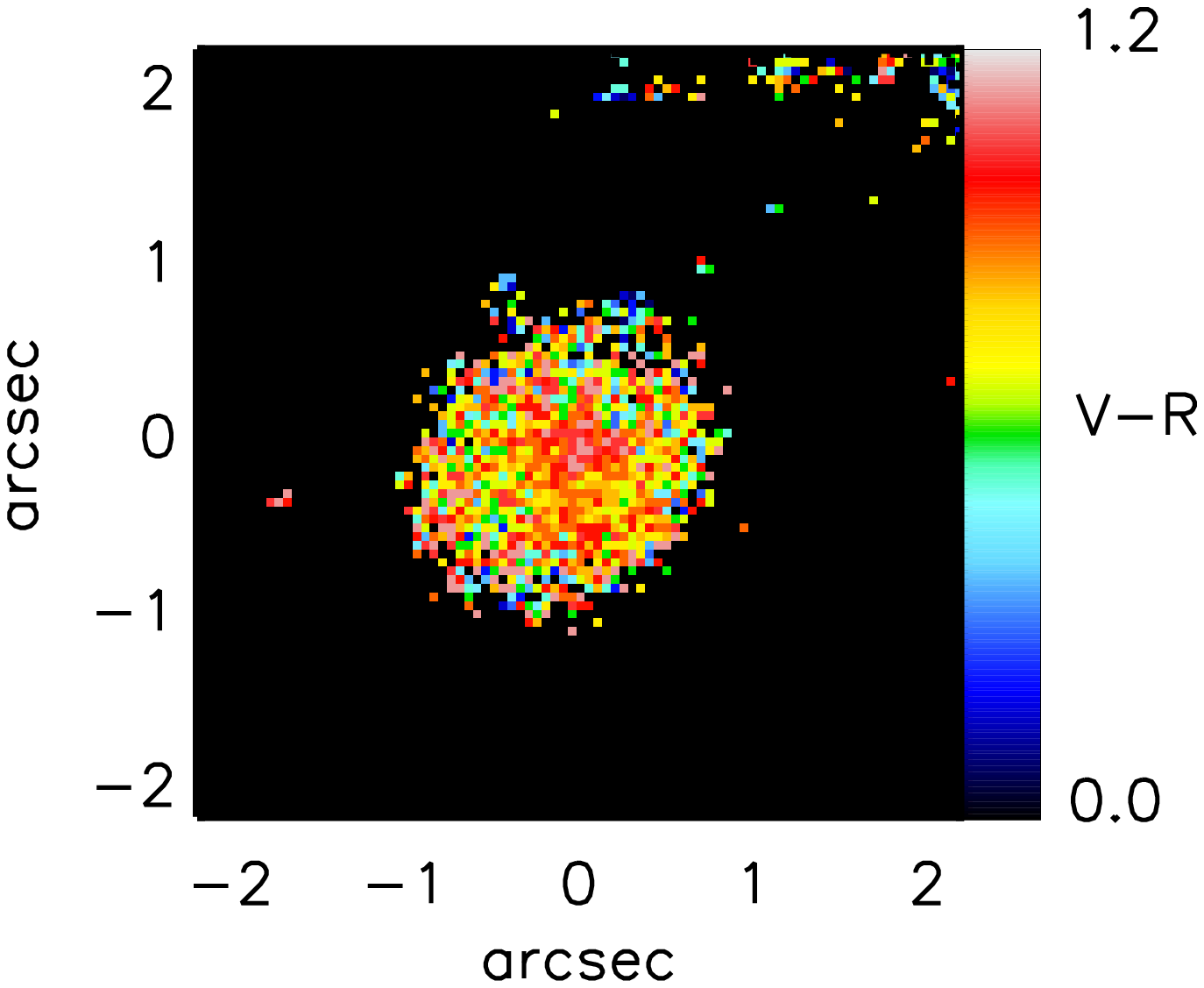}\hspace{-0.7cm}
      \end{minipage}
       \begin{minipage}{0.95\textwidth}
\hspace{-1.3cm}
         \includegraphics[width=4.0cm, angle=0, trim=0 0 0 0]{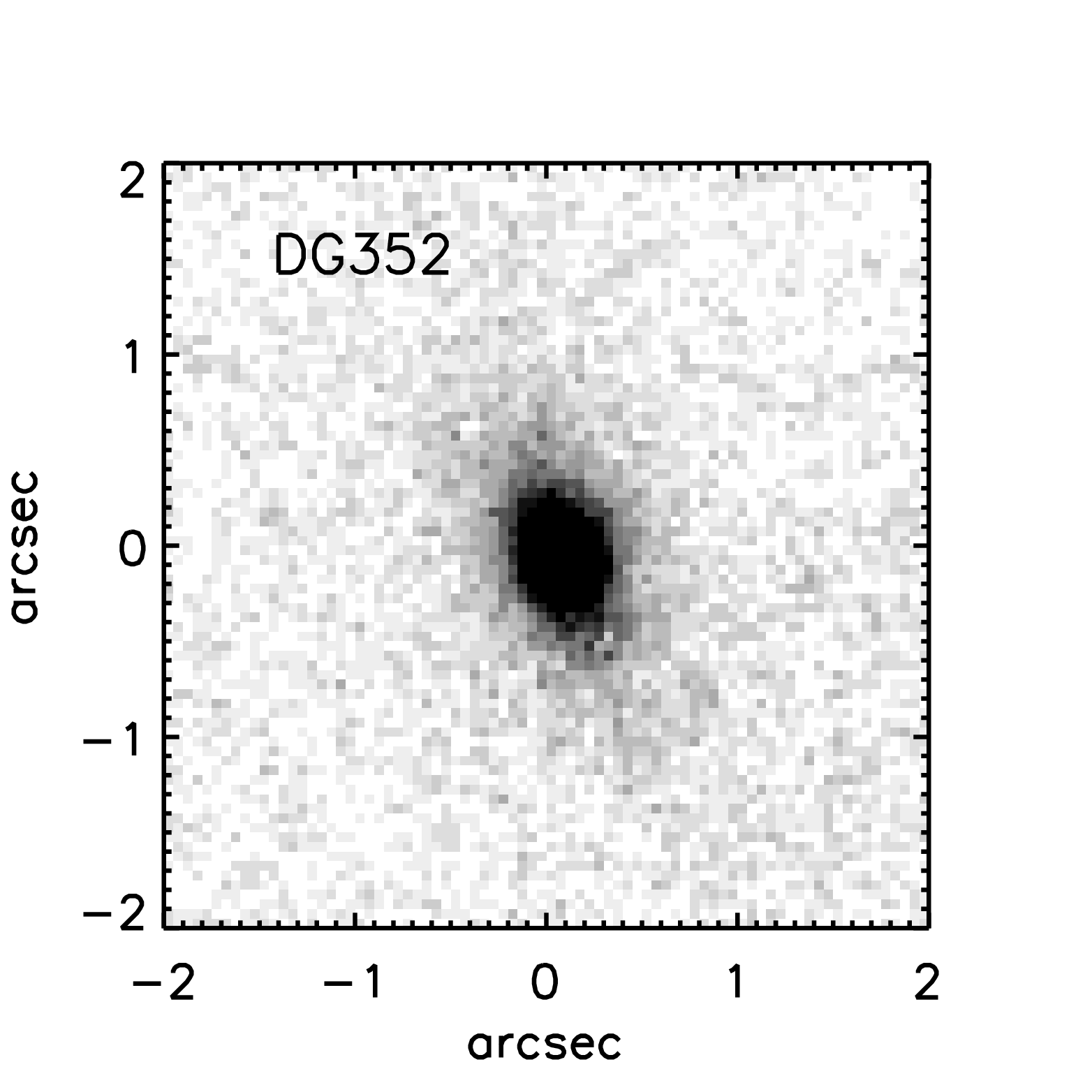}\hspace{-0.7cm}
         \includegraphics[width=4.0cm, angle=0, trim=0 0 0 0]{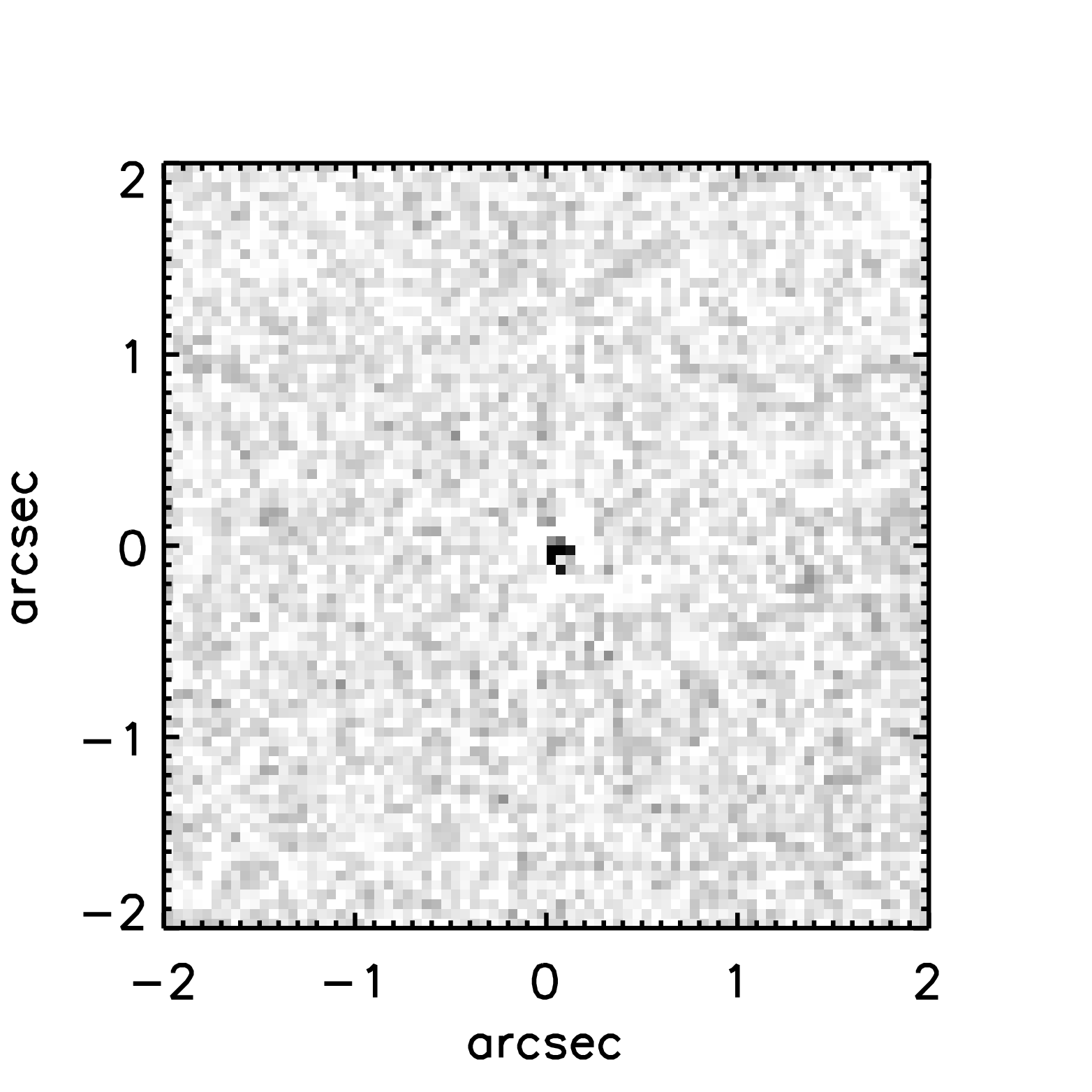}\hspace{-0.7cm}
         \includegraphics[width=3.6cm, angle=90, trim=0 0 0 0]{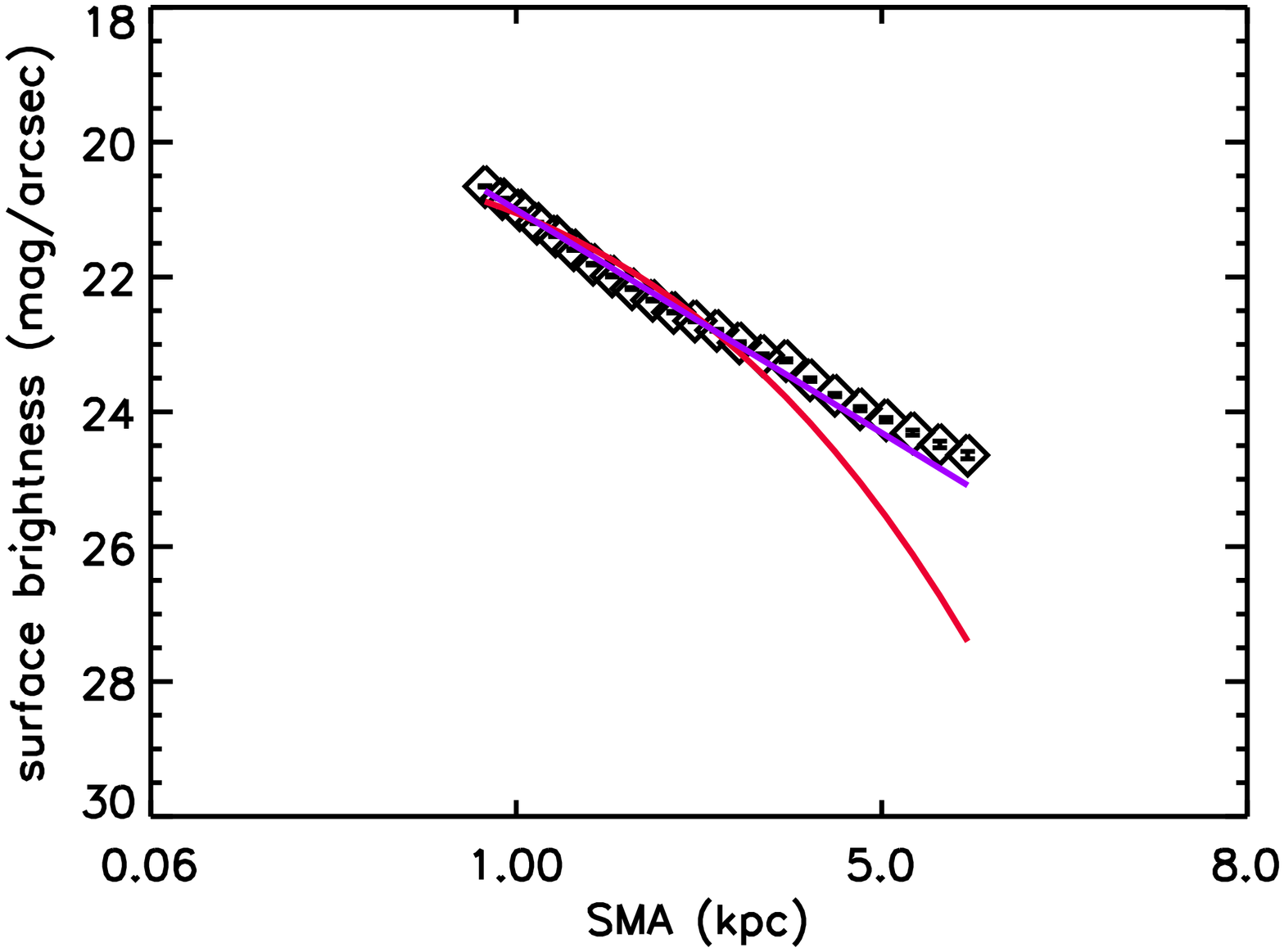}\hspace{-0.7cm}
         \includegraphics[width=5.4cm, angle=0, trim=0 0 0 0]{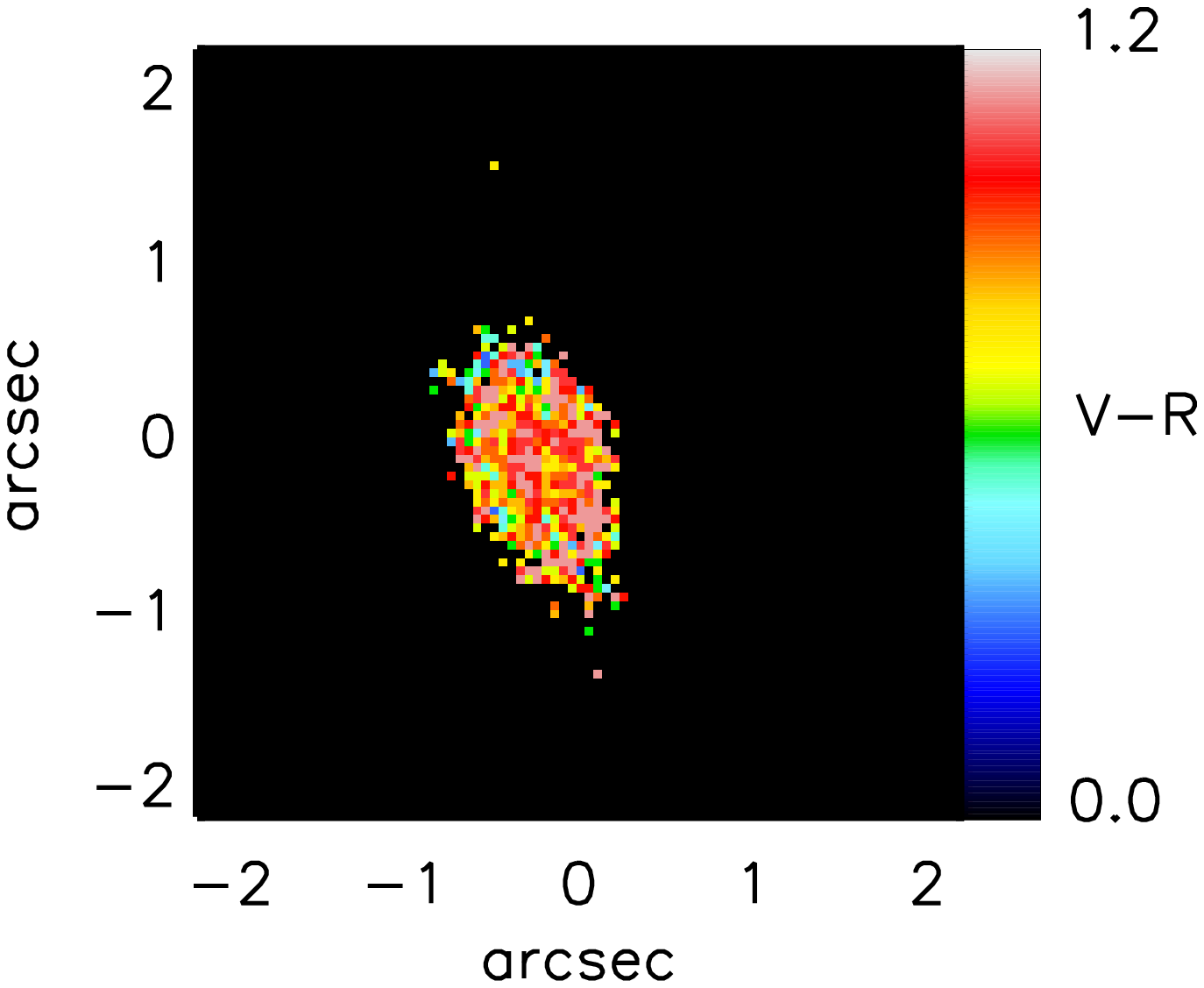}\hspace{-0.7cm}
      \end{minipage}
       \begin{minipage}{0.95\textwidth}
\hspace{-1.3cm}
         \includegraphics[width=4.0cm, angle=0, trim=0 0 0 0]{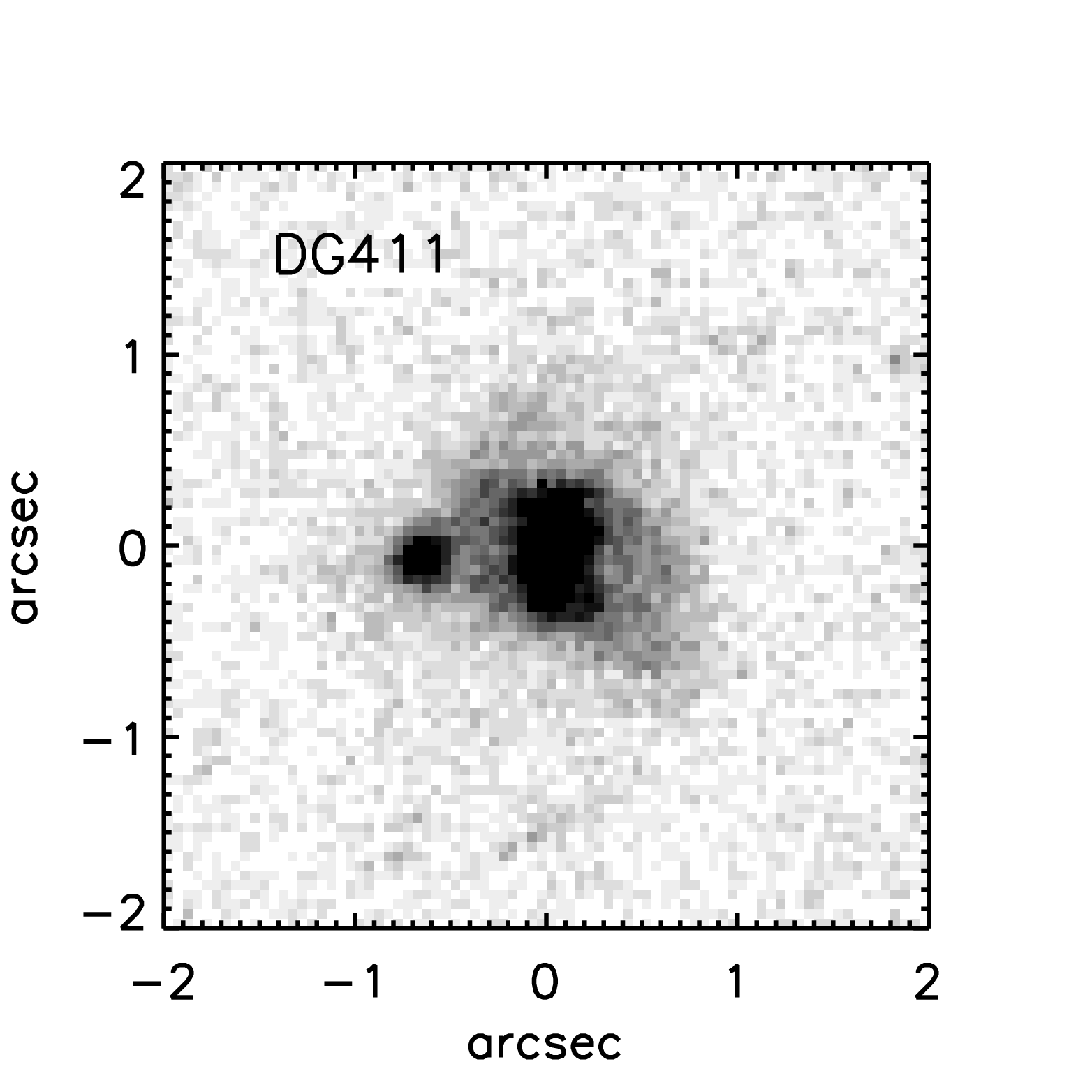}\hspace{-0.7cm}
         \includegraphics[width=4.0cm, angle=0, trim=0 0 0 0]{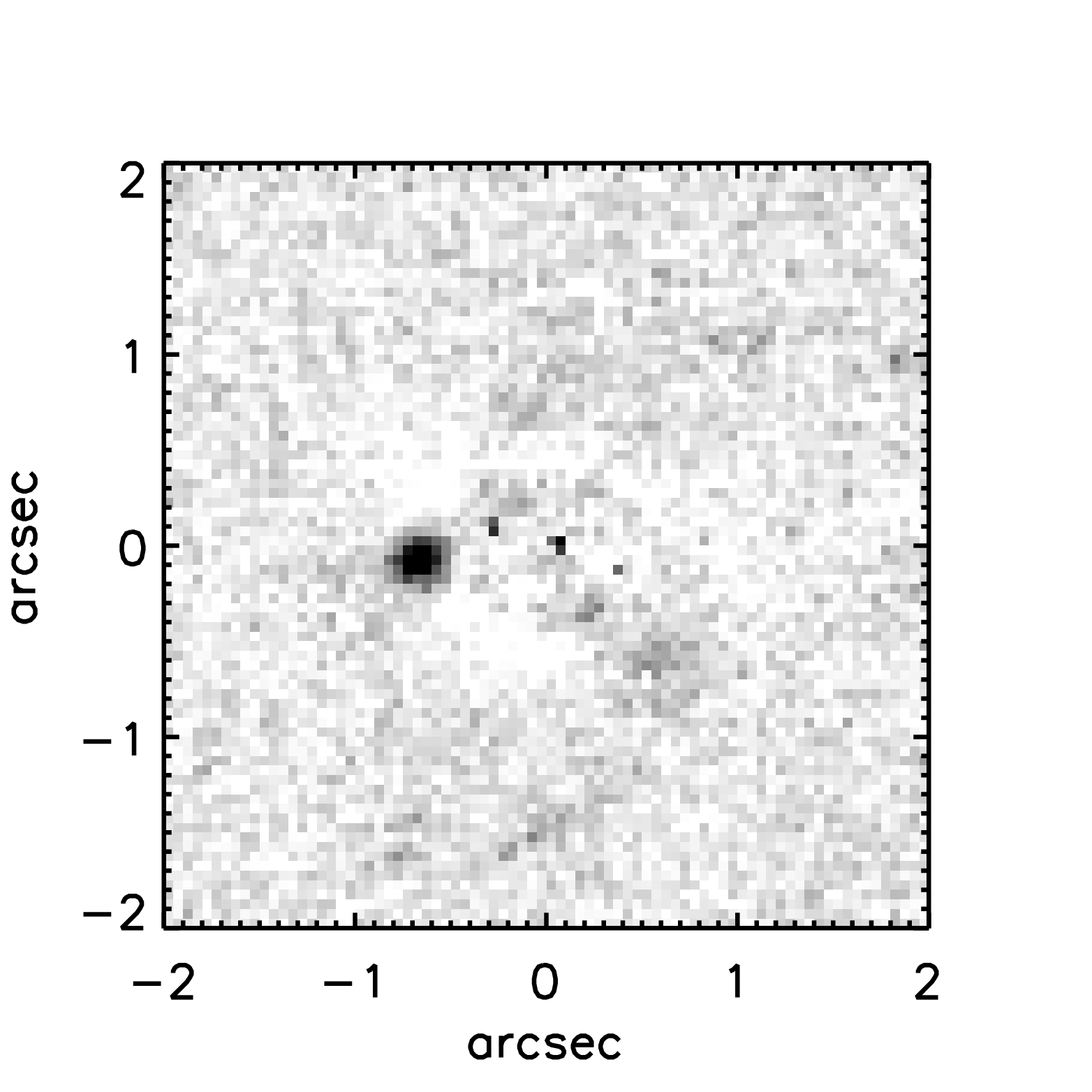}\hspace{-0.7cm}
         \includegraphics[width=3.6cm, angle=90, trim=0 0 0 0]{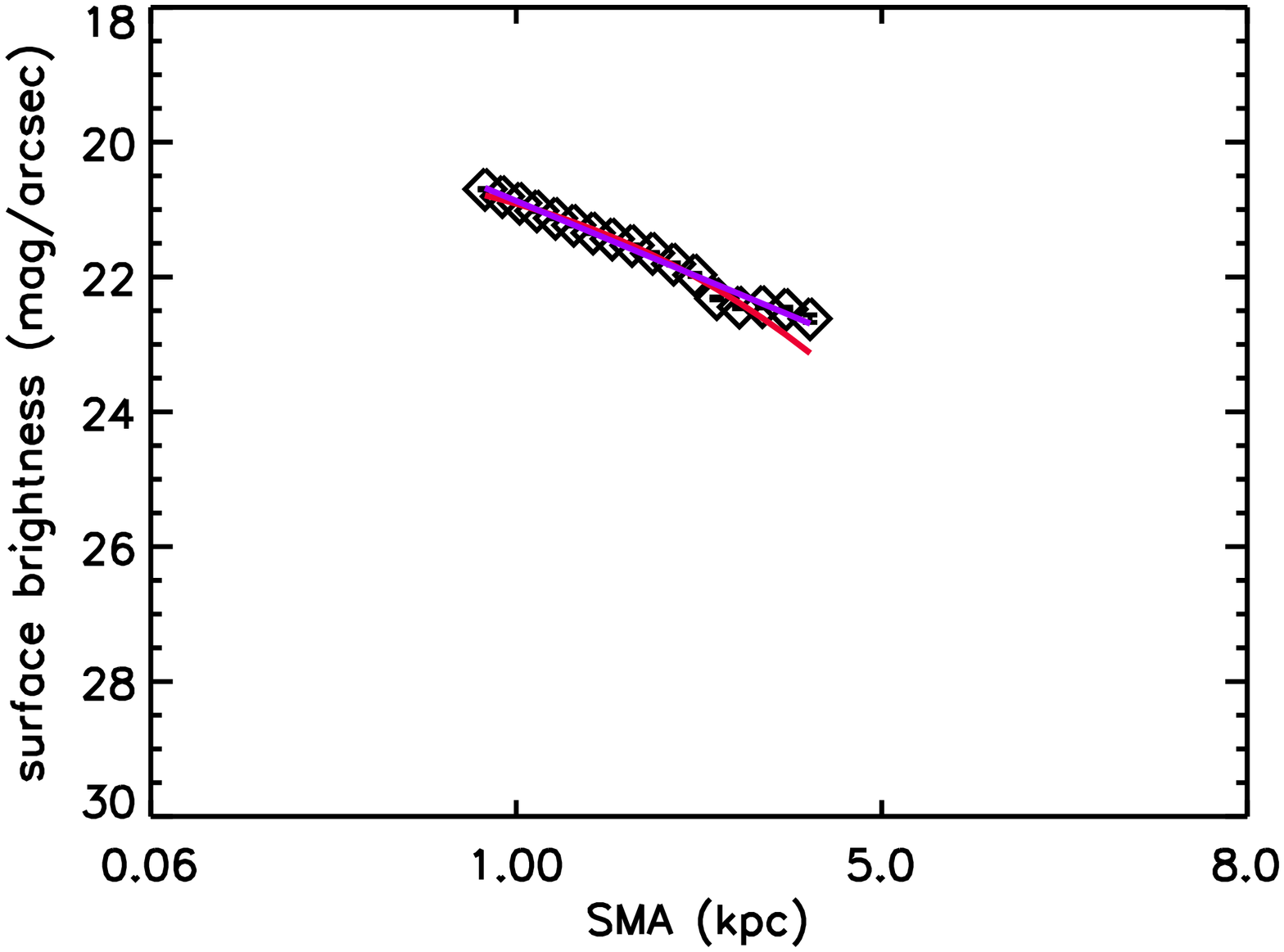}\hspace{-0.7cm}
         \includegraphics[width=5.4cm, angle=0, trim=0 0 0 0]{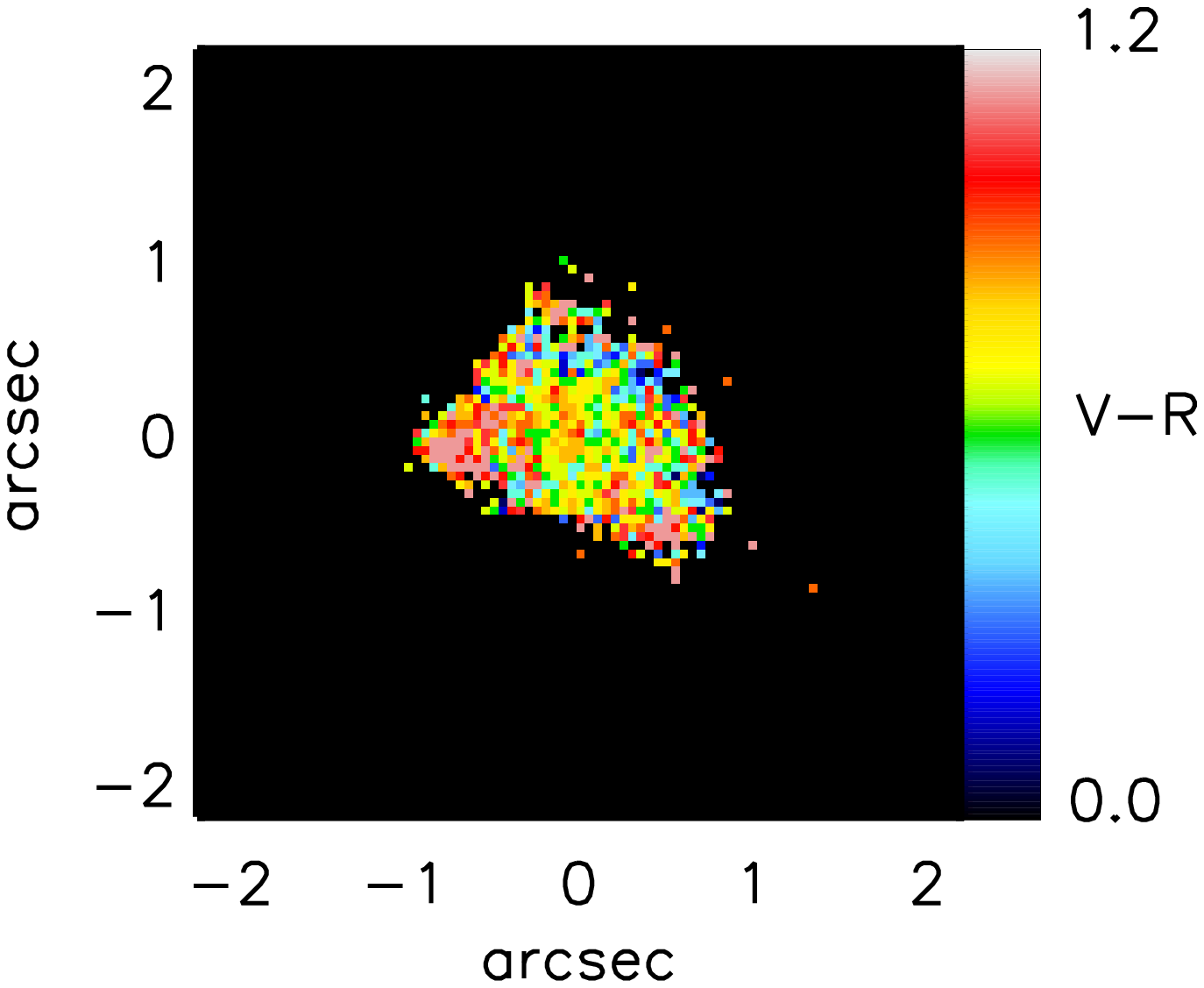}\hspace{-0.7cm}
       \end{minipage}
\end{center}
\vspace{-0.4cm}
\caption{Some morphological and photometric diagnostics of the post--starburst galaxies. The leftmost 
column shows a postage stamp ACS image in the F775W passband. The second column shows a residual image 
after subtraction of a symmetric elliptical model. The third column is the isophotal profile for each
galaxy along with fits of an exponential model ({\it red line}) and a de Vaucouleurs model ({\it blue line}).
The final column shows a V-R colour image (actually F606W-F775W).}
\label{fig:psbmorph}
\end{figure*}

Visual inspection of the R-band postage-stamp images shown in the first column of Figure 
\ref{fig:psb} (and displayed again in the first column of Figure \ref{fig:psbmorph} 
reveals that all of the post-starburst galaxies are bulge-dominated 
early-type systems -- which is true of E+As generally \citep{dressler99}. However,
while E+A galaxies in local field samples commonly display early type morphologies, they
also display a high incidence of interactions, disruption and tidal tails 
\citep{zabludoff96,norton01,blake04,yamauchi05,goto08,yang08,pracy09}. With the exception
of DG\_411, there is no obvious sign of tidal disruption in this sample.  

To investigate this more thoroughly, we used the 
{\sc iraf} tasks {\sc ellipse} and {\sc bmodel} to produce symmetric elliptical model images 
for the E+A galaxies and then subtracted the models from the data in order to look for evidence of 
irregularities or tidal disruption. This technique has been 
used on local E+A field samples and has revealed a high incidence of morphological disturbance \citep{yang08,pracy09}.
Such features have been used as evidence for a  merger or tidal interaction origin in these populations. 
The model-subtracted images are shown in the second column of Figure \ref{fig:psbmorph} and show no evidence
of disruption or underlying tidal features -- again with the exception of DG\_411 which has a companion and tidal tail.
This is in distinct contrast to local field samples and argues in favor of truncation by the ICM over mergers as the
mechanism involved in producing these E+As.

One caveat when comparing our study with local samples is that the ability 
to detect tidal features is sensitive to the surface brightness
limits probed. Our HST/ACS imaging is intrinsically deeper 
than the local HST/ACS study of \citet{yang08} with an integration time of 
$\sim$2200\,s (in the F775W passband) compared with their 900\,s 
integration (in the F625W passband). However, the cosmological surface 
brightness dimming due to the higher redshift
of our targets reduces our ability to detect faint features. Assuming a
redshift for our target galaxies of $z=0.55$  and a typical
redshift of the \citet{yang08} sample of $z\sim 0.1$ the relative effect of 
cosmological dimming is a factor of $\sim 3.9$. Note the different passbands
and redshifts of the observations partially cancel out making the restframe
wavelength range of the imaging similar in both samples, with the high redshift observations 
$\sim$500\AA\, bluer. For the detection of faint features the imaging is in
the background limited regime and the signal-to-noise ratio scales as 
the square root of the ratio of the exposure times. Taking both of these
effects into account, and neglecting the small differences in filter sensitivity 
and restframe wavelength, the true `restframe surface brightness limit' 
of the \citet{yang08} study is $\sim 1$\,mag fainter. Although many of the tidal features in the
low redshift data are generally clear--cut we cannot entirely rule out that some faint tidal
features that would be detected in the low redshift data are missed in our imaging.

In the third column of Figure \ref{fig:psbmorph} we show the isophotal profiles
measured with the {\sc ellipse} task along with the best fitting exponential profile 
({\it red line}) and de Vaucouleurs profile ({\it blue line}).
We only show isophotes at radii larger than the FWHM of the HST ACS point spread function to avoid 
any issues caused by bluring near the resolution limit 
on the galaxy profiles. For the most part the isophotal profile are well fitted by a 
de Vaucouleurs profile profile. DG\_134 and DG\_181 have an
exponential disk--like component. In the final column of Figure \ref{fig:psbmorph} we show V-R colour maps. 
Local field samples show variation in the colour properties of E+As with examples of uniform distributions, 
red centres, blue centres and irregular 
colour distributions \citep{yang08,pracy09}. Although the signal-to-noise ratio of the 
maps in Figure \ref{fig:psbmorph} is moderate there does appear to be diversity among 
the sample with examples of uniform (DG\_106), blue centre (DG\_134), red centre
(DG\_181 and DG\_352) and irregular (DG\_115) colour distributions. In most cases 
these gradients are not very pronounced. 

\subsection{Star forming galaxies}
The remaining three galaxies in our sample are star--forming galaxies with clear [OII]$\lambda 3727$
emission. These three galaxies (DG\_338, DG\_356, DG\_371) are consistent with an e(a) spectral 
classification within their 1$\sigma$ error bars.  

\subsubsection{Spatially resolved spectroscopy}
\begin{figure*}[t]
   \begin{center}
     \begin{minipage}{0.95\textwidth}
\hspace{-1.2cm}
         \includegraphics[width=3.6cm, angle=0]{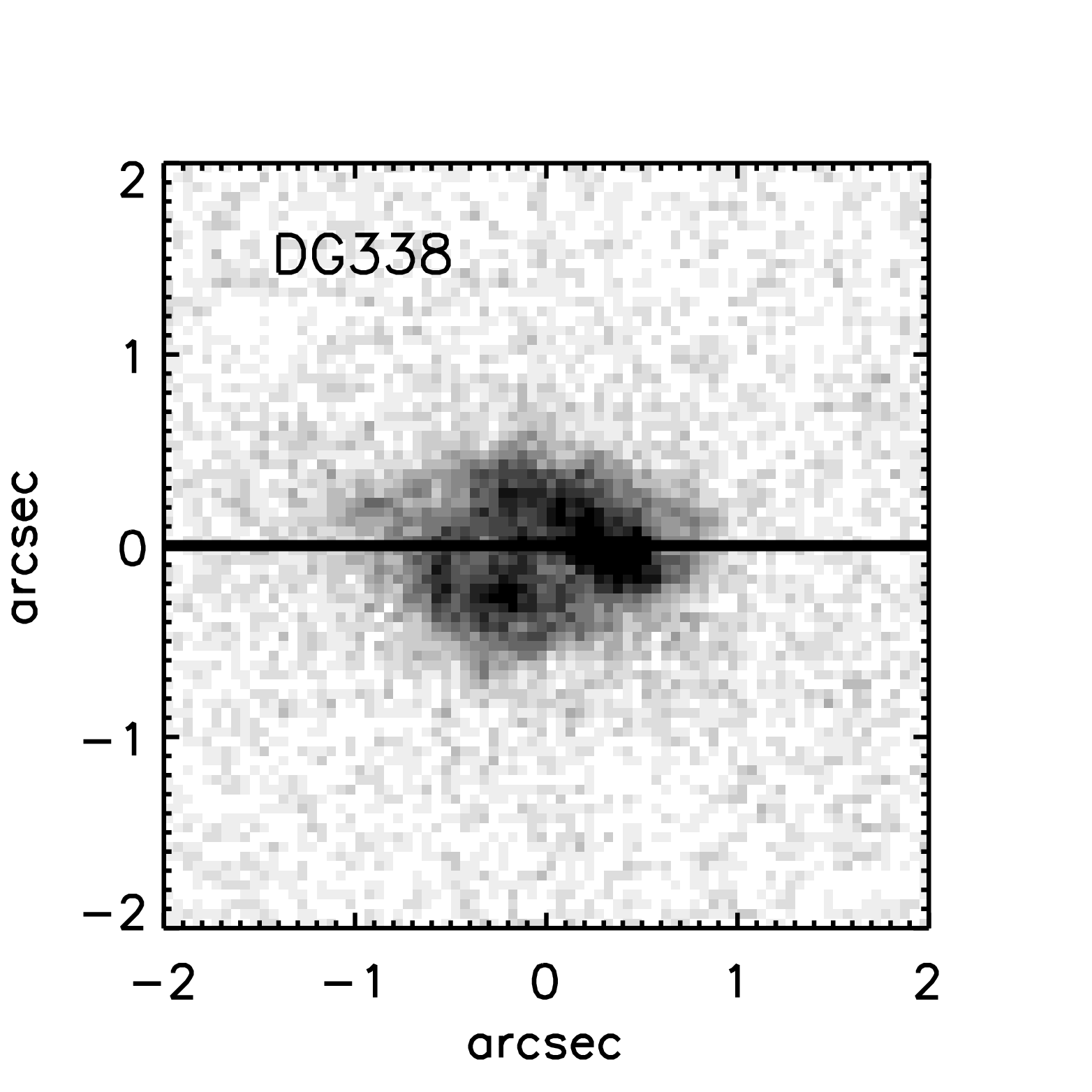}
\hspace{-0.8cm}
        \includegraphics[width=3.4cm, angle=90]{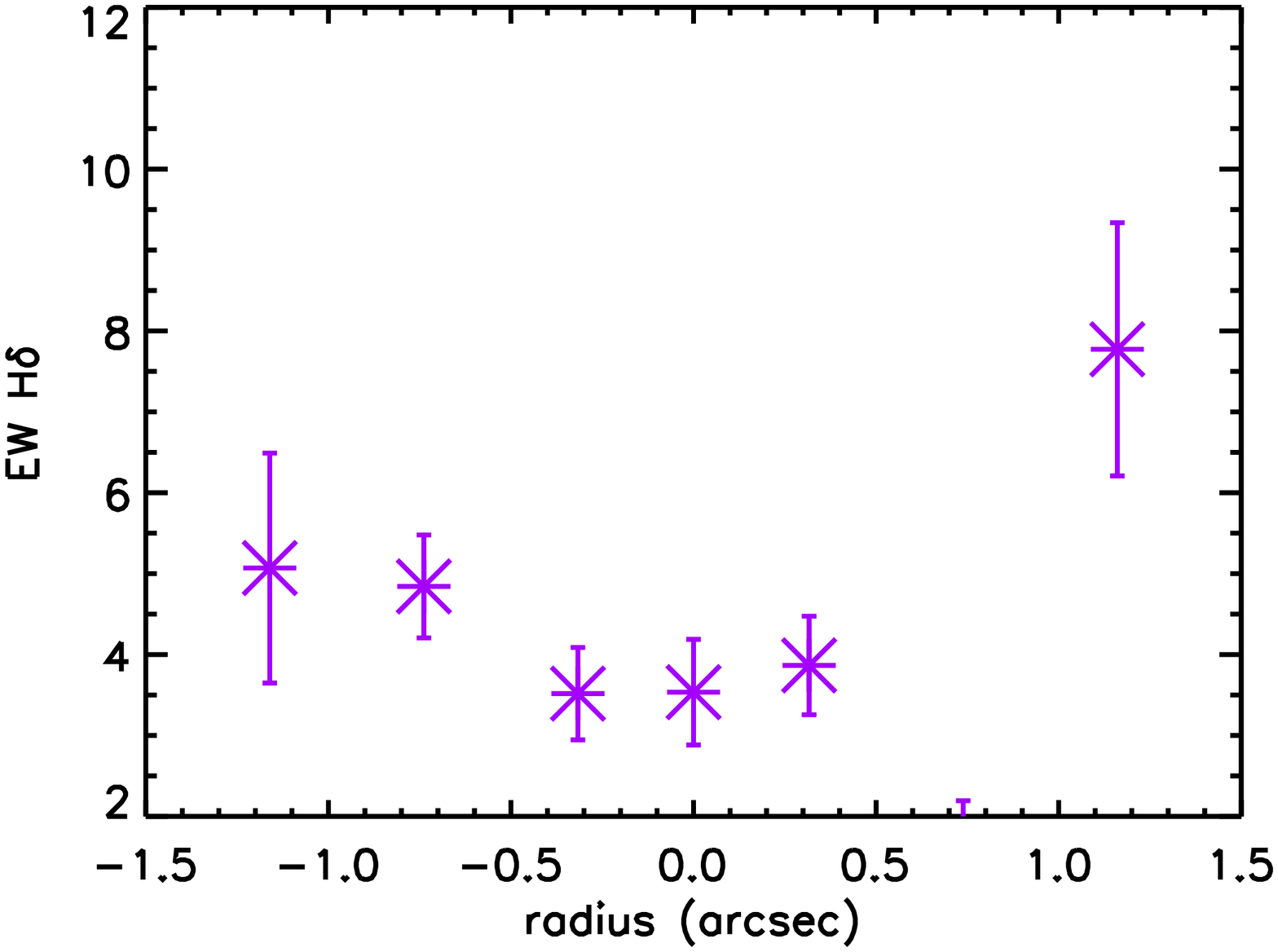}
\hspace{-0.6cm}
         \includegraphics[width=3.4cm, angle=90]{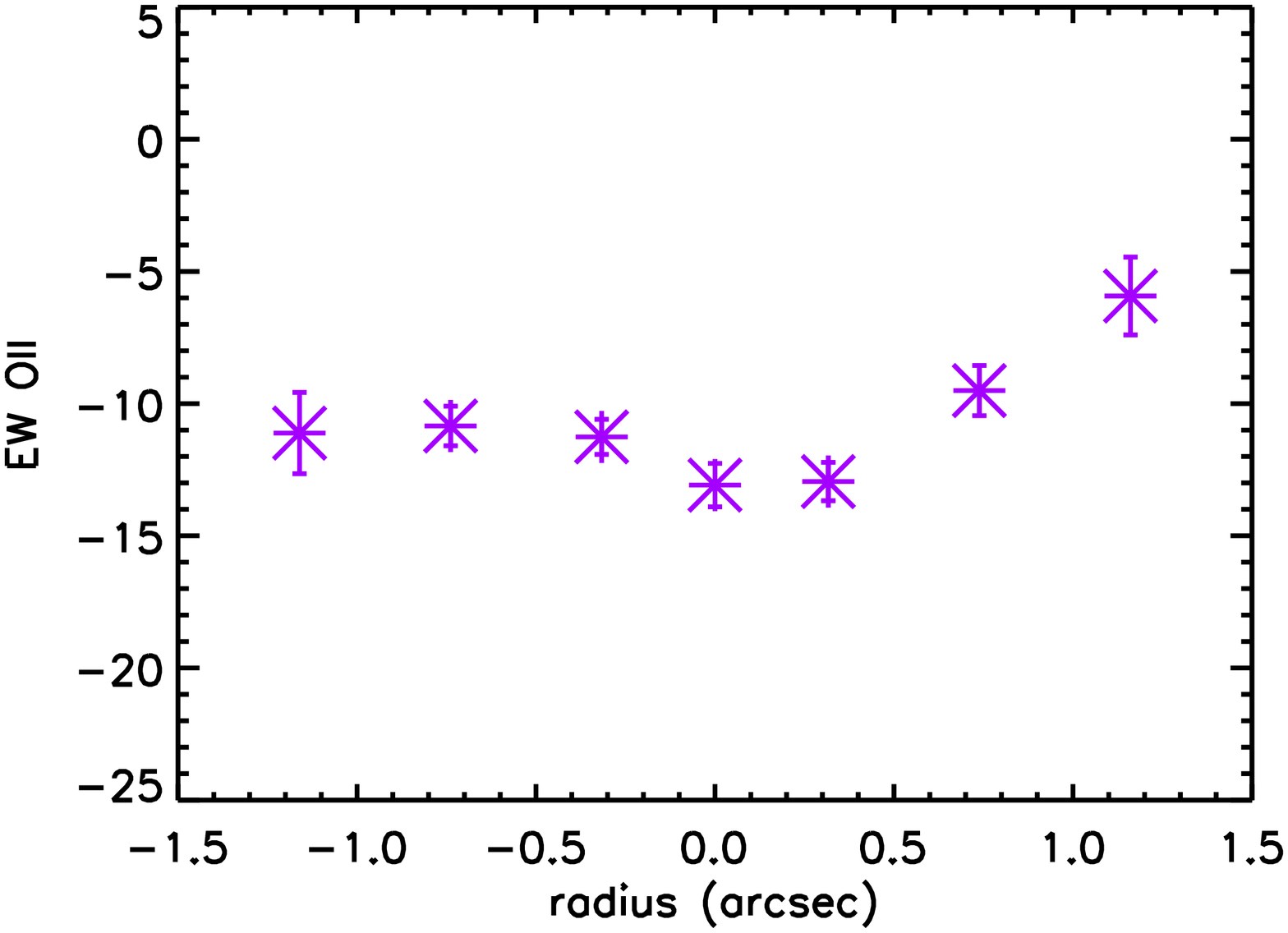}
\hspace{-0.6cm}
         \includegraphics[width=3.4cm, angle=90]{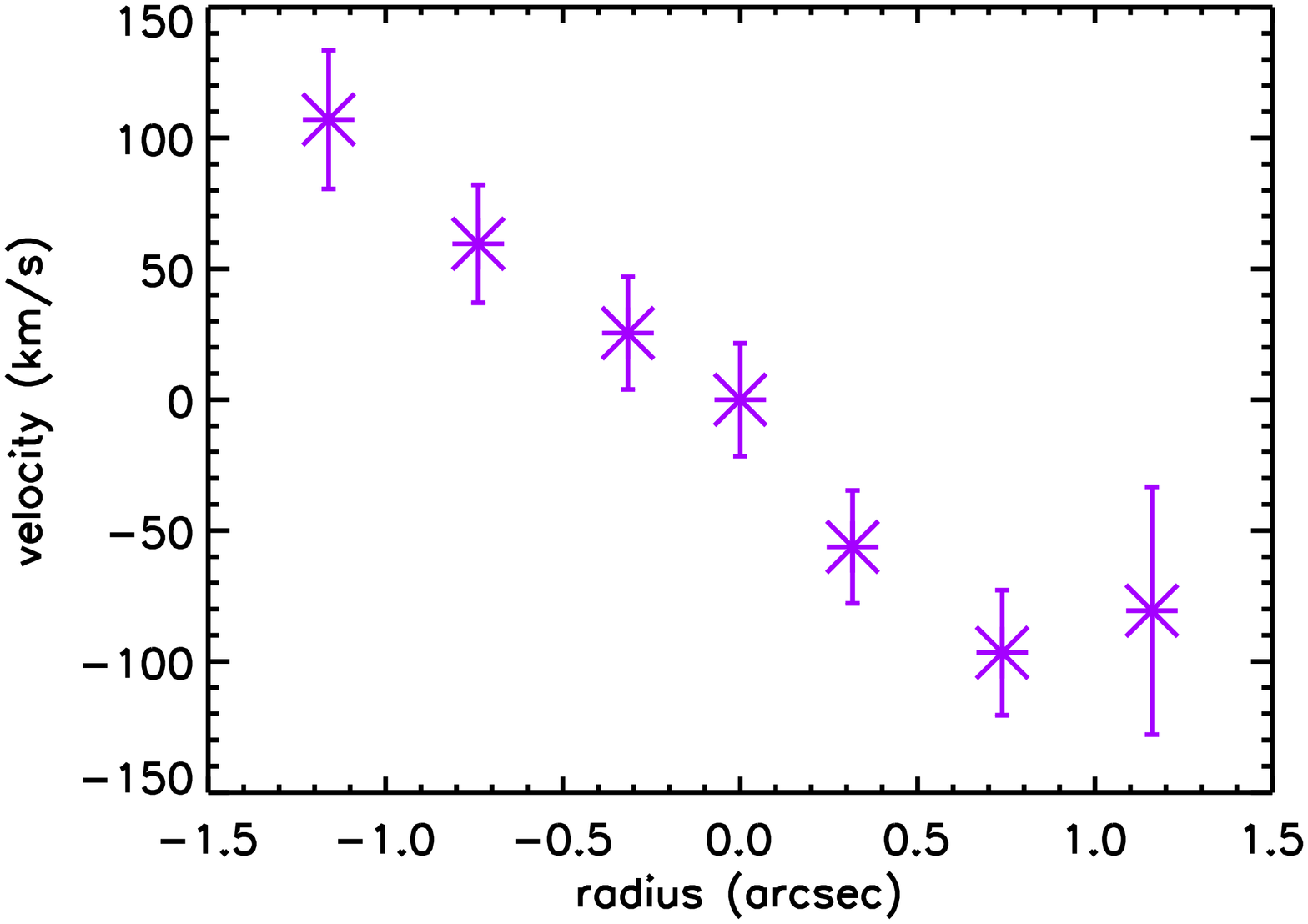}
      \end{minipage}
       \begin{minipage}{0.95\textwidth}
\hspace{-1.2cm}
         \includegraphics[width=3.6cm, angle=0, trim=0 0 0 0]{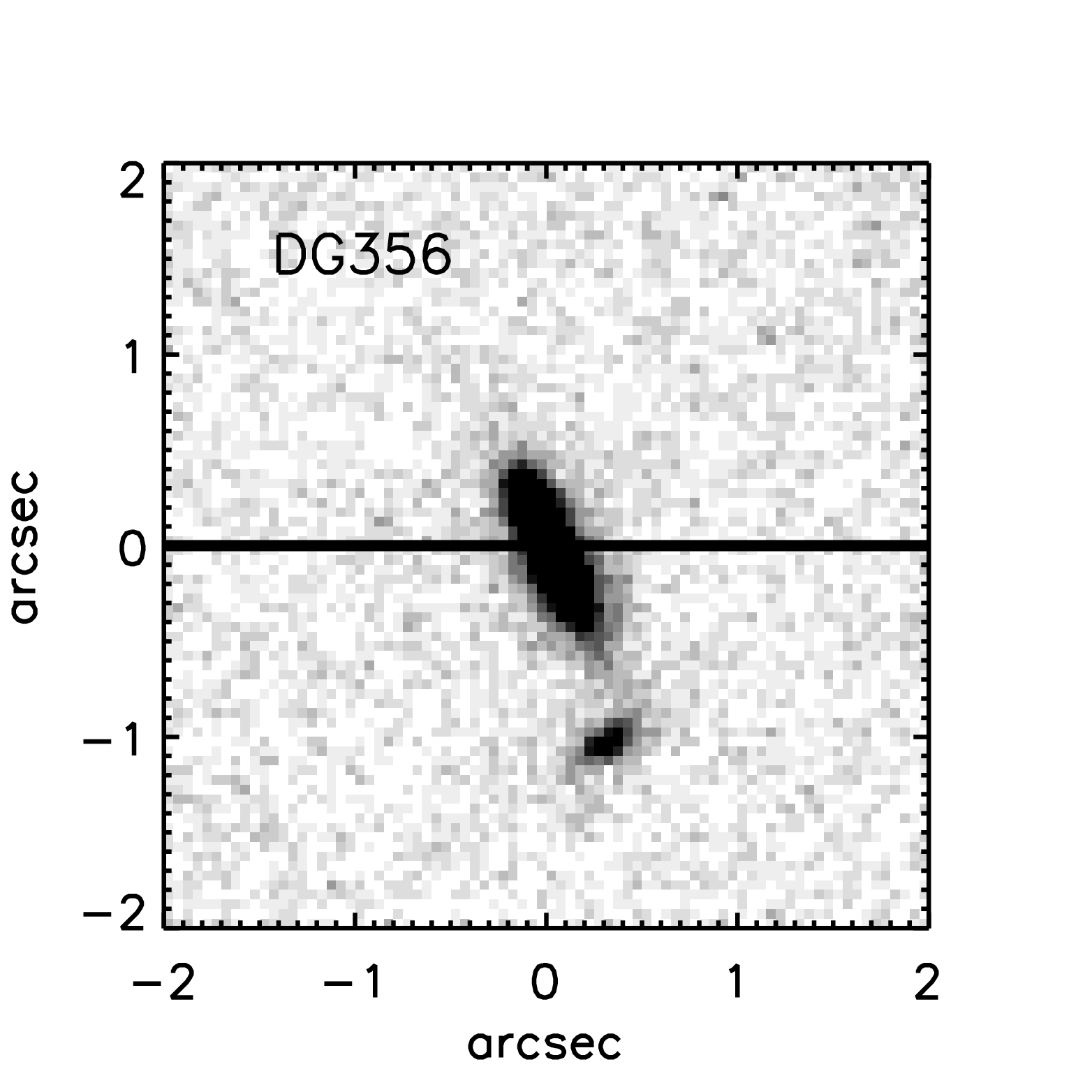}
\hspace{-0.8cm}
        \includegraphics[width=3.4cm, angle=90, trim=0 0 0 0]{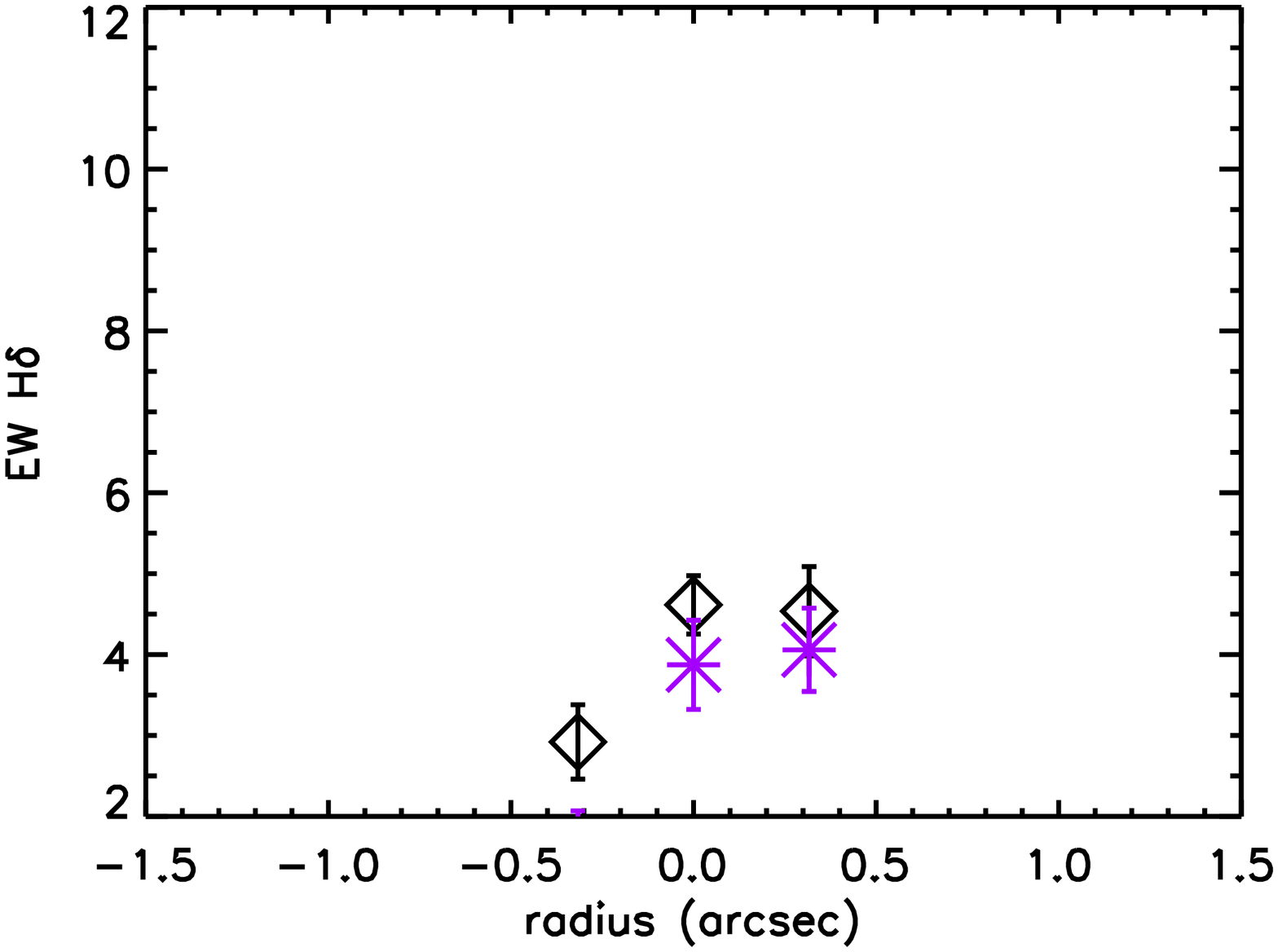}
\hspace{-0.6cm}
         \includegraphics[width=3.4cm, angle=90, trim=0 0 0 0]{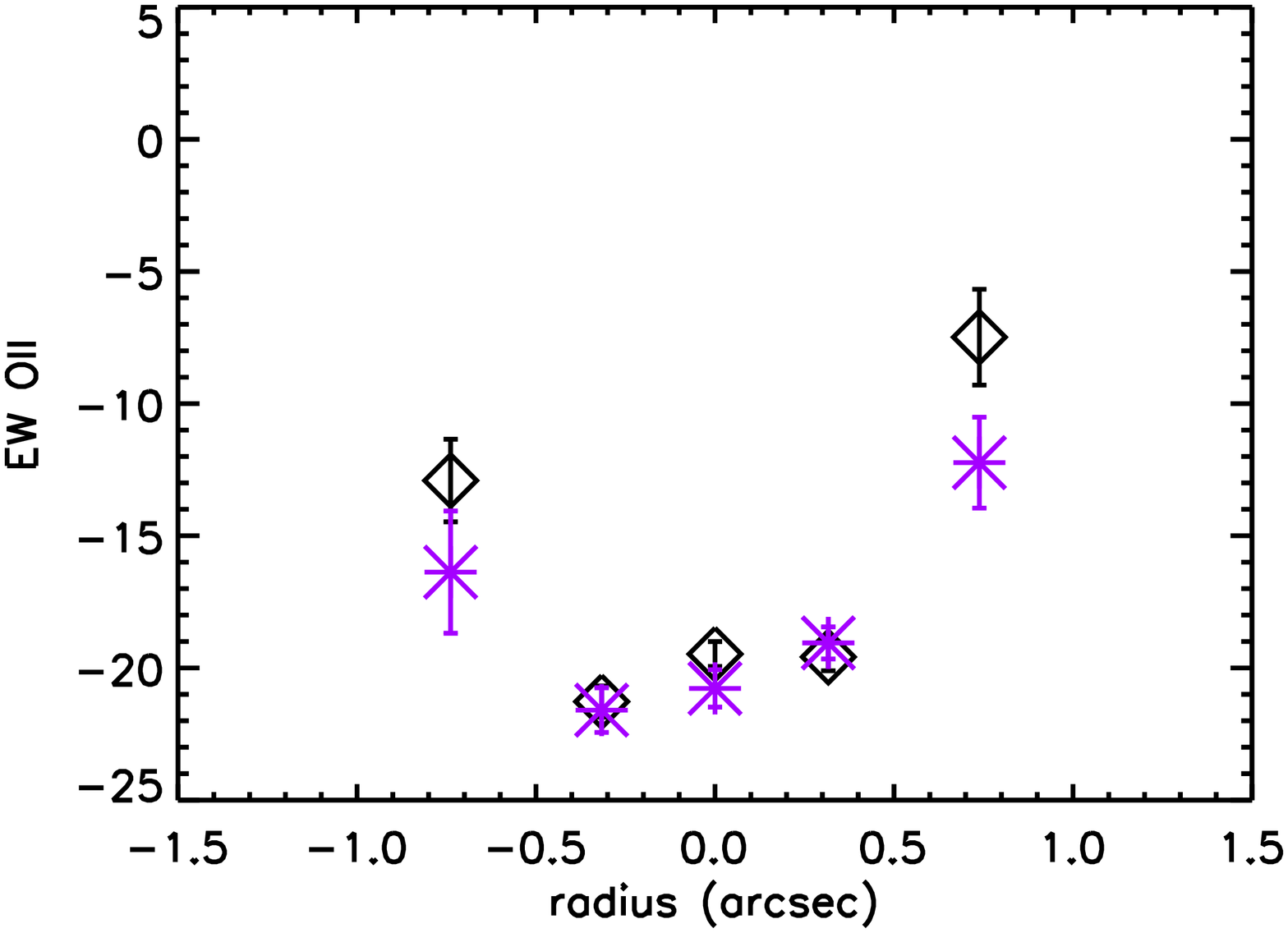}
\hspace{-0.6cm}
         \includegraphics[width=3.4cm, angle=90, trim=0 0 0 0]{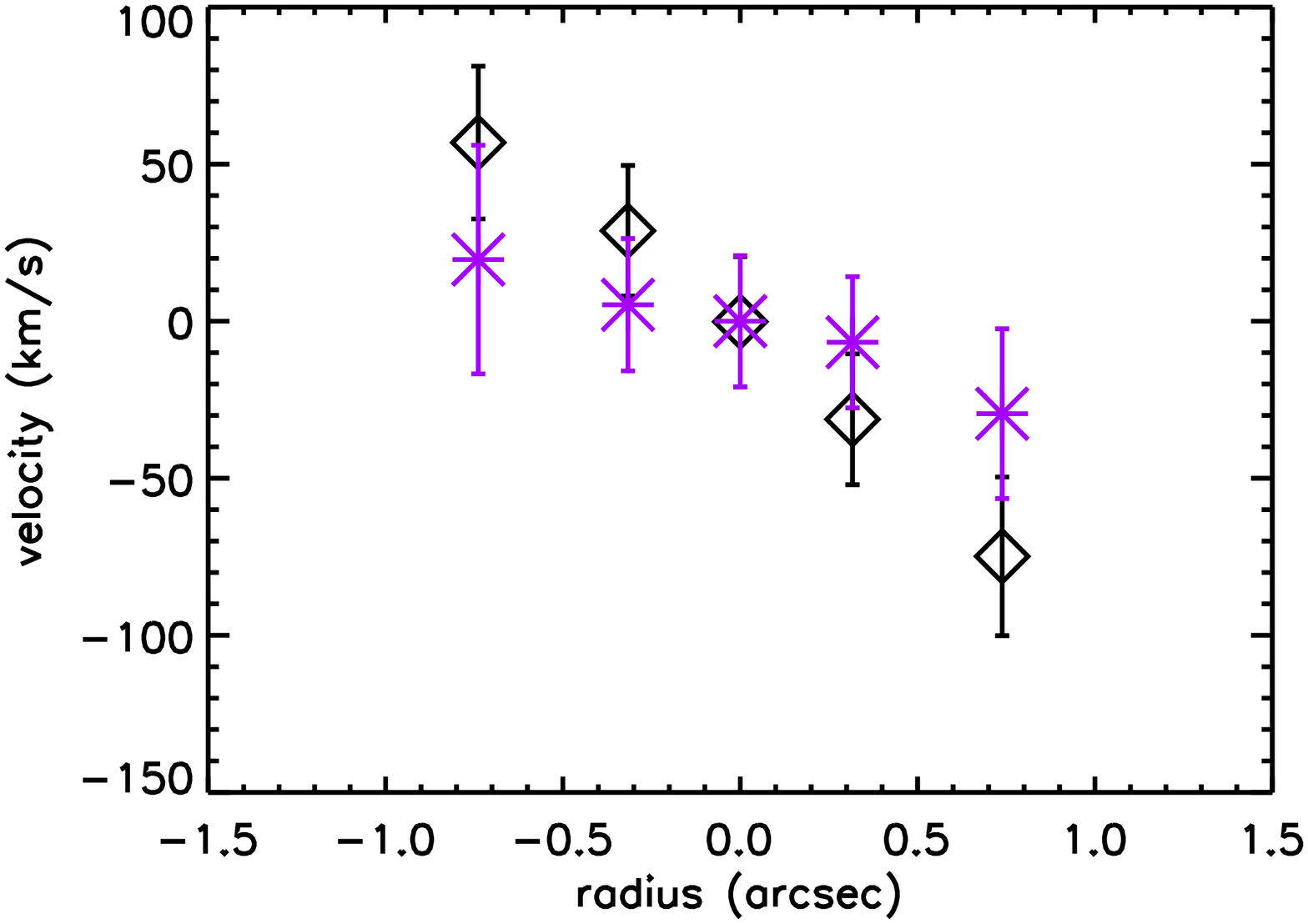}
      \end{minipage}
       \begin{minipage}{0.95\textwidth}
\hspace{-1.2cm}
         \includegraphics[width=3.6cm, angle=0, trim=0 0 0 0]{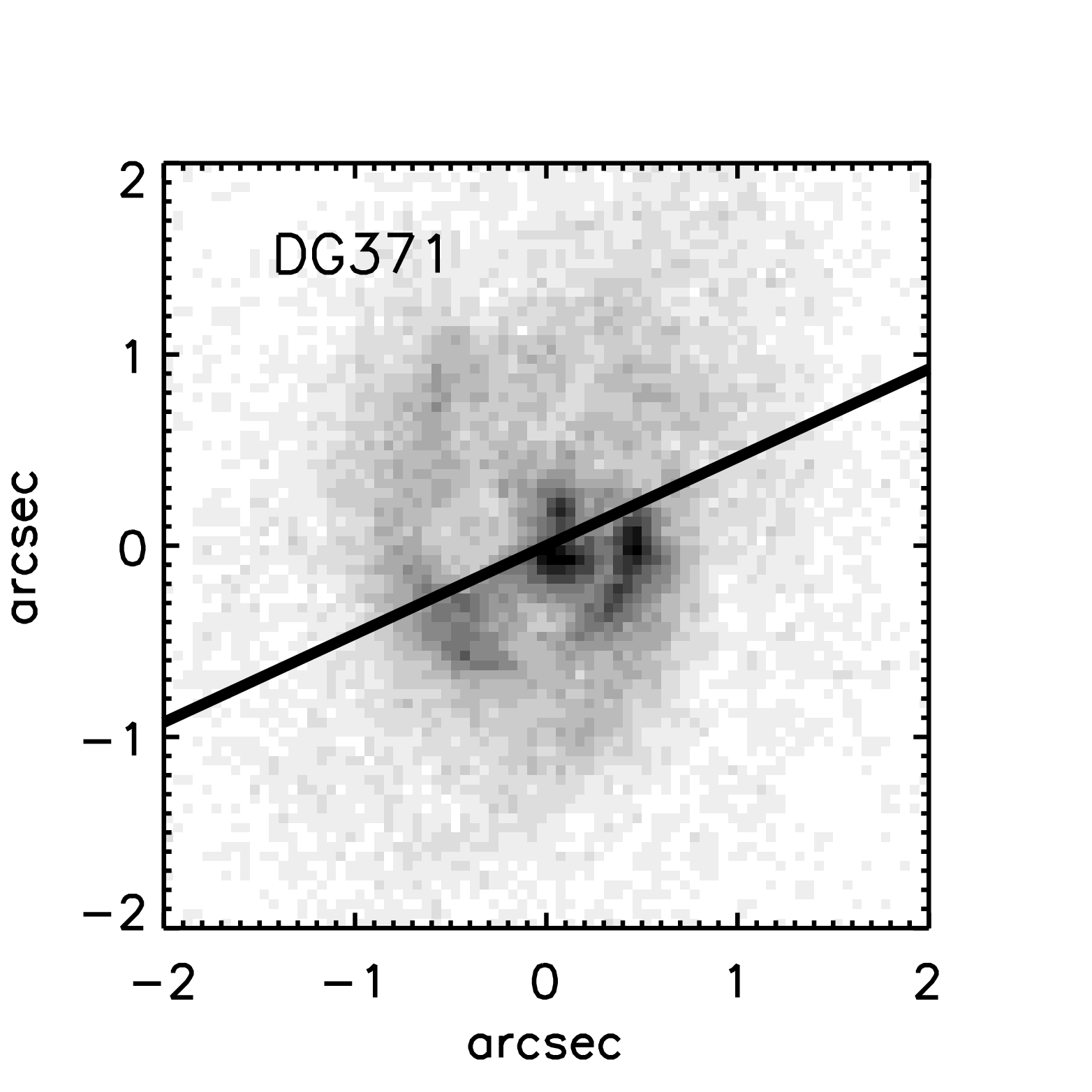}
\hspace{-0.8cm}
        \includegraphics[width=3.4cm, angle=90, trim=0 0 0 0]{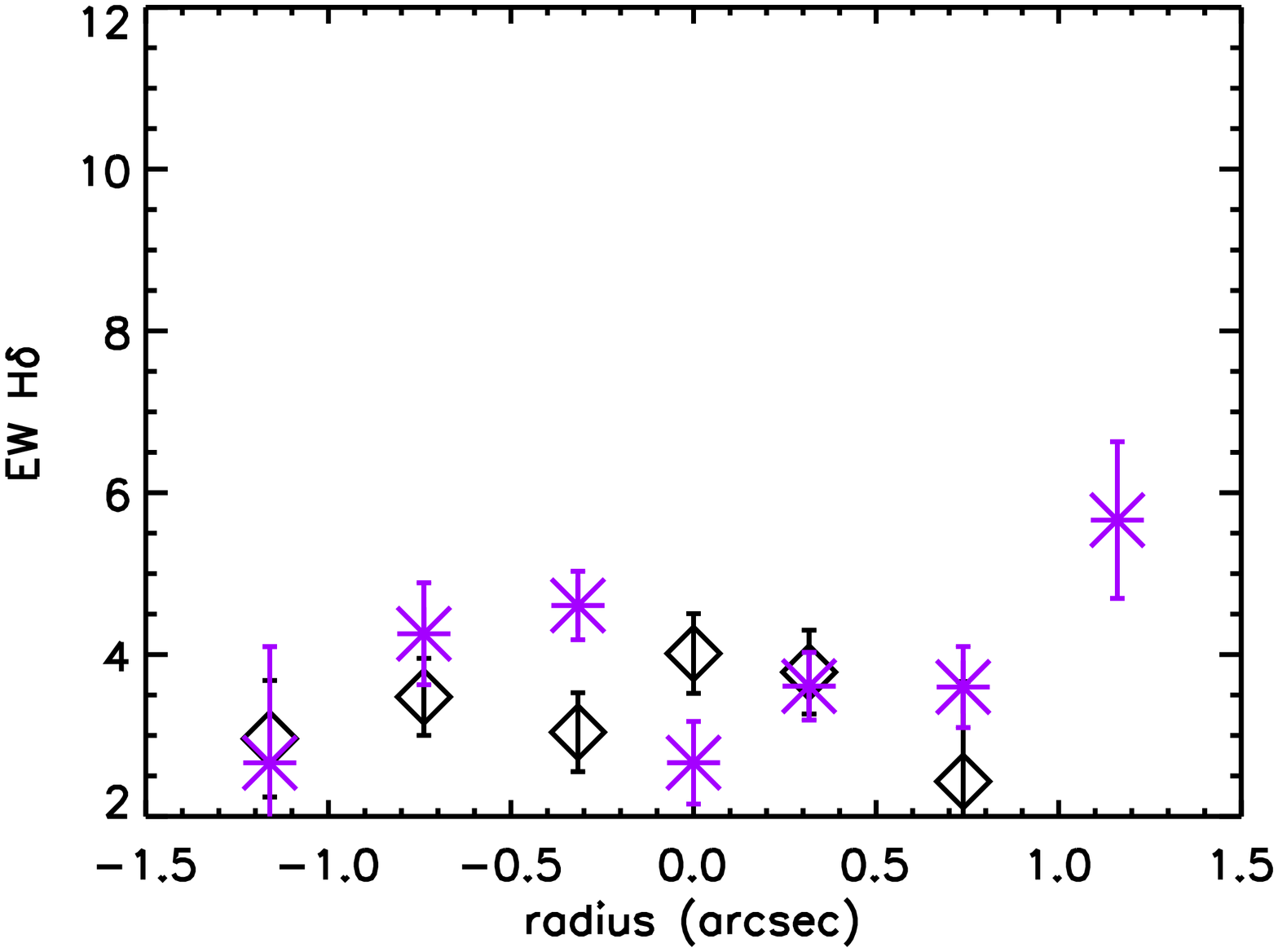}
\hspace{-0.6cm}
         \includegraphics[width=3.4cm, angle=90, trim=0 0 0 0]{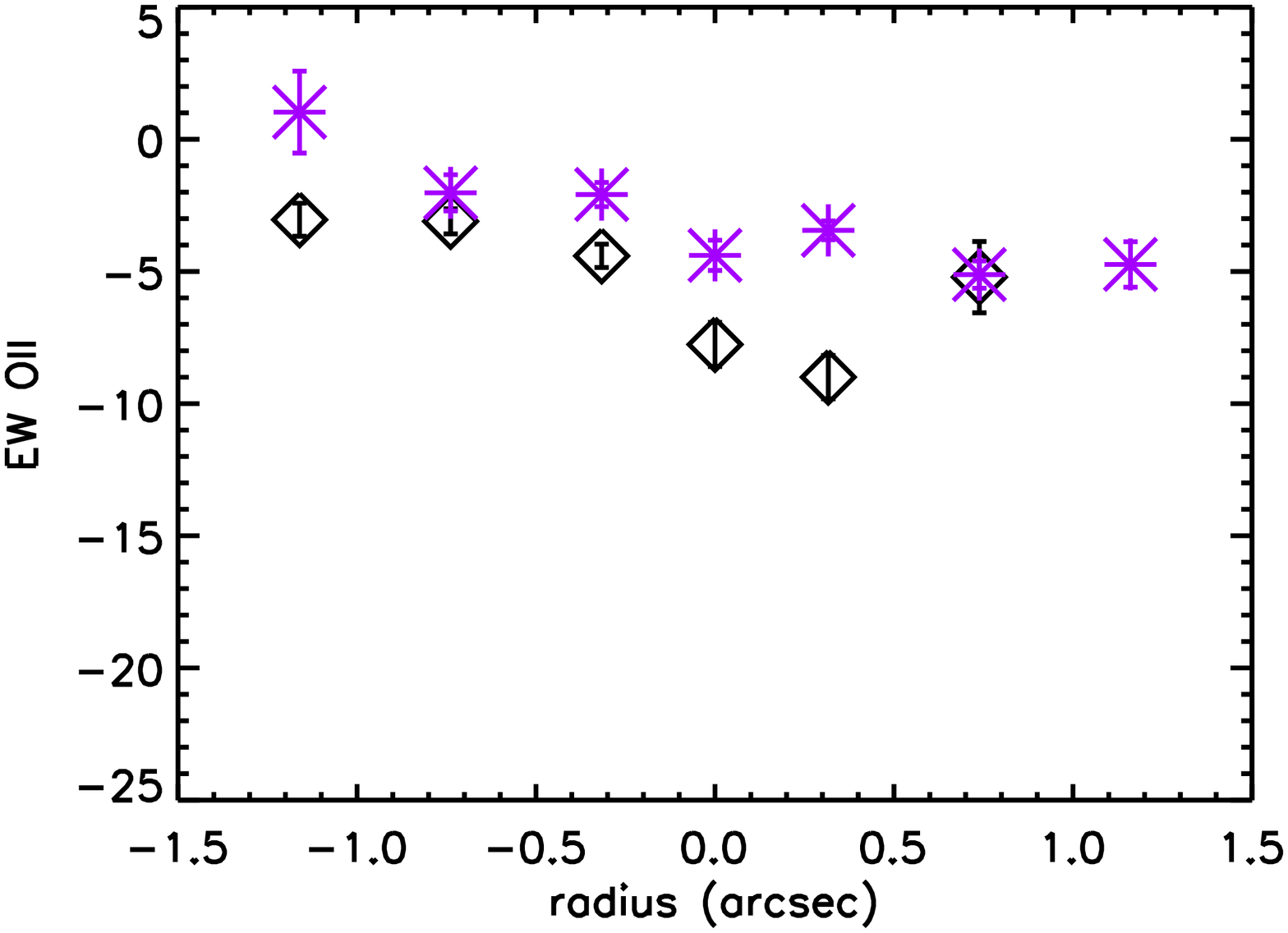}
\hspace{-0.6cm}
         \includegraphics[width=3.4cm, angle=90, trim=0 0 0 0]{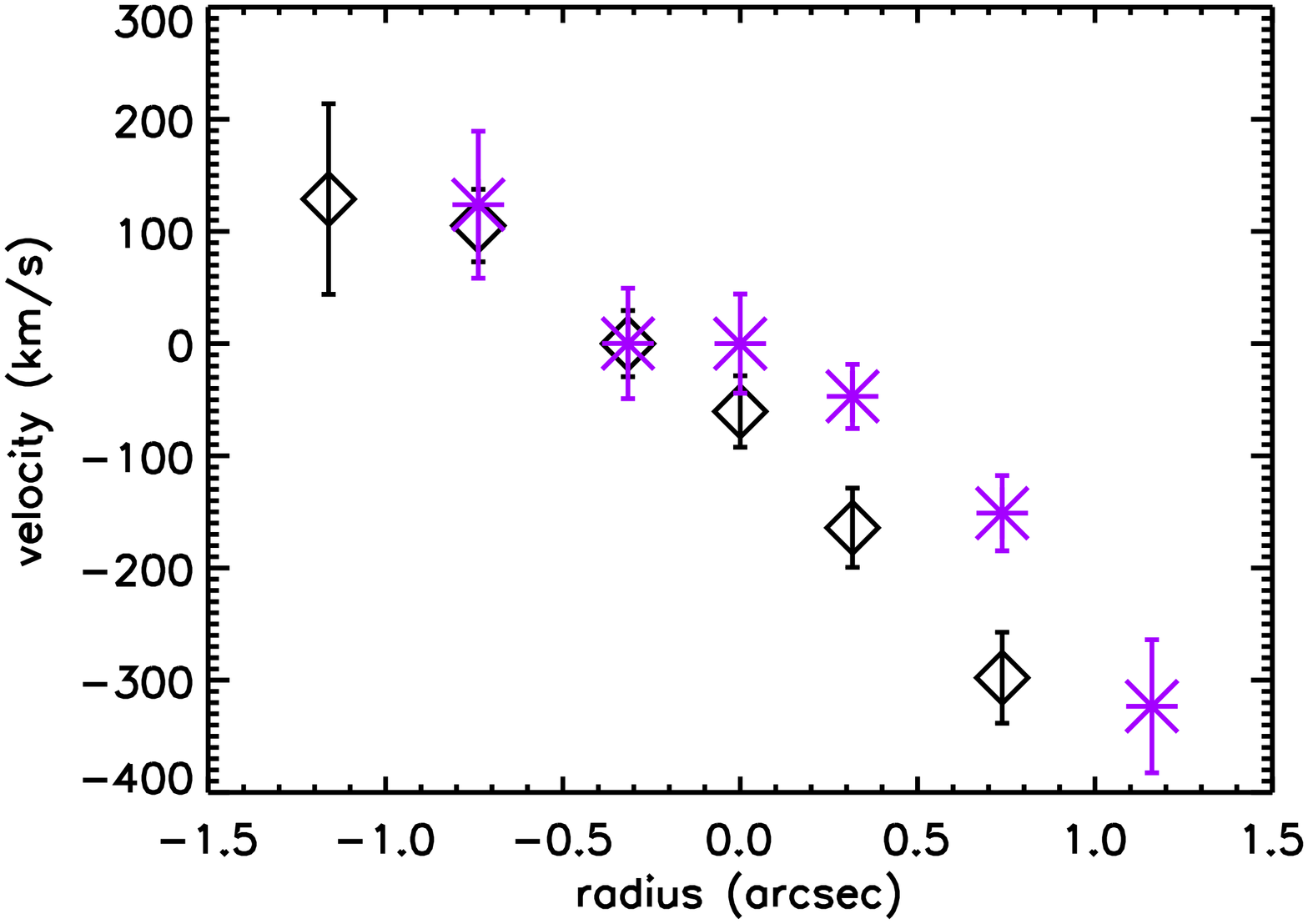}
      \end{minipage}
\hspace{-1.5cm}
\end{center}
\caption{The star--forming galaxies. From {\it left to right}: HST postage stamp image taken with the 
F775W filter and has the angle of the slit superimposed; radial H$\delta_{\rm A}$ profile; radial [OII]$\lambda 3727$ equivalent width profile; 
and projected streaming velocity profile along the slit.  Mask 1 observations
are plotted as {\it black} symbols and Mask 2 observations are  plotted as {\it blue} symbols. At the redshift of the cluster
1\,arcsecond corresponds to $\sim 6.4$\,kpc.}
\label{fig:em}
\end{figure*}

In identical fashion to Figure \ref{fig:psb}, the radial profiles along the slit of the 
H$\delta$ equivalent width, the [OII]$\lambda 3727$ 
equivalent width, and the streaming velocity for these 3 e(a) galaxies are shown in
columns 2--4 of Figure \ref{fig:em}, respectively. As in the case
of the E+A galaxies, we do not detect any radial gradient in the H$\delta$ line. 
We do, however, see a trend in the radial distribution
of the [OII]$\lambda 3727$ equivalent width in the sense that the equivalent width of the emission line
is stronger at small radii (i.e. larger negative equivalent width at small radii). This
trend is apparent in all three galaxies. In the case of DG\_356 (second row in Figure \ref{fig:em})
the slit is near to perpendicular to the major axis of the galaxy meaning the increase in emission
line equivalent width is toward the plane of the disk. In DG\_371 (3rd row in Figure \ref{fig:em}) we
see a more pronounced central concentration in the better quality Mask~1 data than in the Mask~2 data. This
is likely an example of the inferior image quality during the Mask~2 observations smoothing out the profile.
The high signal-to-noise ratio of the emission lines makes it easy to detect rotation in these galaxies. The
projected rotation curves are shown in the final column of Figure \ref{fig:em} -- again the difference in
seeing between the two masks is evident in the damping of the amplitude of the curve in Mask~2.

\subsubsection{Morphological properties}
\begin{figure*}[t]
   \begin{center}
     \begin{minipage}{0.95\textwidth}
\hspace{-1.3cm}
         \includegraphics[width=4.0cm, angle=0]{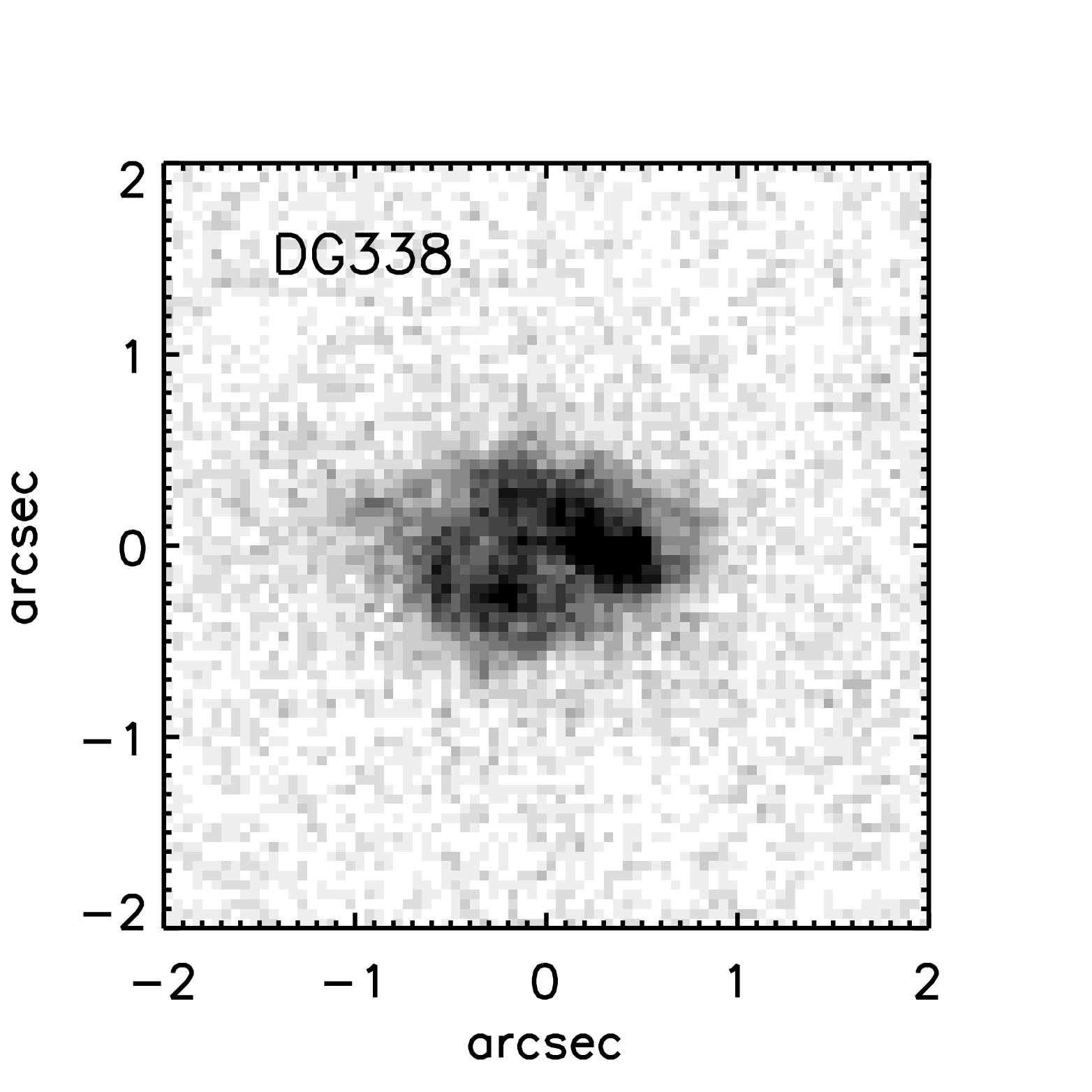}\hspace{-0.7cm}
         \includegraphics[width=4.0cm, angle=0]{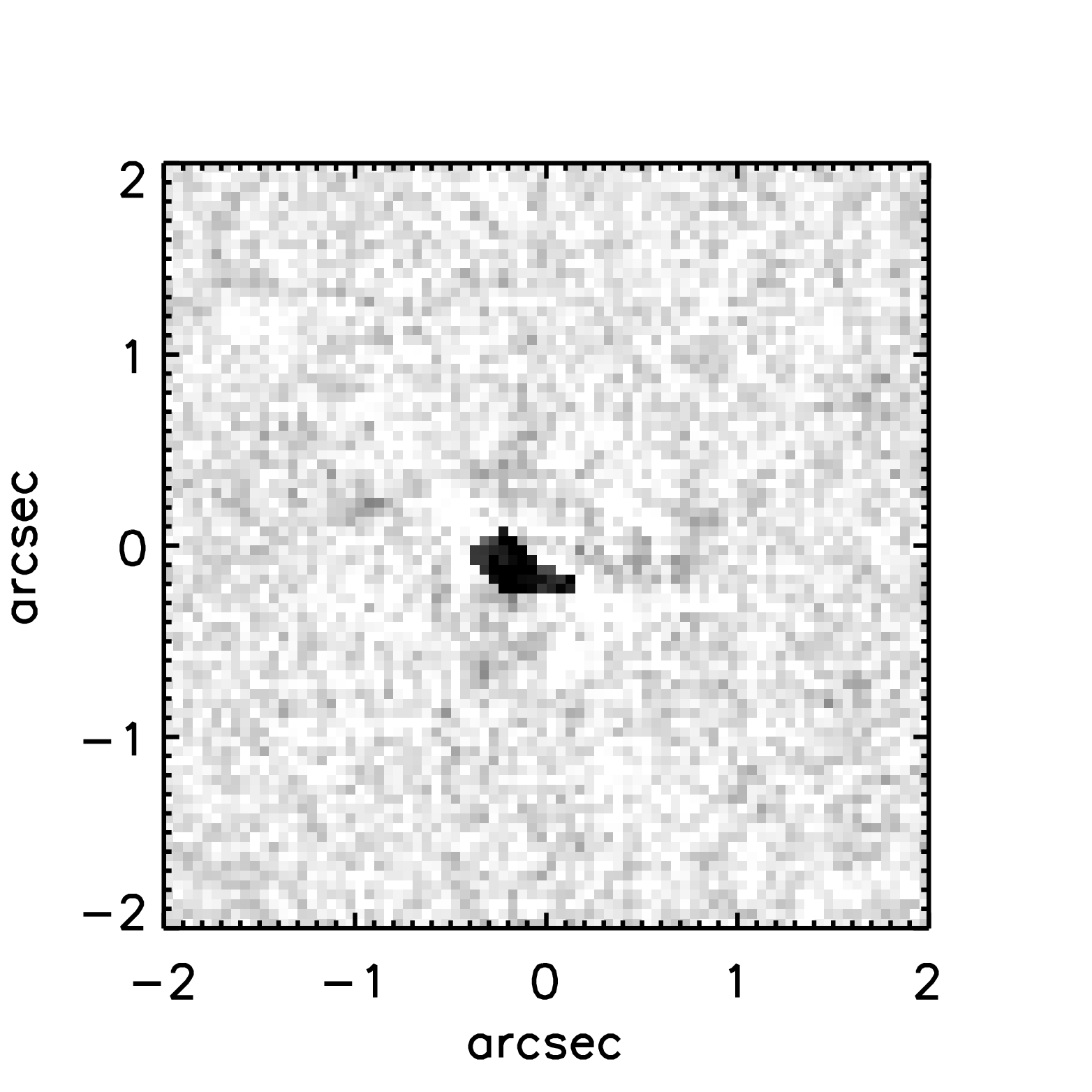}\hspace{-0.7cm}
         \includegraphics[width=3.6cm, angle=90]{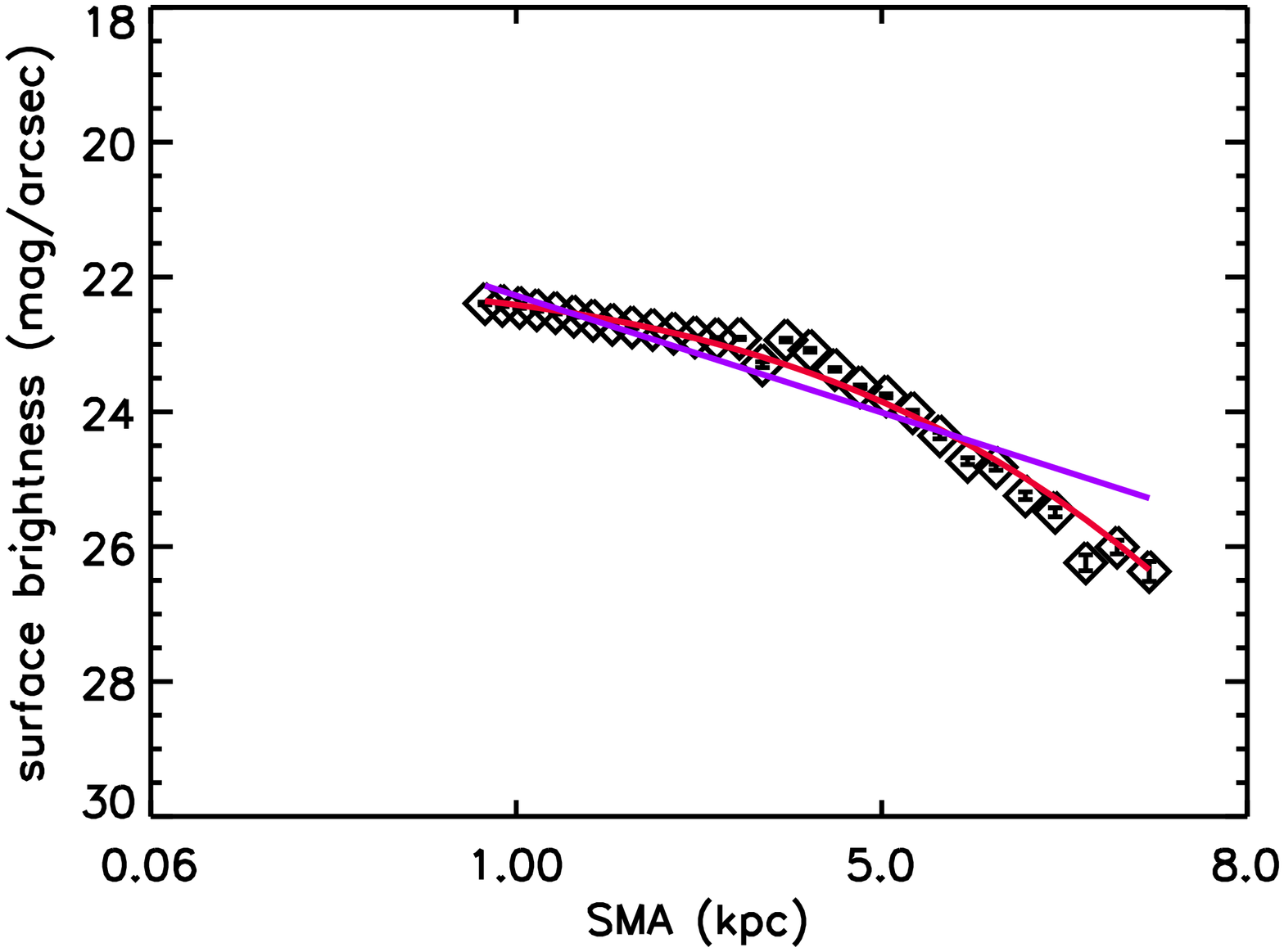}\hspace{-0.7cm}
         \includegraphics[width=5.4cm, angle=0]{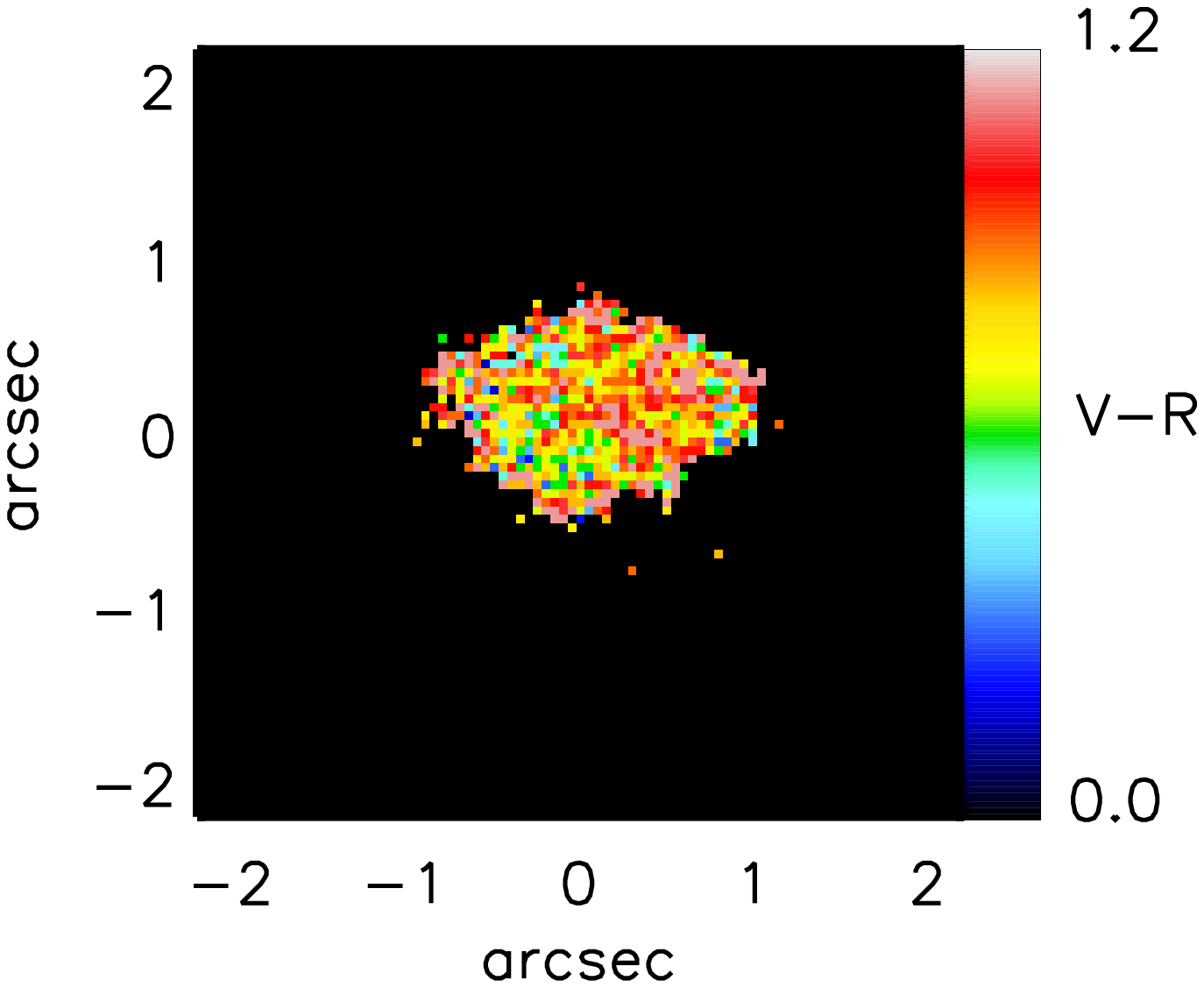}\hspace{-0.7cm}
      \end{minipage}
       \begin{minipage}{0.95\textwidth}
\hspace{-1.3cm}
         \includegraphics[width=4.0cm, angle=0]{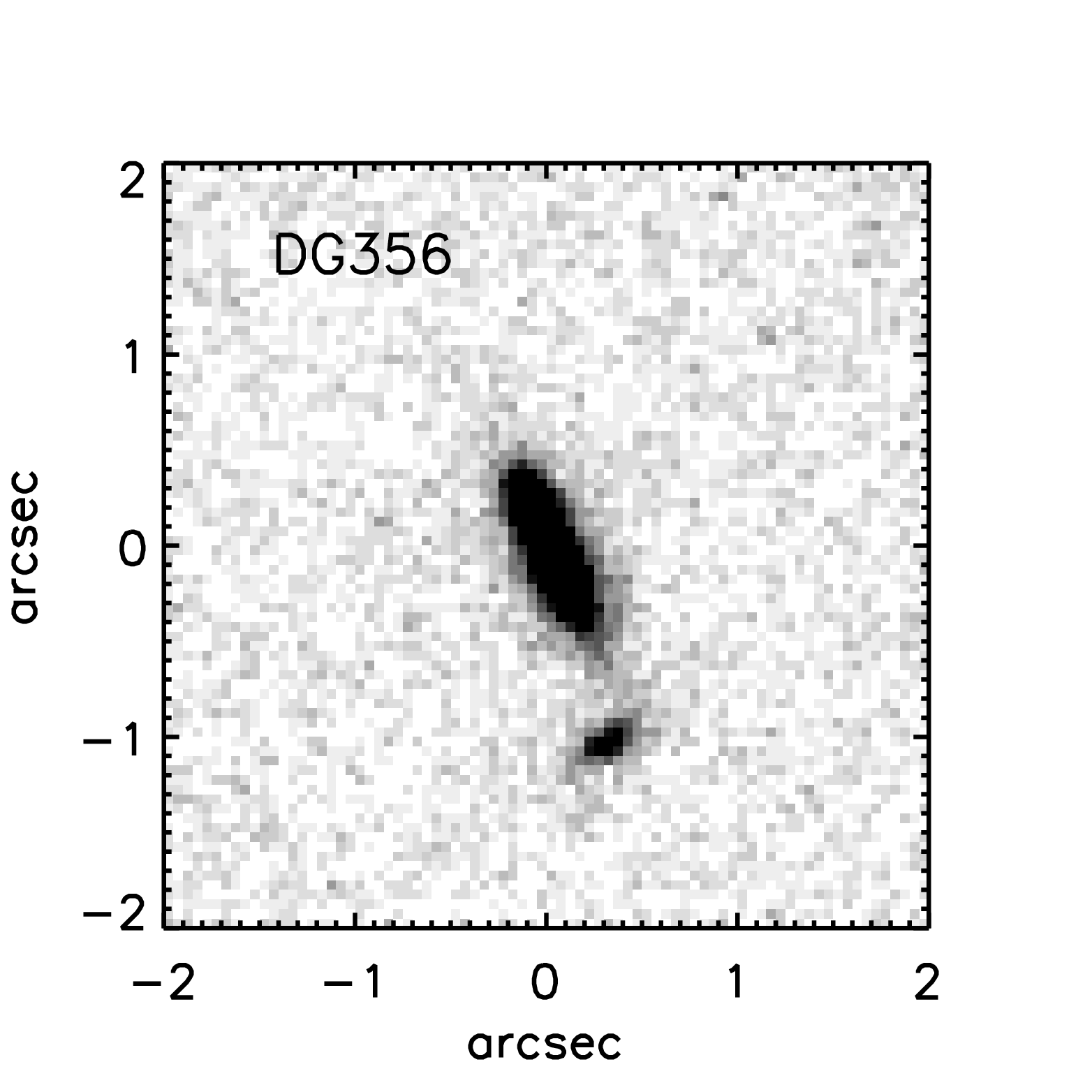}\hspace{-0.7cm}
         \includegraphics[width=4.0cm, angle=0]{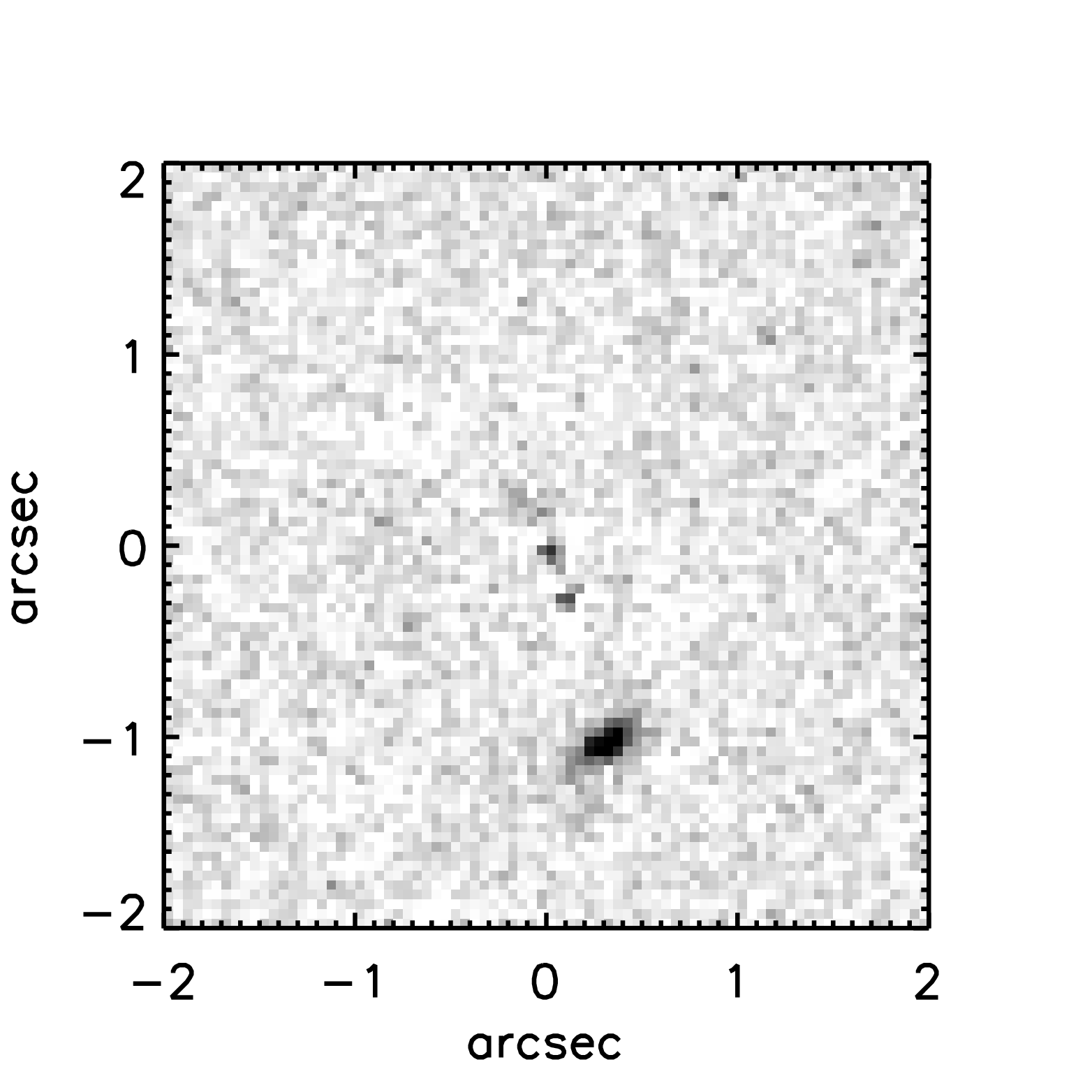}\hspace{-0.7cm}
         \includegraphics[width=3.6cm, angle=90, trim=0 0 0 0]{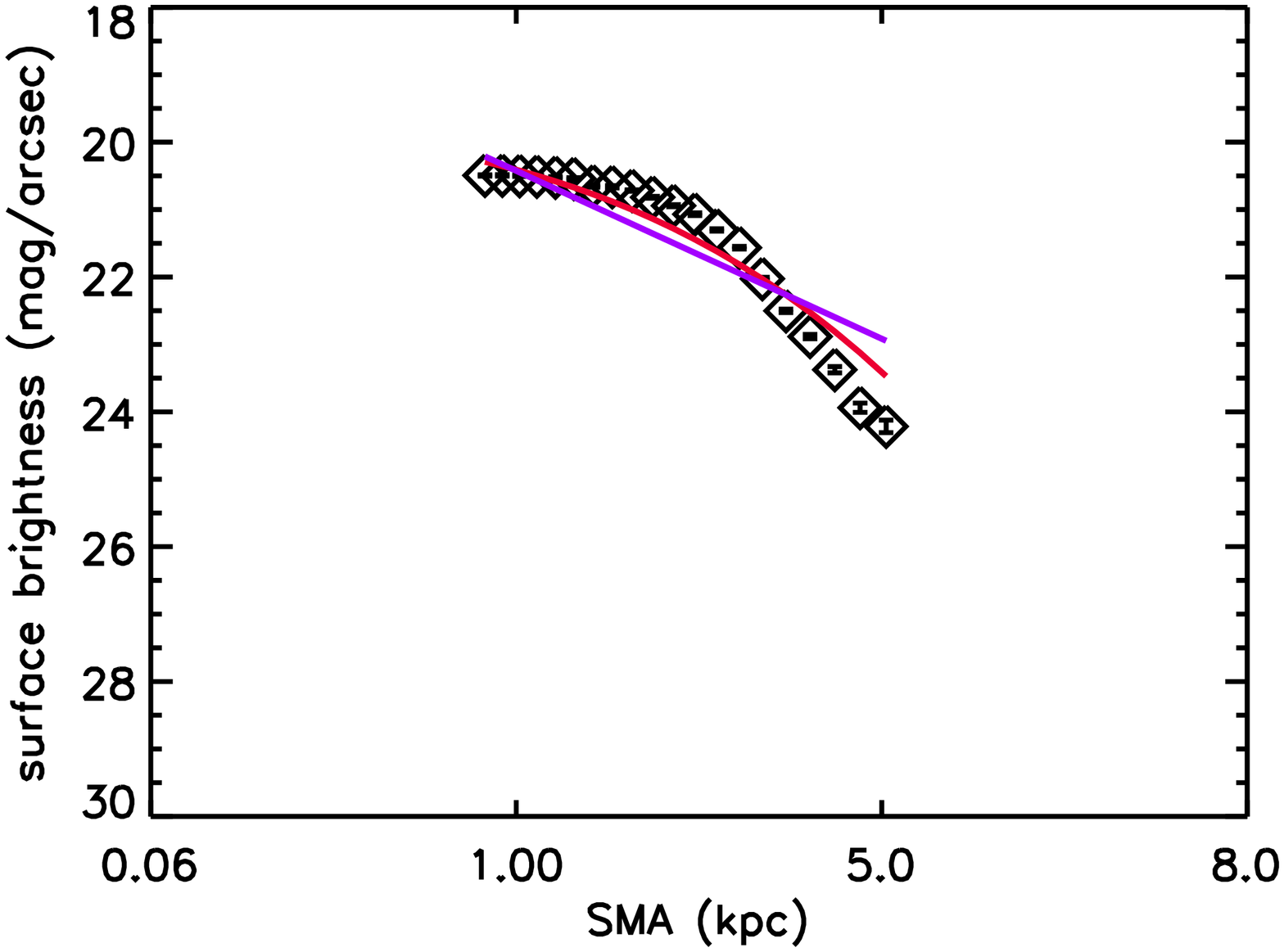}\hspace{-0.7cm}
         \includegraphics[width=5.4cm, angle=0, trim=0 0 0 0]{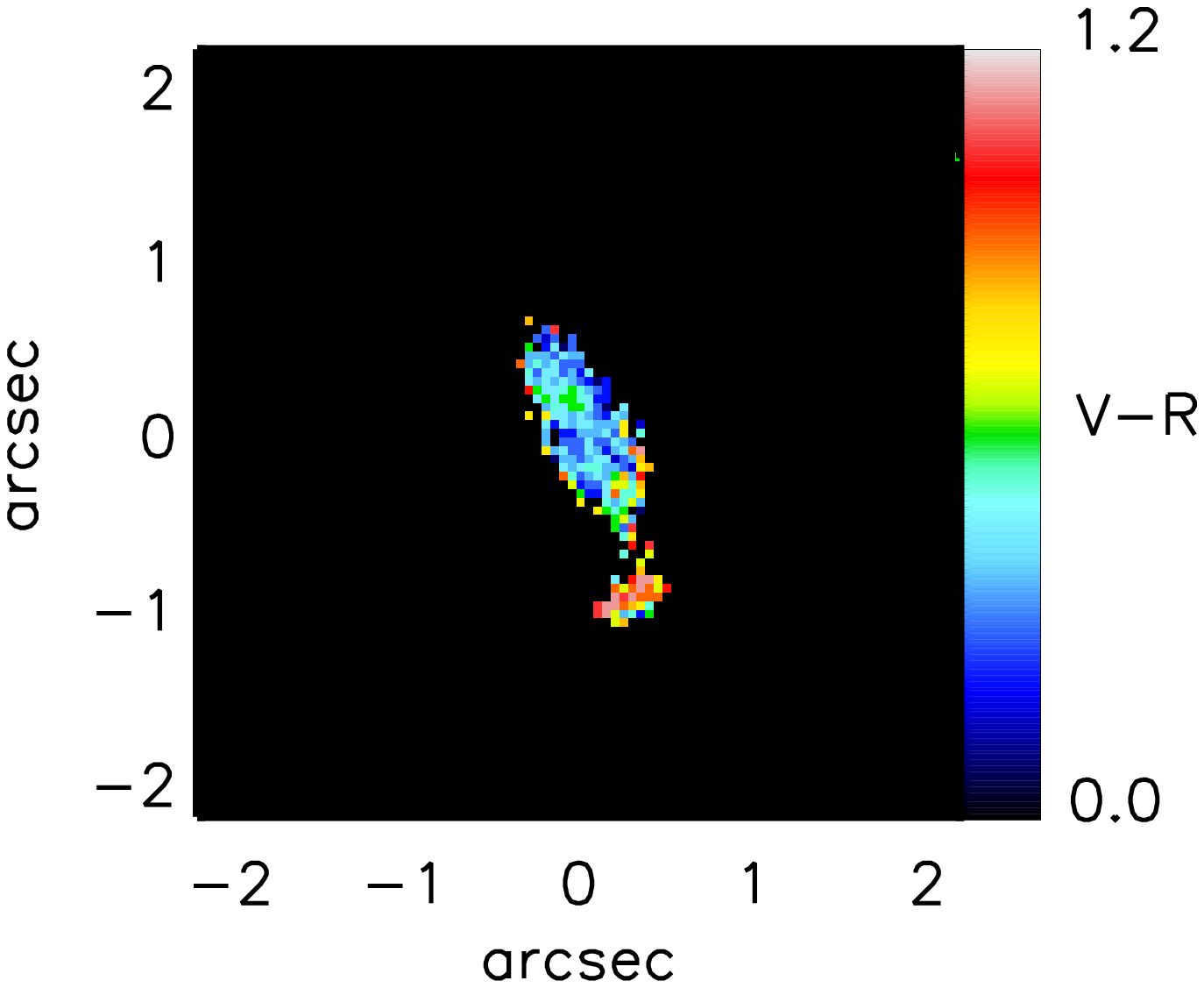}\hspace{-0.7cm}
      \end{minipage}
       \begin{minipage}{0.95\textwidth}
\hspace{-1.3cm}
         \includegraphics[width=4.0cm, angle=0, trim=0 0 0 0]{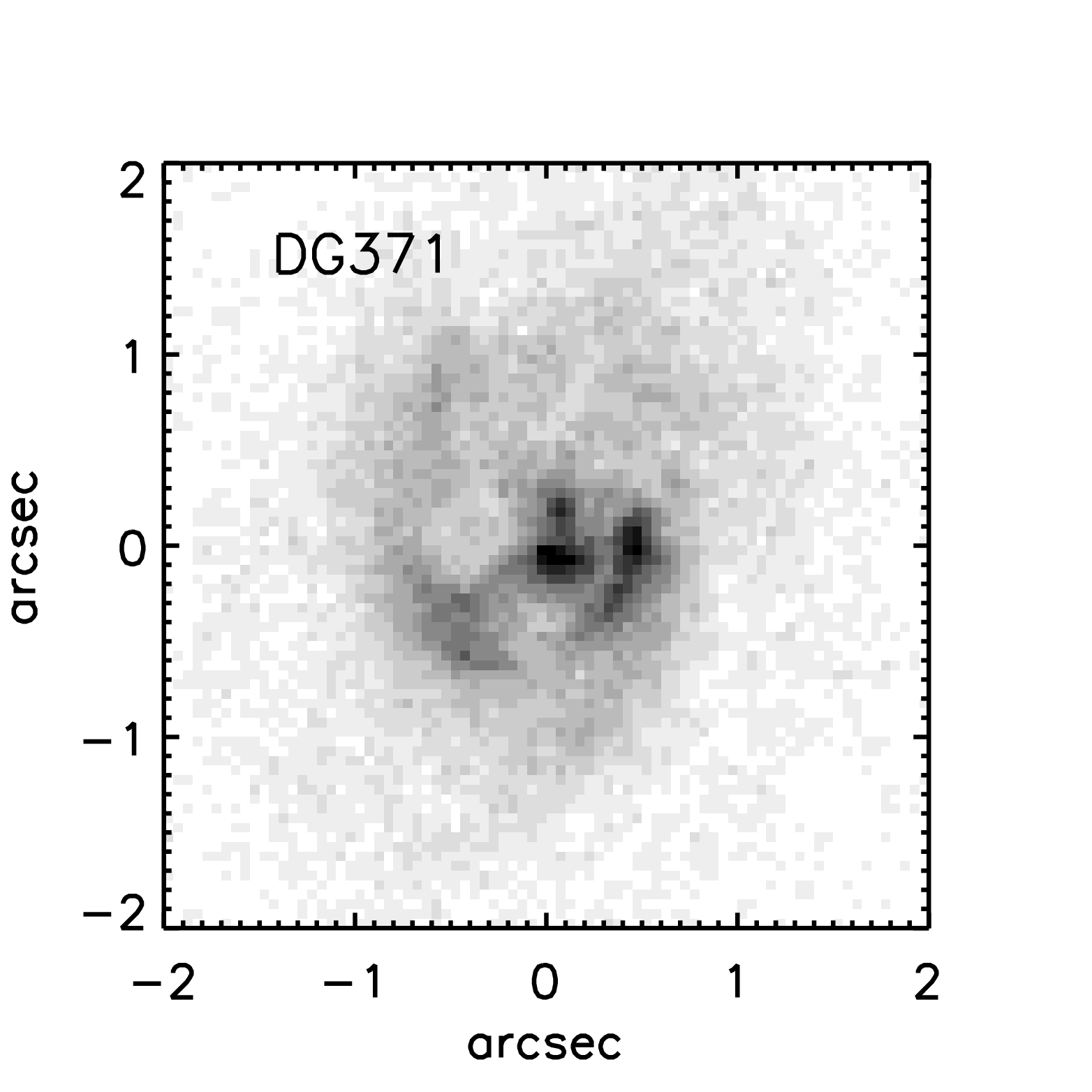}\hspace{-0.7cm}
        \includegraphics[width=4.0cm, angle=0, trim=0 0 0 0]{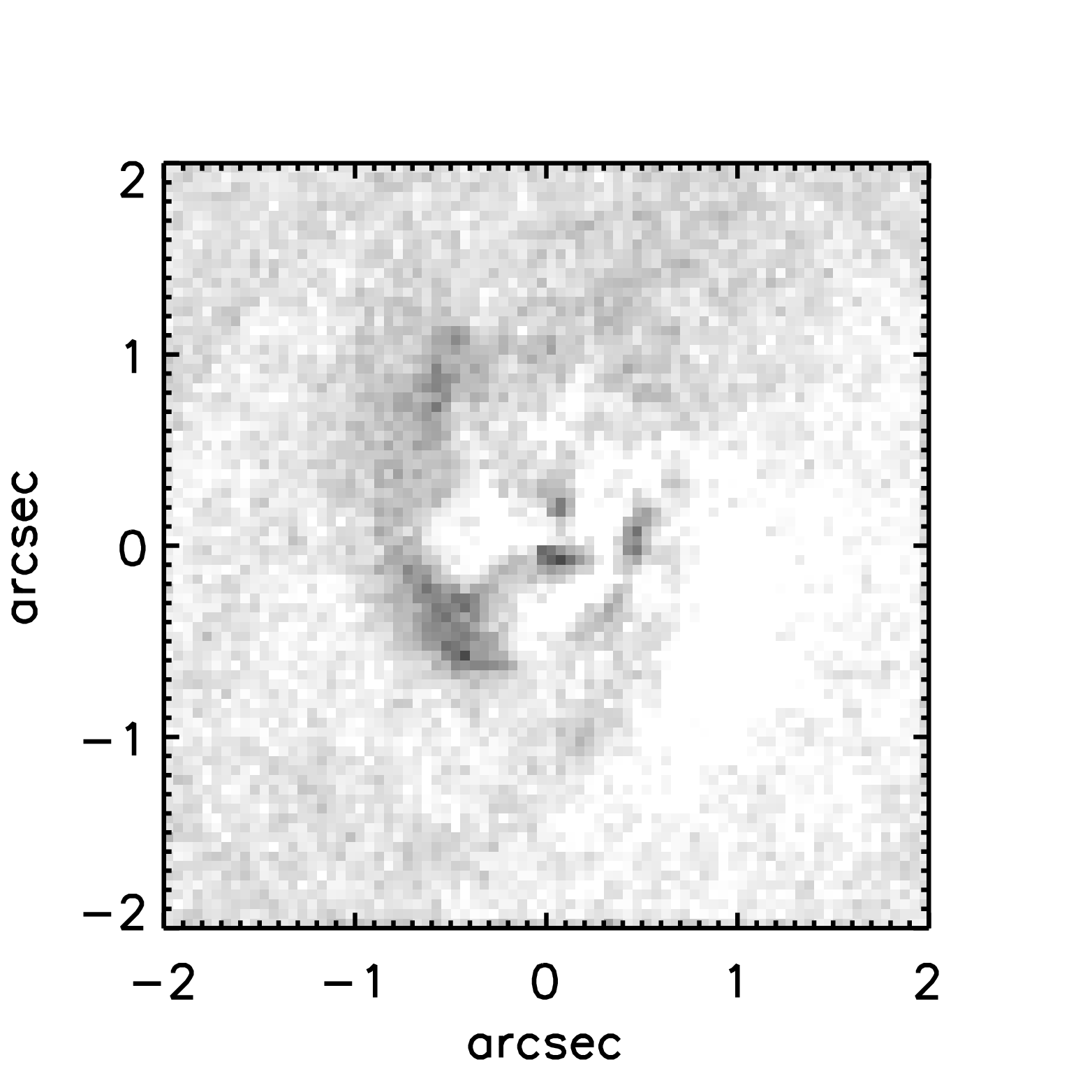}\hspace{-0.7cm}
         \includegraphics[width=3.6cm, angle=90, trim=0 0 0 0]{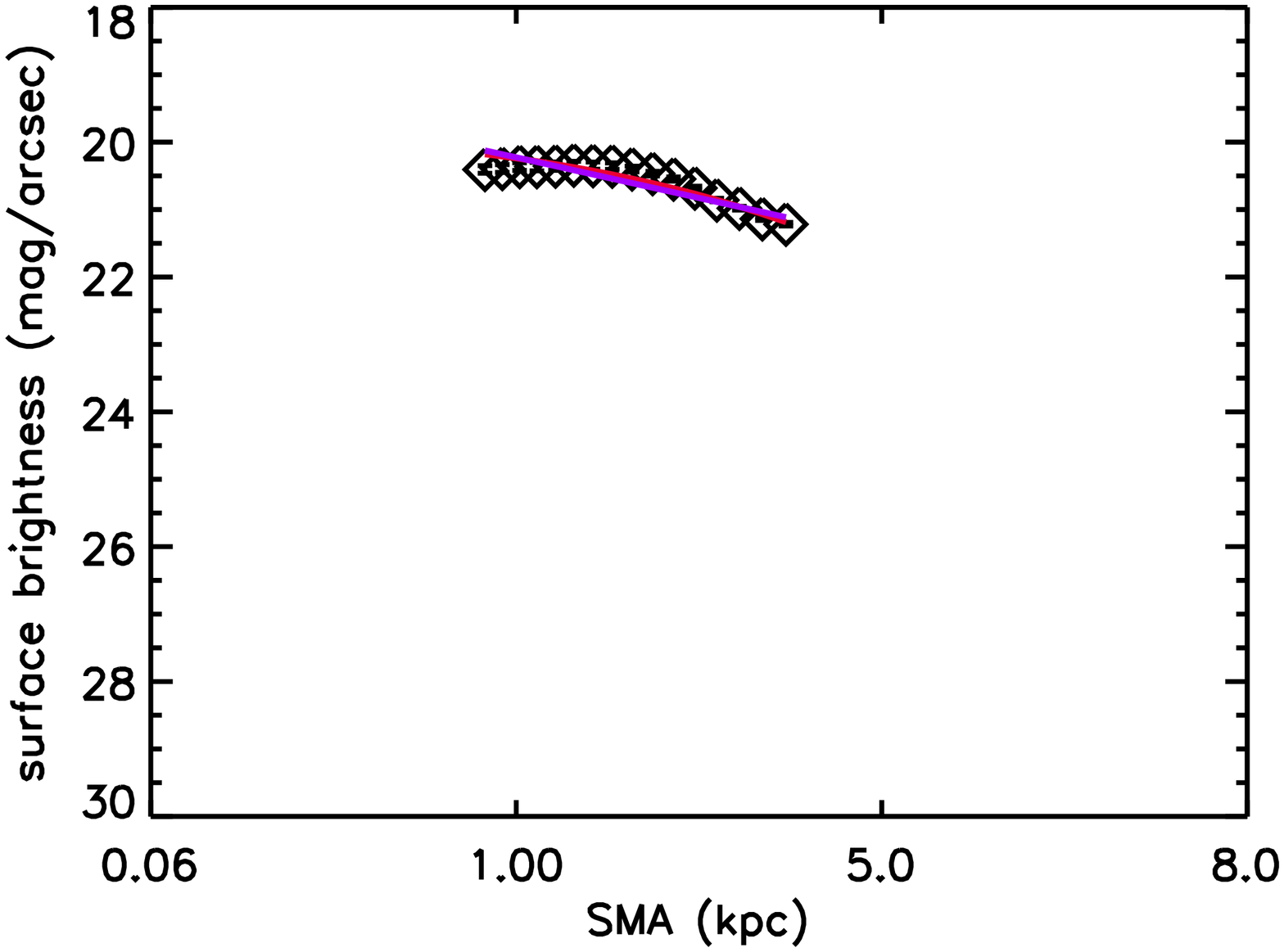}\hspace{-0.7cm}
         \includegraphics[width=5.4cm, angle=0, trim=0 0 0 0]{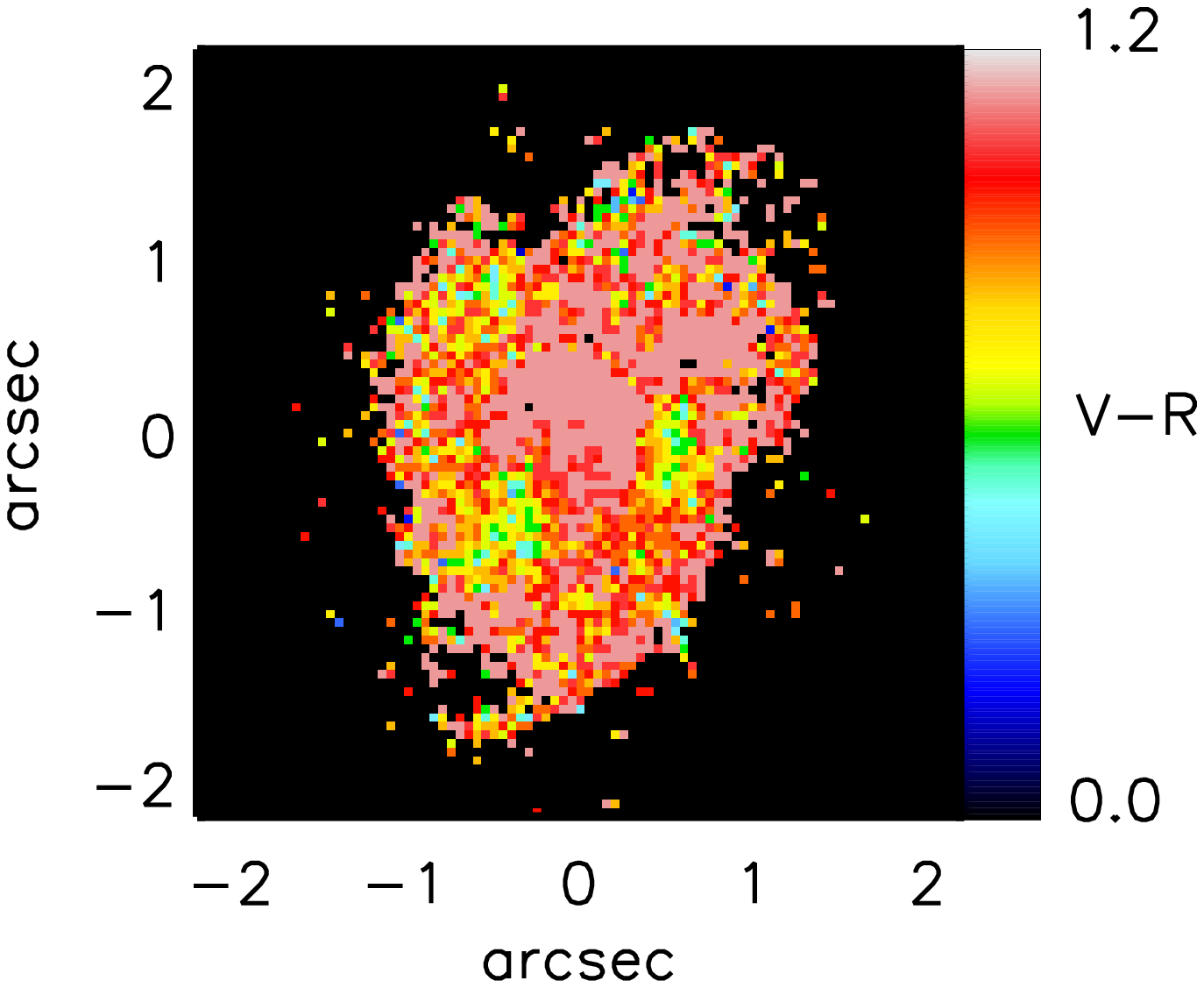}\hspace{-0.7cm}
      \end{minipage}
\hspace{-1.0cm}
\end{center}
\caption{Some morphological and photometric diagnostics of the starburst galaxies. The leftmost 
column shows a postage stamp ACS image in the F775W passband. The second column shows a residual image 
after subtraction of a symmetric elliptical model. The third column is the isophotal profile for each
galaxy along with fits of an exponential model ({\it red line}) and a de Vaucouleurs model ({\it blue line}). 
The final column shows a V-R colour image (actually F606W-F775W).}
\label{fig:emmorph}
\end{figure*}

In Figure \ref{fig:emmorph} we plot the same set of morphological and photometric
diagnostics for the e(a) galaxies as we did for the E+As in Figure \ref{fig:psbmorph}.
The e(a) galaxies are all of late and irregular morphological type. The model subtracted
images show significant residuals for two of the three (DG\_338 and DG\_371) galaxies and the 
asymmetry is also illustrated by the irregular isophotal profiles (third column of Figure \ref{fig:emmorph}). 
Although irregular, the isophotal profiles of the e(a)s are closer to exponential than de Vaucouleur profiles
-- compatible with their late type status. The morphological properties of the three e(a) galaxies 
is consistent with them being irregular and/or interacting systems.

\section{The effects of seeing}
As mentioned above, attempting to measure radial gradients in equivalent width is hampered by 
the convolution of the galaxy light with the seeing disk. This problem is a natural consequence
of the expected angular scale of equivalent width gradients in E+A galaxies at this redshift
being comparable to the seeing scale \citep{pracy05}. The observed radial gradients in line
index strength depend on both the intrinsic spatial distribution of the line equivalent width and 
the overall distribution of the continuum light, after having suffered the smearing effects
of the seeing 
\citep[see][for an in-depth discussion of the convolution effects on spatially resolved spectroscopy, with
an emphasis on IFU data]{pracy05}. The convolution of the intrinsic radial gradient with the
seeing disk always acts to flatten out any gradients.

Simulations of the expected H$\delta$ equivalent width gradients in E+A galaxies were performed 
by \citet{pracy05} for both merger and truncation scenarios (see also \citet{bekki05} for more
detailed modeling of E+A galaxy formation and its observable effects). For each case, \citet{pracy05} simulated
the radial H$\delta$ gradients at three different time steps. In the case of the merger model, 
the remnant galaxy possesses a strong negative H$\delta$ equivalent width gradient (stronger absorption
toward the centre) contained within the central $\sim 2$\,kpc in radius. As time progresses the 
strong central H$\delta$ equivalent
width and gradient decrease as the A--star population dies out until the signature is gone after $\sim$1.5\,Gyr.
A different pattern is seen  in the truncation model which begins with uniformly strong H$\delta$ and 
little gradient, and evolves with time to have a positive radial gradient. These models represent 
the only predictions in the literature of radial H$\delta$ gradients in E+A galaxies.
\begin{figure*}
   \begin{center}
     \begin{minipage}{0.95\textwidth}
\hspace{-0.5cm}
        \includegraphics[width=3.7cm, angle=90]{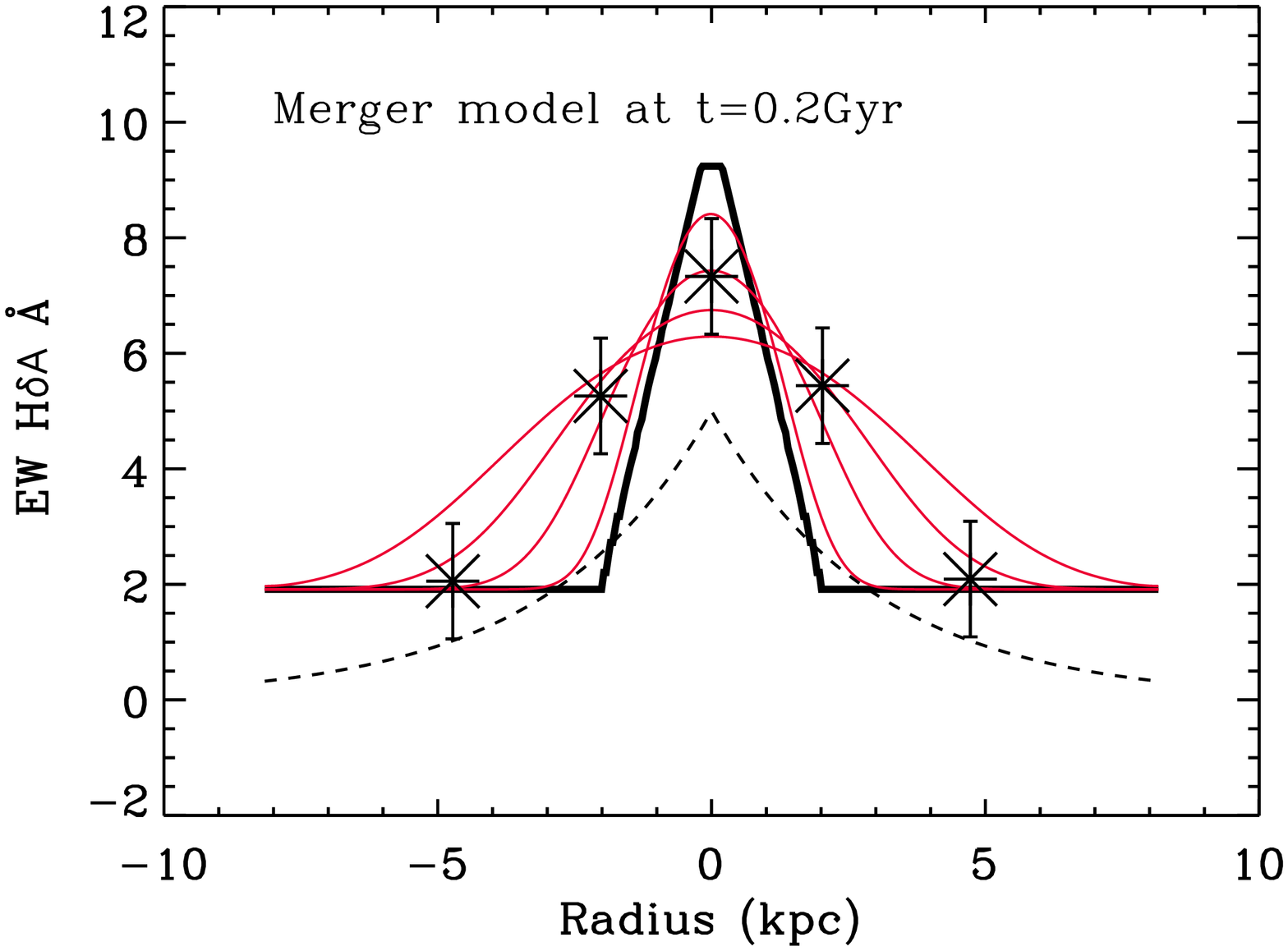}
\hspace{-0.5cm}
         \includegraphics[width=3.7cm, angle=90]{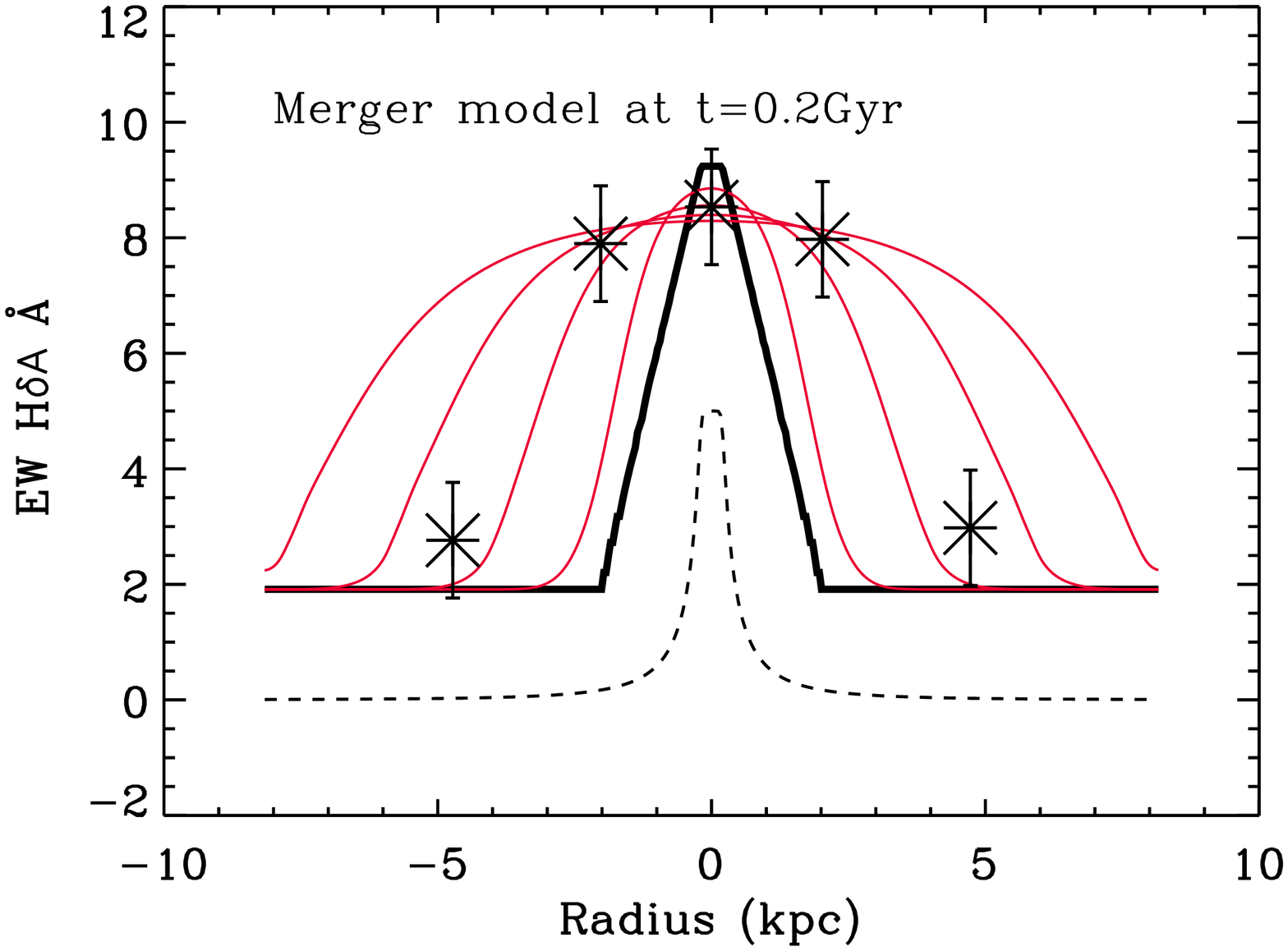}
\hspace{-0.5cm}
         \includegraphics[width=3.7cm, angle=90]{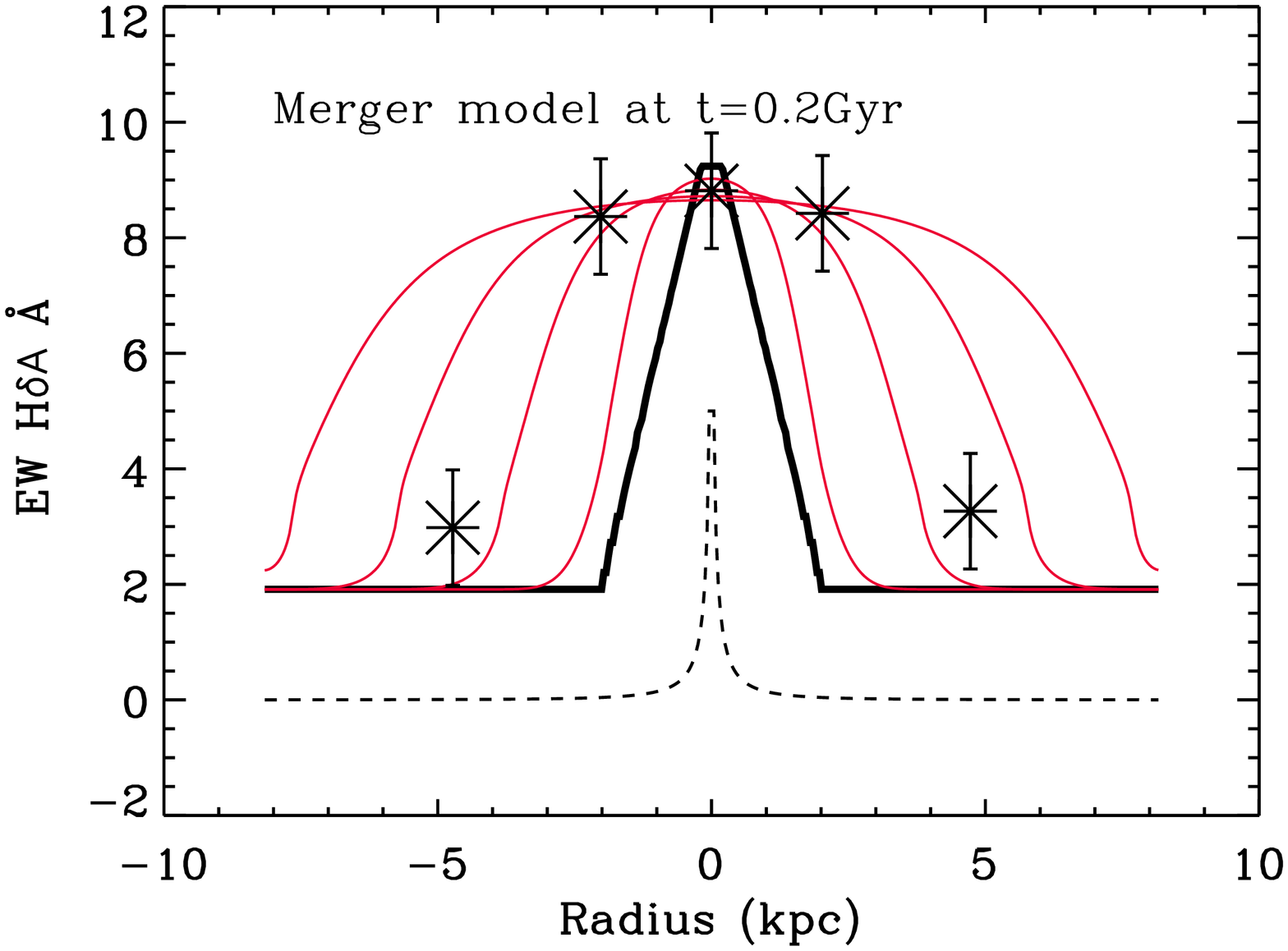}
      \end{minipage}
       \begin{minipage}{0.95\textwidth}
\hspace{-0.5cm}
        \includegraphics[width=3.7cm, angle=90]{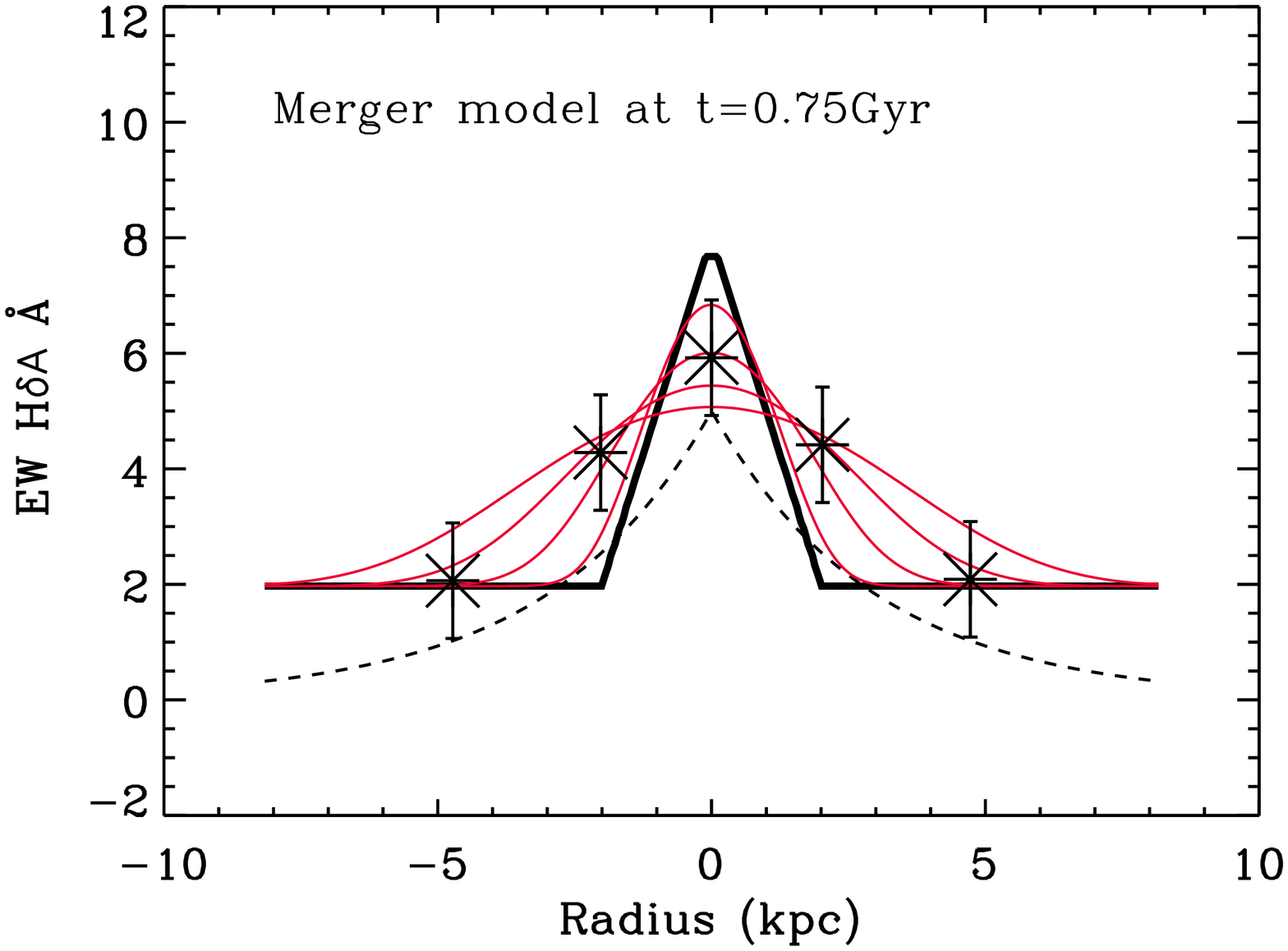}
\hspace{-0.5cm}
         \includegraphics[width=3.7cm, angle=90]{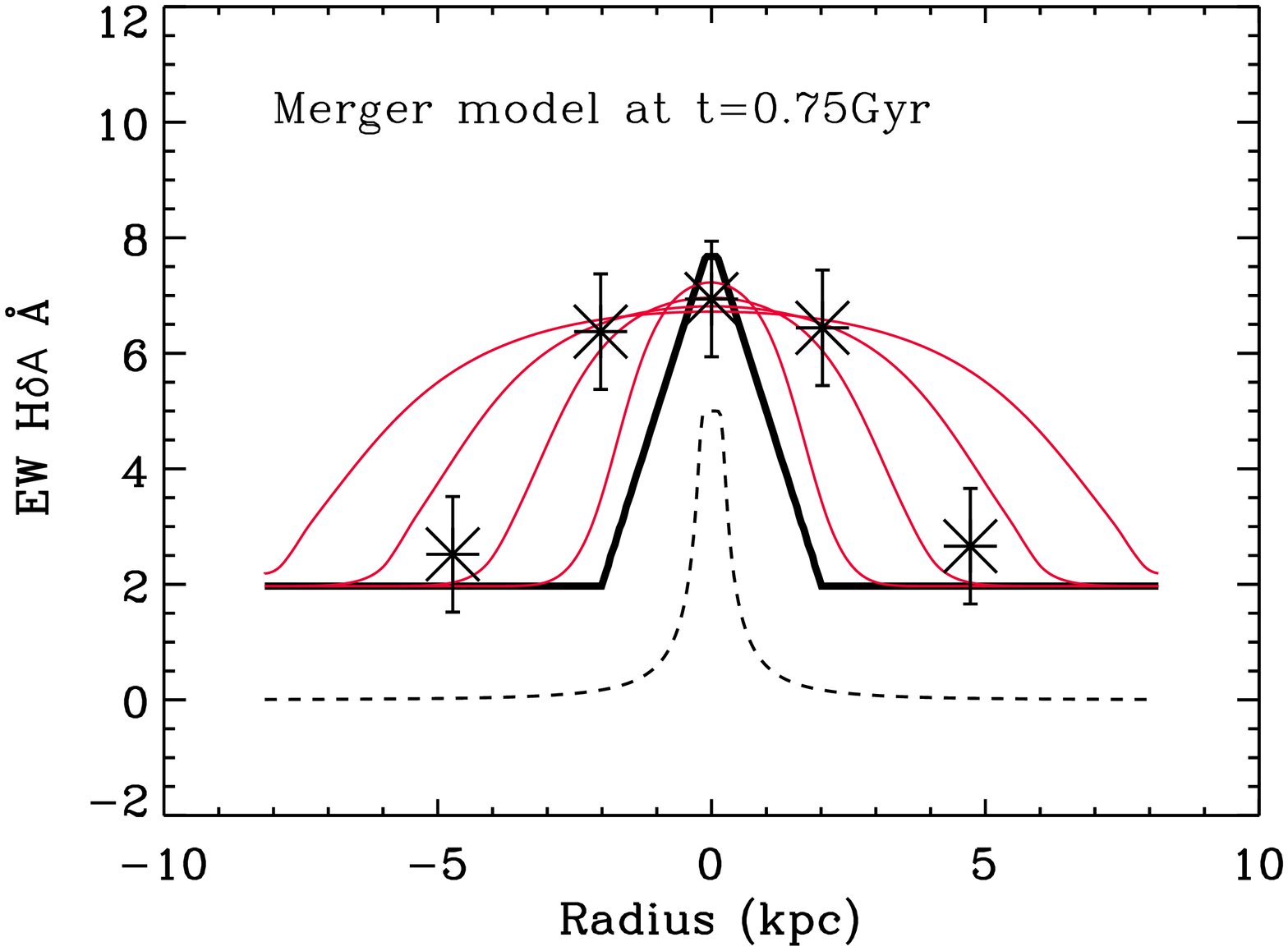}
\hspace{-0.5cm}
         \includegraphics[width=3.7cm, angle=90]{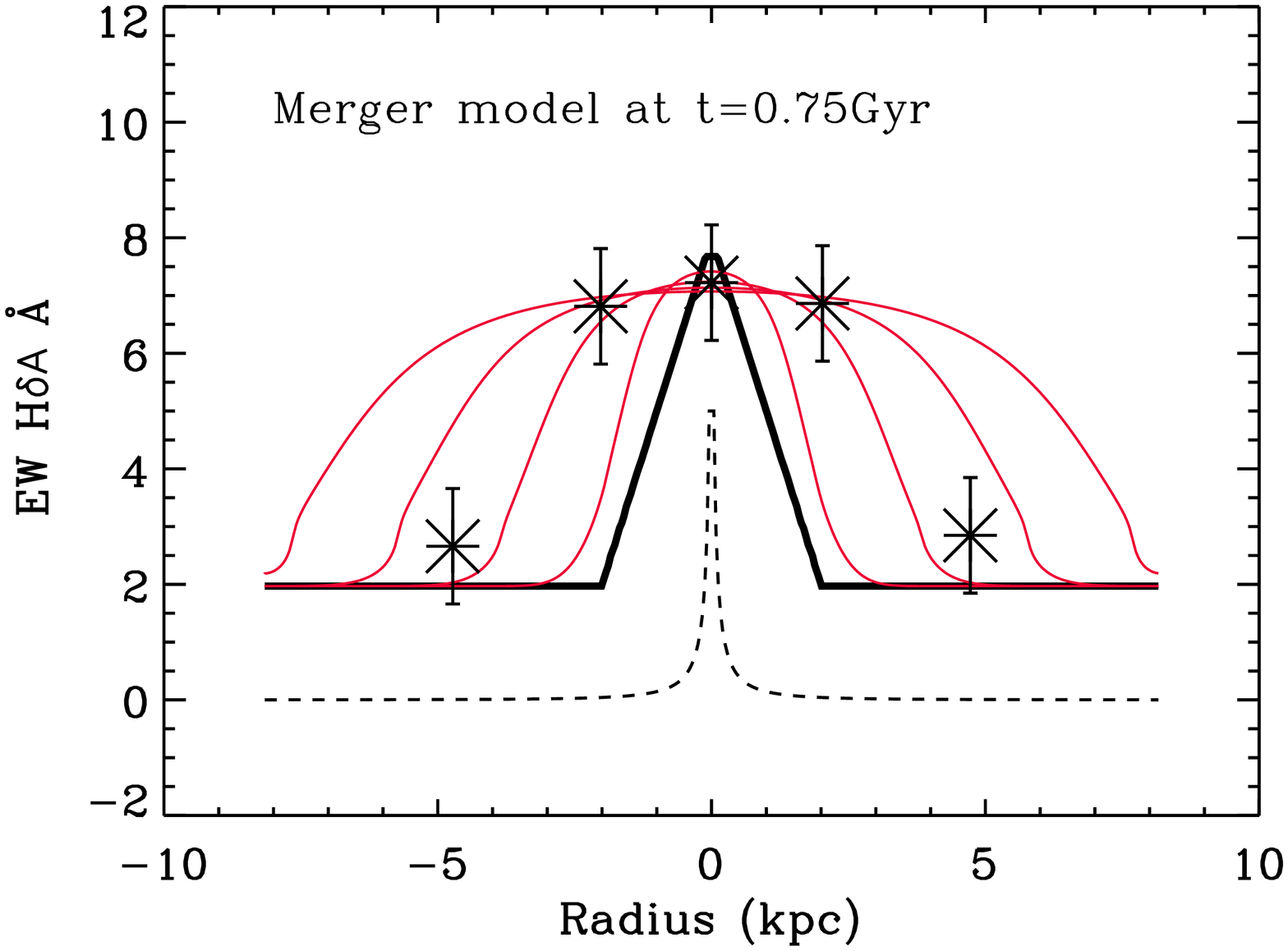}
      \end{minipage}
       \begin{minipage}{0.95\textwidth}
\hspace{-0.5cm}
        \includegraphics[width=3.7cm, angle=90]{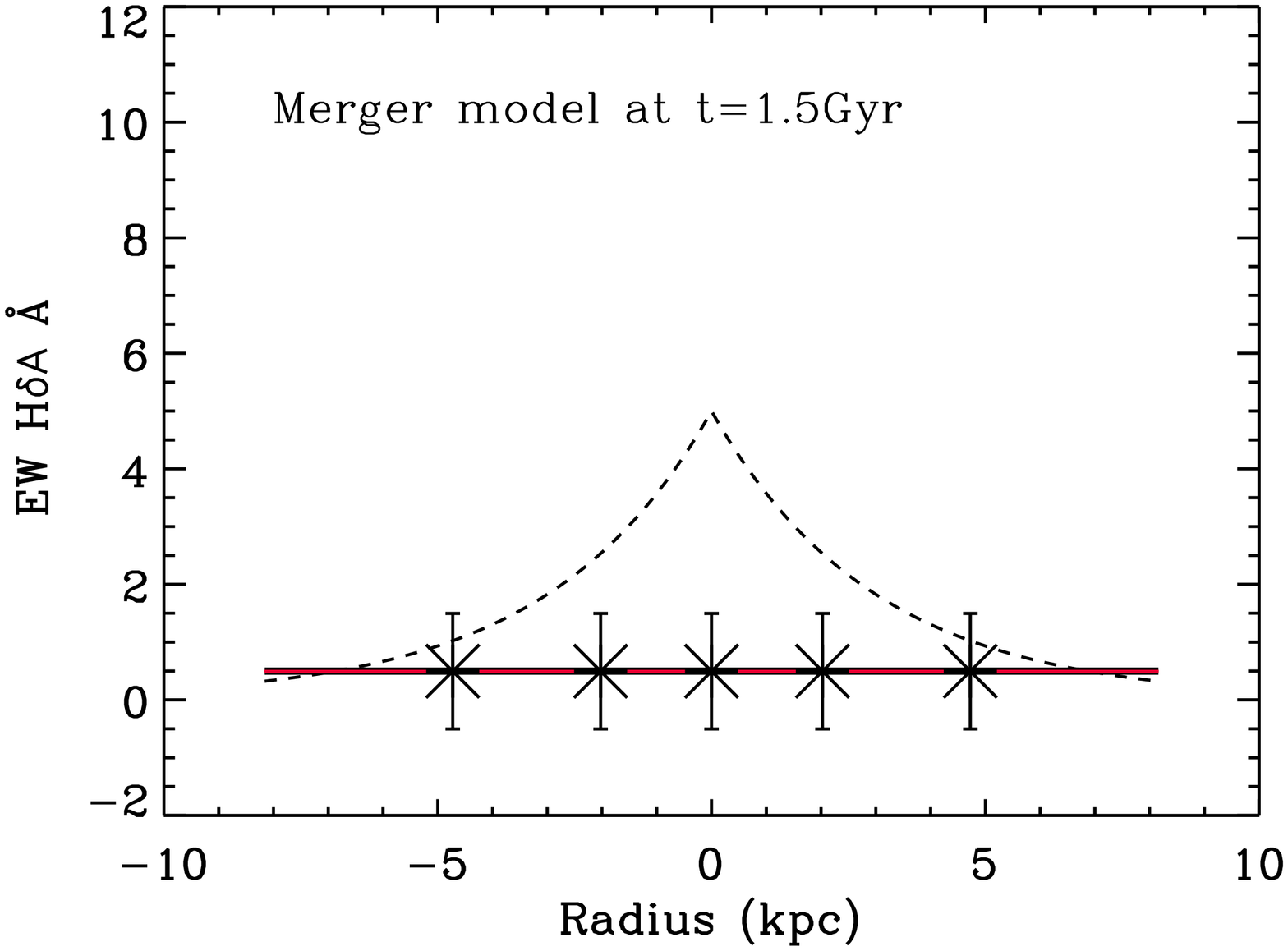}
\hspace{-0.5cm}
         \includegraphics[width=3.7cm, angle=90]{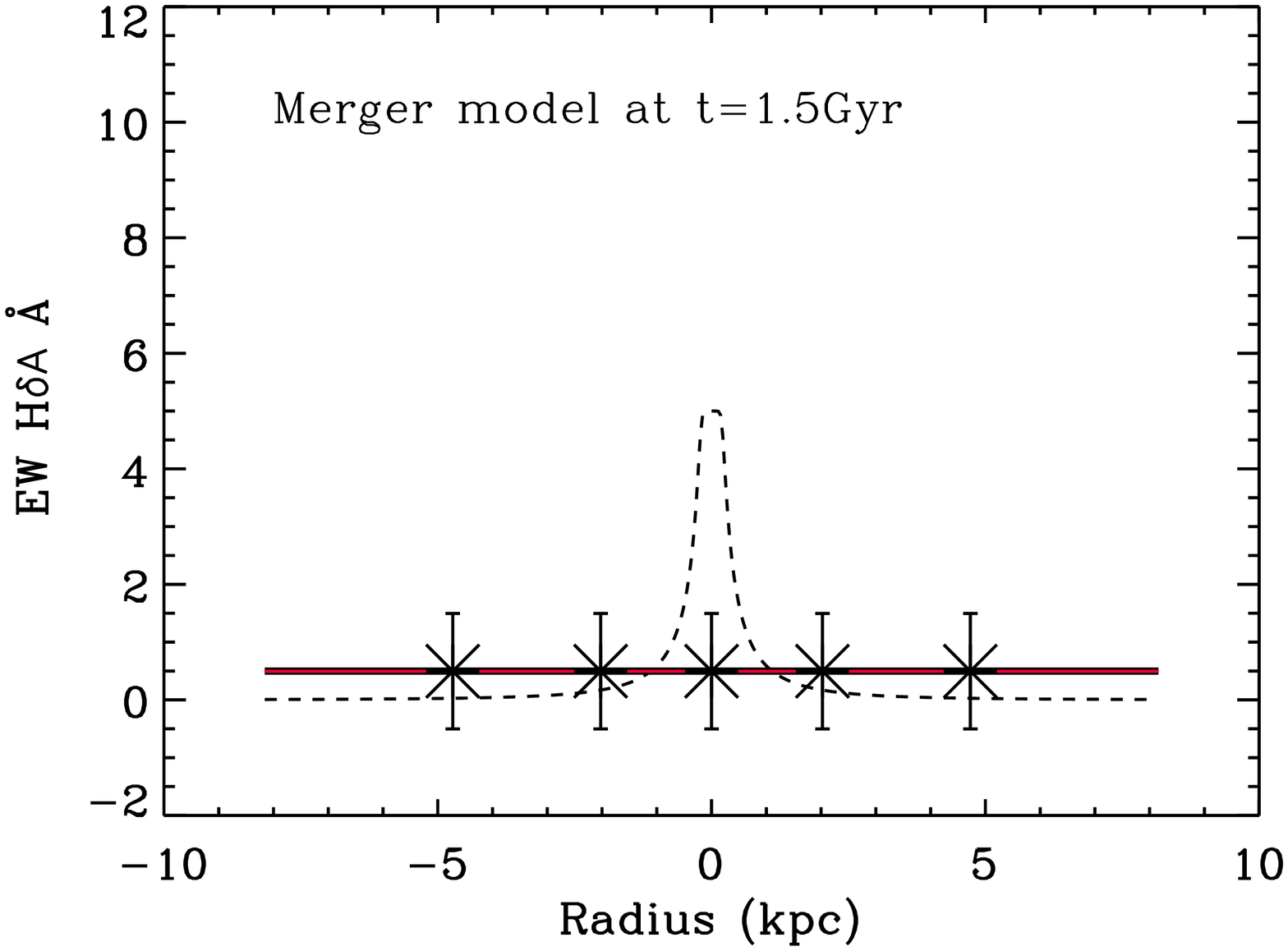}
\hspace{-0.5cm}
         \includegraphics[width=3.7cm, angle=90]{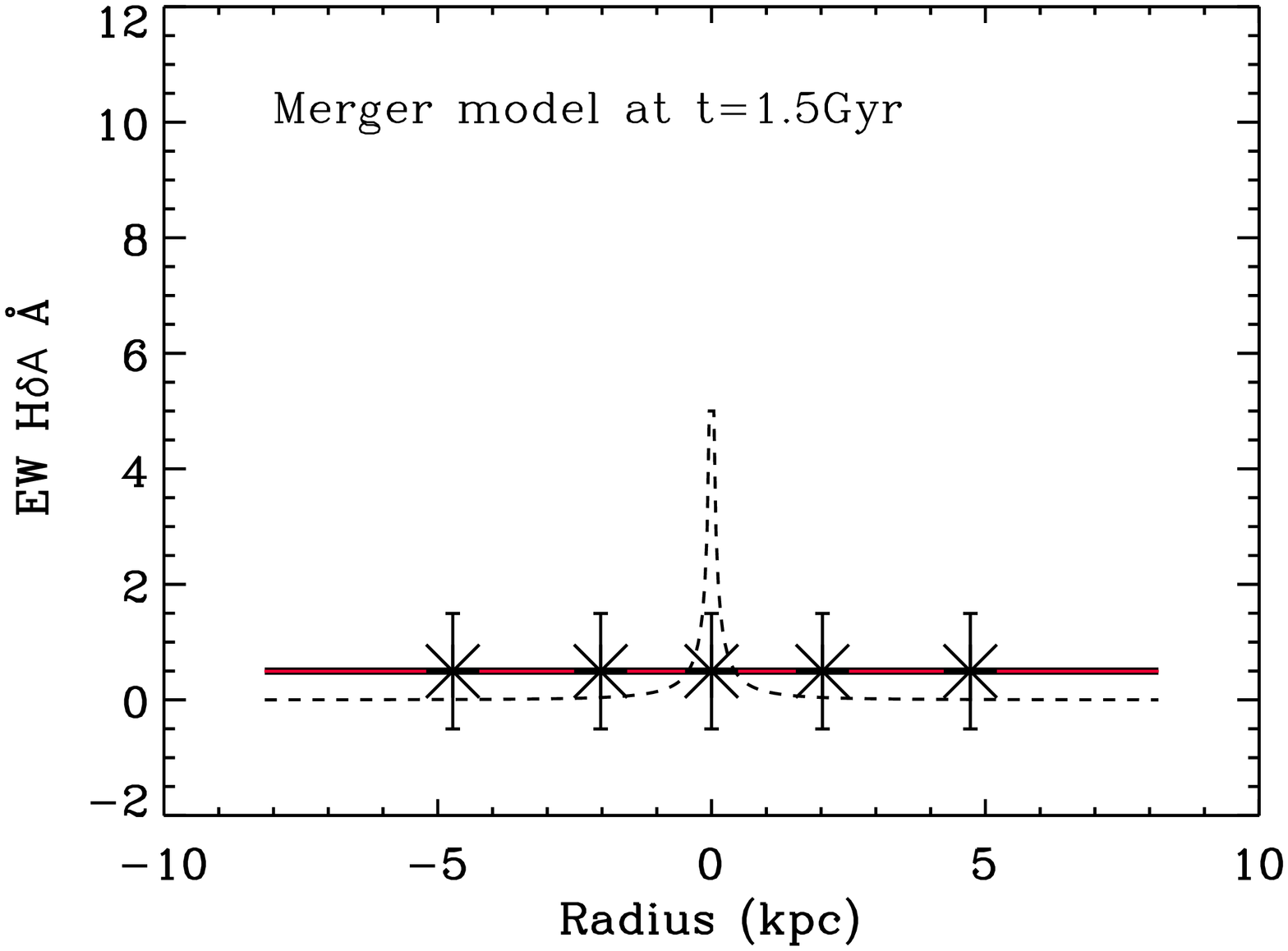}
      \end{minipage}
     \begin{minipage}{0.95\textwidth}
\hspace{-0.5cm}
        \includegraphics[width=3.7cm, angle=90]{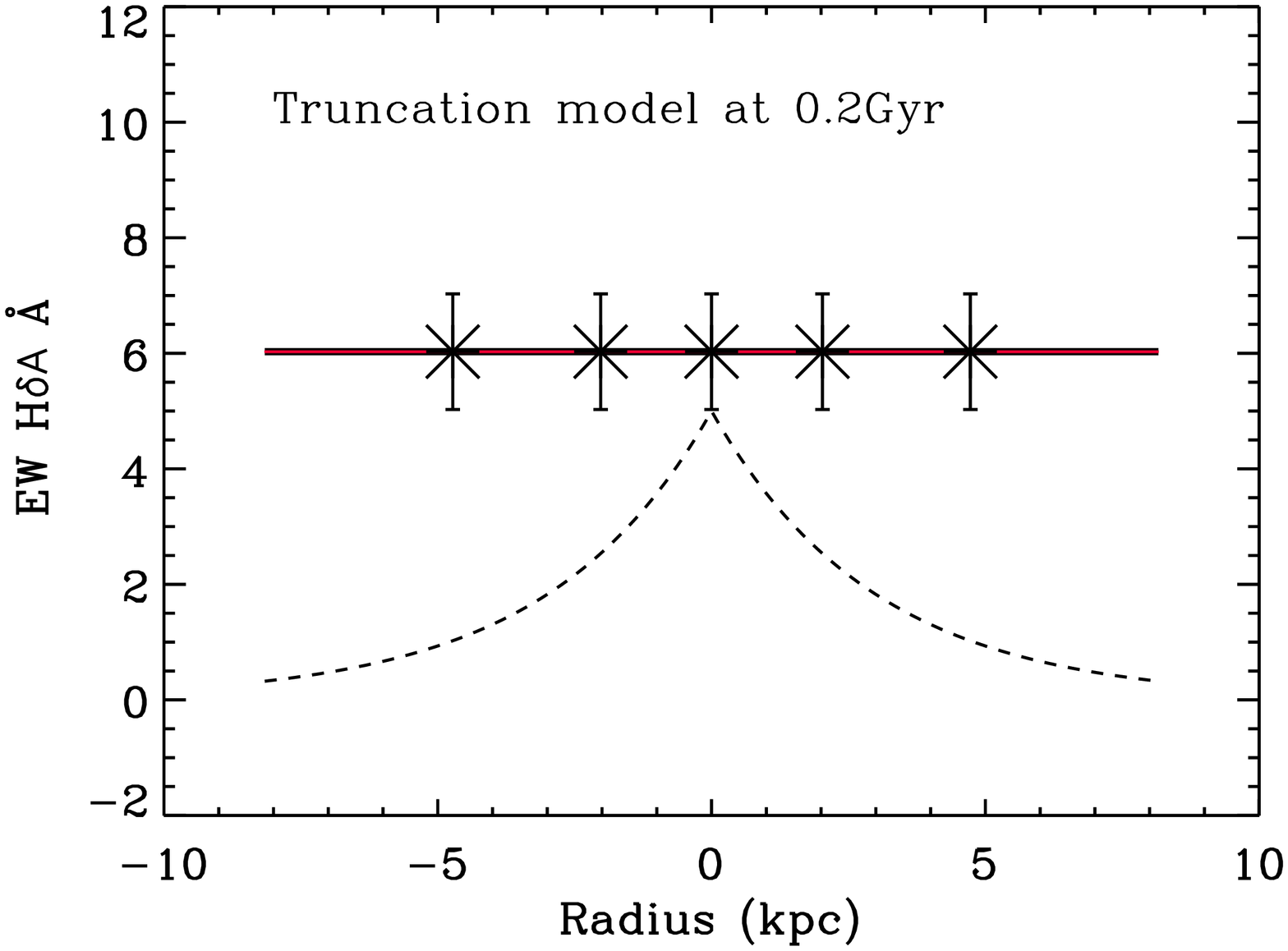}
\hspace{-0.5cm}
         \includegraphics[width=3.7cm, angle=90]{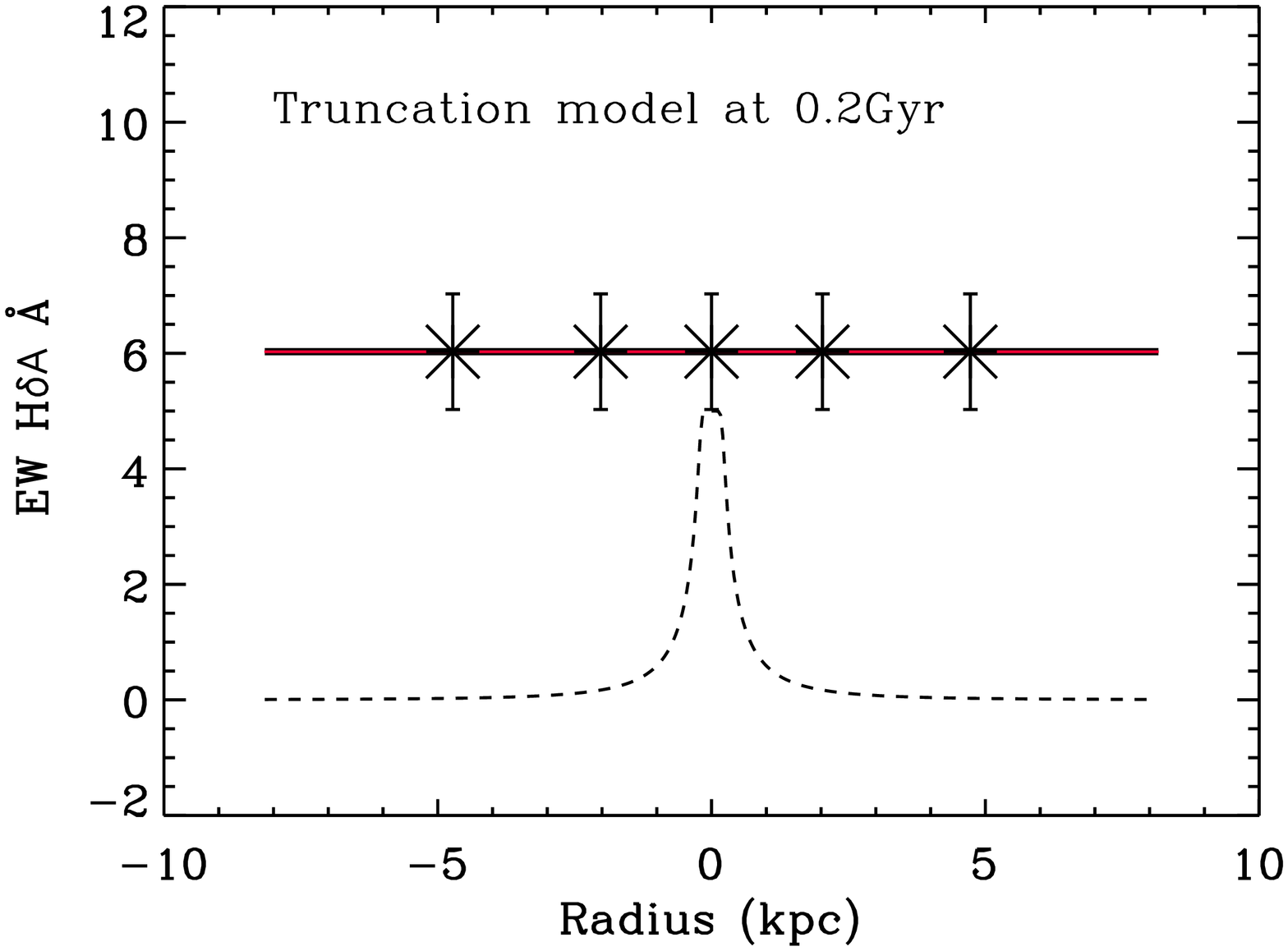}
\hspace{-0.5cm}
         \includegraphics[width=3.7cm, angle=90]{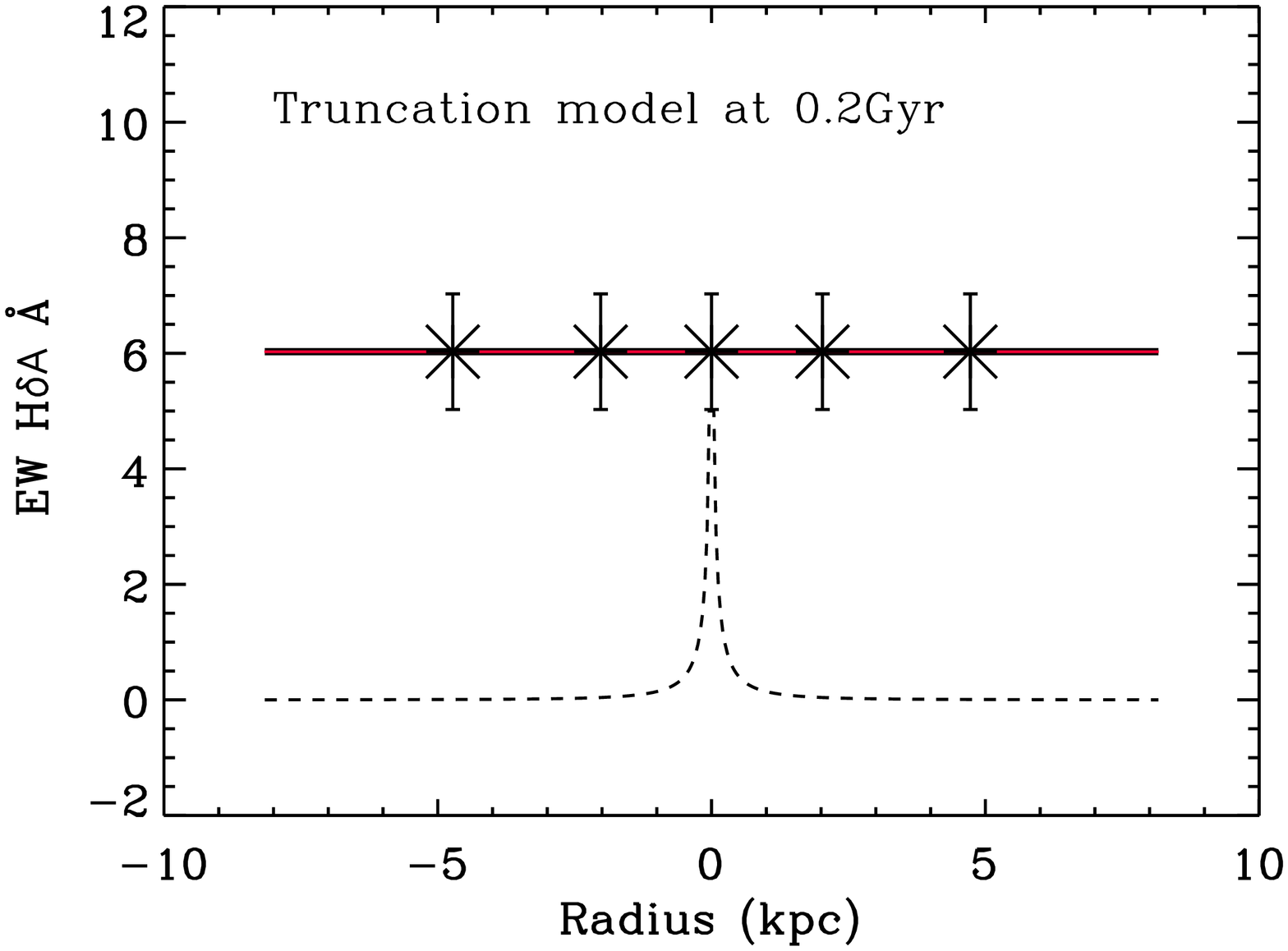}
      \end{minipage}
       \begin{minipage}{0.95\textwidth}
\hspace{-0.5cm}
        \includegraphics[width=3.7cm, angle=90]{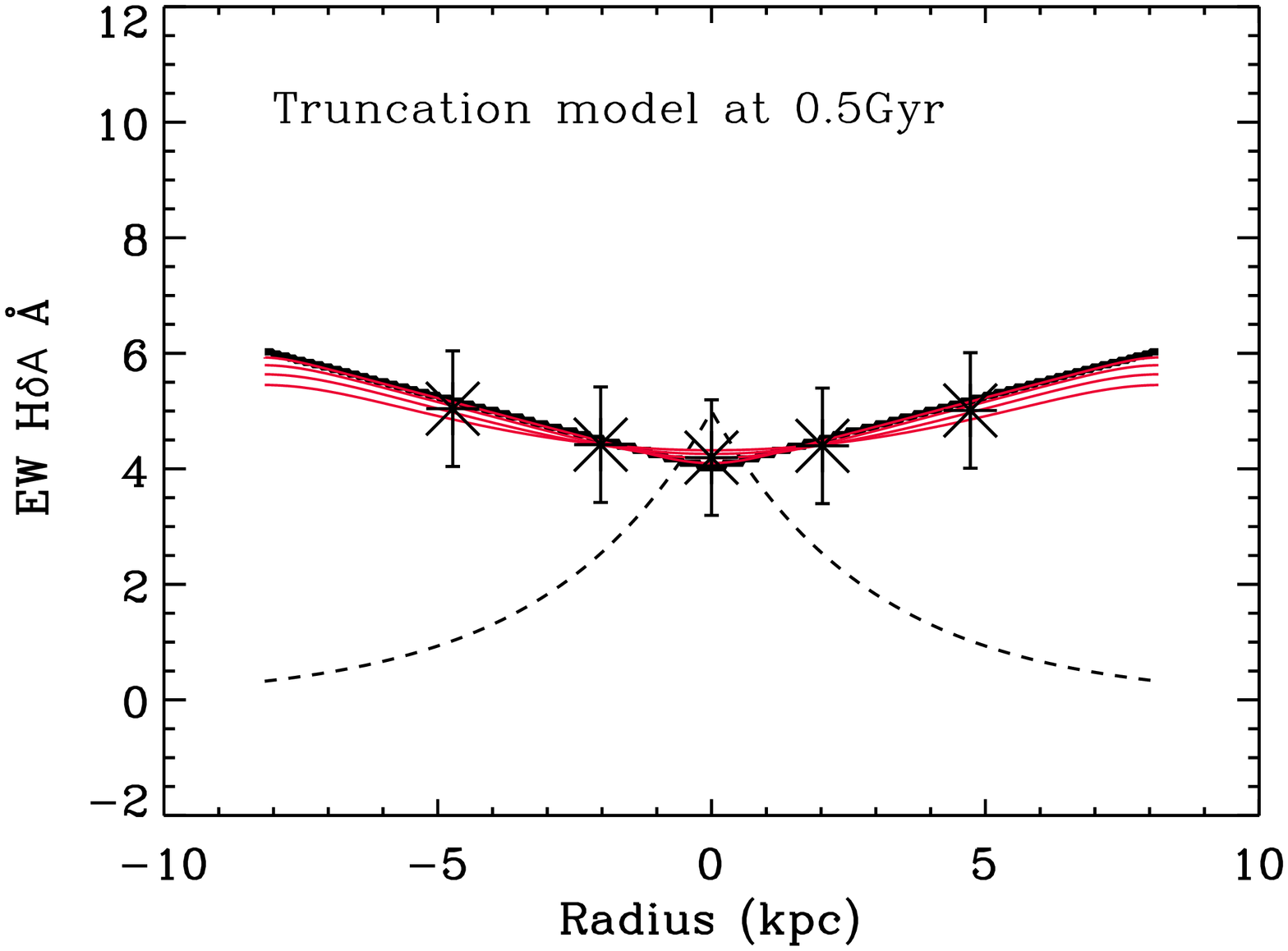}
\hspace{-0.5cm}
         \includegraphics[width=3.7cm, angle=90]{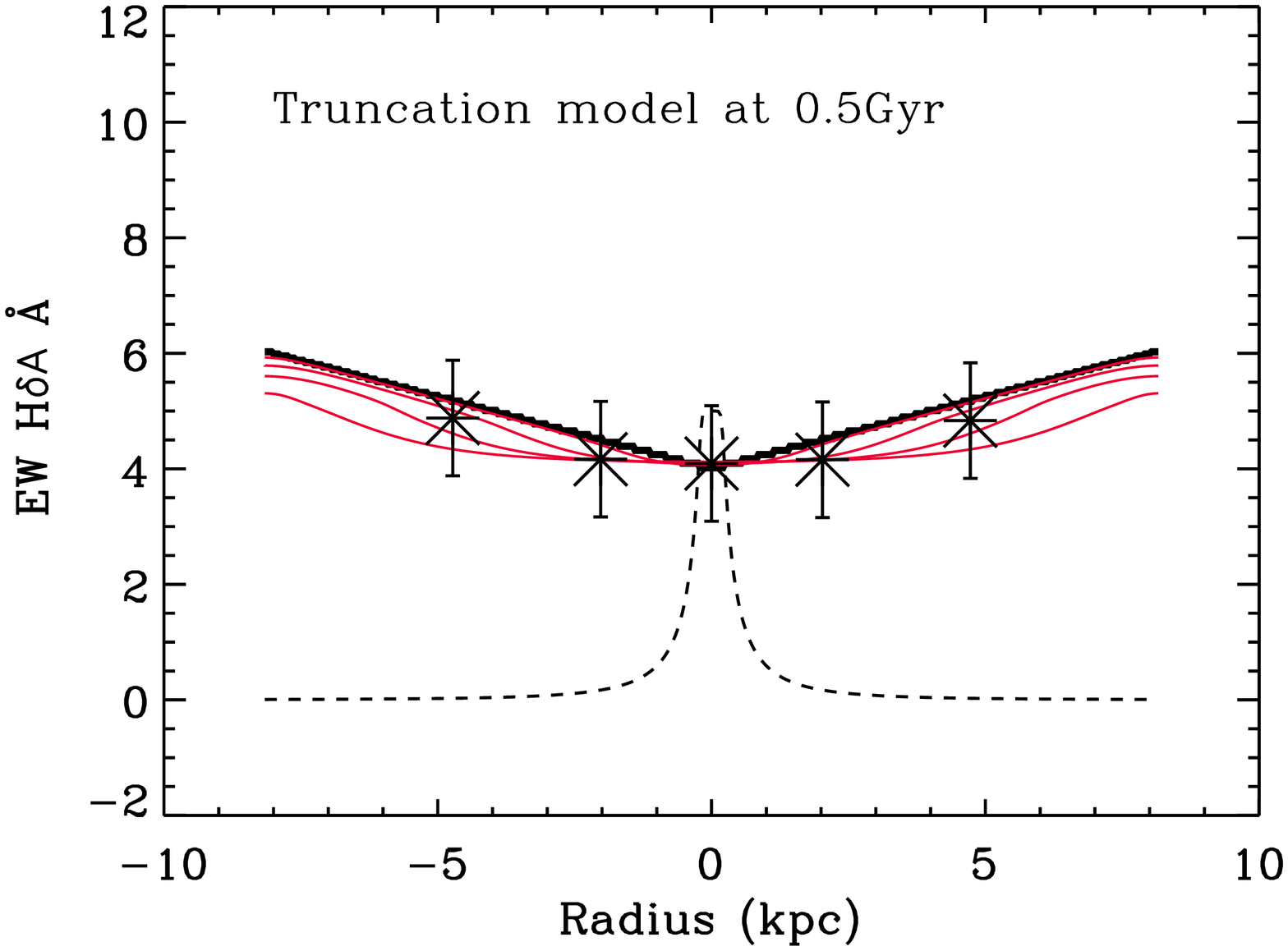}
\hspace{-0.5cm}
         \includegraphics[width=3.7cm, angle=90]{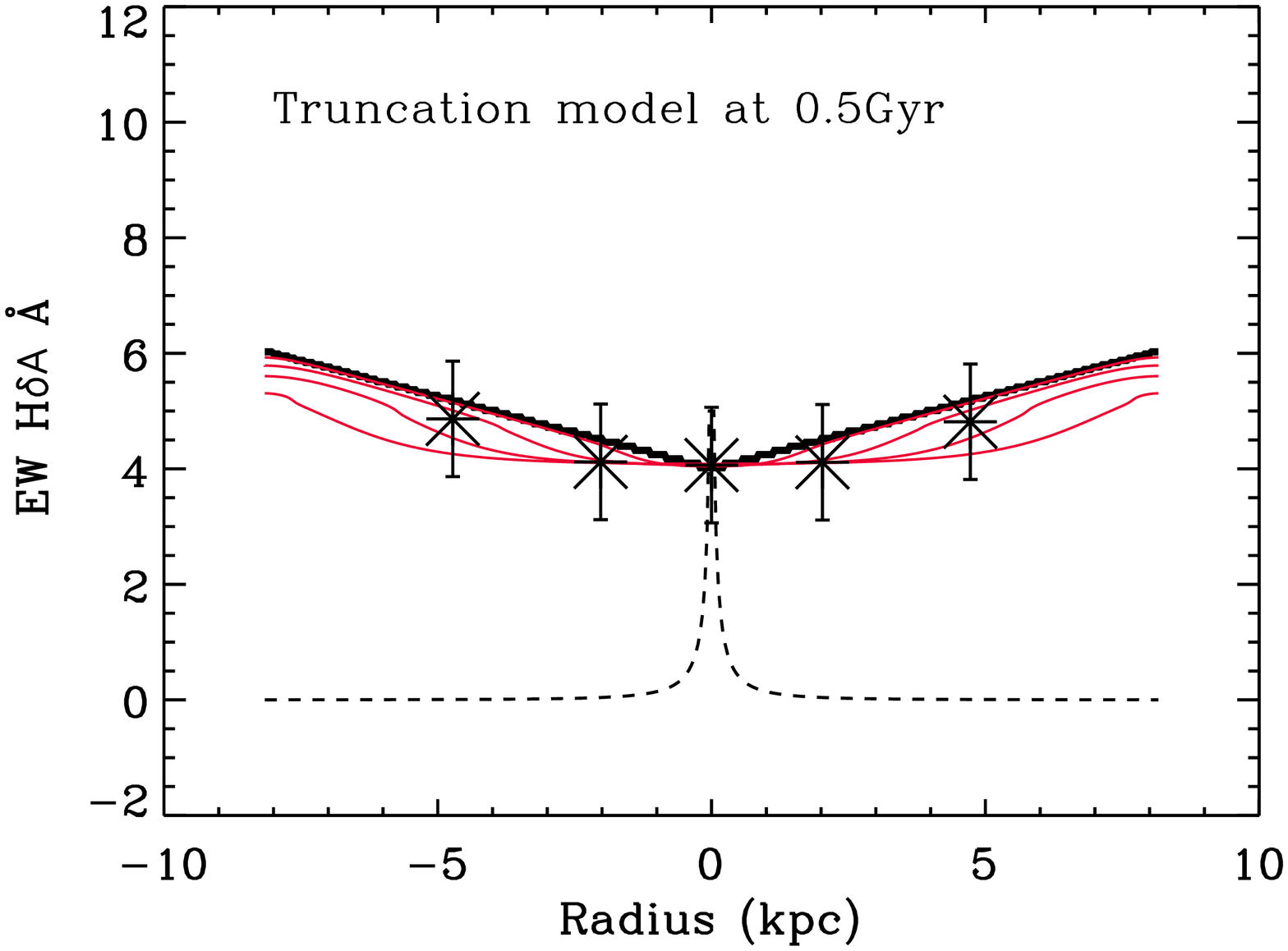}
      \end{minipage}
       \begin{minipage}{0.95\textwidth}
\hspace{-0.5cm}
        \includegraphics[width=3.7cm, angle=90]{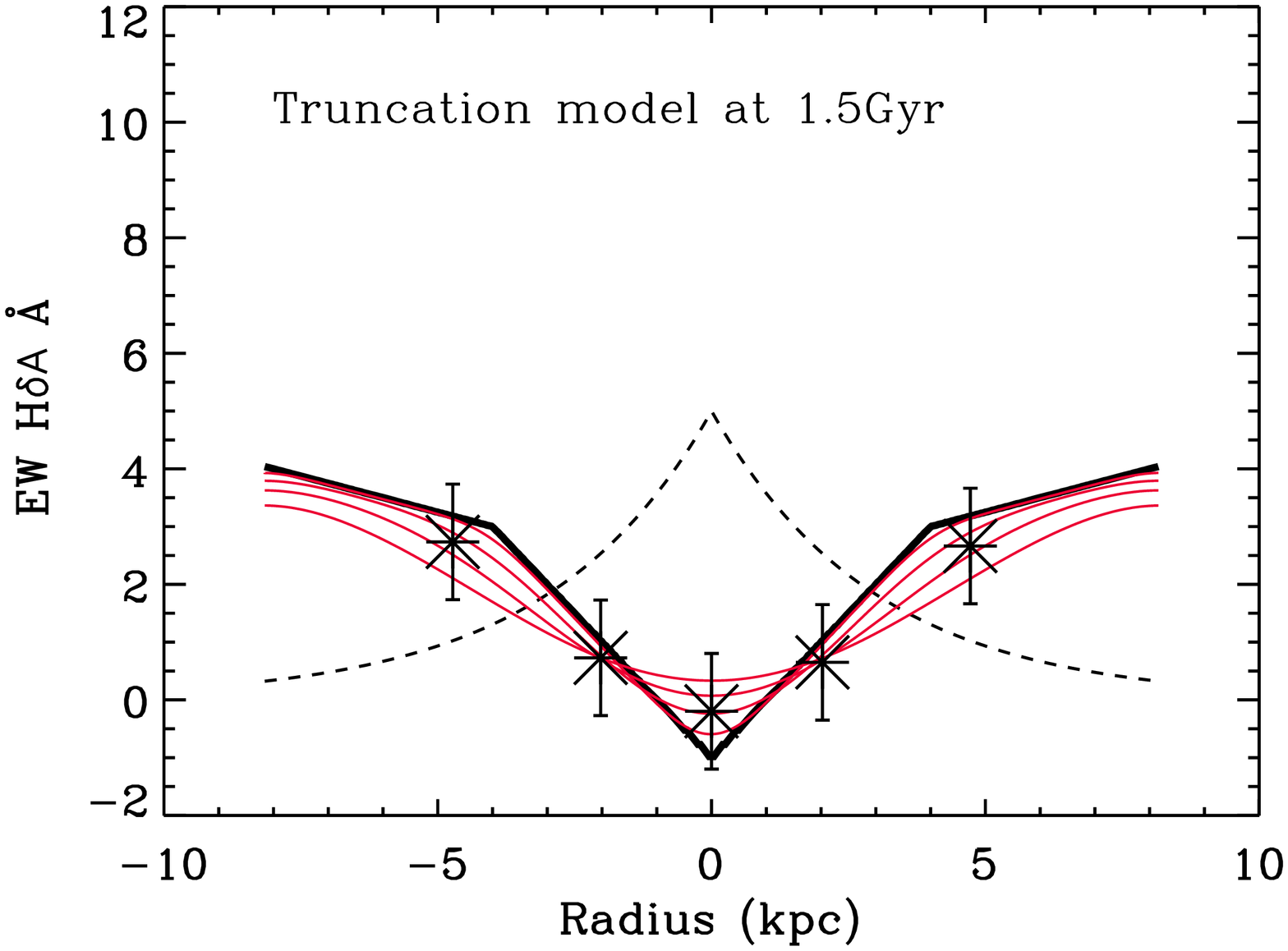}
\hspace{-0.5cm}
         \includegraphics[width=3.7cm, angle=90]{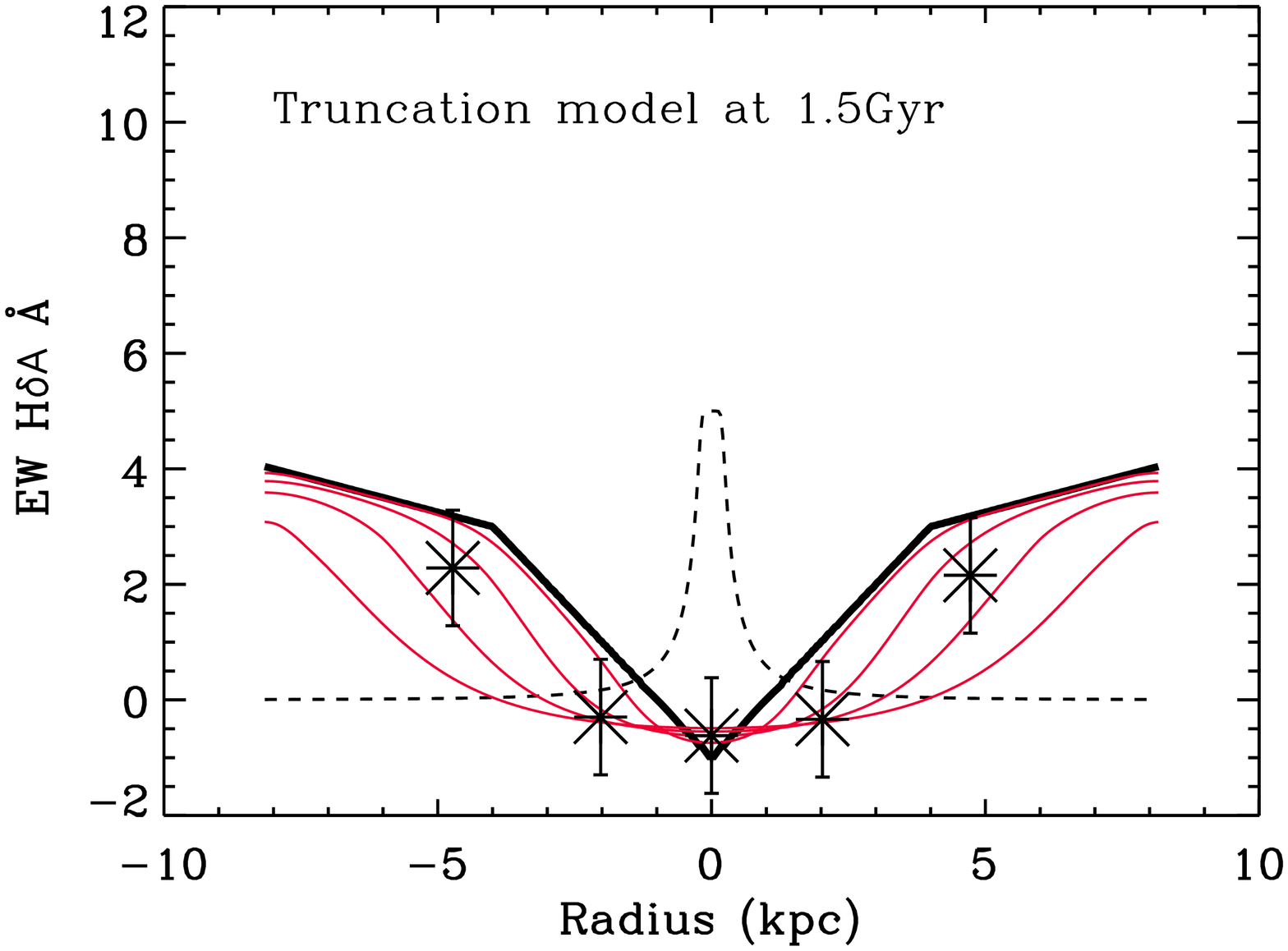}
\hspace{-0.5cm}
         \includegraphics[width=3.7cm, angle=90]{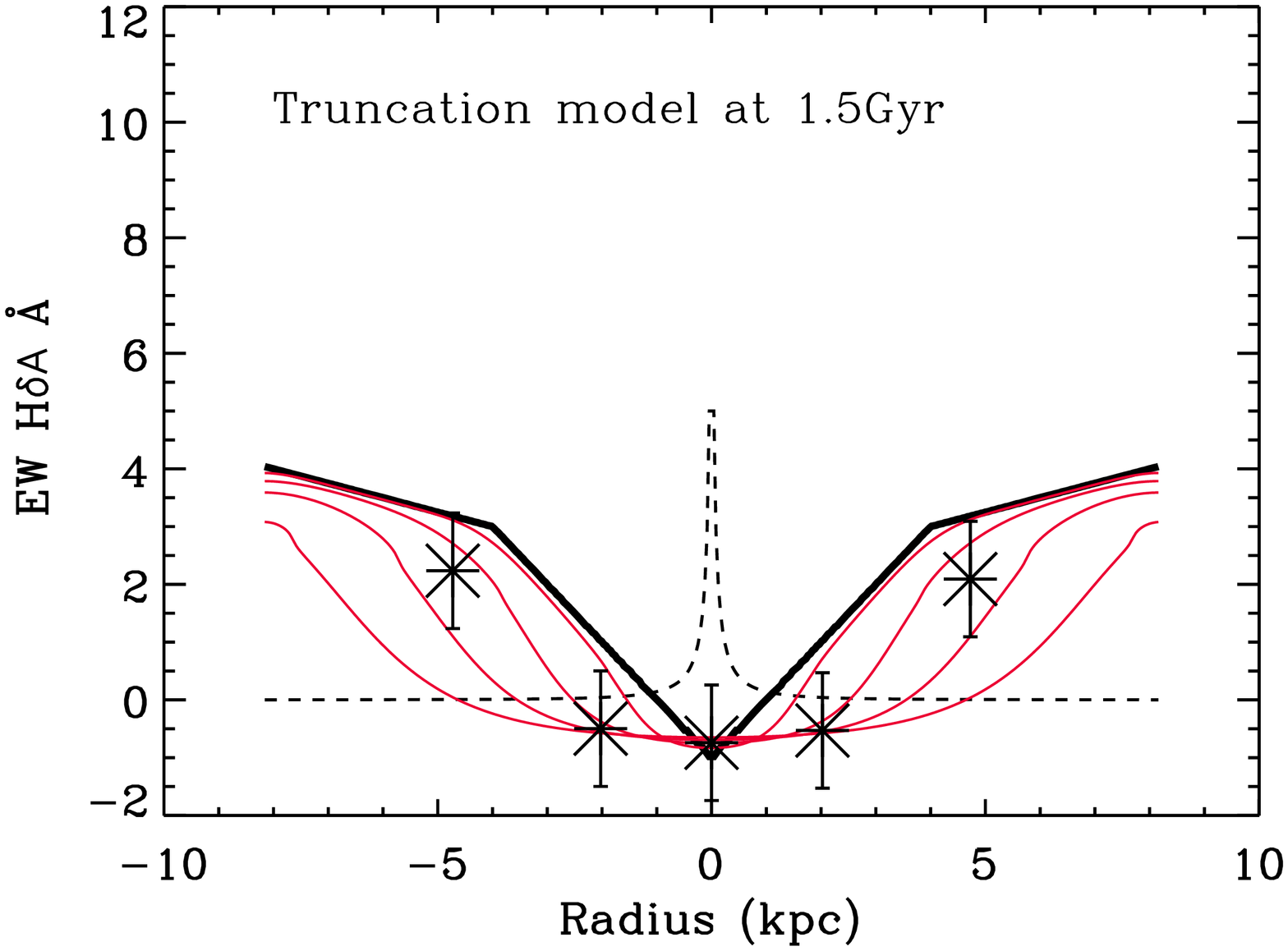}
      \end{minipage}
\hspace{-1.0cm}
\end{center}
\caption{Simple models of the effects of seeing on observed equivalent width distributions of H$\delta$ for
various intrinsic distributions and galaxy light profile types. Moving down a column for different intrinsic
H$\delta$ equivalent width gradients ({\it solid black lines}) and across a row for different galaxy light
profiles ({\it dashed lines}), left to right: exponential; core--Sersic; and de Vaucouleurs. The solid {\it red lines}
shows the observed H$\delta$ gradients after convolution with seeing of 0.2, 0.4, 0.6 and 0.8 arcseconds. The {\it black
stars} show what the H$\delta$ profiles would look like, for the 0.4 arcsecond case, if binned in the same manner as the observations.
See text for more details.}
\label{fig:seeingmod}
\end{figure*}

We used the model predictions of \citet{pracy05} as templates for investigating the effect of seeing on
the intrinsic radial equivalent width profiles. We first approximate the \citet{pracy05} model H$\delta$ 
gradients as either
simple linear functions of radius or piecewise two component linear functions. These are reasonable approximations --
see Figures 9 and 10 of \citet{pracy05}. We then created two dimensional spectra having these H$\delta$ equivalent
width radial profiles. We did this by using a combination of two single--age stellar population (SSP) models
from the MILES library \citep{vazdekis07}. In the combination we used an old ($\sim 8$\,Gyr) SSP to represent 
the old stellar 
population and a young SSP (with the nearest age to the time since the starburst in the 
\citet{pracy05} models) to represent
the burst population, with the fractional
contribution of each set so as to produce the required H$\delta$ equivalent width. These mock H$\delta$ gradients
are shown as the {\it solid black lines} in Figure \ref{fig:seeingmod} for the merger model after 0.2\,Gyr (top row);
merger model after 0.75\,Gyr (second row); merger model after 1.5\,Gyr (third row); truncation model
after 0.2\,Gyr (4th row); truncation model after 0.5\,Gyr (5th row); and truncation model after 1.5\,Gyr (last row).
Each one dimensional spectrum in the two dimensional array is then weighted depending on it's radial position to
represent the overall galaxy light profile. This is done for three different profiles: exponential disk with
an effective radius of 5\,kpc (1st column); a core--Sersic profile \citep{graham03} with $n=4$, a flat 200\,pc core and an
effective radius of 2\,kpc (middle column); and a de Vaucouleurs profile with an effective radius of 2\,kpc 
(third column). 
Since the de Vaucouleurs surface brightness becomes unrealistically large at small $r$ it
was truncated within the central 40\,pc. Using these three profiles gives a series of increasing central concentrations. The effective
radius values are typical for such galaxies and are similar to the values obtained by the fits in Figure \ref{fig:psbmorph}. 
These galaxy light profiles are shown by the {\it dashed lines} in Figure \ref{fig:seeingmod}. We then convolved our two dimensional spectra in
the spatial direction with a Gaussian of widths representing seeing of 0.2, 0.4, 0.6, and 0.8\,arcseconds. 
Note that the seeing in our observations ranged from 0.4 to 0.8\,arcseconds. These seeing convolved profiles are shown
as the {\it solid red curves} in Figure \ref{fig:seeingmod}. Since binning the data is also affected by the galaxy light
profile we reproduce our observational binning for the 0.4\,arcsecond case on 
Figure \ref{fig:seeingmod} as the {\it black stars}. The error bars are 1\AA\, and are representative 
of the errors on the line indices in our observational data. 

Immediately evident from Figure \ref{fig:seeingmod} is that the central concentration of galaxy continuum light has
a significant effect on how much the radial gradients are smoothed out. This is because the 
more centrally concentrated the overall
galaxy continuum light is the more it's spectral signature contaminates the light at 
other radii when the galaxy image is smeared
by the seeing. This can be seen in the progression across a row in Figure \ref{fig:seeingmod} going 
from the least centrally 
concentrated exponential profile to the most concentrated de Vaucouleur profile. Note that the exponential 
profile is typical
of disk galaxies whilst the core--Sersic and de Vaucouleur profiles are typical of 
early type galaxies -- which, in general,  includes 
E+A samples and the majority of our sample. Of course in cases (such as the 3rd and 4th rows 
in Figure \ref{fig:seeingmod}) where the intrinsic profiles are
flat they remain so after smearing. Another point to note from Figure \ref{fig:seeingmod} is 
that at the spatial resolution 
of our observations the central burst of a merger model can be flattened to appear similar 
to a truncation model, for example, compare
the Merger model at 0.75\,Gyr and de Vaucouleur profile with that of the 0.2\,Gyr truncation model 
for the cases of 0.6 and 0.8\,arcsecond seeing.
It is also important to note how sensitive the gradients are to seeing given the radial 
coverage of our data. While the models suggest that
we should be able to detect these kinds of gradients in our outermost bins for the 
0.4\,arcsecond seeing case, this becomes marginal at 0.6\,arcseconds and
the merger profiles are close to flat for 0.8\,arcsecond seeing.
Qualitatively, Figure \ref{fig:seeingmod} suggests that {\it the resolution and depth of our 
observations are around that where we should expect
to make marginal detections of radial H$\delta$ equivalent width gradients 
in galaxies if they were present.} This is particularly so for the `younger' 
($t_{burst}\leq 0.6$\,Gyr) galaxies in our sample (DG\_134, DG\_106,DG\_411) 
where more extreme gradients are expected.

\begin{figure*}[t!]
   \begin{center}
     \begin{minipage}{0.95\textwidth}
         \includegraphics[width=3.5cm, angle=90]{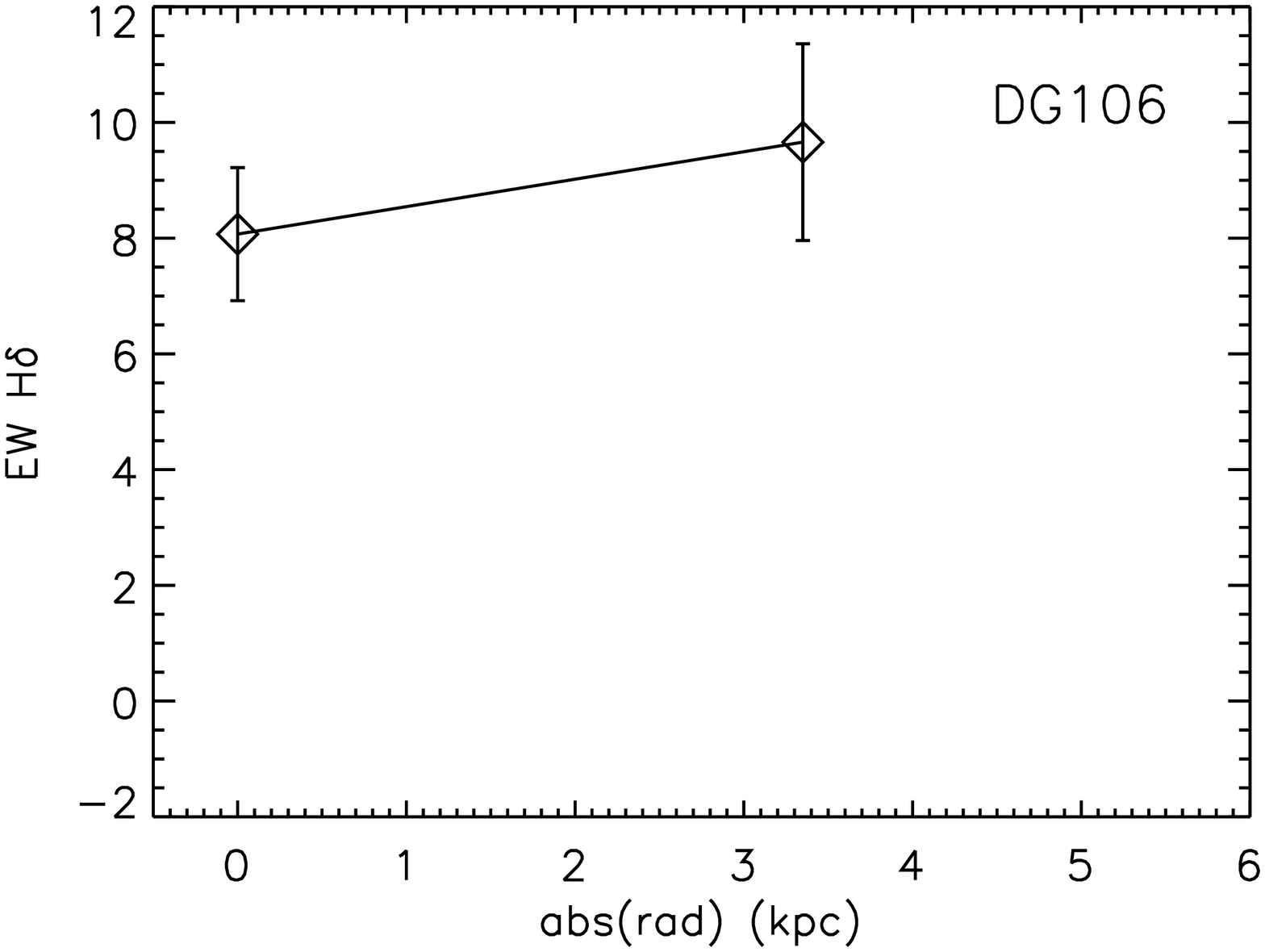}
         \includegraphics[width=3.5cm, angle=90]{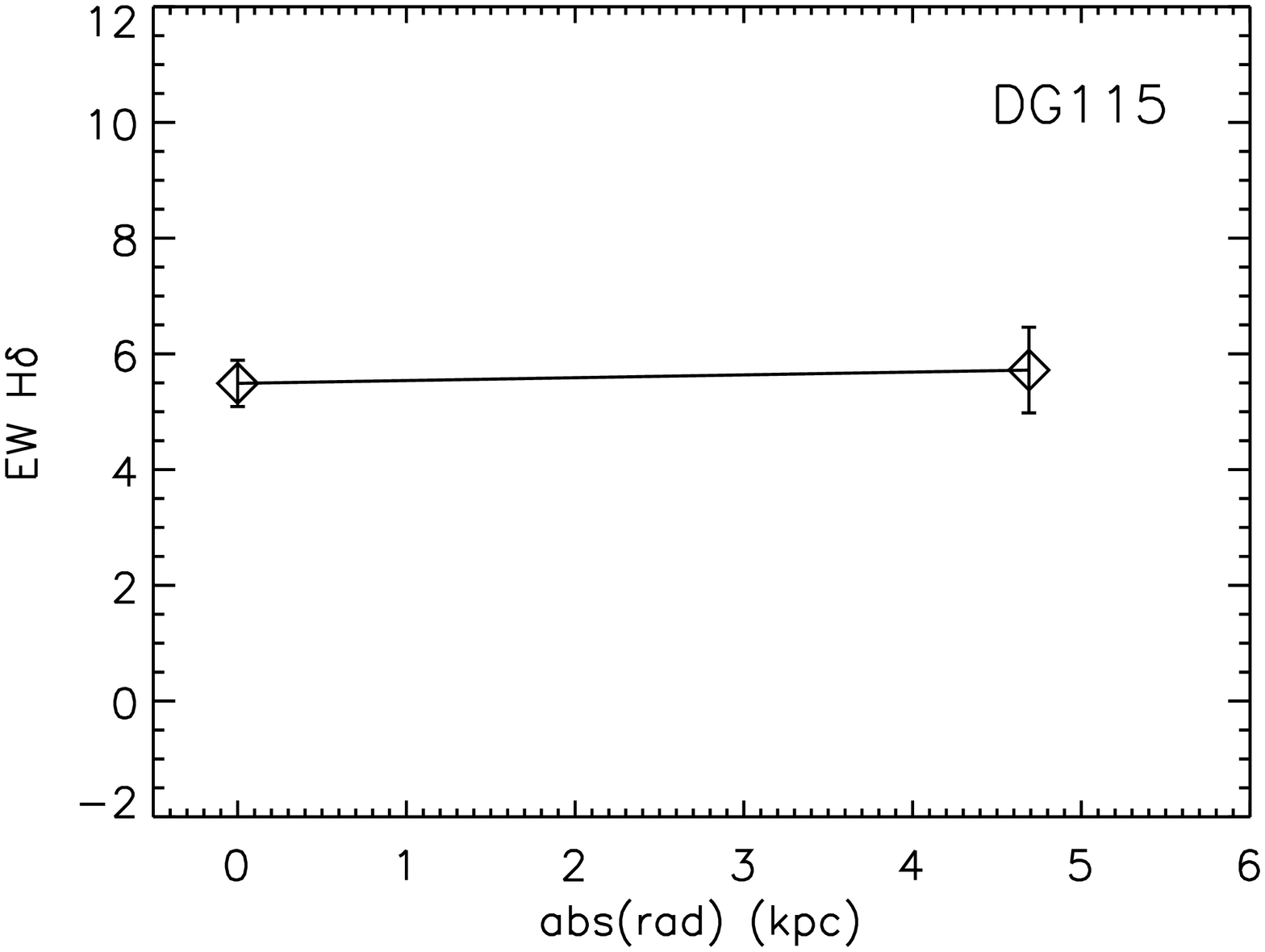}
        \includegraphics[width=3.5cm, angle=90, trim=0 0 0 0]{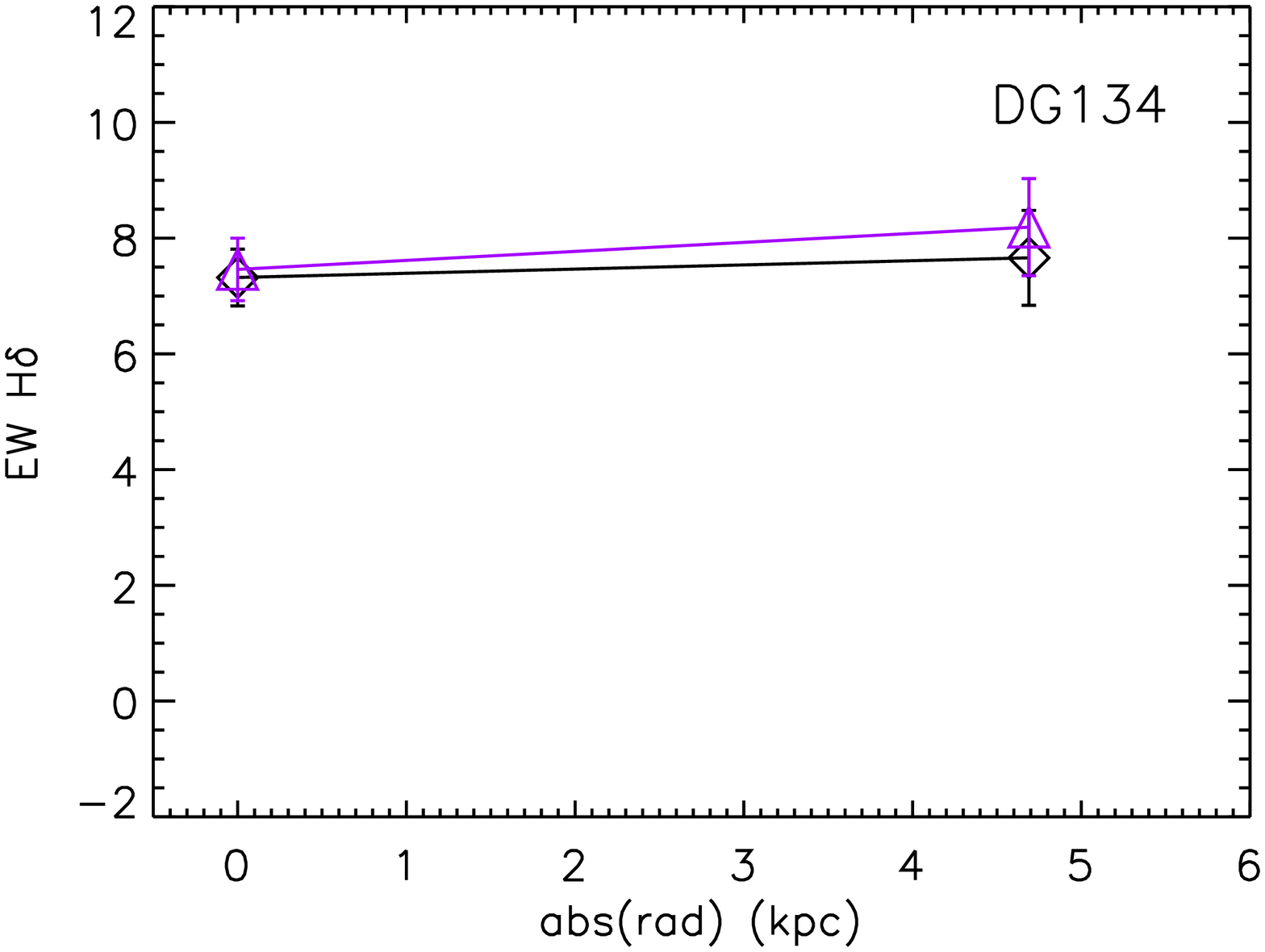}
       \end{minipage}
       \begin{minipage}{0.95\textwidth}
          \includegraphics[width=3.5cm, angle=90, trim=0 0 0 0]{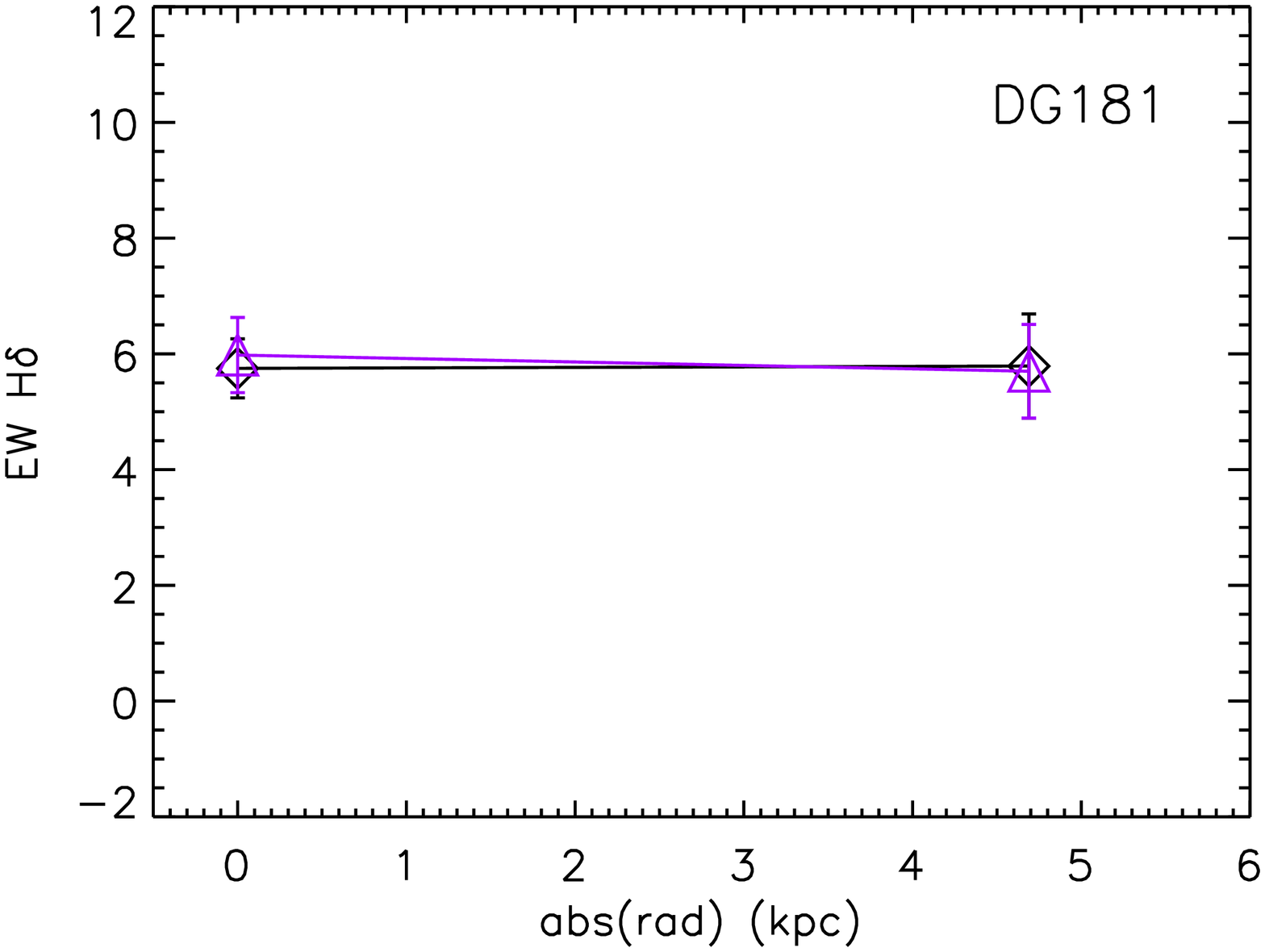}
        \includegraphics[width=3.5cm, angle=90, trim=0 0 0 0]{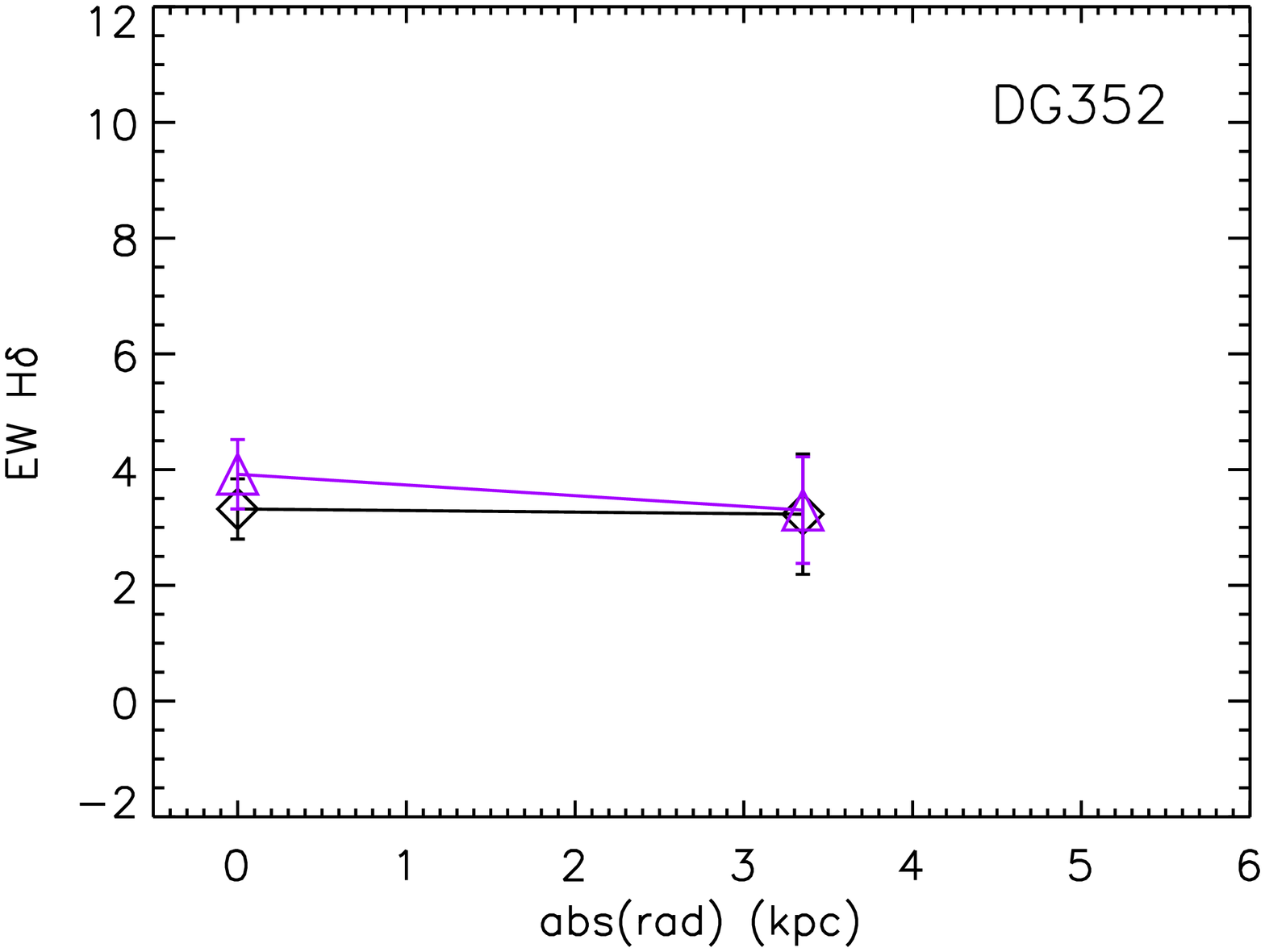}
        \includegraphics[width=3.5cm, angle=90, trim=0 0 0 0]{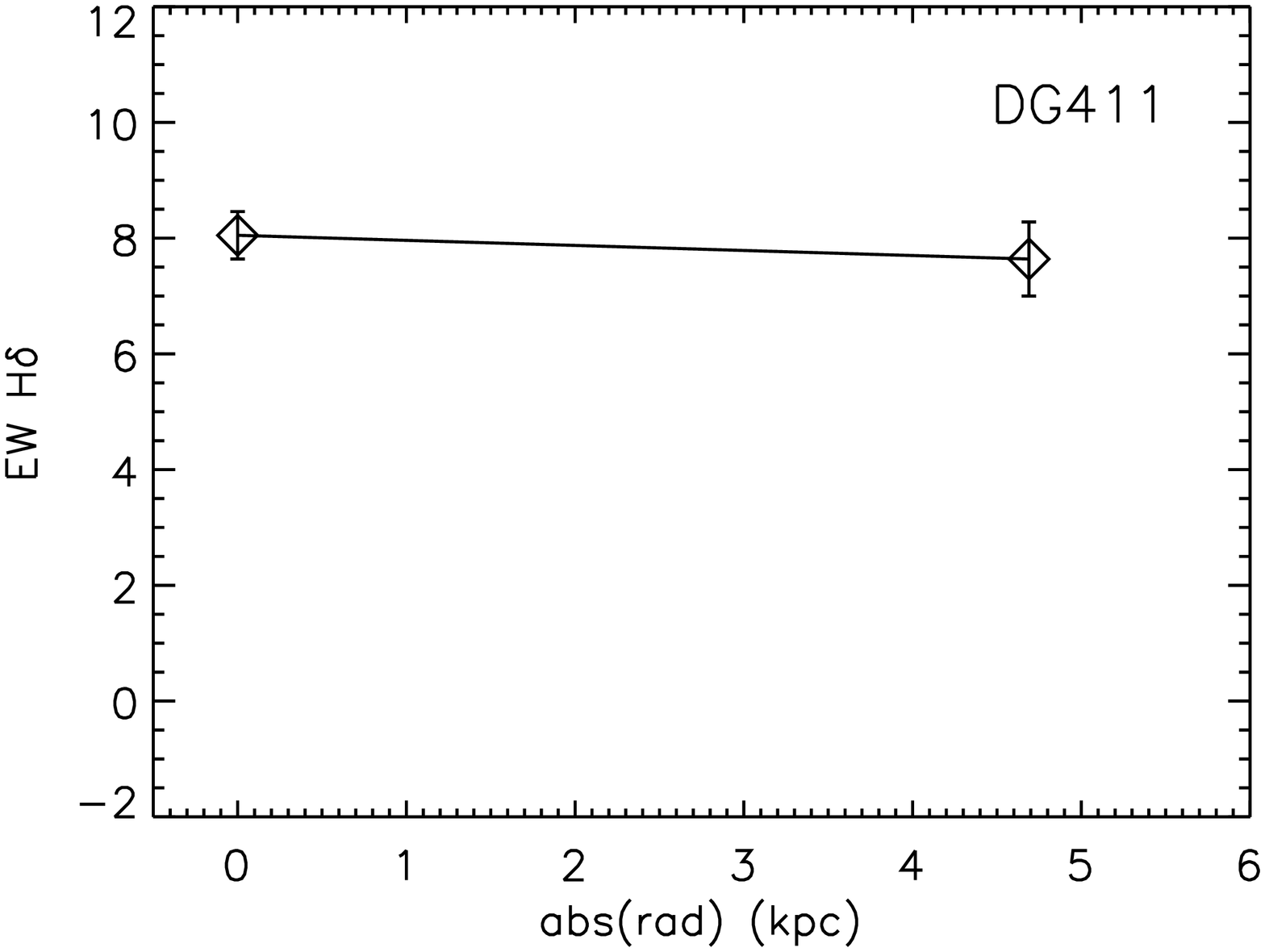}
       \end{minipage}
     \end{center}
\caption{Radial H$\delta_{\rm A}$ profile based on the binning of data in two radial regions: 
(i)$r\leq2$\,kpc, and (ii)$r>2$\,kpc. Data were binned in both spatial directions along the
slit, and only when of usable quality in the outer region. The {\it black symbols} represent
the Mask~1 data while the {\it blue symbols} represent the Mask~2 data.}
\label{fig:twopoints}
\end{figure*}

Another important feature of our radial H$\delta$ equivalent width profile predictions
in Figure \ref{fig:seeingmod} is that they show the central enhancement (in the case
of the merger scenario) or deficit (in the case of the truncation scenario) is 
contained within the central $\sim$4\,kpc (2\,kpc in radius). Therefore, in order to
maximize our chances of detecting any gradient in H$\delta$ strength that might
exist, we have extracted and combined our spatially resolved spectra of the E+A 
sample in this manner (i.e. inside and outside the central
4\,kpc) and remeasured the H$\delta$ equivalent widths. These 
two point equivalent width radial profiles are shown in Figure \ref{fig:twopoints}. 
For all six galaxies, no significant gradient if found to be present.

\section{Discussion and Summary}
The enhancement in the frequency of E+A galaxies in intermediate redshift clusters relative to the field implies 
that a cluster-specific phenomenon, such as interaction with the ICM, should be responsible for the 
truncation of the star--formation in a significant fraction of the E+A galaxies in these environments 
\citep{poggianti99}.  At face value our observations are in agreement with this expectation. The lack of radial 
gradients in H$\delta$ equivalent width in our E+A sample is consistent with the ICM being 
responsible for the truncation of the star formation via gas stripping processes \citep{rose01,bekki05}.
However, the spatial resolution imposed on our observation by the seeing is similar to the spatial scale
of the gradients expected in merger scenarios \citep{pracy05,bekki05}. Based on simple models of the effects
of seeing marginal detection of H$\delta$ gradients in merger models could be expected from our observations 
(c.f. 2nd column of Figure \ref{fig:psb} with the 2nd column of Figure \ref{fig:seeingmod}). In 
addition our expectations
for the expected gradients are based on just one set of merger simulations \citep{pracy05,bekki05} and 
so whilst our spectroscopic observations
are more consistent with a truncation model the conclusion is not robust and merger models cannot be ruled out. 
Given that our observations were taken on a 10-m telescope in nearly the best conditions that 
can be expected for natural seeing observations from the ground
another strategy is required to advance these studies. Better spatial resolution spectroscopic observations of 
cluster E+As could be obtained  by observing low-redshift galaxies where the physical size to angular size ratio 
is larger (although bright cluster E+A galaxies are extremely rare at the present epoch). 

Further support for truncation by the ICM being an important source of the intermediate cluster E+A population is the
distinct lack of evidence for tidal disturbance in our sample. A high incidence of tidal disruption and interaction
is ubiquitous in low redshift E+A samples \citep{zabludoff96,norton01,blake04,yamauchi05,goto08,yang08,pracy09}. 
While observing at higher redshift makes detection of such features 
more difficult, it should be noted that as a result of the superb image quality delivered by ACS/HST, our physical scale
resolution is better than the low redshift studies with the exception of \citet{yang08}s low redshift HST observations.

The e(a) galaxies we observed do show a tendency to have centrally concentrated emission. This is likely 
evidence for a centralized starburst. A high fraction of e(a) galaxies are expected to be the result of mergers
or tidal interactions \citep{poggianti00} and these processes are expected to give rise to a centralized starburst
as gas is funnelled toward the galaxy centre \citep{noguchi88,barnes91,mihos92,mihos96,bekki05,bournaud08}. 
However, these galaxies are also expected to have uneven dust and age distributions \citep{poggianti00} and
it is difficult to draw conclusions about the distribution of star-formation in these objects without understanding
the  distribution of dust. The morphology of these galaxies are of very late type with evidence of disturbance or
interactions as expected for e(a) galaxies and starbursts in general.

\noindent In Summary:
\begin{list}{$\bullet$}{\itemsep=0.1cm}

\item We do not detect any significant radial variation in the strength of the H$\delta$ line in any of 
the six post--starburst galaxies in our sample. However, we do detect a continuous gradient across the face 
of one galaxy in our sample, DG\_411, which is also the one galaxy with evidence of a current interaction. 
The lack of H$\delta$ equivalent width gradients is consistent with their production being via interaction 
with the ICM. However, the scale of the expected gradients in merger models
is similar to our spatial resolution element, making this conclusion uncertain.

\item In contrast to the local field there is little evidence for tidal disturbance, mergers or interactions
in this intermediate redshift cluster sample of E+As. At low redshift such features have been interpreted
as evidence for mergers in production of the E+A population. The lack of such features is consistent
with ICM truncation of star formation.

\item The post--starburst signature is present at all radial points in each of the sample galaxies
with uniformly strong Balmer line absorption and no [OII]$\lambda 3727$ emission present in any radial bin.

\item The three e(a) galaxies all show some evidence of centrally concentrated emission with the 
equivalent width of the  [OII]$\lambda 3727$ line increasing toward the galactic centre. While one
explanation for this is a centralized burst of star--formation the interpretation is clouded by
not knowing the spatial distribution of the extinction.

\item The morphological properties of the e(a) galaxies are consistent with a merger/interaction origin.

\end{list}

\section*{Acknowledgments}
The authors wish to recognize and acknowledge the very significant cultural role and reverence that 
the summit of Mauna Kea has always had within the indigenous Hawaiian community.  We are most 
fortunate to have the opportunity to conduct observations from this mountain. We would like to 
thank Matt Auger for providing us with his pipeline reduction software for the LRIS instrument. 
Parts of this work are based on observations made with the NASA/ESA Hubble Space Telescope, and 
obtained from the Hubble Legacy Archive, which is a collaboration between the Space Telescope 
Science Institute (STScI/NASA), the Space Telescope European Coordinating 
Facility (ST-ECF/ESA) and the Canadian Astronomy Data Centre (CADC/NRC/CSA).
M.B.P. and W.J.C. acknowledge the generous financial support of the Australian Research
Council throughout the course of this work.

\bibliographystyle{apj}
\bibliography{references}

\begin{thebibliography}{39}
\expandafter\ifx\csname natexlab\endcsname\relax\def\natexlab#1{#1}\fi

\bibitem[{{Barnes} \& {Hernquist}(1991)}]{barnes91}
{Barnes}, J.~E., \& {Hernquist}, L.~E. 1991, \apjl, 370, L65

\bibitem[{{Bekki}(1999)}]{bekki99}
{Bekki}, K. 1999, \apjl, 510, L15

\bibitem[{{Bekki} {et~al.}(2005){Bekki}, {Couch}, {Shioya}, \&
  {Vazdekis}}]{bekki05}
{Bekki}, K., {Couch}, W.~J., {Shioya}, Y., \& {Vazdekis}, A. 2005, \mnras, 359,
  949

\bibitem[{{Bekki} {et~al.}(2001){Bekki}, {Shioya}, \& {Couch}}]{bekki01}
{Bekki}, K., {Shioya}, Y., \& {Couch}, W.~J. 2001, \apjl, 547, L17

\bibitem[{{Blake} {et~al.}(2004){Blake}, {Pracy}, {Couch}, {Bekki}, {Lewis}, \&
  {and 26 other authours}}]{blake04}
{Blake}, C., {Pracy}, M.~B., {Couch}, W.~J., {Bekki}, K., {Lewis}, I., \& {and
  26 other authours}. 2004, \mnras, 355, 713

\bibitem[{{Bothun} \& {Dressler}(1986)}]{bothun86}
{Bothun}, G.~D., \& {Dressler}, A. 1986, \apj, 301, 57

\bibitem[{{Bournaud} {et~al.}(2008){Bournaud}, {Bois}, {Emsellem}, \&
  {Duc}}]{bournaud08}
{Bournaud}, F., {Bois}, M., {Emsellem}, E., \& {Duc}, P.~A. 2008, Astron.
  Nachr., in press

\bibitem[{{Butcher} \& {Oemler}(1978)}]{butcher78}
{Butcher}, H., \& {Oemler}, Jr., A. 1978, \apj, 226, 559

\bibitem[{{Butcher} \& {Oemler}(1984)}]{butcher84}
---. 1984, \apj, 285, 426

\bibitem[{{Caldwell} {et~al.}(1996){Caldwell}, {Rose}, {Franx}, \&
  {Leonardi}}]{caldwell96}
{Caldwell}, N., {Rose}, J.~A., {Franx}, M., \& {Leonardi}, A.~J. 1996, \aj,
  111, 78

\bibitem[{{Cappellari} \& {Emsellem}(2004)}]{cappellari04}
{Cappellari}, M., \& {Emsellem}, E. 2004, \pasp, 116, 138

\bibitem[{{Couch} {et~al.}(1994){Couch}, {Ellis}, {Sharples}, \&
  {Smail}}]{couch94}
{Couch}, W.~J., {Ellis}, R.~S., {Sharples}, R.~M., \& {Smail}, I. 1994, \apj,
  430, 121

\bibitem[{{Couch} \& {Sharples}(1987)}]{couch87}
{Couch}, W.~J., \& {Sharples}, R.~M. 1987, \mnras, 229, 423

\bibitem[{{Dressler} \& {Gunn}(1982)}]{dressler82}
{Dressler}, A., \& {Gunn}, J.~E. 1982, \apj, 263, 533

\bibitem[{{Dressler} \& {Gunn}(1983)}]{dressler83}
---. 1983, \apj, 270, 7

\bibitem[{{Dressler} \& {Gunn}(1992)}]{dressler92}
---. 1992, \apjs, 78, 1

\bibitem[{{Dressler} {et~al.}(2004){Dressler}, {Oemler}, {Poggianti}, {Smail},
  {Trager}, {Shectman}, {Couch}, \& {Ellis}}]{dressler04}
{Dressler}, A., {Oemler}, Jr., A., {Poggianti}, B.~M., {Smail}, I., {Trager},
  S., {Shectman}, S.~A., {Couch}, W.~J., \& {Ellis}, R.~S. 2004, \apj, 617, 867

\bibitem[{{Dressler} {et~al.}(1999){Dressler}, {Smail}, {Poggianti}, {Butcher},
  {Couch}, {Ellis}, \& {Oemler}}]{dressler99}
{Dressler}, A., {Smail}, I., {Poggianti}, B.~M., {Butcher}, H., {Couch}, W.~J.,
  {Ellis}, R.~S., \& {Oemler}, A.~J. 1999, \apjs, 122, 51

\bibitem[{{Goto}(2007)}]{goto07}
{Goto}, T. 2007, \mnras, 381, 187

\bibitem[{{Goto} {et~al.}(2008){Goto}, {Kawai}, {Shimono}, {Sugai}, {Yagi}, \&
  {Hattori}}]{goto08}
{Goto}, T., {Kawai}, A., {Shimono}, A., {Sugai}, H., {Yagi}, M., \& {Hattori},
  T. 2008, ArXiv e-prints, 801

\bibitem[{{Graham} {et~al.}(2003){Graham}, {Erwin}, {Trujillo}, \& {Asensio
  Ramos}}]{graham03}
{Graham}, A.~W., {Erwin}, P., {Trujillo}, I., \& {Asensio Ramos}, A. 2003, \aj,
  125, 2951

\bibitem[{{Gunn} \& {Gott}(1972)}]{gunn72}
{Gunn}, J.~E., \& {Gott}, J.~R.~I. 1972, \apj, 176, 1

\bibitem[{{Mihos} \& {Hernquist}(1996)}]{mihos96}
{Mihos}, J.~C., \& {Hernquist}, L. 1996, \apj, 464, 641

\bibitem[{{Mihos} {et~al.}(1992){Mihos}, {Richstone}, \& {Bothun}}]{mihos92}
{Mihos}, J.~C., {Richstone}, D.~O., \& {Bothun}, G.~D. 1992, \apj, 400, 153

\bibitem[{{Moore} {et~al.}(1996){Moore}, {Katz}, {Lake}, {Dressler}, \&
  {Oemler}}]{moore96}
{Moore}, B., {Katz}, N., {Lake}, G., {Dressler}, A., \& {Oemler}, A. 1996,
  \nat, 379, 613

\bibitem[{{Noguchi}(1988)}]{noguchi88}
{Noguchi}, M. 1988, \aap, 203, 259

\bibitem[{{Norton} {et~al.}(2001){Norton}, {Gebhardt}, {Zabludoff}, \&
  {Zaritsky}}]{norton01}
{Norton}, S.~A., {Gebhardt}, K., {Zabludoff}, A.~I., \& {Zaritsky}, D. 2001,
  \apj, 557, 150

\bibitem[{{Poggianti} {et~al.}(1999){Poggianti}, {Smail}, {Dressler}, {Couch},
  {Barger}, {Butcher}, {Ellis}, \& {Oemler}}]{poggianti99}
{Poggianti}, B.~M., {Smail}, I., {Dressler}, A., {Couch}, W.~J., {Barger},
  A.~J., {Butcher}, H., {Ellis}, R.~S., \& {Oemler}, A.~J. 1999, \apj, 518, 576

\bibitem[{{Poggianti} \& {Wu}(2000)}]{poggianti00}
{Poggianti}, B.~M., \& {Wu}, H. 2000, \apj, 529, 157

\bibitem[{{Poggianti} {et~al.}(2009){Poggianti}, {Arag{\'o}n-Salamanca},
  {Zaritsky}, {De Lucia}, {Milvang-Jensen}, {Desai}, {Jablonka}, {Halliday}, \&
  {10 other authors}}]{poggianti09}
{Poggianti}, B.~M., {et~al.} 2009, \apj, 693, 112

\bibitem[{{Pracy} {et~al.}(2005){Pracy}, {Couch}, {Blake}, {Bekki}, {Harrison},
  {Colless}, {Kuntschner}, \& {de Propris}}]{pracy05}
{Pracy}, M.~B., {Couch}, W.~J., {Blake}, C., {Bekki}, K., {Harrison}, C.,
  {Colless}, M., {Kuntschner}, H., \& {de Propris}, R. 2005, \mnras, 359, 1421

\bibitem[{{Pracy} {et~al.}(2009){Pracy}, {Kuntschner}, {Couch}, {Blake},
  {Bekki}, \& {Briggs}}]{pracy09}
{Pracy}, M.~B., {Kuntschner}, H., {Couch}, W.~J., {Blake}, C., {Bekki}, K., \&
  {Briggs}, F. 2009, \mnras, 396, 1349

\bibitem[{{Rose} {et~al.}(2001){Rose}, {Gaba}, {Caldwell}, \&
  {Chaboyer}}]{rose01}
{Rose}, J.~A., {Gaba}, A.~E., {Caldwell}, N., \& {Chaboyer}, B. 2001, \aj, 121,
  793

\bibitem[{{Tran} {et~al.}(2004){Tran}, {Franx}, {Illingworth}, {van Dokkum},
  {Kelson}, \& {Magee}}]{tran04}
{Tran}, K., {Franx}, M., {Illingworth}, G.~D., {van Dokkum}, P., {Kelson},
  D.~D., \& {Magee}, D. 2004, \apj, 609, 683

\bibitem[{{Vazdekis} {et~al.}(2007){Vazdekis}, {Cardiel}, {Cenarro},
  {Cervantes}, {Falc{\'o}n-Barroso}, {Gorgas}, {Jim{\'e}nez-Vicente},
  {Mart{\'{\i}}n-Hern{\'a}ndez}, {Peletier}, {S{\'a}nchez-Bl{\'a}zquez},
  {Selam}, \& {Toloba}}]{vazdekis07}
{Vazdekis}, A., {et~al.} 2007, in IAU Symposium, Vol. 241, IAU Symposium, ed.
  A.~{Vazdekis} \& R.~F. {Peletier}, 133--137

\bibitem[{{Worthey} \& {Ottaviani}(1997)}]{worthey97}
{Worthey}, G., \& {Ottaviani}, D.~L. 1997, \apjs, 111, 377

\bibitem[{{Yamauchi} \& {Goto}(2005)}]{yamauchi05}
{Yamauchi}, C., \& {Goto}, T. 2005, \mnras, 359, 1557

\bibitem[{{Yang} {et~al.}(2008){Yang}, {Zabludoff}, {Zaritsky}, \&
  {Mihos}}]{yang08}
{Yang}, Y., {Zabludoff}, A.~I., {Zaritsky}, D., \& {Mihos}, J.~C. 2008, \apj,
  688, 945

\bibitem[{{Zabludoff} {et~al.}(1996){Zabludoff}, {Zaritsky}, {Lin}, {Tucker},
  {Hashimoto}, {Shectman}, {Oemler}, \& {Kirshner}}]{zabludoff96}
{Zabludoff}, A.~I., {Zaritsky}, D., {Lin}, H., {Tucker}, D., {Hashimoto}, Y.,
  {Shectman}, S.~A., {Oemler}, A., \& {Kirshner}, R.~P. 1996, \apj, 466, 104

\end{thebibliography}

\end{document}